\documentclass[11pt, final]{ucthesis}

\usepackage{import} 

\usepackage{epsfig, graphicx, color, amsmath, amsthm, amssymb}
\usepackage[sort]{cite} 
\usepackage{cancel}
\usepackage{caption, subfig}
\input{qcircuit}

\usepackage{diagrams}
\def\ket #1{\vert #1\rangle}
\def\bra #1{\langle #1\vert}
\def\ketbra #1#2{\ket{#1}\!\bra{#2}}

\def\abs #1{\lvert #1\rvert}

\DeclareMathOperator{\tr}{Tr}
\newcommand{\beq}{\begin{equation}}
\newcommand{\eeq}{\end{equation}}
\newcommand{\binomial}[2]{\ensuremath{\left(\begin{smallmatrix}#1 \\ #2 \end{smallmatrix}\right)}}
\newcommand\pr{{\bf {P}}}

\newcommand{\comment}[1]{\emph{\color{red}Comment:\color{black} #1}} 
\newlength{\commentslength}
\newcommand{\comments}[1]{
\hspace{-2\parindent}
\addtolength{\commentslength}{-\commentslength}
\addtolength{\commentslength}{\linewidth}
\addtolength{\commentslength}{-\parindent}
\fcolorbox{red}{white}{\smallskip\begin{minipage}[c]{\commentslength}
\emph{Comments:}\begin{itemize}#1\end{itemize}\end{minipage}}\bigskip
}
\renewcommand{\comment}[1]{}
\renewcommand{\comments}[1]{}

\def\Ref{}
\def\Refs{}

\theoremstyle{plain} 
\newtheorem{theorem}{Theorem}  

\newtheorem{lemma}{Lemma}

\newtheoremstyle{note}{}{}{\slshape}{}{\bfseries}{.}{ }{}
\theoremstyle{remark} 
\newtheorem{remark}{Remark}
\theoremstyle{definition} 
\newtheorem{definition}{Definition}

\DeclareMathOperator{\poly}{\operatorname{poly}}

\begin{document}


\title{Error-detection-based quantum fault tolerance against discrete Pauli noise}

\author{Benjamin W. Reichardt}

\prevdegrees{B.S. (Stanford University) 2001}

\degreemonth{Fall} \degreeyear{2006} \degreename{Doctor of Philosophy}
\defensemonth{Fall} \defenseyear{2006}

\numberofmembers{3} 
	\chair{Professor Umesh Vazirani} 
	\othermemberA{Professor Christos Papadimitriou} 
	\othermemberB{Professor K.~Birgitta Whaley}
	
\field{Computer Science} 

\campus{Berkeley}

\begin{frontmatter}

\maketitle

\approvalpage

%
\copyrightpage

%
\abstract


A quantum computer -- i.e., a computer capable of manipulating data in quantum superposition -- would find applications including factoring, quantum simulation and tests of basic quantum theory. 
Since quantum superpositions are fragile, the major hurdle in building such a computer is overcoming noise. 

Developed over the last couple of years, new schemes for achieving fault tolerance based on error detection, rather than error correction, appear to tolerate as much as 3-6\% noise per gate -- an order of magnitude better than previous procedures.  But proof techniques could not show that these promising fault-tolerance schemes tolerated any noise at all.  

With an analysis based on decomposing complicated probability distributions into mixtures of simpler ones, we rigorously prove the existence of constant tolerable noise rates (``noise thresholds") for error-detection-based schemes.  
Numerical calculations indicate that the actual noise threshold this method yields is lower-bounded by 0.1\% noise per gate.

	\abstractsignature	
\endabstract

\end{frontmatter}

\begin{optionalFrontMatter}


\tableofcontents

\clearpage 

\pagestyle{plain} 





\end{optionalFrontMatter}


\begin{dissertationText}
\renewcommand{\baselinestretch}{1.66}

\chapter{Introduction}    
\section{Overview: Background, techniques and new results}

\subsection{Background}

Fault tolerance is the study of reliable computation using unreliable, noisy components.  For example, one can run several copies of a calculation in parallel, periodically using majority gates to catch and correct faults.  

Historically, fault tolerance was particularly important before the development of the transistor and the integrated circuit, which are much more reliable than vacuum tubes.  As transistors continue to shrink, becoming more vulnerable to errors, fault tolerance will become more important in classical computing.  

In quantum computing, the focus of this dissertation, the basic computational elements are already physically very small, and therefore intrinsically vulnerable to noise.  For example, the spin of a single electron or atomic nucleus, the orbital of an electron, and the polarization of a photon have all been proposed as practical instantiations of a quantum bit (``qubit").  Fundamentally, quantum systems are inherently fragile because of entanglement; a single unintended interaction with a single qubit can collapse, or decohere, the entire system.  Fault-tolerance techniques will be essential for quantum computers. 

The key result in fault tolerance is the existence of a noise \emph{threshold}.  The noise threshold is a positive, constant noise rate (or set of noise model parameters) such that below this rate, reliable computation is possible.  To continue our initial example, when the majority gates themselves are faulty, it is not clear at all that interspersing them should improve overall reliability.  Only if the noise rate is low enough, below the threshold, will reliability improve.  Slightly more precisely, a threshold theorem states that provided the noise rate is low enough, one can compute reliably and also efficiently -- i.e., only moderately slower, and using only moderately more space, than the original computation.  (Still more precisely, a threshold theorem specifies a particular fault-tolerance scheme, or class of schemes, which protects against a certain form of noise in a certain computational model.)

In classical computing, the existence of a constant threshold was first shown by Von Neumann~\cite{VonNeumann56}.  (He was motivated not just by unreliable vacuum tubes, but also to try to understand how the brain might handle with aplomb unreliably firing neurons.)  Roughly, in his model each circuit element failed, flipping its output bit $0 \leftrightarrow 1$, \emph{independently} and with probability $p$.

Quantum computers have enormous potential.  Besides allowing for efficient factoring of large numbers, they should for example also be capable tools for simulating (quantum) physical systems beyond the limits of classical techniques.  But this potential cannot be realized without fault-tolerance techniques for controlling noise. 

In the quantum setting, the first proofs of the \emph{existence} of a constant tolerable noise threshold, by Aharonov and Ben-Or and independently by Kitaev~\cite{AharonovBenOr99, Kitaev96b}, in 1997, were for a noise model very similar to that Von Neumann considered.  The main difference is that, in addition to bit flip X errors which swap $0$ and $1$, there can also be phase flip Z errors which swap $\ket{+}=\tfrac{1}{\sqrt{2}}(\ket{0}+\ket{1})$ and $\ket{-}=\tfrac{1}{\sqrt{2}}(\ket{0}-\ket{1})$ (Fig.~\ref{f:xandzerrors}).  A noisy gate is modeled as a perfect gate followed by independent introduction of X, Z or Y (both X and Z) errors with respective probabilities $p_X, p_Z, p_Y$.\footnote{For two-qubit gates, there are fifteen different failure parameters -- $p_{IX}, p_{IZ}, \ldots, p_{YZ}, p_{YY}$ -- one for each setting of errors to be applied to the qubits after the gate, $I$ being the identity.}  Even though realistic models of quantum noise can be more complicated -- for example affecting continuous, rather than discrete, changes in a quantum state -- the essential features can be captured by this simple ``probabilistic Pauli" noise model.  (X, Z, Y are the Pauli operators.)  

\begin{figure*}
\begin{center}
\includegraphics[scale=1]{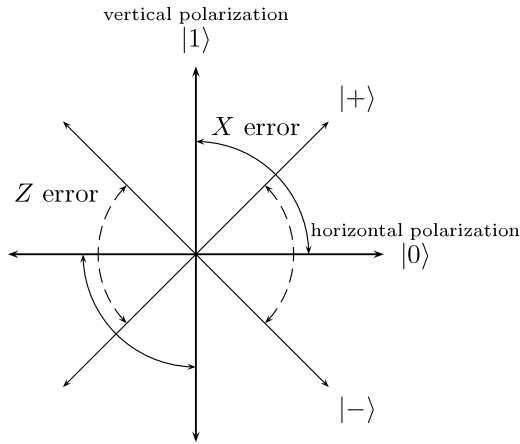}
\end{center}
\caption{
Bit flip X errors flip $0$ and $1$.  In a qubit, $\ket{0}$ and $\ket{1}$ might be represented by horizontal and vertical polarization of a photon, respectively.  Phase flip Z errors flip the $\pm 45^\circ$ polarized states.  Z errors have no classical analog, but are the same as X errors in a different, \emph{dual} basis -- reflect the diagram about the $22.5^\circ$ line to relate $\ket{0}$ to $\ket{+}$, $\ket{1}$ to $\ket{-}$, and X to Z errors.
}
\label{f:xandzerrors}
\end{figure*}

Aharonov/Ben-Or and Kitaev's threshold existence results established that building a working quantum computer is possible \emph{in principle}.  Physicists need only engineer quantum systems with a low enough constant noise rate.  
But realizing the potential of a quantum computer will require \emph{practical} fault-tolerance schemes for dealing with noise.  This means that schemes will have to tolerate a high noise rate (not just some constant), and do so with low overhead (not just polynomial).  Rough estimates of the maximum tolerable noise rates allowed by the early threshold existence proofs, however, were not promising -- below $10^{-6}$ per gate.  It is extremely difficult for physicists to engineer quantum systems with low error rates, and it was clear that a quantum computer would be nearly impossible to build if the tolerable noise rate were only $10^{-6}$.

Is the tolerable noise rate high enough to be practical?  Quantum fault-tolerance researchers have followed three different approaches in order to answer this question:
\begin{enumerate}
\item Develop a new proof technique which allows for a less conservative analysis, to compute a higher rigorous noise threshold lower bound.  More efficient analyses require more complete control over the state of the quantum system, with less worst-case slack.
\item Optimize an existing fault-tolerance scheme, squeezing out inefficiences, or come up with an entirely new fault-tolerance scheme.
\item Estimate the tolerable noise rate of a fault-tolerance scheme with Monte Carlo simulations.  Noise threshold estimates are higher -- more optimistic -- than rigorous threshold lower bounds, and simulations may track typical system behavior better than a necessarily conservative rigorous analysis.  However, extrapolating the results of small simulations to (asymptotically) large systems may be unreliable.  Various intuitive assumptions made in the simulations are only approximately valid. 
\end{enumerate}

Over the last two years, there has been dramatic progress along these lines.  First, new proof techniques have allowed the rigorous analysis of schemes for which estimated noise thresholds were high, but for which previously no positive threshold at all had been proven to exist.  The gap between rigorously proven threshold lower bounds and estimates from simulations has narrowed substantially thanks to more efficient analysis.  Second, Knill has developed a new fault-tolerance scheme which according to estimates can tolerate substantially higher noise rates than previous schemes.  However, existing proof techniques again were insufficient to prove the existence of any positive threshold at all for Knill's scheme.

\subsubsection*{More efficient proof techniques}

Tolerable noise rates for different fault-tolerance schemes have typically been investigated through simulations, and not rigorously lower-bounded, because the original threshold proof techniques of Aharonov/Ben-Or and Kitaev are too inefficient.  They give lower bounds which are thought to be far too conservative, by five orders of magnitude or more.\footnote{In a few very simple noise models, like detected qubit loss, high threshold lower bounds had been established~\cite{KnillLaflammeMilburn00, Knill03erasure}.}  In fact, most simulations evaluated fault-tolerance schemes for which analysis techniques could not prove any positive threshold at all (schemes using distance-three quantum error-correcting codes; analysis required codes of distance at least five).

In 2005, Aliferis, Gottesman and Preskill, and separately this author, gave more efficient analysis techniques which could each handle schemes using distance-three codes \cite{AliferisGottesmanPreskill05, Reichardt05distancethree}.  Also given were the first explicit, rigorous threshold lower bounds, which are now approaching a $10^{-4}$ noise rate (unpublished).  The gap between simulation-based estimates and what could be proved closed to less than two orders of magnitude, placing the estimates on firmer footing.

\subsubsection*{Knill's postselection-based fault-tolerance scheme}

In the meantime, though, Knill has constructed a breakthrough scheme based on very efficient distance-two codes~\cite{Knill05}.  Being of distance two, his codes allow for error \emph{detection}, not error correction, and the scheme uses extensive \emph{postselection} on no detected errors -- i.e., on detecting an error, the enclosing subroutine is restarted.  Using simulations and heuristic analysis, Knill has estimated that the threshold for his scheme is perhaps as high as 3-6\%, 
about an order of magnitude higher than threshold estimates for schemes not using postselection.  (Let us remark immediately that the overhead required for computing at such high error rates is far from practical; we will discuss this further below.)

Despite Knill's high noise threshold estimate, though, once again it was not known if his scheme gave any positive threshold at all.  Postselection is a key factor allowing computation in the face of high error rates, but the rigorous threshold proof techniques did not accommodate it, being limited to more standard fault-tolerance schemes based on error correction.  

\subsection{Existence of noise threshold with postselection}

Our main result here is a proof of the existence of a positive constant noise threshold for a postselection-based fault-tolerance scheme similar, but not identical, to Knill's scheme.  
The proof shows that the strong error correlations one worries about in considering a scheme using postselection can be avoided.

\subsection{Numerical noise threshold calculations}

Postselection-based schemes are attractive because of their high tolerable noise estimates, but can such estimates be rigorously verified?  After proving the existence of a noise threshold, we apply the same technique more carefully to attempt to calculate explicit numerical lower bounds.  
\begin{enumerate}
\item In a noise model which is known \emph{exactly} to the experimenter, who can adjust the fault-tolerance scheme accordingly, we show that $0.7\%$ noise can be tolerated.  
\item With the more realistic assumption that the experimenter knows only upper bounds on the noise parameters, we 
show that $0.1\%$ noise can be tolerated.  (Better analysis can likely improve this number.)
\end{enumerate}
For comparison, the highest current threshold lower bound, for an error-correction scheme in a comparable model, is a $5.36 \times 10^{-5}$ noise rate (DiVincenzo and Aliferis~\cite{DiVincenzoAliferis06slow}).\footnote{Unpublished lower bounds are closer to $10^{-4}$.}

These calculations are subject to some minor caveats, however, including a familiar caveat for numerical work: we use high-precision arithmetic but do not maintain rigorous upper and lower bounds on every number.  For a few specific small problems, we also make mild monotonicity assumptions -- that, roughly, behavior improves as error rates drop.  More careful numerical work could remove all these caveats, but tediously.

The calculations show that the proof technique is efficient enough to use for obtaining fairly high rigorous threshold lower bounds, and is not just useful for threshold existence proofs.  
By narrowing the gap to noise threshold estimates, our higher lower bounds support the estimates' validity, at least in the idealized models for which they were made.  

\subsection{Proof technique} \label{s:introductionprooftechnique}

A difficulty in proving noise thresholds is that the system's error distribution is complex, with many correlations.  Proofs typically handle this complexity by characterizing only a large portion of the probability mass of the system, losing control of some small remainder.  

The intuitive problem for proving a noise threshold with postselection is that renormalizing the error distribution, to condition on no detected errors, can allow the uncontrolled remainder to become exponentially more likely.  

Even if the computer starts with bitwise-independent errors, all well bounded, after applying a ``logical" gate (a gate sequence designed to affect data encoded into an error-correcting code), errors will no longer be independent.  However, if the logical gate is applied properly, then errors in the output will be close to independent.  Close is not enough, since any deviations from independence might be amplified by renormalization.  However, in fact the output error distribution can be rewritten as a \emph{mixture} of well-bounded distributions in which errors are independent.  The analysis can be continued on to the next logical gate by picking one of these distributions from the mixture.  

Rewriting probability distributions with small correlations as mixtures of ``nice" probability distributions with bounded-probability independent events is the main technical tool of this thesis.  In Chapter~\ref{s:postselectchapter}, we give a Mixing Lemma which characterizes the convex hull of the set of ``nice" distributions.  For now, a few simple examples should clarify the idea.  Say we have the following distribution over two events, $A$ and $B$:
\begin{center}\begin{tabular}{r c c c c}
Event: & $\neg (A \vee B)$ & $A$ & $B$ & $A \wedge B$   \\
Probability: & $1-4p+5p^2$ & $2p$ & $2p$ & $5p^2$ \\
\end{tabular}\end{center}
(Notation: $\neg$, $\wedge$ and $\vee$ are {\textsc{NOT}}, {\textsc{AND}} and {\textsc{OR}}, respectively.)
Here $A$ and $B$ are not independent events -- $\pr[A \wedge B] = 5p^2 > (2p)^2 = \pr[A] \pr[B]$.  However, $\pr[A \wedge B]$ is $O(\pr[A] \pr[B])$.  We can rewrite this distribution as
\beq \label{e:introductionmixtureexample}
\left(\begin{array}{c}
1-4p+5p^2 \\ 2p \\ 2p \\ 5 p^2
\end{array}\right)
= 
\frac{1}{2}
\left(\begin{array}{c}
(1-p)^2 \\ p \\ p \\ p^2
\end{array}\right)
+ \frac{1}{2}
\left(\begin{array}{c}
(1-3p)^2 \\ 3p \\ 3p \\ 9p^2
\end{array}\right) \enspace ,
\eeq
a mixture of two distributions in each of which $A$ and $B$ are independent and also still $O(p)$.

Two more simple examples will foreshadow issues which will arise during the analysis.  First, 
$$
\left(\begin{array}{c}
1-2p \\ p \\ p \\ 0
\end{array}\right)
= 
\frac{1}{2}
\left(\begin{array}{c}
1-2p \\ 2p \\ 0 \\ 0
\end{array}\right)
+ \frac{1}{2}
\left(\begin{array}{c}
1-2p \\ 0 \\ 2p \\ 0
\end{array}\right) \enspace .
$$
By rewriting the distribution on the left-hand side as a mixture of two distributions in which $A$ and $B$ are independent, we lose a factor of two in our upper bounds on $\pr[A]$ and $\pr[B]$.  A constant factor loss is acceptable if we aim just to prove the existence of a noise threshold.  For an efficient analysis, though, we want to minimize such losses, so it is helpful if the initial distribution is already very ``close" to having the desired independence properties.  But ``close" does not just mean in total variation distance.

Consider the following distribution:
\begin{center}\begin{tabular}{r c c c c}
Event: & $\neg (A \vee B)$ & $A\wedge\neg B$ & $B\wedge\neg A$ & $A \wedge B$   \\
Probability: & $1-p^2$ & $0$ & $0$ & $p^2$ \\
\end{tabular}\end{center}
This distribution cannot be rewritten as a mixture of distributions in which $A$ and $B$ are independent and $O(p)$ events.  It can be rewritten as $(1-p^2)(\pr[A]=0,\pr[B]=0) + p^2 (\pr[A]=1,\pr[B]=1)$, but this mixture is not interesting.  With care, one can ensure that distributions like this one do not arise.  However, it is convenient to allow for them, but then to \emph{deliberately} introduce first-order errors.  For example, if the initial distribution is known exactly, errors can be added so $A$ and $B$ are exactly independent.  More commonly, errors can be added so the new distribution can be rewritten as a mixture of bounded, independent distributions, as in the first two examples.

Our mixing proof technique, for the fairly simple probabilistic Pauli noise model described above, is completely classical -- just a way of controlling and manipulating probability distributions.  Indeed most of the work in proving a postselection noise threshold is classical.  (One main exception is the extension to a quantum universal gate set, in Ch.~\ref{s:magicchapter}.)  However, the original noise threshold existence proofs of Aharonov and Ben-Or and of Kitaev have since been extended to show the existence of thresholds for more general and realistic noise models -- for example, for continuous Hamiltonian interactions with a non-Markovian bath~\cite{TerhalBurkard04, AliferisGottesmanPreskill05, AharonovKitaevPreskill05}.  An open problem is similarly to extend the mixing technique to show postselection noise thresholds for more general noise models -- and a purely classical technique will not suffice.  

\subsection{Application to error-correction schemes}

In practice, for someone trying to build a quantum computer, these high noise threshold estimates for postselection-based fault-tolerance schemes may be rather too optimistic.  
Problematical assumptions include the simple noise model, and the assumption of nonlocal gates (i.e., any qubit can interact with any other equally well).  Additionally, the space \emph{overhead} required for implementing the schemes tolerating the highest noise rates is greatly impractical.  (Although the overhead can be made to be theoretically efficient, it should not be surprising that a scheme based on restarting whenever an error is detected does not scale well in practice.)  

Knill has given a scheme using a certain error-correction method on concatenated distance-two codes which has considerably less overhead than his postselection-based scheme (but still more than one would like), and for which he estimates the tolerable noise rate remains above $1\%$.  

A distance-two code cannot correct any errors, but concatenation with itself yields a distance-four code.  Roughly, Knill's scheme therefore works on two code concatenation levels at a time: levels one and two, two and three, three and four, etc.  Instead of dropping like $(c p)^{2^k}$ in $k$ the number of levels of concatenation, one expects the effective error rate to drop with an exponent growing like the Fibonacci sequence $F(k) = F(k-1) + F(k-2) \sim 1.6^k$.  No positive threshold had been proven to exist for this method of error correction, because analyzing overlapping concatenation levels is difficult.  By applying the mixing technique, we prove the existence of a positive ``Fibonacci-type" threshold.

\section{Organization}

In the remainder of this chapter, we give high-level intuition for fault tolerance.  We explain the fault-tolerance techniques of Steane and Knill, and give a more detailed history of quantum fault tolerance to place the current work in context.  
In the next chapter, we explain more of the details of particular fault-tolerance schemes and techniques -- for example, the choice of the error-detecting or error-correcting code to protect the data, and the important step of fault-tolerantly encoding into this code.  (A reader well-versed in quantum fault-tolerance can skim through the first two chapters.)  

Chapter~\ref{s:overviewchapter} gives an overview of the postselection noise threshold proof, using classical fault-tolerance examples.  
It starts with more intuition for what might go wrong in fault-tolerant computing with postselection.  

In Chapter~\ref{s:postselectchapter}, 
we apply a Mixing Lemma to prove the existence of a positive noise threshold for a postselection-based fault-tolerance scheme, in a toy noise model in which only bit flip $X$ errors occur.  The next two chapters extend this basic result to more general noise, and, using so-called ``magic states" distillation, to a larger, fully universal gate set.  This completes the main result, the proof of the \emph{existence} of a noise threshold for a class of postselection-based fault tolerance schemes against probabilistic Pauli noise.

In Chapter~\ref{s:correctionchapter}, we will show how to apply the same mixing technique to fault-tolerance schemes not based on postselection -- using error correction instead of error detection.  We show the existence of a so-called ``Fibonacci-type" threshold.  
We also briefly explain how to concatenate an error-correction-based fault-tolerance scheme on top of a postselection-based scheme in order to make the overhead theoretically efficient, while still tolerating higher noise rates thanks to postselection.  

Finally, we apply the mixing technique to attempt to calculate explicit numerical noise threshold lower bounds for postselection-based fault-tolerance schemes in Chapter~\ref{s:numericalchapter}.  

\section{Fault-tolerance intuition: Goals and methods}

\subsection{Reliable simulation idea}

In quantum computing, we have some quantum circuit $\mathcal{C}$ which we want to compute.  $\mathcal{C}$ is made up of $N$ gates from a universal gate set; for example including controlled-NOT (CNOT) gates, single qubit unitaries U, preparation of ancillas, and measurement.  

The problem is that any physical noise quickly disrupts a large computation.  In the presence of random errors, for example, a single random gate failure can a priori randomize the entire output.  Therefore, if we want to have confidence in the result, we need the gate error rate to be $\lesssim 1/N$.  

But this is completely impractical.  Errors appear to be intrinsic to controlled quantum systems.  For one thing, a classical bit is discrete, 0 or 1, and a classical state is just a bit string.  However, a quantum bit -- qubit -- is continuous.  Since quantum states are continuous, at the very least we'll always have precision errors in our operations (unitary rotations).  Also, there is a tradeoff between stability and controllability.  States which we can control are also vulnerable to errors coming from the environment, and states which are well-isolated from the environment are very difficult to control.  This tradeoff appears across the range of proposed quantum computer architectures, from the microscopic to the atomic scale.  

Therefore, we compile the ideal circuit $\mathcal{C}$ into a larger simulating circuit, $\text{FT}\mathcal{C}$, which is so-called ``fault tolerant": meaning that even though its individual physical gates are still perhaps faulty, the effective logical gates are nearly perfect.  
To allow for practical quantum computers, we want to optimize the compilation procedure so that:
\begin{enumerate}
\item $\text{FT}\mathcal{C}$ should tolerate as much noise as possible.
\item $\text{FT}\mathcal{C}$ should be as efficient as possible; the blowup in circuit size should be small.
\end{enumerate}

\subsection{Fault-tolerance targets: algorithms and error rates}

Interesting physical simulations of quantum systems can be run even on fairly small, specially tailored quantum ``computers" \cite{DengPorrasCirac05simulation}.  
To run other interesting algorithms on a general purpose quantum computer, we might need at least a few thousand qubits.  
For example, a circuit factoring a $K= 1024$ bit number might use about $38 K^3 \approx 4 \cdot 10^{10}$ Toffoli gates on about $5K \approx 5000$ qubits \cite{Preskill97,BeckmanChariDevabhaktuniPreskill96efficientfactoring}, so we would need a gate error rate of less than $10^{-10}$ (and an even lower memory error rate) to have a reasonable probability of success.  

Physically reasonable error rates are much larger than this, probably $10^{-2}$ to $10^{-3}$ per gate (Fig.~\ref{f:physicaliontrapparameters}), but not much smaller.  

\begin{table*} 
\caption{Some experimentally determined parameters for ion traps, taken from \Ref\cite{MetodievCrossThakerBrownCopseyChongChuang04}.  (Especially for larger systems, these parameters may be optimistic.)} \label{f:physicaliontrapparameters}
\begin{center}
\begin{tabular}{r c c}
\hline \hline
Operation      & Prob. failure & Time        \\
\hline
One-qubit gate & 0.0001        & $1 \mu s$   \\
Two-qubit gate & 0.03          & $10 \mu s$  \\
Measurement    & 0.01          & $100 \mu s$ \\
Memory time    &               & 100s        \\
\hline \hline
\end{tabular}
\end{center}
\end{table*}

Therefore, fault-tolerance techniques will be essential to building a quantum computer.

\subsection{Fault-tolerance methods} \label{s:introductionfaulttolerancemethods}

To make an ideal quantum circuit $\mathcal{C}$ fault tolerant, encode each qubit of the data into a quantum error-correcting code, and work directly on the encoded or logical qubits.  So for example, as shown in Fig.~\ref{f:faulttoleranceoverview}, an ideal CNOT gate between two qubits might be compiled into, or substituted by, transversal physical CNOT gates on the corresponding code blocks -- meaning CNOT gates from bit 1 of the first block to bit 1 of the second block, 2 to 2, 3 to 3, etc.  Follow these transversal gates by error correction steps on each output block.  Altogether, this affects a logical CNOT gate on the encoded data, and does so without allowing local errors to spread out of control either within or between the two data blocks.  If, say, we had instead decoded the data, computed on it with a single physical CNOT gate, and then reencoded, the logical qubits would have been exposed to local noise.

\begin{figure*}
\begin{center}
\includegraphics[scale=.5]{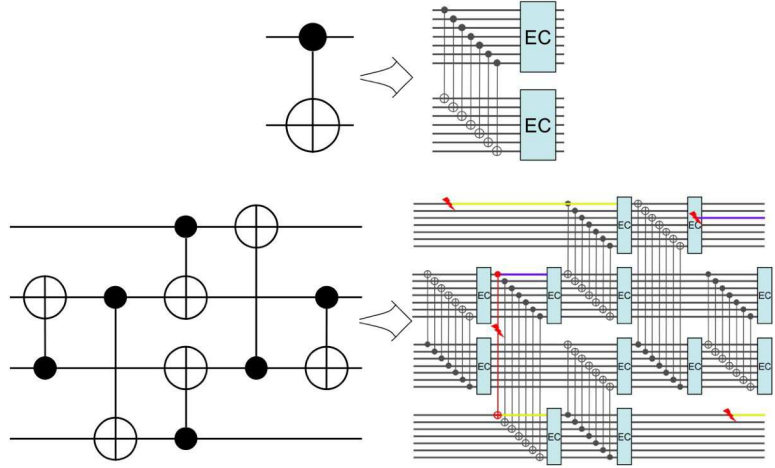}
\end{center}
\caption{
In fault-tolerant computing (classical or quantum), the logical (qu)bits are encoded into an error-correcting code.  The ideal circuit's gates are compiled into gates acting directly on the encoded data.  For example, an encoded CNOT gate can often be implemented as transversal physical CNOT gates (bit 1 to 1, 2 to 2, etc.).  To prevent errors from spreading and accumulating, error-correction modules are placed between encoded gates.  
Corrections are themselves possibly faulty but 
errors remain under control.
}
\label{f:faulttoleranceoverview}
\end{figure*}

Of course the error corrections are themselves faulty, and can introduce new errors, but when implemented carefully -- ``fault-tolerantly'' -- errors will overall remain under control.  Figure~\ref{f:errorcorrectionoverview} gives an example of a fault-tolerant error-correction procedure -- we will explain this and other error correction schemes in Sec.~\ref{s:introfaulttoleranceschemes} and Ch.~\ref{s:ftconstructionschapter}.

\begin{figure*}
\begin{center}
\includegraphics[scale=.5]{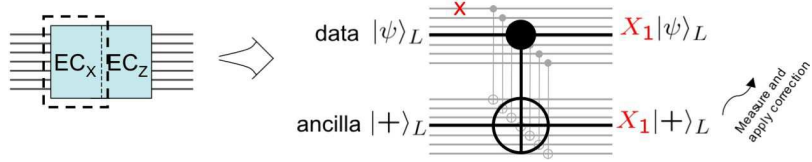}
\end{center}
\caption{
Steane-type error correction consists of X (bit flip) error correction, followed by Z (phase flip, the dual) error correction.  
To correct X errors, prepare an ancilla in the encoded state $\ket{+}_L = \tfrac{1}{\sqrt{2}} (\ket{0}_L + \ket{1}_L)$.  Then apply transversal CNOTs into the ancilla, measure the ancilla, and apply any necessary correction.  
}
\label{f:errorcorrectionoverview}
\end{figure*}

Using an $m$-qubit code of distance $d=2t+1$ (i.e., correcting $t$ errors), one intuitively expects the ``effective/logical error rate'' of an encoded CNOT gate to be 
\begin{equation} \label{e:introductionthresholdlevelone}
\eta_1 \leq c \eta_0^{t+1} = O(\eta_0^{t+1}) \enspace ,
\end{equation}
if $\eta_0$ is the physical error rate.  This is because $t+1$ errors might be corrected in the wrong direction, causing a logical error.\footnote{The constant $c$ should come from counting the number of ``malignant" sets of $t+1$ error locations in the encoded implementation.}  
Below a break-even point, $\eta_0 < (1/c)^{1/t}$, the effective gate error rate is reduced.  

However, one can't directly achieve an arbitrarily low effective error rate $1/N$ just by choosing a large code -- setting $t = \Omega(\log N)$ -- because then even the initial encoding would likely fail (encoding is also often used in fault-tolerant quantum error correction).  
Instead, in order to get the effective error rate arbitrarily low, use a smaller code and repeat, or \emph{concatenate}, the whole procedure.  
Each of the qubits on the right-hand side of Fig.~\ref{f:faulttoleranceoverview} is itself encoded, and the CNOT gates are themselves replaced by the same substitution rule -- transversal CNOT gates followed by error correction at the lower level.  Letting $\eta_j$ be the probability that a level-$j$-encoded CNOT gate ``fails," it should be the case that 
\begin{equation} \label{e:introductionthresholdlevelj}
\eta_j \leq c (\eta_{j-1})^{t+1} \enspace .
\end{equation}
Therefore, provided the initial error rate is beneath a constant threshold, the effective error rate can be reduced arbitrarily with sufficient concatenation~\cite{NielsenChuang00}.  

While this argument may be intuitive, it falls apart upon further examination.  Logical gate ``failures" on blocks of qubits are not just analogous to physical gate failures, and so Eq.~\eqref{e:introductionthresholdlevelj} may not be a valid extension of Eq.~\eqref{e:introductionthresholdlevelone}.  The most natural definition for success of a logical gate (level-$j$ for $j \geq 1$) may be the commutativity of the diagram:
\begin{equation} \label{e:introductionlogicalfailure}
\begin{diagram}[balance,width=4em,height=4em,tight]
\, & \rTo{\text{logical gate}}  & \, \\
\dTo>{\rotatebox{-90}{\makebox[0pt]{ideal decoding}}} &  & \dTo>{\rotatebox{-90}{\makebox[0pt]{ideal decoding}}} \\
\, & \rTo_{\text{ideal gate}} & \,
\end{diagram} 
\end{equation}
I.e., define a logical gate on an encoded qubit to have succeeded if following that logical gate by an ideal (faultless) decoding procedure would have the same effect as first decoding and then applying the ideal gate.  Otherwise, the gate has failed.  However, with this definition or others, it is not clear at all that the logical error model should have the same properties as the initial physical error model.  Why for example should different logical gates still fail independently?  E.g., with two consecutive logical gates, mightn't the failure of the first, by the definition of Eq.~\eqref{e:introductionlogicalfailure}, leave more bit errors in the block, in turn making the second logical gate more likely to fail?

The major problem in \emph{proving} fault-tolerance thresholds is in understanding and managing the behavior of ``failures" on encoded blocks of qubits.  
It isn't even clear that any constant-sized definition of logical failure will suffice to describe the behavior of asymptotically large blocks of qubits.
Still, it is a useful fiction to assume that encoded gate failures can be treated completely analogously to physical gate failures, so let us follow it through a bit longer to develop more intuition.

Assuming Eq.~\eqref{e:introductionthresholdlevelj} holds, and provided the initial error rate is beneath a constant threshold, the effective error rate will drop as 
$$
\eta_j \leq \frac{1}{c^{1/t}} (c^{1/t} \eta_0)^{(t+1)^j} \enspace ;
$$
by concatenating the compilation procedure on top of itself $k = \log_{t+1} \log \tfrac{N}{\epsilon}$ times, the effective error rate of a level-$k$-encoded gate will be $\eta_k \leq \epsilon/N$.  

The overhead is growing very quickly, like $(c' m)^k$ if the code has $m$ qubits.  But since the effective error rate is dropping doubly exponentially fast in the number of concatenation levels, the overhead in terms of $N$ is only 
$$
\left(\log \frac{N}{\epsilon}\right)^{\tfrac{\ln (c' m)}{\ln (t+1)}} ,
$$
or polylogarithmic in $N/\epsilon$; the scheme is efficient.  

The constant threshold beneath which the effective error rate can be improved -- $\eta_0 < (1/c)^{1/t}$, above -- is known as the fault-tolerance threshold.  The threshold together with the overhead roughly determine how hard it is to build a quantum computer.  In reality, these two parameters \emph{trade off} against each other, and are highly model-dependent and typically multi-dimensional.  Understanding them well gives resource tradeoffs of how much noise can be tolerated for a given computation with a given qubit budget for overhead.  

\section{Quantum fault tolerance history} \label{s:history} \label{s:faulttolerancetheorems}

\subsection{Noise threshold existence proofs}

Quantum states are inherently fragile, and quantum operations inherently noisy, so developing fault-tolerance techniques is essential for progress toward a quantum computer.  A quantum circuit with $N$ gates can a priori tolerate only $O(1/N)$ error per gate.  In 1996, Shor showed how to tolerate $O(1/\poly(\log N))$ error by encoding each qubit into a $\poly(\log N)$-sized quantum error-correcting code, then implementing each gate of the desired circuit directly on the encoded qubits, alternating computation and error-correction steps as in Fig.~\ref{f:faulttoleranceoverview} \cite{Shor96}.  

Several groups -- Aharonov and Ben-Or \cite{AharonovBenOr99}, Kitaev \cite{Kitaev96b}, and Knill, Laflamme and Zurek \cite{KnillLaflammeZurekProcRSocLondA98} -- each came up with the idea of using smaller codes, and concatenating the procedure repeatedly on top of itself.  Aharonov/Ben-Or and Kitaev in 1997 gave independent proofs of the existence of a positive constant noise threshold, or maximum tolerable gate error rate allowing reliable quantum computation.  Intuitively, more frequent error correction at the lower levels allows one to tolerate a constant error rate, while still reducing the effective logical error rate at high levels to be arbitrary small.  
Again, the challenge in making this rigorous is that logical gates on blocks of qubits do not ``fail" independently, and their behavior must be carefully controlled.

The existence of a tolerable noise threshold is important, because it means that quantum computers can in principle be built.  Physicists only need to put in a constant amount of engineering work to lower error rates enough.  But in practice, the value of the noise threshold, together with the overhead required to attain it, roughly determine how hard it is to build a useful quantum computer.  One would like the tolerable noise rate to be high, because reducing error rates below 1\%, or perhaps 0.1\%, could be nigh well impossible.  However, rough estimates of the noise rate tolerated by the original existence proofs are not promising: perhaps $10^{-7}$ to $10^{-6}$ noise per gate.

Broadly speaking, there has since been progress on two fronts of the fault-tolerance problem.

First, there has been substantial work on optimizing fault-tolerance schemes primarily in order to improve the tolerable noise rate.  These optimizations are typically evaluated with simulations and heuristic analytical models, and will be discussed in Sec.~\ref{s:introductionhistorysimulationresults} below.  

Second, work has proceeded on rigorous fault-tolerance threshold results.  This includes extending the set of noise and computation models in which a noise threshold is known to exist.  For example, correlated noise, leakage errors, and non-Markovian noise have all been shown to be tolerable, even in a computation model allowing only geometrically constrained local gates.
\begin{description} 
\item[Correlated noise] Knill, Laflamme and Zurek \cite{KnillLaflammeZurek96} show that the independent noise assumption of the original threshold proofs can be weakened to allow for weak \emph{correlations} both spatially and temporally -- instead of the error strength on each location being at most $\eta$ with independent noise, the weaker condition is that the total error strength on any $k$ locations must be upper-bounded by $\eta^k$.  
\item[Leakage errors] Knill/Laflamme/Zurek also consider \emph{leakage} errors -- errors which take the system out of the computational Hilbert space.  For example, a physical qubit might just disappear; or, if a qubit (two-level quantum system) is represented by the location of an electron in two orbitals, the electron might move to a third orbital.  Leakage errors propagate through the quantum computer differently from errors within the computational Hilbert space, so require a slightly more careful analysis -- which has now been made rigorous by Aliferis and Terhal \cite{AliferisTerhal05}.  (When a leakage error is reliably detectable, it is known as an \emph{erasure} error.  Very high thresholds can be shown for error models with only erasure errors \cite{KnillLaflammeMilburn00, Knill03erasure}.) 
\item[Non-Markovian noise] \emph{Non-Markovian} noise models -- in which the environment is allowed to have a memory of previous interactions instead of starting fresh after every gate -- are considered by Terhal and Burkard, who extend Aharonov/Ben-Or's threshold proof to allow for non-Markovian noise with environment baths local to each qubit \cite{TerhalBurkard04}.  Aliferis, Gottesman and Preskill show the existence of a threshold even with a single, non-local bath \cite{AliferisGottesmanPreskill05}.  Whereas these results require that noise affects pairs of system qubits only when the experimentalist deliberately interacts them to apply a gate, Aharonov, Kitaev and Preskill have recently proved the existence of a threshold even for always-on non-Markovian noise on pairs of system qubits \cite{AharonovKitaevPreskill05}.
\item[Local gates] The original threshold results assumed a non-local gate model.  That is, arbitrary pairs of qubits could interact equally well.  For some proposed quantum computer implementations, e.g., using photons, this is a reasonable assumption.  But if qubits are arranged spatially and are not very mobile (or can't move at all), then this is a poor assumption.  Gottesman has shown that thresholds exist even for non-mobile qubits arranged in a two-dimensional grid with only \emph{local}, nearest-neighbor interactions, or arranged in one dimension with next-nearest-neighbor interactions \cite{Gottesman00local}.  His argument is general enough to apply with care to most threshold proofs.  Most simulations have also assumed non-local gates.  Other simulations, though, have indicated that adding two-dimensional locality constraints imposes only a modest threshold penalty \cite{SvoreTerhalDiVincenzo04} (much worse in one-dimension \cite{SzkopekFanRoychowdhuryYablonovitchBoykinSimmsGyureFong04local}).  However, this is an area which still needs to be evaluated (see Sec.~\ref{s:architectures}), particularly for fault-tolerance schemes based on postselection.
\end{description}

\subsection{Simulation results} \label{s:introductionhistorysimulationresults}

The effectiveness -- noise threshold and overhead -- of optimized fault-tolerance schemes can be estimated efficiently by simulation, because the limited set of ``stabilizer" operations used in quantum fault tolerance with probabilistic Pauli noise models can be efficiently classically simulated, by the Gottesman-Knill theorem (Sec.~\ref{s:gottesmanknilltheorem}).

Most threshold estimates have used Steane's seven-qubit, distance-three code, from basic estimates \cite{Gottesman97thesis, Preskill97, KnillLaflammeZurekScience98, AharonovBenOr99}, to estimates using optimized fault-tolerance schemes \cite{Zalka97,Reichardt04,SvoreCrossChuangAho}, to a threshold estimate with a two-dimensional locality constraint \cite{SvoreTerhalDiVincenzo04}.  

Steane has developed an optimized fault-tolerance scheme, and used simulations and a heuristic model to evaluate its performance using different codes \cite{Steane02, Steane03}.\footnote{
There are many details to optimize in fault-tolerance schemes, including the method of fault-tolerant encoded ancilla preparation, which we will address in Sec.~\ref{s:encodingverification}.  But one also has to answer questions like, in Steane's scheme, should X or Z error correction come first?  To minimize interblock error propagation, according to the rules of Fig.~\ref{f:introductionerrorpropagation}, one might prefer to have X error correction on the control block immediately before transversal CNOTs, and on the target block immediately after transversal CNOTs -- and contrariwise for Z error correction.
}
Steane estimates that, from among a large set of codes, the seven-qubit code comes in only third behind the twenty-three-qubit Golay code (distance seven) and a forty-seven-qubit quadratic residue code (distance eleven), which offer better efficiency compromises.  To give some numbers, Steane estimates a threshold noise rate of around $2 \times 10^{-3}$ for the seven- and forty-seven-qubit codes (in a certain precise error model), and $3 \times 10^{-3}$ for the Golay code.  The $3 \times 10^{-3}$ threshold estimate is three times higher than Zalka's earlier estimate \cite{Zalka97}, and five times higher than an estimate of Gottesman and Preskill \cite{Preskill97, Gottesman97thesis}.  

Steane finds that at noise rates well below the threshold it is best to start with a small code at the lowest concatenation level, then switch to a larger code.  Intuitively, small codes can be more efficient in fault-tolerance schemes because encoding into the quantum code is a threshold bottleneck.  Quick encoding allows for frequent, rapid error correction.  But this intuition needs to be qualified.  Larger codes can offer more efficient protection with higher distances, and even multiple encoded qubits per code block, so there is potentially a tradeoff.  And if measurements are slow, then small codes can no longer perform rapid error correction, because they are waiting for measurements to complete.  

The main feature of Steane's techniques is a different error-correction method using encoded ancilla states (Sec.~\ref{s:introfaulttoleranceschemessteane}), and an optimized encoded ancilla preparation procedure (which we will discuss in Sec.~\ref{s:stabilizerstatepreparation}).  Reliable ancilla states are important for tolerating high noise.  In order to tolerate the most noise, without regard to overhead, this author considered careful testing of encoded ancilla states during and after preparation \cite{Reichardt04}.  If a single test fails -- any error detected -- then throw the whole ancilla away and start over (or, prepare many in parallel).  In other words, we \emph{postselect}, or condition, on no detected errors.  Intuitively, this should improve performance because, e.g., a distance-three code can correct only one error, but detect up to two errors.  In simulations, the estimated tolerable noise rate roughly doubled, but at the cost of increased overhead.

In a breakthrough, Knill has constructed a novel fault-tolerance scheme based on very efficient distance-\emph{two} codes \cite{Knill05}.  
His codes cannot correct any errors 
and the scheme uses extensive postselection on no detected errors -- i.e., on detecting an error, the enclosing subroutine is restarted.  This leads to an enormous overhead at high error rates, limiting practicality.  However, Knill has estimated that the threshold for his scheme is perhaps as high as 3-6\% (independent depolarizing noise in a nonlocal gate model), an order of magnitude higher than Steane's estimates not using postselection.  (Knill also gives schemes using less postselection, and thus having more reasonable overhead but tolerating less error, too.)

\subsection{Rigorous noise threshold bounds}

Noise threshold proofs trade off control with worst-case analysis.  For example, an analyst constructing a proof might decide to call a block ``controlled" if it lies exactly in the codespace.  Within the set of controlled blocks, he would try to maintain bounds on logical error rates, with a definition like that of Eq.~\eqref{e:introductionlogicalfailure} and probabilistic analysis.  Outside the set of controlled blocks, however, he would have to assume worst-case behavior.  Error correction at higher encoding levels needs not only to correct encoded errors, but also to restore control to the affected block.  In this example, almost all blocks would be uncontrolled, and it would be very difficult to regain any control -- the proof would fail.  

However, valid proofs can be constructed maintaining different levels of control.  Stronger control requires a better understanding of the behavior of errors in the fault-tolerance procedure.  Stronger control is difficult to maintain, and the necessary definitions may be more complex -- but a more efficient analysis, with less worst-case slack, can give the reward of better noise threshold lower bounds and broader applicability.  

The threshold proof in this thesis can be viewed as part of a progression of proof techniques each maintaining stronger control over the quantum computer.  We sketch this progression below, starting with Aharonov and Ben-Or's threshold proof.

\subsubsection*{Weak control: 1-good}

The classic noise threshold proof of Aharonov and Ben-Or \cite{AharonovBenOr99} can be reformulated to rely on a key control definition of ``1-goodness."  Roughly, define a code block to be 1-good if it has at most one subblock which is not itself 1-good.  (The allowed bad subblock can have worst-case behavior.)  The definition is recursive; at the base encoding level 0, every physical qubit can be said to be 1-good.  

For the CNOT substitution rule of Fig.~\ref{f:faulttoleranceoverview}, if the two input blocks are both 1-good and at most one error occurs within the block, then the output blocks will be 1-good.  This is provided the code has distance seven or higher, for then the three total errors (one from each input block, and one during the CNOT) can be corrected in the proper direction.  Thus two errors occurring during the CNOT implementation is the bad event, so the error rate drops quadratically, giving a positive threshold.  The difficulty lies in formalizing and making rigorous this intuition.  

Aharonov and Ben-Or's threshold proof can be made to work for the repeated concatenation of codes of distance-five or higher.  
The 1-good definition, however, is very strong, so much of the system falls out of control and must be assumed to have worst-case behavior.  
With a concatenated distance-five code, effective error rates ``should'' drop cubically at each level of encoding (according to the intuition of Sec.~\ref{s:introductionfaulttolerancemethods}, because three errors can be corrected in the wrong direction).  However, the analysis is inefficient, only giving a \emph{quadratic} error-rate reduction.   

Most fault-tolerance simulations have been run using Steane's seven-qubit, distance-three code, probably because the code is easy to simulate and heuristically analyze, apparently efficient, and implementable in small experimental systems.  
With a distance-three code, 
we intuitively expect the effective error rate to drop quadratically at each concatenation level.  But with the 1-good definition, the system is too poorly controlled for the proof to apply at all.  
For, take a 1-good block with the allowed one erroneous (worst-case) subblock, and apply a logical gate to it.  If a single subblock failure occurs while applying the logical gate, there can be two bad subblocks total, enough to flip the state of the whole block (since the distance is only three).  Therefore, the block failure rate is first-order in the subblock failure rate.  Logical behavior is not necessarily improved by encoding, and the basic premise of fault tolerance, controlling errors even with imperfect controls, is violated.  

\subsubsection*{More control: 1-well}

This author has extended Aharonov and Ben-Or's proof to work for the popular, concatenated distance-three, seven-qubit Steane code, by using a stronger induction assumption \cite{Reichardt05distancethree}.  With 1-goodness, we are assuming that the block entering a computation step has no more than one bad subblock.  Intuitively, though, most of the time there should be no bad subblocks at all, and the stronger definition of ``1-wellness" captures this intuition.  In a 1-well block, not only is there at most one bad subblock, but also the \emph{probability} of a bad (worst-case) subblock is small.  Using 1-wellness as the controlling induction assumption, the problem for distance-three codes sketched above does not occur; the probability of there being an erroneous subblock in the input is already first-order, so a logical failure is still a second-order event.  (For the argument to go through, though, the definition must be carefully stated, and we need to carefully define what is required by each logical gate and how the logical gates will be implemented.)  The proof therefore relies on recursively controlling the probability distribution of errors in the system's code blocks as the computation progresses.  

Aliferis, Gottesman and Preskill independently proved a threshold for concatenated distance-three codes \cite{AliferisGottesmanPreskill05}, based instead on formalizing the ``overlapping steps" threshold argument of Knill, Laflamme and Zurek \cite{KnillLaflammeZurekScience98}.  This argument extends the basic units of Aharonov and Ben-Or's analysis to include also the previous error correction (Fig.~\ref{f:idealcircuit}).  Aliferis et al.~define a logical gate to have failed if two errors occur in the ``extended rectangle."  This is somewhat analogous to ``wellness" because every error in an input code block can be accounted for by some error in the previous error correction.  The analogy breaks down, though, in the technical definitions.

\begin{figure*}
\begin{center}
\includegraphics[scale=.37]{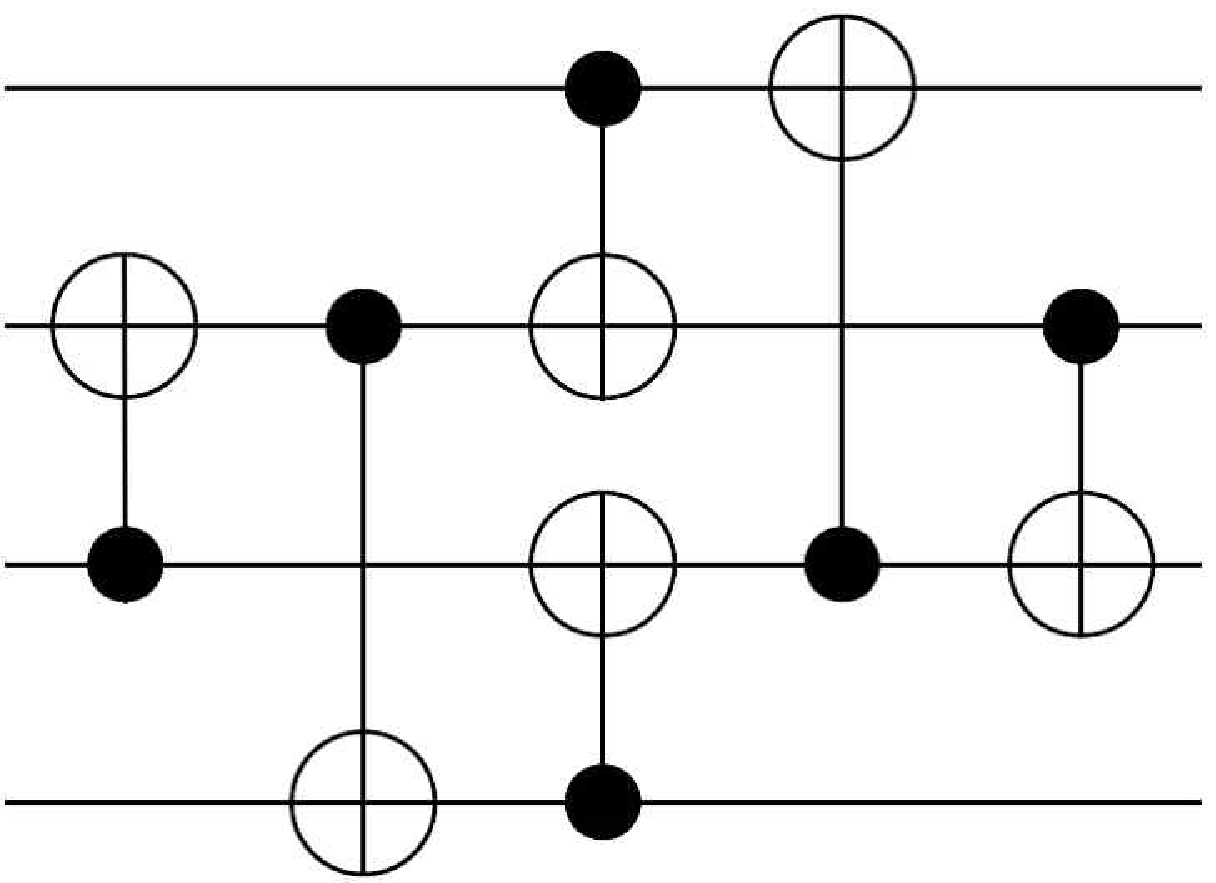} $\qquad$
\includegraphics[scale=.37]{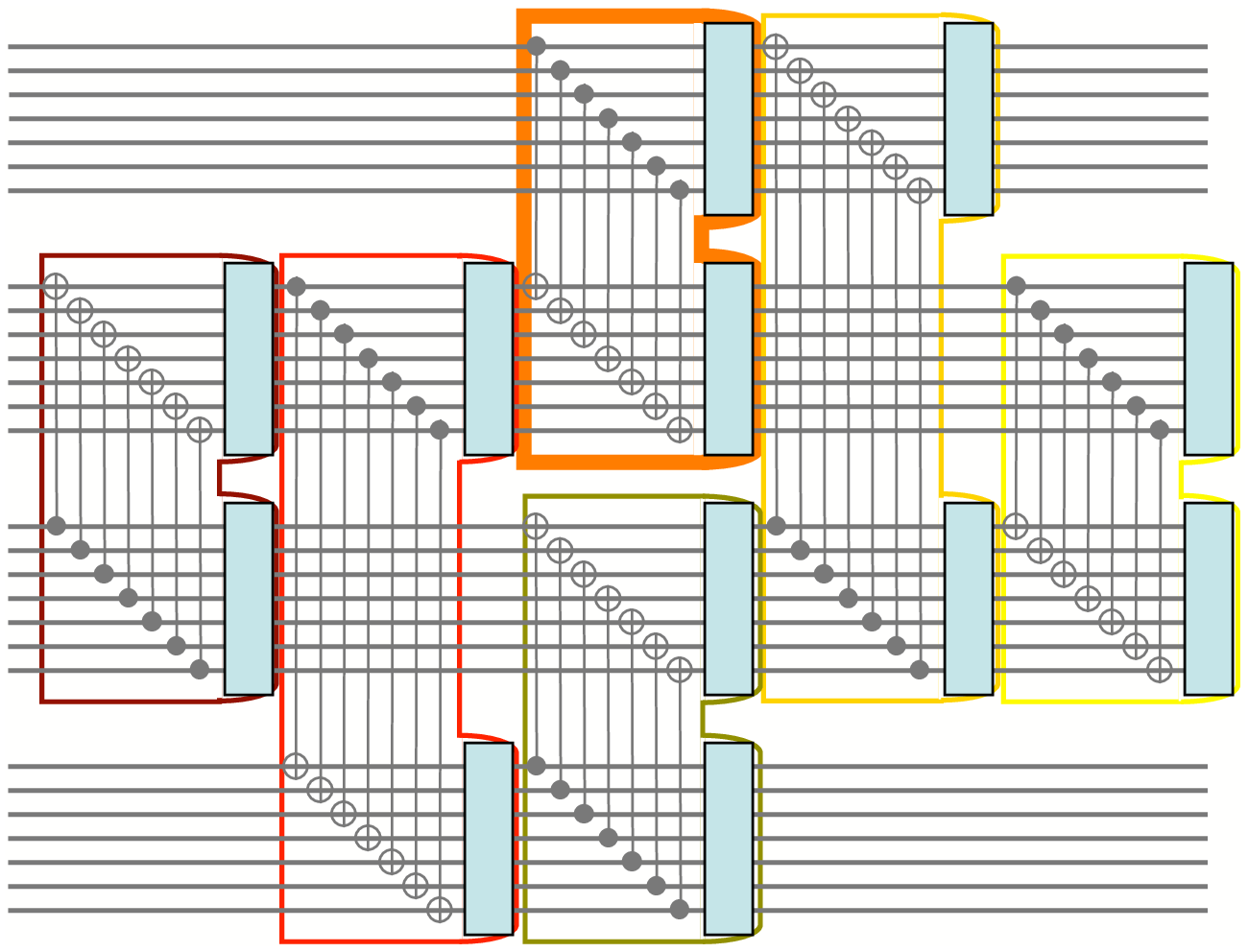} $\quad$
\includegraphics[scale=.37]{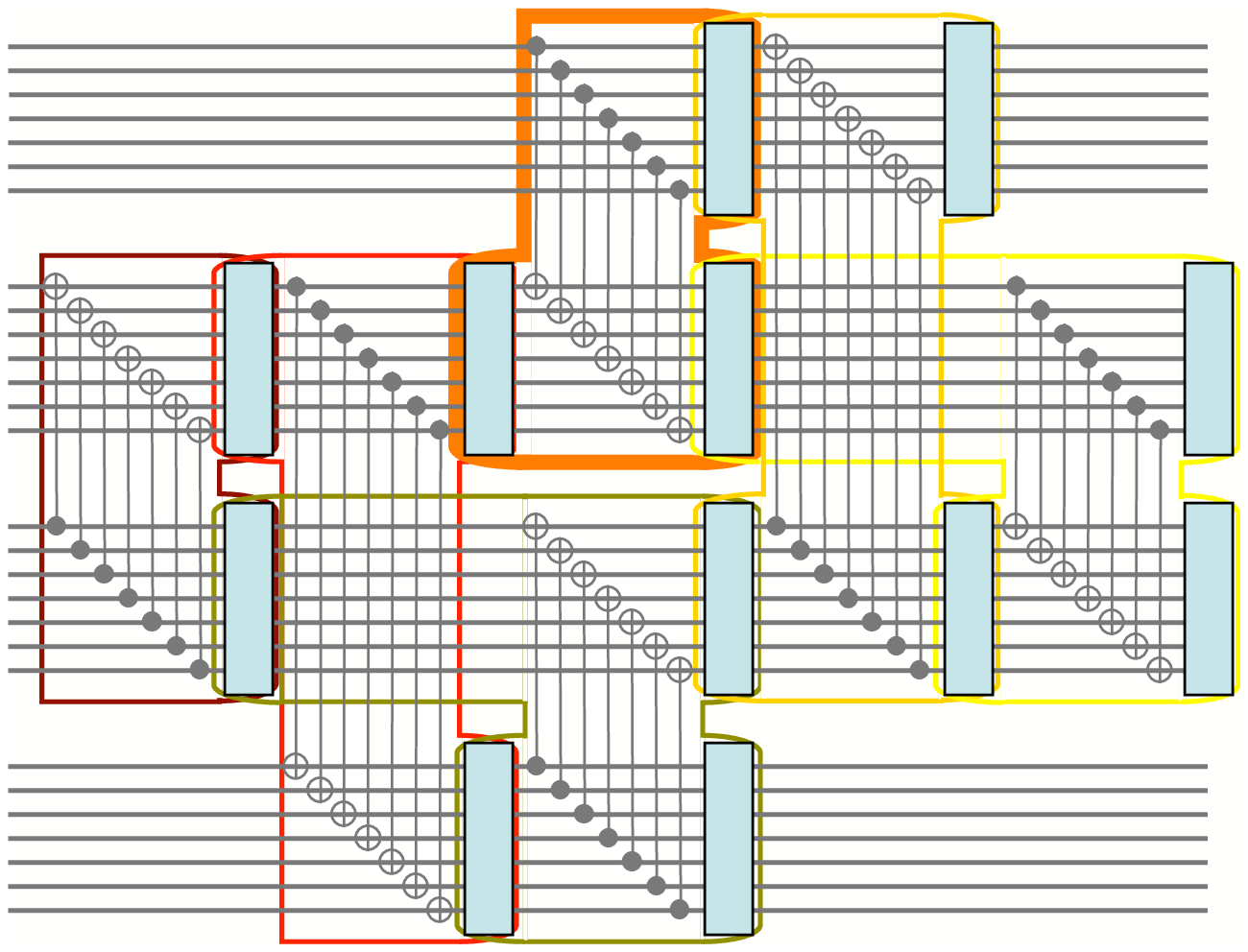}
\end{center}
\caption{Left: A certain ideal circuit using CNOT gates.  Right: The same circuit compiled with one application of the substitution rule of Fig.~\ref{f:faulttoleranceoverview}.  The ``rectangles," of physical gates corresponding to an ideal CNOT, are highlighted.  Bottom: Used in the proof of Aliferis, Gottesman and Preskill \cite{AliferisGottesmanPreskill05}, ``extended rectangles" -- highlighted -- overlap on the leading error correction.
} \label{f:idealcircuit}
\end{figure*}

In addition to giving the first proofs of the existence of a threshold for concatenated distance-three codes, Refs.~\cite{Reichardt05distancethree, AliferisGottesmanPreskill05} include the first proven explicit numerical threshold lower bounds.  Rigorous threshold lower bounds based on previous existence proofs had not been carefully evaluated, probably because rough analyses did not seem promising compared to simulation-based estimates.  The more efficient techniques required to prove the existence of a threshold for distance-three codes also made careful numerical threshold evaluations more worthwhile.  The highest current rigorous threshold lower bound, for an error-correction scheme in a comparable model, is a $5.36 \times 10^{-5}$ noise rate \cite{DiVincenzoAliferis06slow}, obtained by applying the technique of \Ref\cite{AliferisGottesmanPreskill05} to an optimized fault-tolerance scheme.

\subsubsection*{This thesis: Total control via mixing}

Knill's fault-tolerance scheme has the highest estimates for the amount of noise it can tolerate.  Despite Knill's high estimated noise threshold, though, it was not known if his scheme gave \emph{any} positive threshold.  Postselection is a key element of his scheme, allowing for computation in the face of high error rates.  But threshold proof techniques have not accommodated it, being limited to more standard fault-tolerance schemes based on error correction.  We here prove the existence of a positive constant noise threshold for a postselection-based fault-tolerance scheme using concatenated distance-two, error-detecting codes. 

The method, as sketched in Sec.~\ref{s:introductionprooftechnique}, is based on controlling the probability distribution of errors in the system.  But whereas in \Ref\cite{Reichardt05distancethree} it sufficed to control the errors within ``well" code blocks (and allow for worst-case errors in other blocks), here we need strong control over errors in all blocks, in order to prevent postselection from amplifying correlations.  
A block is never allowed to have worst-case behavior.
At all times, we know that the true distribution of errors in the system can be rewritten as a mixture of distributions with independent, bounded-rate errors, as in Eq.~\eqref{e:introductionmixtureexample}.  Still, the analysis is conservative because at each time step we must choose the worst of the mixed distributions (e.g., the rightmost distribution of Eq.~\eqref{e:introductionmixtureexample}) to advance control to the next time step.  

(In Sec.~\ref{s:numericalexact}, we will consider a simple noise model for which the error distribution can be tracked exactly throughout the computation, even by the experimenter.  Code blocks are never allowed to have worst-case error behavior, and there is no mixing of distributions.  The analysis is still very slightly conservative because in order to maintain such complete control, errors must be deliberately introduced into the computation, although only at a very low rate.)  

More efficient analyses, based on controlling more of the system and letting less worst-case behavior slip through the cracks, 
have been required to extend threshold existence proofs to smaller concatenated codes, from distance $d \geq 5$ to $d=3$ to $d=2$.   
Calculations in Ch.~\ref{s:numericalchapter} show that the new mixing technique developed here has promise to give high rigorous threshold lower bounds.  Speculatively, perhaps the mixing technique, by offering even stronger control of errors in the system, will lead to more efficient analysis even of schemes which do not use postselection.

\subsubsection*{Extension to more realistic noise models}

The proof technique of \Ref\cite{Reichardt05distancethree}, like this one, is probabilistic, but analogies can be drawn between it and \Ref\cite{AliferisGottesmanPreskill05}, which does not require a probabilistic error model.  
Unfortunately, it seems less likely that our new technique can be extended to more general, coherent errors.  Writing the error distribution as a mixture of nice distributions is a classical idea which does not work for general quantum states.  In quantum mechanics language, this rewriting is equivalent to saying that the environment (in this case, the analyst!) 
can measure which element of the mixture the system is in -- but with coherent errors, that is simply not possible.  Proving the existence of a postselection noise threshold for more general noise models therefore might require us to give up some analytical control over the system.

\section{Classical versus quantum fault tolerance} \label{s:classicalversusquantum}

We will now explain the building blocks of quantum fault-tolerance.  For a gentler, intuitive introduction, we begin with the classical case.  

Faulty vacuum tubes in early classical computers such as the ENIAC were a major limitation to scalability, and one impetus to develop a theoretical understanding of how to compute reliably in the presence of errors.  

Von Neumann considered a model in which each gate fails independently with some probability \cite{VonNeumann56}.\footnote{Von Neumann originally considered nonlocal AND gates -- arbitrary bits allowed to interact -- but G{\'a}cs later considered the situation with geometrical locality constraints \cite{Gacs83}.}  Certain other operations, including introduction of fresh bits and splitting a wire (fan-out), can be applied with zero error.  

To compute in the presence of this noise, run the ideal circuit on bits each encoded into the $n$-bit repetition code $0 \mapsto 0^n$, $1 \mapsto 1^n$.  To apply e.g. a logical AND gate between two encoded blocks, simply use transversal AND gates 
-- meaning an AND gate from bit $i$ of the first block to bit $i$ of the second block, $1 \leq i \leq n$.  
Between logical gates, a recovery operation prevents errors from accumulating.  Error recovery can be accomplished for example with majority gates, as shown in Fig.~\ref{f:vonneumannmajorityscheme}.  (Von Neumann actually used two rounds of $n/2$ NAND gates between random pairs of bits.)  Provided the physical error rate is below a constant threshold, the probability of a logical error can be made arbitrarily small for $n$ large enough.

\begin{figure*}
\begin{center}
\includegraphics[scale=.5]{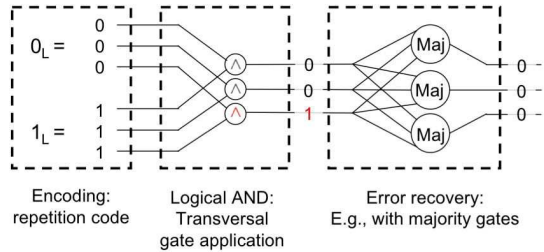}
\end{center}
\caption{A classical fault-tolerance scheme, using encoding into the repetition code.  Logical gates can be applied transversally, and are followed by an error recovery step using majority gates.}
\label{f:vonneumannmajorityscheme}
\end{figure*}

Careful definitions of fault tolerance differ, according to what one is trying to prove.  Intuitively, fault tolerance means that local errors have only local effect.  This is perhaps most easily seen with counterexamples.  Figure~\ref{f:classicalschemenotft} gives two examples of logical AND gate implementations which are not fault tolerant.

\begin{figure*}
\begin{center}
\includegraphics[scale=.5]{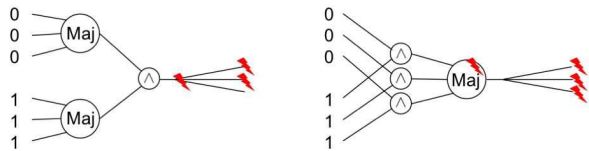}
\end{center}
\caption{It is \emph{not} fault tolerant to decode the data, apply an AND gate, then reencode it -- for unencoded data is not protected against errors.  Error recovery with a single majority gate and fan-out is also not fault tolerant, since gates used in error recovery are themselves potentially faulty.}
\label{f:classicalschemenotft}
\end{figure*}

The quantum fault-tolerance scheme sketched in Fig.~\ref{f:faulttoleranceoverview} is analogous to Von Neumann's scheme, in terms of applying an encoded gate followed by error correction.  But the analogy doesn't go much further, since quantumly we can't apply the same methods.  
For example, the bit-copying operation (fan-out) which Von Neumann assumed works perfectly doesn't even exist quantumly.  Classical gates often used for error correction, like the majority or NAND gates useful with the repetition code, are similarly non-unitary, so can't be implemented quantumly.  

Another important difference is the encoding step.  If we were somehow given encoded qubits with bitwise independent errors for free, then quantum fault-tolerance would be more analogous to Von Neumann's scheme.  However, the quantum states for large codes are highly entangled -- for example the repetition code maps $\ket{0}+\ket{1} \mapsto \ket{0^n} + \ket{1^n}$ a GHZ or cat state -- and we cannot assume that they can be prepared with bitwise independent errors.  It is for this reason that quantum fault-tolerance schemes use code concatenation.  They start with smaller codes, for which it is possible to prepare reliable encoded states, and use them to bootstrap into larger codes.  By using this bootstrapping procedure, we will ultimately show how to prepare large encoded states which, except for a tiny probability, have only bitwise independent errors (Sec.~\ref{s:postselectefficient}, and see \Ref\cite{Knill03erasure}).

Despite these quantum disadvantages, there are several quantum advantages, too.  For example, we can assume perfect classical computers controlling our quantum computer.\footnote{Purely unitary error correction methods exist \cite{AharonovBenOr99}, but adaptive classical control based on measurements is more efficient.  In some proposed quantum computer implementations, though -- for example globally controlled arrays of qubits -- fully adaptive classical control is impossible.}
Perhaps the biggest quantum advantage is that we can defer exposure of the data to operations, allowing us to catch errors in the computation before they touch the data.  There is no classical analog to this fact, which, as we will see in the next section, is based on teleportation.

\section{Quantum fault-tolerance schemes} \label{s:introfaulttoleranceschemes}

The two currently most successful quantum fault-tolerance schemes, according to simulations, are due to Steane \cite{Steane03, Steane04computer} and Knill \cite{Knill05}.  
We here briefly explain the error-correction parts of these schemes, omitting many of the details and optimizations.  
We then detail some similarities between the two schemes, and give a few other useful ingredients for a fault-tolerance scheme.

\subsection{Steane-type error correction} \label{s:introfaulttoleranceschemessteane}

In Steane's fault-tolerance scheme, shown in Figs.~\ref{f:faulttoleranceoverview} and~\ref{f:errorcorrectionoverview}, 
application of logical gates alternates with error correction.  

Error correction is split into two parts: bit flip (or X) error correction, and phase flip (Z) error correction.  
Even though quantum errors can be continuous, it suffices to correct discrete X and Z errors. 
The justification is that the four Pauli matrices, written in the computational 0/1 basis as $I = \left(\begin{smallmatrix}1&0\\0&1\end{smallmatrix}\right)$, $X = \left(\begin{smallmatrix}0&1\\1&0\end{smallmatrix}\right)$, $Z = \left(\begin{smallmatrix}1&0\\0&-1\end{smallmatrix}\right)$ and $Y = i X Z = \left(\begin{smallmatrix}0&-i\\i&0\end{smallmatrix}\right)$, form a basis over $\mathbf{C}$ for all $2\times 2$ matrices.  Any error in the qubits' Hilbert space can therefore be expanded out as a sum of discrete errors.  \emph{Leakage} errors, leaving the Hilbert space, can be treated separately.

Quantumly, one can't just look at the data to determine where the errors are.  Measuring the data block will collapse its quantum state.  Fortunately, with a little care, error locations can be extracted fault-tolerantly and without collapsing the encoded state.  To see how, first recall the definition of the controlled-NOT (CNOT) gate; in the computational 0/1 basis, it exors the control qubit into the target:
$$
\begin{array}{c}
\Qcircuit @C=.5em @R=.85em {
\lstick{a}&\ctrl{1}&\rstick{a}\qw \\
\lstick{b}&\targ   &\rstick{a \oplus b}\qw
}
\end{array}
$$

Therefore, an X error on the control wire preceding a CNOT gate (giving $a \oplus 1$ instead of $a$) has the same effect as a CNOT gate followed by X errors on both wires:
\begin{equation} \label{e:introductioncnotforward}
\begin{array}{c}
\Qcircuit @C=.5em @R=.85em {
\lstick{X} & \ctrl{1} & \qw \\
               & \targ   & \qw 
}
\end{array}
\mspace{13mu}
=
\mspace{13mu}
\begin{array}{c}
\Qcircuit @C=.5em @R=.85em {
& \ctrl{1} & \qw & X \\
& \targ   & \qw & X
}
\end{array}
\end{equation}
In short, CNOT gates copy X errors forward.  

Also, note that a CNOT gate targeted on $\ket{+} = \tfrac{1}{\sqrt{2}}(\ket{0}+\ket{1})$ has no effect:
\beq \label{e:introductioncnotplus}
\begin{array}{c}
\Qcircuit @C=.5em @R=.85em {
\lstick{\ket{\psi}}&\ctrl{1}&\rstick{\ket{\psi}}\qw \\
\lstick{\ket{+}}&\targ   &\rstick{\ket{+}}\qw
}
\end{array}
\eeq
Indeed, if the control wire is $0$, then the gate isn't triggered.  If the control wire is $1$, the target is flipped, to $\tfrac{1}{\sqrt{2}}(\ket{1}+\ket{0})$, which is still $\ket{+}$.  By linearity, the gate has no effect regardless of the state $\ket{\psi} = \alpha \ket{0} + \beta \ket{1}$ on the control wire.  The state $\ket{+}$ is the $+1$ eigenstate of the NOT gate.

Using Eqs.~\eqref{e:introductioncnotforward} and~\eqref{e:introductioncnotplus} together, we can understand X error correction (Fig.~\ref{f:errorcorrectionoverview}).
First prepare an encoded $\ket{+}_L =  \tfrac{1}{\sqrt{2}}(\ket{0}_L+\ket{1}_L)$ state.  Then apply a logical CNOT from the data block into this ancilla.  Implemented as transversal physical CNOT gates, this will copy any X errors on the data into the ancilla, by Eq.~\eqref{e:introductioncnotforward}.  E.g., if there is originally an X error on the data block's first qubit, after the transversal CNOT gates, there will now be an X error on the first qubit in each of the data and ancilla blocks.  Now measure each qubit of the ancilla, and use the measurement results and a classical computer to diagnose the errors and determine what correction to apply to the data.  

It is safe to measure the ancilla block by Eq.~\eqref{e:introductioncnotplus}.  The transversal CNOT gates implement a logical CNOT, which has no logical effect on $\ket{+}_L$.  Therefore, no entanglement is created with the encoded data, so measuring the ancilla block does not collapse the data.  (If we had instead used an ancilla prepared as $\ket{0}_L$, then measuring the ancilla would certainly have collapsed the data.)

Phase flip Z error correction, not shown in Fig.~\ref{f:errorcorrectionoverview}, is implemented just in a dual manner, since Z errors are dual to X errors under the Hadamard transform $H$: $\ket{0} \leftrightarrow \ket{+}$, $\ket{1} \leftrightarrow \ket{-} = \tfrac{1}{\sqrt{2}}(\ket{0}-\ket{1})$.  Since 
$$
\begin{array}{c}
\Qcircuit @C=.5em @R=.85em {
\lstick{H} & \ctrl{1} & \rstick{H} \qw \\
\lstick{H} & \targ    & \rstick{H} \qw 
}
\end{array}
\mspace{26mu}
=
\mspace{13mu}
\begin{array}{c}
\Qcircuit @C=.5em @R=.85em {
& \targ     & \qw \\
& \ctrl{-1} & \qw 
}
\end{array} \enspace ,
$$
CNOT gates copy Z errors backward -- see Fig.~\ref{f:introductionerrorpropagation}.  Therefore, in Z error correction, an ancillary encoded state $\ket{0}_L$ is prepared, and CNOT gates applied from it into the data.  Then the ancilla is measured transversally in the +/- basis, and any necessary corrections determined.

\begin{figure}
\begin{center}
\includegraphics{images/propagation}
\end{center}
\caption{Propagation of X and Z Paulis through a CNOT gate; bit flips X are copied forward and phase flips Z copied backward.} \label{f:introductionerrorpropagation}
\end{figure}

\subsection{Knill-type error correction}

\subsubsection*{Error correction through teleportation}

Knill-type error correction is based on the logical operation of quantum \emph{teleportation} (Fig.~\ref{f:steaneknilltypeerrorcorrection}).  An ancillary encoded Bell pair $\tfrac{1}{\sqrt{2}}(\ket{00}_L+\ket{11}_L)$ is prepared, and a logical Bell measurement, implemented by transversal physical Bell measurements, is applied to the data block and one half of the encoded Bell pair (Fig.~\ref{f:steaneknilltypeerrorcorrection}).  

In the absence of any errors, the data block is teleported into the output block.  

In the presence of bit errors, some of the physical Bell measurements may be incorrect.  Following the error propagation rules of Fig.~\ref{f:introductionerrorpropagation}, X errors on the data will be copied forward causing mistakes in the 0/1 measurements, while data Z errors will remain in place to cause mistakes in the +/- measurements.  Bit errors cannot propagate to the output block!  However, if there are too many bit errors, the determined logical Bell measurement result will be incorrect, so a logical Pauli error on the output will be introduced.  

\begin{figure*}
\begin{center}
\includegraphics[scale=.5]{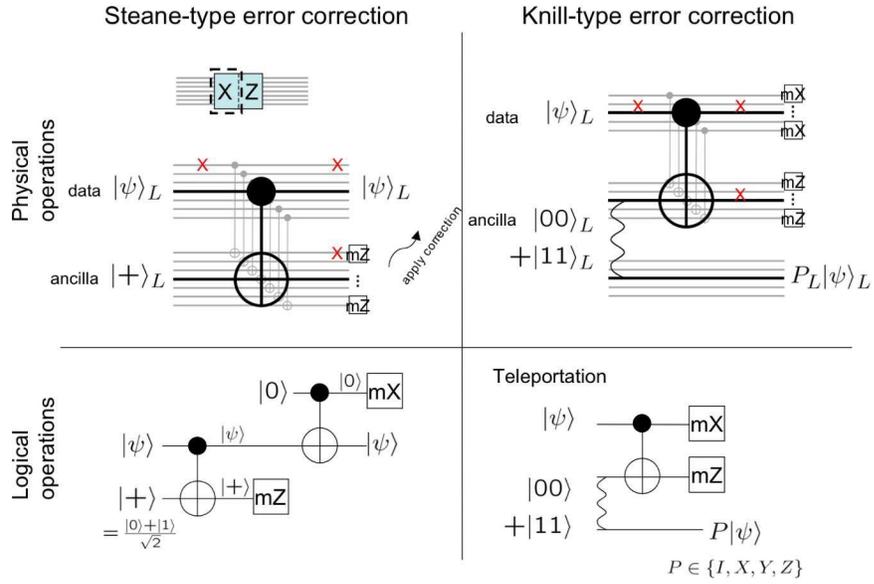}
\end{center}
\caption{Steane-type error correction is based on the logical operation of CNOTs into $\ket{+}$ and from $\ket{0}$ for X and Z error correction, respectively.  Knill's error-correction method corrects X and Z errors simultaneously, and is based on the logical operation of teleportation.  Here mZ denotes a measurement in the computational 0/1 basis (Z eigenbasis), while mX denotes a measurement in the +/- X eigenbasis.  The latter can be implemented with a Hadamard gate followed by a 0/1 measurement.}
\label{f:steaneknilltypeerrorcorrection}
\end{figure*}

By simultaneously correcting X and Z errors in a single step, error correction through teleportation may be able to tolerate higher noise rates than Steane-type error correction.  However, it also uses a larger ancilla state, requiring more overhead -- at least for a given physical error rate, if not necessarily for a given effective logical error rate.  
For threshold analyses, Knill-type error correction is convenient to work with because understanding error correlations in the system reduces to understanding correlations in the encoded Bell pair (Sec.~\ref{s:reliablebellstate}). 

\subsubsection*{Teleportation plus computation}

Teleportation-based error correction becomes more efficient when combined with computation.  

As shown in Fig.~\ref{f:teleportationcomputation}, teleportation followed by a computation $U$ on the output is equivalent to applying $U$ on the second half of the Bell pair, and then teleporting into the computation.  (Not shown is the Pauli correction $P$.  If a nontrivial $P$ is required, then after applying $U$, the required correction becomes $U P U^\dagger$.  For a large class of $U$ -- stabilizer operations -- the conjugated operator remains some Pauli [Sec.~\ref{s:gottesmanknilltheorem}].  Also, although a single-qubit unitary is shown here, one can also teleport into a multi-qubit unitary -- see Eq.~\ref{e:cnotreducestobellpairs} of Sec.~\ref{s:cnotteleport}.)

\begin{figure*}
\begin{center}
\includegraphics[scale=.5]{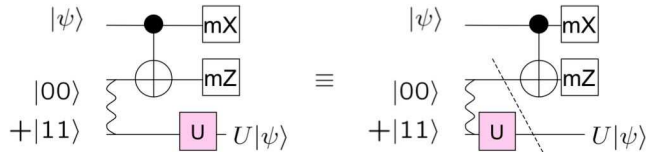}
\end{center}
\caption{Combining computation with teleportation.}
\label{f:teleportationcomputation}
\end{figure*}

In a fault-tolerance scheme, we do this all at an encoded level, using an ancilla state of $(I_L \otimes U_L) \tfrac{1}{\sqrt{2}} (\ket{00}_L + \ket{11}_L)$.  The advantage over splitting computation and error correction into two steps, as in Fig.~\ref{f:faulttoleranceoverview}, is that the ancilla here can be carefully \emph{tested} before it is allowed to touch the data.  If it fails the test(s), we can simply throw it away and prepare another copy.  This should allow the quantum computer to tolerate higher error rates, but of course testing the ancilla carefully trades off the overhead.  

\subsection{Commonalities of the Steane- and Knill-type fault-tolerance schemes}

\subsubsection{Ancilla preparation} \label{s:introductionancillapreparation}
In both schemes, it is important to be able to prepare reliable encoded ancilla states.  This is particularly true in Knill's scheme, in which even computation is done through ancilla preparation.  Only during ancilla preparation do qubits within the same code block interact with each other, possibly multiplying errors.  Additionally, encoding an $n$-qubit ancilla typically requires $\sim n^2$ gates.  High-fidelity encoded ancillas do not suffice.  We need a stronger requirement, that the errors not be correlated.  So for example, a weight-two error needs to occur with second-order probability in a probabilistic error model (or second-order weight in a coherent error model).  Careful fault-tolerant procedures, which will be described in Sec.~\ref{s:encodingverification}, must be used.  Therefore, encoded ancilla preparation is a bottleneck in the threshold tolerable noise rate.  

\subsubsection{Virtual corrections} \label{s:virtualcorrections}
In our description of Steane's scheme, we said that Pauli error corrections are applied to the data; and in Knill's scheme a logical Pauli correction $P_L$, or $(U P U^\dagger)_L$, is applied to the output.  However, any Pauli corrections do not actually have to be applied to the data.  Instead, a classical computer can track the corrections through the computation by linearity (the Gottesman-Knill theorem -- Sec.~\ref{s:gottesmanknilltheorem}), and only after a qubit is finally measured is the correction applied to the measurement outcome.  Tracking the ``Pauli frame" in this way has two advantages.  It removes one possible point of failure in the computation (since applying a real correction could itself introduce errors).  It also can reduce delays in the quantum computer.  For example, if measurements are much slower than other operations, the rest of the computation can continue without waiting for the measurements to complete, and any required corrections can later be propagated forward classically.  

The exception is when the correction $U P U^\dagger$ is not a Pauli operator, so it cannot be tracked classically (see Sec.~\ref{s:intromagic} below).  Then the computation must be delayed for measurements to finish so the correction can be determined.  However, for all the gates used in encoding ancillas (``stabilizer operations"), the correction is a Pauli.  For a fault-tolerance scheme based on ancilla preparation, the majority of gates in the computer will be stabilizer operations.  (Necessarily some gates $U$ will not be stabilizer operations, but these can be chosen so the correction $U P U^\dagger$ is a stabilizer operation.)

\subsubsection{Assistance of classical computers}
Both schemes take advantage of classical computers to interpret measured syndromes.  (For a constant-size code, locating errors based on measurement results is an efficient classical computation.)  Fault tolerance is possible without measurements, but both the overhead and the tolerable noise rate are then significantly worse.

\subsubsection{Detected/erasure errors} \label{s:detectederasureerrors}
It is worth remarking that detected errors, also known as erasures, are much easier to correct than undetected errors.  A distance-$d$ error-correcting code can correct up to $d-1$ erasures, but only $(d-1)/2$ undetected errors.  (In particular, a distance-2 code can only correct detected errors.)  Both schemes will therefore benefit from any physical techniques allowing one to judge the reliability of a qubit.  

\subsection{Other fault-tolerance scheme ingredients}

\subsubsection{Postselection}

In both Steane- and Knill-type fault-tolerance, most of the work is in preparing reliable encoded ancilla states.  Assume for a moment an erasure error model, that all errors are detected when they occur.  Then if an erasure is detected during ancilla preparation, the whole ancilla can be thrown away (since it has not yet interacted with any data), and a new one prepared.  Then we can assume that the states $\ket{0}_L$, $\ket{+}_L$ and $(I_L \otimes U_L) \tfrac{1}{\sqrt{2}} (\ket{00}_L + \ket{11}_L)$ are in fact perfect when they are used.  

In more interesting error models, errors are not necessarily detected, but a similar strategy can be applied.  Work to try to detect any errors.  For example, one might use a small error-detecting code and prepare the ancilla on top of this code (two levels of encoding total).  During preparation, continually check the lower code for any errors, and if any are found then start over.  Finally, decode out the lower error-detecting code.  Another technique is to prepare two copies of the ancilla state and then check one against the other, looking for both bit errors and logical errors -- this \emph{purification} method will be discussed and optimized in Sec.~\ref{s:encodingverification}.

Very high noise rates can be tolerated in this manner, but the overhead from failed preparations can of course be substantial.  The challenge in using postselection is in deploying it in an efficient and limited manner, to find a compromise between tolerable noise and overhead.  Also, in physical models with locality constraints -- only nearby qubits allowed to interact -- then postselection might become particularly difficult to implement.  Even if there is a ``state factory" carrying out many ancilla preparations in parallel so one is always ready when needed, there is a challenge in quickly getting the ancilla to \emph{where} it is needed in the quantum computer.  

Once again, extensive use of postselection is safe because most of the work in fault-tolerant quantum computation is in preparing reliable ancilla states.  It is safe to throw away ancilla states, before the data is exposed to errors.  

\subsubsection{Universality via magic states distillation} \label{s:intromagic}

The operations required for error correction, including encoded ancilla preparation, are known as stabilizer operations (consisting of preparation of fresh qubits as $\ket{0}$, measurement in the computational basis, and application of Clifford group unitaries like the CNOT gate -- see Sec.~\ref{s:gottesmanknilltheorem}).  Stabilizer operations are easy to implement fault-tolerantly, and are particularly easy to analyze because bit flips and phase flips -- Pauli errors -- propagate through linearly.  

However, a circuit consisting only of stabilizer operations can be efficiently classically simulated.  (For this reason, probabilistic Pauli noise models for stabilizer operations can be efficiently simulated to obtain threshold estimates.)  Stabilizer operations do not form a universal gate set; something more is needed.  For example, it suffices to add a Toffoli gate~\cite{Aharonov03universal}:
$$
\begin{array}{c}
\Qcircuit @C=.5em @R=.85em @!R {
\lstick{a}&\ctrl{1}&\rstick{a}\qw \\
\lstick{b}&\ctrl{1}&\rstick{b}\qw \\
\lstick{c}&\targ   &\rstick{c \oplus (a \cdot b)}\qw
}
\end{array}
$$

It turns out that in most constructions, achieving fault-tolerant universality is not much harder than achieving fault-tolerant stabilizer operations.  The bottleneck for the noise threshold typically comes from getting reliable stabilizer operations.\footnote{This is not the case in some constructions; for example, in \Ref\cite{RaussendorfHarringtonGoyal05oneway}, the estimated noise threshold for stabilizer operations is higher than that for universality.  Even here, one suspects that the threshold for universality can be increased to match that for stabilizer operations, possibly at some cost to overhead.}  For some constructions, this is simply because the implementation of the logical CNOT gate is the most complicated fault-tolerant operation.  But there is a rough reduction, known as magic states distillation, which implies that this should always be the case.  

Adaptive stabilizer operations together with the ability to prepare the ancilla state
$$
\tfrac{1}{2}(\ket{000}+\ket{010}+\ket{100}+\ket{111})
$$
(a Toffoli applied to $\ket{\!+\!+0}$) allow for implementing a Toffoli, and hence give universality~\cite{Shor96}.  
Like the Toffoli ancilla, 
certain other ``magic" ancilla states similarly allow for universality with stabilizer operations.  Some give universality even if they can only be prepared with relatively high noise rates; stabilizer operations can be used to \emph{distill} a noiseless ancilla from multiple noisy copies.  Magic states distillation is a method for obtaining universality, given reliable encoding, stabilizer operations and decoding, together with unreliable (noisy) preparation of certain ancilla states.  Below a noise threshold, (encoded) stabilizer operations can be assumed to be perfect, so the threshold for universal computation is (roughly) the smaller of the stabilizer operation threshold and the state distillation noise threshold.

Magic states distillation also lets us skip the fault-tolerance hierarchy for universal quantum computing operations, considerably
simplifying proofs.  
We explain the technique in more detail in Ch.~\ref{s:magicchapter}.

\subsubsection{Fault-tolerance architecture} \label{s:architectures}

Locality constraints in a physical quantum computer mean that fully specifying a fault-tolerance scheme also requires laying out the qubits and specifying how information moves through the computer (e.g., using swap gates, or by moving the qubits themselves).  Using an entanglement purification technique, it suffices to build a number of small but very reliable systems \cite{DurBriegel03purification}.  Still, though, even using fault-tolerance to get say five very reliable logical qubits requires a large apparatus of physical qubits.  Fortunately, the concatenated structure of standard fault-tolerance schemes gives a structure to the gates -- most gates are used for error correction, and at low levels of encoding.  Qubits need to be arranged in order to take advantage of this structure.  Szkopek et al.~have studied, through simulations, the effect of a one-dimensional geometry constraint on the noise threshold \cite{SzkopekFanRoychowdhuryYablonovitchBoykinSimmsGyureFong04local}.  Steane \cite{Steane02architecture} and Svore, Terhal and DiVincenzo \cite{SvoreTerhalDiVincenzo04} have considered two-dimensional layouts.  Detailed simulations of two-dimensional ion trap architectures have been run by Metodiev et al.~\cite{MetodievCrossThakerBrownCopseyChongChuang04, CopseyOskinImpensMetodievCrossChongChuangKubiatowicz03}, and by Cross and Chuang \cite{CrossChuang06}, and they have observed that ancilla preparation costs more than expected in a local model.  More efficient ancilla preparation and verification techniques are probably required for practical quantum computing with locality constraints.  

\section{Complementary and alternative approaches to fault tolerance}

\subsection{Low-level error-avoidance and error-correction techniques} \label{s:lowlevelft}

The high overhead required for general fault-tolerance schemes, particularly at high error rates, will require the use of more efficient, specialized low-level techniques.  Examples include decoherence-free subspaces \cite{KempeBaconLidarWhaley01}, dynamical decoupling and concatenated dynamical decoupling \cite{RottelerWocjan04,KhodjastehLidar04}, and composite pulse sequences \cite{BrownHarrowChuang04,ReichardtGrover05} and other NMR techniques \cite{VandersypenChuang04nmr}.  These specialized schemes each take advantage of some known structure to the errors in the system.  But the exploitable structure only goes so far, so quantum computers for large calculations (hundreds to thousands of qubits) will still require high-level fault tolerance.  

In the intermediate range, it is not clear what fault-tolerance techniques can most efficiently be brought to bear, and this will depend on the directions taken by experiments.  One  suggestion is to use continuous-time feedback control for error correction \cite{SarovarAhnJacobsMilburn04continuous}.

\subsection{Topological quantum computing}

It is possible that, like the transistor for classical computing, an inherently reliable quantum-computing device will be engineered or discovered -- and this is in fact the idea behind topological quantum computing \cite{Kitaev97topological, FreedmanKitaevLarsenWang02topological, DennisKitaevLandahlPreskill01topological, Bacon05operator}.  
We are now only at a very early stage in building quantum computers -- with eight ions trapped and manipulated to create entanglement by H{\"a}ffner et al.~\cite{Blatt05iontrap}.  Controllable qubits are scarce, and it seems that the vast majority of qubits in a general-purpose quantum computer will have to be dedicated to achieving fault tolerance.  
Perhaps physical processes which strongly suppress errors in quantum systems, but still allow for control, exist in nature -- by changing the model, they might spare us the daunting task of building a full fault-tolerant architecture.

\section{Open problems}

Dealing with noise may turn out to be the most daunting task in building a quantum computer.  At the moment, physicists' low-end estimates of achievable noise rates are only slightly below theorists' high-end (simulation-based) estimates of tolerable noise rates, at reasonable levels of overhead.  But this is with different noise models -- most simulations are based on a simple independent depolarizing noise model, and threshold estimates for more general noise are much lower.  And both communities may be being too optimistic.  Unanticipated noise sources may well appear as experiments progress.  The probabilistic noise models used by theorists in simulations may not match reality closely enough, the overhead/threshold tradeoff may be impractical, or locality constraints may harm the threshold more than now thought.
It is not clear if fault-tolerant quantum computing will work in practice, unless inefficiencies are wrung out of the system.  Developing more efficient fault-tolerance techniques is a major open problem.  

The gaps between threshold upper bounds, threshold estimates and rigorously proven threshold lower bounds are closing at least for simple noise models, like independent depolarizing noise.   Our understanding of what to expect with more general noise models is less developed, though.  Rigorous threshold lower bounds in the more general noise models may still be far too conservative (according to arguments, for the most part only intuitive, known as ``twirling" \cite{BennettDivincenzoSmolinWootters96, KernAlberShepelyansky04randomize, AschaurerDurBriegel05purify}), and new ideas are needed for more efficient analyses.

Open problems specific to this thesis are discussed in Sec.~\ref{s:postselectopenproblems}.  Among those problems, dealing with locality constraints, and giving better explicit threshold lower bounds are certainly tractable.  Significant work or even new analysis techniques may be required to obtain the most efficient threshold lower bounds.  It is less clear how even to begin to extend the approach to more general noise models, and doing so is a major challenge.

\chapter{Fault-tolerant constructions} \label{s:ftconstructionschapter}
An operation is said to be fault tolerant if it does not cause correlated errors within a code block.  For example, decoding a code, applying a unitary to the unencoded qubit, then reencoding, is certainly not fault tolerant.  As a more subtle example, when qubits are arranged on a line with constrained interactions -- say, only nearest-neighbor interactions -- then swapping qubits is useful to allow communication.  However, directly swapping two adjacent qubits in a single code block is not fault tolerant, since a single failure could then cause two errors in the block~\cite{Gottesman00local}.  

Applying a one-qubit gate transversally in a code block -- i.e., applying the same gate to each qubit -- is always fault tolerant.  (However, this will only be a valid operation for certain gates, depending on the code.)  Transversal gates between code blocks -- meaning the same two-qubit gate is applied between the first qubits of each block, the second qubits, etc. -- are also always fault tolerant.


The circuit in Fig.~\ref{f:steanenotft} extracts the parity of the last four qubits of a seven-qubit code block (the IIIZZZZ stabilizer's syndrome for the Steane code, which we will describe in Sec.~\ref{s:sevenqubitsteanecode}).  However, this operation is not fault tolerant, because a single Z error on the ancilla qubit can be copied backward into multiple locations on the code block.  (Following the rules of Fig.~\ref{f:introductionerrorpropagation}, a Z error on the ancilla after the second CNOT will be copied back onto the last two qubits of the code.)  This could possibly cause a logical error with first-order probability, whereas for the existence of a noise threshold we need logical errors to be quadratically suppressed.  Figure~\ref{f:steaneft} sketches a \emph{fault-tolerant} circuit for computing the same parity check, presuming the initial GHZ/cat state is prepared fault tolerantly.  The basic idea is to avoid reusing qubits, so that errors never have the chance to spread.


\begin{figure}
\begin{center}
\includegraphics{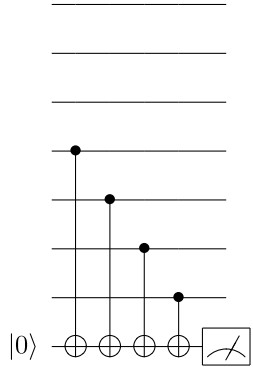}
\end{center}
\caption{A circuit for extracting the last four qubits' parity which is not fault tolerant, because a single Z error on the ancilla can cause multiple Z errors in the code block.}
\label{f:steanenotft}
\end{figure}

\begin{figure}
\begin{center}
\includegraphics{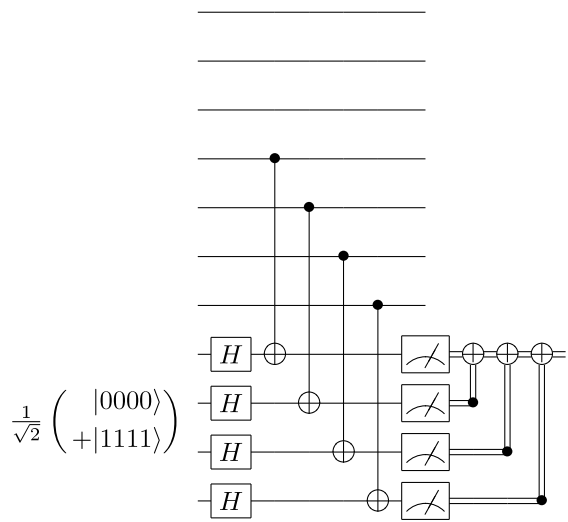}
\end{center}
\caption{A fault-tolerant circuit for extracting the last four qubits' parity -- assuming the initial GHZ/cat state is prepared fault tolerantly.}
\label{f:steaneft}
\end{figure}

This is not a hard-and-fast rule, though.  Certain codes, including the four-qubit code we will use for numerical threshold calculations in Ch.~\ref{s:numericalchapter}, may allow ancilla qubits to be efficiently reused, while still maintaining fault tolerance.  Figure~\ref{f:fourqubitcodeft} shows how to extract both ZZZZ and XXXX syndromes of the four-qubit code, 
using just \emph{one} ancilla qubit for each syndrome.  See Sec.~\ref{s:fourqubitcode} for more details.

\begin{figure}
\begin{center}
\includegraphics{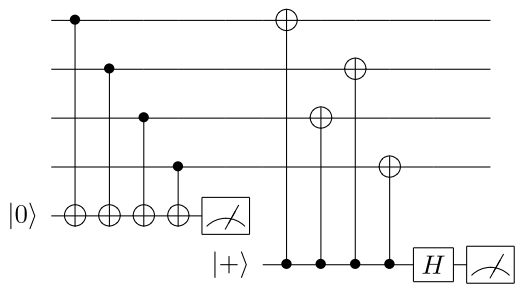}
\end{center}
\caption{A circuit for extracting the ZZZZ and XXXX syndromes for the four-qubit code which protects one of the two encoded qubits.}
\label{f:fourqubitcodeft}
\end{figure}

In this chapter, we will describe some of the most useful fault-tolerant constructions.\footnote{Fault-tolerant constructions for general stabilizer codes are given in \Ref\cite{Gottesman97}.}  
After a brief explanation of the Gottesman-Knill theorem and the stabilizer algebra formalism, we present Shor's original method for fault-tolerant error correction.  Two other fault-tolerance schemes, Steane-type and Knill-type, were described in Sec.~\ref{s:introfaulttoleranceschemes}.  We also present some of the most popular codes used in quantum fault tolerance, in Sec.~\ref{s:quantumerrorcorrectingcodes}.  

Shor-, Steane- and Knill-type fault-tolerance schemes all rely on the ability to prepare fault tolerantly certain multi-qubit ancilla states (like the cat state of Fig.~\ref{f:steaneft}).  For the two- and four-qubit codes used in the remainder of this thesis, fault-tolerant ancilla preparation is easy.  But for the sake of completeness, we finally present na{\"i}ve and newly optimized fault tolerant preparation procedures, useful for computing with larger codes (Sec.~\ref{s:encodingverification}).



\section{Stabilizer algebra} \label{s:gottesmanknilltheorem}

We assume basic familiarity with quantum computing, as can be found for example in the textbooks of Nielsen and Chuang~\cite{NielsenChuang00} or Kitaev, Shen and Vyalyi~\cite{KitaevShenVyalyi02textbook}.  

Stabilizer algebra is an extremely useful tool for understanding fault-tolerant constructions, so 
we will give a hands-on account of it.
(For an exposition with all the formal details, also oriented around fault tolerance, we recommend \Ref\cite{Gottesman97}.)

\begin{definition}{\ } \label{t:faulttolerantpaulistabilizerdefinition}
\begin{itemize}
\item The one-qubit Pauli operators are the identity $I$, $X = \left(\begin{smallmatrix}0&1\\1&0\end{smallmatrix}\right)$ in the computational $\ket{0}$, $\ket{1}$ basis, $Z = \left(\begin{smallmatrix}1&0\\ 0&-1\end{smallmatrix}\right)$ and $Y = i X Z = \left(\begin{smallmatrix}0&-i\\i&0\end{smallmatrix}\right)$.  The Pauli group consists of tensor products of one-qubit Pauli operators with phase $\pm 1$ or $\pm i$.  

\item The Clifford group is the set of unitaries which conjugate Paulis to Paulis.  Clifford group unitaries are generated by the Hadamard gate $H = \tfrac{1}{\sqrt{2}}\left(\begin{smallmatrix}1&1\\1&-1\end{smallmatrix}\right)$, the phase gate $Z^{1/2} = \left(\begin{smallmatrix}1&0\\0&i\end{smallmatrix}\right)$ and the controlled-NOT gate, CNOT $\ket{a} \ket{b} = \ket{a} \ket{a + b \mod 2}$ for $a, b \in \{0,1\}$.  

\item Stabilizer operations consist of Clifford group unitaries, preparation of $\ket{0}$ and measurement in the computational $\ket{0}, \ket{1}$ basis.  
\end{itemize}
\end{definition}

A few basic facts: Note that every nontrivial Pauli operator squares to $\pm I$, and has half its eigenvalues $+1$, the other half $-1$ (or half $+i$, half $-i$).  Any two Pauli operators either commute or anticommute; for example, $X\otimes Z\otimes X\otimes Y$ -- which we'll write just XZXY for short -- commutes with YXZZ and anticommutes with YZYI.  This follows because in the former case, the number of locations where the two operators differ nontrivially is even (four), and the latter case it is odd (three).  

The Gottesman-Knill theorem says that an inputless circuit with only stabilizer operations and adaptive classical control is efficiently classically simulatable \cite{AaronsonGottesman04}.  Since the Clifford unitaries are essentially the quantum analog of classical linear gates, the Gottesman-Knill theorem is essentially the quantum analog of linearity; see Fig.~\ref{f:classicalgottesmanknill}.  The simulation procedure is the stabilizer algebra technique.

\begin{figure*}
\begin{center}
\includegraphics[scale=.45]{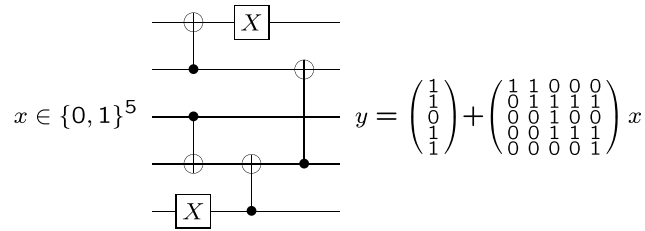}
\end{center}
\caption{
The Gottesman-Knill theorem is the quantum analog of an obvious classical fact.  A classical linear circuit can use CNOT gates and X operators (NOT gates).  The output is always an affine function of the input and the precise equation can be determined by first determining the circuit's functionality on $0^n$ the all-zeros input, then checking the effects of flipping each of the input bits.  In the quantum case, we also have to consider phase (Z) information.  X and Z flips propagate through CNOT gates following the rules of Fig.~\ref{f:propagation}.
}
\label{f:classicalgottesmanknill}
\end{figure*}

We say the operator $S$ \emph{stabilizes} the state $\ket{\psi}$ if $S \ket{\psi} = \ket{\psi}$.  The set of Pauli stabilizers for a state forms a group, since $S$ and $T$ stabilizing $\ket{\psi}$ implies that $S \cdot T$ does as well.  After applying a unitary $U$ to $\ket{\psi}$, the set of stabilizers is updated by conjugation, since 
\begin{equation*}
(U S U^\dagger) (U \ket{\psi}) = U S \ket{\psi} = U \ket{\psi} \enspace .
\end{equation*}
Figure~\ref{f:propagation} shows how the CNOT gate conjugates Pauli operators (the rules for Y follow since $Y  = i X Z$).  Another common Clifford operator is the Hadamard gate $H = H^\dagger$, which conjugates X to Z -- $H X H = Z$ -- and vice versa.  

\begin{figure}
\begin{center}
\includegraphics{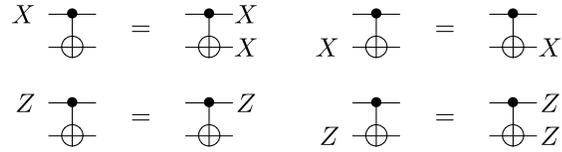}
\end{center}
\caption{Propagation of X and Z 
Paulis through a CNOT gate; bit flips X are copied forward and phase flips Z copied backward.} \label{f:propagation}
\end{figure}

At its simplest, stabilizer algebra is the technique of commuting Pauli operators past Clifford gates.  For example, an X operator on the last qubit preceding the circuit of Fig.~\ref{f:classicalgottesmanknill}, is the same as that circuit followed by IXIXX (corresponding to the last column of the matrix in Fig.~\ref{f:classicalgottesmanknill}), as one can see by repeatedly applying the conjugation rules of Fig.~\ref{f:propagation}:
\begin{equation*}
\mathrm{IIIIX} \overset{\substack{\text{X}_{5}}}{\longrightarrow} \mathrm{IIIIX} \overset{\substack{\text{CNOT}_{5,4}}}{\longrightarrow} \mathrm{IIIXX} \overset{\substack{\text{CNOT}_{4,2}}}{\longrightarrow} \mathrm{IXIXX} \enspace .
\end{equation*}
Similarly, preceding that circuit with Z operators on the first two qubits is equivalent to applying the circuit followed by $-\text{Z}$.  (The CNOT copies the first Z back, canceling the second Z.  The minus sign comes from pulling the Z past the X gate.)

Below, we will develop the stabilizer algebra formalism incrementally, using a series of more substantial examples.  We start by describing the useful class of ``stabilizer states," states fixed by their Pauli stabilizers.  We then use quantum teleportation to describe how to incorporate measurements into stabilizer algebra, and generalize teleportation to the ``one-way" quantum computation model.  Finally, we introduce ``stabilizer codes," which are \emph{subspaces} fixed by Pauli stabilizers -- the remaining degrees of freedom are used for encoding qubits.  In the next section, we will present some of the most popular stabilizer codes used for fault tolerance.

\subsection{Example 1: Stabilizer states}

An $n$-qubit \emph{stabilizer state} is a pure state stabilized by $n$ independent Pauli operators.  An equivalent operational definition is that stabilizer states are exactly the pure states which can be created using only the stabilizer operations of Def.~\ref{t:faulttolerantpaulistabilizerdefinition}.
(In fact, measurements are never needed, and so nor is adaptive control of the circuit using a classical computer.)  

For example, consider the quantum circuit:
\begin{equation} \label{e:cnotteleportancilla1}
\begin{array}{c} 
\includegraphics{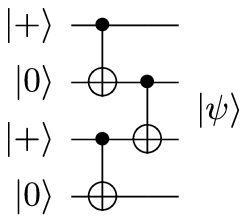}
\end{array}
\end{equation}
Here, $\ket{+} \equiv \tfrac{1}{\sqrt{2}}(\ket{0}+\ket{1})$, the $+1$ eigenstate of X.  ($\ket{-} \equiv \tfrac{1}{\sqrt{2}}(\ket{0}-\ket{1})$.)
One can compute that the output state is 
$\ket{\psi} = \text{CNOT}_{2,3}\left(\tfrac{1}{2}(\ket{00}+\ket{11})\otimes(\ket{00}+\ket{11})\right) = 
\tfrac{1}{2}(\ket{0000} + \ket{0011} + \ket{1110} + \ket{1101})$.  

In terms of stabilizer algebra, the initial state is stabilized by $X_1$, $Z_2$, $X_3$ and $Z_4$, as well as all products of these operators.  Using the conjugating rules of Fig.~\ref{f:propagation}, 
\begin{equation} \label{e:stabilizerstateex1}
\begin{array}{c@{\!}c@{\!}c@{\!}c}
X&I&I&I\\
I&Z&I&I\\
I&I&X&I\\
I&I&I&Z
\end{array}
\overset{\substack{\text{CNOT}_{1,2}\\ \text{CNOT}_{3,4}}}{\longrightarrow}
\begin{array}{c@{\!}c@{\!}c@{\!}c}
X&X&I&I\\
Z&Z&I&I\\
I&I&X&X\\
I&I&Z&Z
\end{array}
\overset{\substack{\text{CNOT}_{2,3}}}{\longrightarrow}
\begin{array}{c@{\!}c@{\!}c@{\!}c}
X&X&X&I\\
Z&Z&I&I\\
I&I&X&X\\
I&Z&Z&Z
\end{array} \enspace ,
\end{equation}
and indeed it is easy to verify that $\ket{\psi}$ has these stabilizers.  

As another example, consider the circuit 
\begin{equation} \label{e:cnotteleportancilla2}
\begin{array}{c} 
\includegraphics{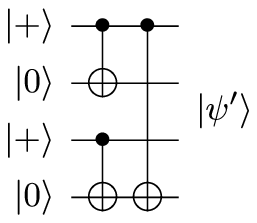}
\end{array}
\end{equation}
The stabilizers of the output state $\ket{\psi'}$ are 
$$
\begin{array}{c@{\!}c@{\!}c@{\!}c}
X&X&I&I\\
Z&Z&I&I\\
I&I&X&X\\
I&I&Z&Z
\end{array}
\overset{\substack{\text{CNOT}_{1,4}}}{\longrightarrow}
\begin{array}{c@{\!}c@{\!}c@{\!}c}
X&X&I&X\\
Z&Z&I&I\\
I&I&X&X\\
Z&I&Z&Z
\end{array} \enspace .
$$
Though these stabilizers are different, the group they generate is the same as the stabilizers in Eq.~\ref{e:stabilizerstateex1}; multiplying the first by the third gives XXXI, and multiplying the fourth by the second gives IZZZ.  Since $\ket{\psi}$ and $\ket{\psi'}$ have the same stabilizer group, $\ket{\psi} = \ket{\psi'}$; the two circuits are equivalent.

This equivalence will slightly simplify the analysis of encoded CNOT gates in Ch.~\ref{s:postselectchapter}.  In our numerical noise threshold analysis (Ch.~\ref{s:numericalchapter}), however, it will allow for substantially faster threshold calculations.  (It allows the mixing technique to be applied to just two encoded qubits at a time, instead of four encoded qubits -- or eight versus sixteen physical qubits.)

\subsection{Example 2: Teleportation}

Now let's throw in a logical degree of freedom, and measurements.  Consider the teleportation circuit \cite{BennettBrassardCrepeauJozsaPeresWooters93teleport}:
\begin{equation} \label{e:teleportationcircuit}
\begin{array}{c} 
\includegraphics{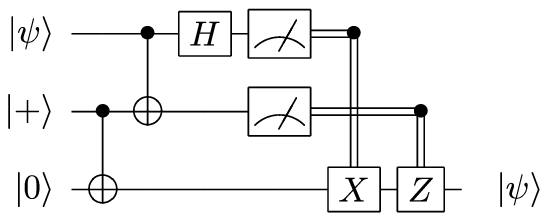}
\end{array}
\end{equation}
(The doubled wires represent classical bits, the 0/1 measurement outcomes used to control the final X and Z corrections.)

Even though $\ket{\psi}$ is not necessarily a stabilizer state (and in fact need not be pure, and may be entangled with the environment), the circuit's behavior can be derived using stabilizer algebra.  The degrees of freedom in $\ket{\psi}$ are tracked, as in the circuit of Fig.~\ref{f:classicalgottesmanknill}, by propagating both X and Z operators from the free qubit.  We represent the initial state as
$$
\begin{array}{c c@{\!}c@{\!}c}
&I&I&Z\\
&I&X&I\\
\cline{2-4}
X_{L} = &X&I&I\\
Z_{L} = &Z&I&I
\end{array} \enspace ,
$$
by convention writing the stabilizers \emph{above} the dividing line and the degrees of freedom, which we call $X_L$ and $Z_L$, below.

The stabilizers and degrees of freedom can be tracked with the Pauli commutation rules:
$$
\begin{array}{c c@{\!}c@{\!}c}
&I&I&Z\\
&I&X&I\\
\cline{2-4}
X_{L} = &X&I&I\\
Z_{L} = &Z&I&I
\end{array}
\overset{\substack{\text{CNOT}_{2,3}}}{\longrightarrow}
\begin{array}{c@{\!}c@{\!}c}
I&Z&Z\\
I&X&X\\
\cline{1-3}
X&I&I\\
Z&I&I
\end{array}
\overset{\substack{\text{CNOT}_{1,2}}}{\longrightarrow}
\begin{array}{c@{\!}c@{\!}c}
I&Z&Z\\
I&X&X\\
\hline
X&X&I\\
Z&I&I
\end{array} \enspace .
$$
Since $X_L$ has an X on the second qubit, it might not be safe to measure that qubit in the computational basis (Z eigenbasis), without collapsing the degree of freedom.  
However, by multiplying $X_L$, $Z_L$ by stabilizers, this stabilizer system is equivalently represented by
$$
\begin{array}{c c@{\!}c@{\!}c}
&Z&Z&Z\\
&I&X&X\\
\cline{2-4}
X_{L} = &X&I&X\\
Z_{L} = &I&Z&Z
\end{array} \enspace .
$$
Now it clearly safe to measure the first qubit in the X eigenbasis, and the second qubit in the Z eigenbasis, without collapsing the degrees of freedom (since they commute).  The measurement outcomes are random (since neither is in the stabilizer).  After measurement, the system becomes 
$$
\begin{array}{c}
\begin{array}{c c@{\!}c@{\!}c@{\!}c}
&\pm_1&X&I&I\\
&\pm_2&I&Z&I\\
\cline{3-5}
X_{L} = &&X&I&X\\
Z_{L} = &&I&Z&Z
\end{array}
\end{array}
\sim
\begin{array}{c}
\begin{array}{c@{\!}c@{\!}c@{\!}c}
\pm_1&X&I&I\\
\pm_2&I&Z&I\\
\cline{2-4}
\pm_1&I&I&X\\
\pm_2&I&I&Z
\end{array}
\end{array} \enspace ,
$$
where the signs $\pm_1$ and $\pm_2$ are the random measurement results; and where we have multiplied $X_L$ and $Z_L$ by the stabilizers to obtain the second representation, in which $X_L$ and $Z_L$ are supported on just the third qubit.
The corrections in Eq.~\eqref{e:teleportationcircuit} are in order to eliminate the random signs $\pm_1$ and $\pm_2$.

One can also teleport into a Clifford computation~\cite{GottesmanChuang99teleportation,Leung02teleport}.  For example, the commutation rules of Fig.~\ref{f:propagation} imply 
\begin{center}
\includegraphics{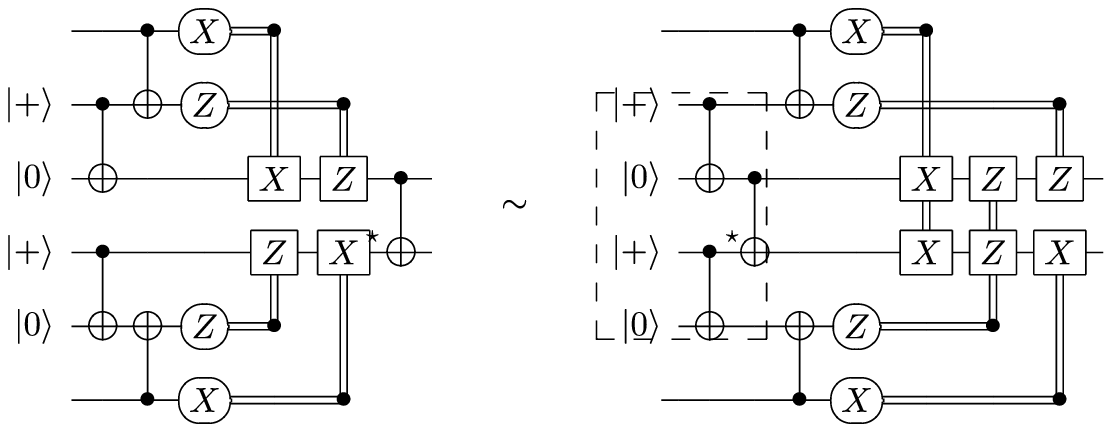}
\end{center}
by commuting the CNOT marked $\star$ past the corrections ($\sim$ here means that the circuits have the same effect on all inputs).  The boxed part of the right circuit is the same as in Eq.~\eqref{e:cnotteleportancilla1} and can be replaced with the circuit from Eq.~\eqref{e:cnotteleportancilla2}.


\subsection{Example 3: One-way/cluster computation}

Consider the circuit of Fig.~\ref{f:compact} (read from top to bottom), where bits 4 and 6 have been prepared in the $+1$ eigenstate of Z and X, respectively.  Then measure bits 1 and 4 in the X and Z bases, respectively; and apply Pauli corrections (not shown) to the other bits according to the measurement results.
Overall, 
$\Lambda_1(X_7 X_8) \Lambda_5(X_1) \Lambda_1(X_2 X_3)$
is applied (where $\Lambda_i(U)$ denotes application of $U$ controlled by qubit $i$ -- $\Lambda(X)$ is a CNOT), and the first bit moves to the sixth position.   

\begin{figure}
\begin{center}
\includegraphics{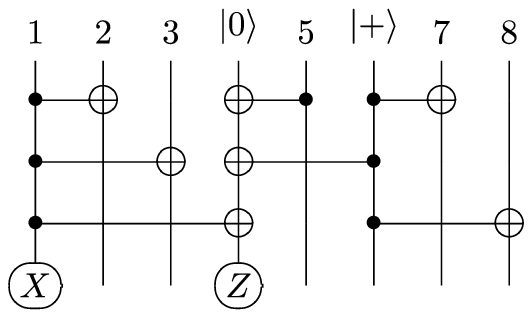}
\end{center}
\caption{}\label{f:compact}
\end{figure}

In five time steps (for three gate applications, a measurement, and the corrections), we have applied five CNOT gates.  The same technique, though, suffices to apply an arbitrary number of CNOT gates in just five time steps, at the cost of using additional qubits.  Figure~\ref{f:cluster} shows the graph on which we based this circuit -- think of there being a qubit for each node, with wires going into the page.  We can add as many other nodes as we like.  The time required to implement the circuit is the maximum degree of a node, which can be kept at three by extending the chain to the right, plus the time for measurements and the corrections.  
For example, $m$ CNOT gates, half controlled by and half targeting some single qubit, can be applied in five time steps using $m-3$ additional qubits.  In general, arbitrary stabilizer operations can be applied in five time steps (the classical processing to determine the corrections, though, is not constant depth).  
This quantum phenomenon is a special case of the ``one-way quantum computer" introduced by Raussendorf and Briegel \cite{RaussendorfBriegel01}; the initial circuit establishes a ``cluster state," into which we can teleport the computation.\footnote{The one-way quantum computer model has interesting properties, but is encompassed by the quantum circuit model.  Fault-tolerance schemes and results for one-way computing, like \Ref\cite{RaussendorfHarringtonGoyal05oneway}, therefore carry over into the quantum circuit model.}  The cluster state for a graph $G$ is a stabilizer state with a stabilizer generator for each vertex $v$ (having neighbors $N(v)$), either $X_v \otimes \left(\bigotimes_{w \in N(v)} X_w\right)$ or $Z_v \otimes \left(\bigotimes_{w \in N(v)} Z_w\right)$.


\begin{figure}
\begin{center}
\includegraphics*[bb=0 0 509 449,scale=.2]{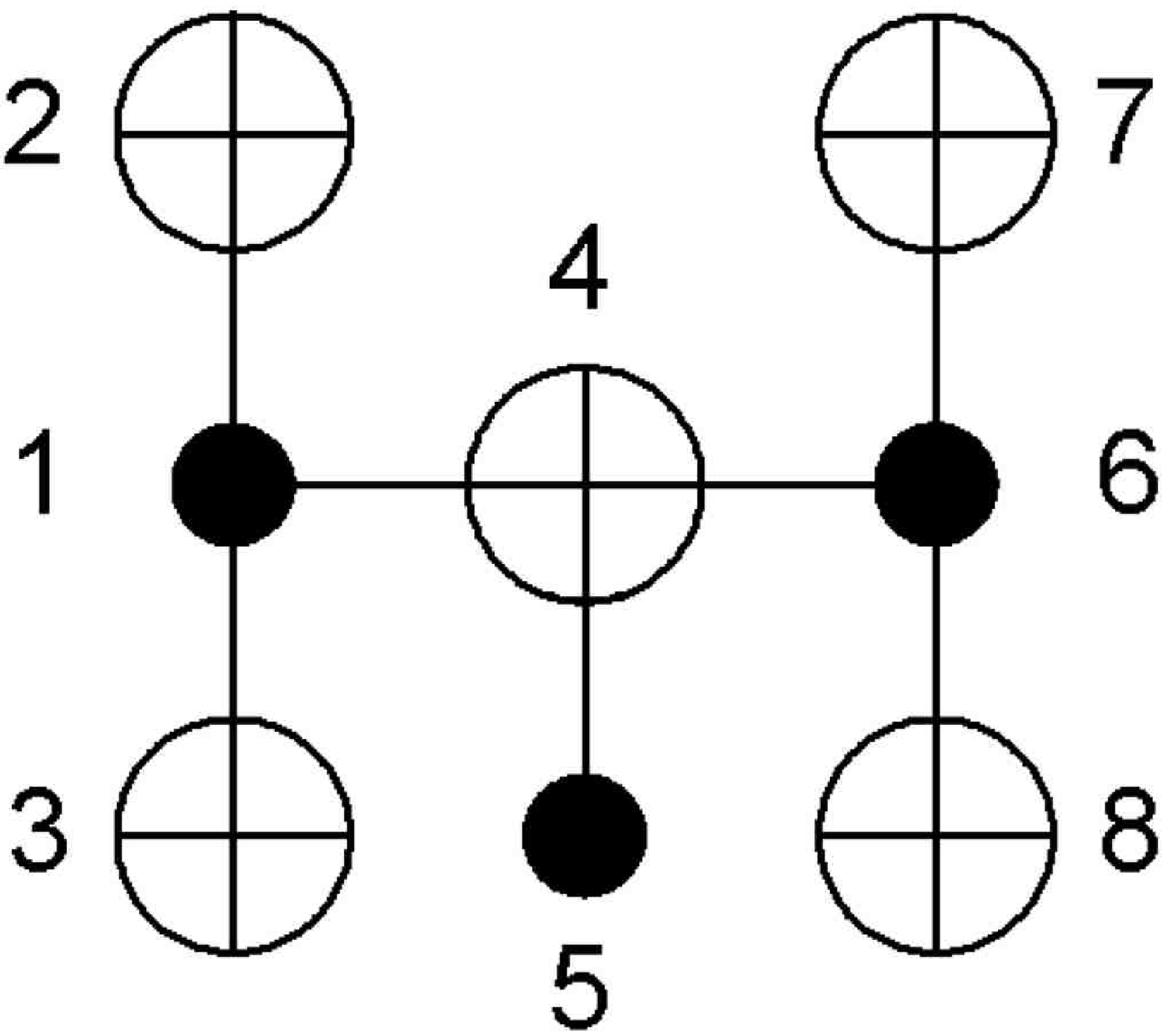}
\end{center}
\caption{}\label{f:cluster}
\end{figure}

\subsection{Example 4: CSS stabilizer codes, fault-tolerant CNOT and error correction}

\def\Ham{\H}
\def\C{\mathbf{C}}
\def\F{\mathbf{F}}
\def\tensor{\otimes}
\def\Aut{\operatorname{Aut}}
\renewcommand{\setminus}{\smallsetminus}
\def\matrix#1{\left(\begin{smallmatrix}#1\end{smallmatrix}\right)}

A \emph{stabilizer code}, encoding $k$ logical qubits into $n$ physical qubits, is the simultaneous $+1$ eigenspace of $n-k$ independent Pauli stabilizers.
Calderbank-Shor-Steane (CSS) codes \cite{CalderbankShor96,Steane96css,Steane96css2} have the property that every codeword is an equal superposition of classical codewords, in both the computational (Z, or $\ket{0}$/$\ket{1}$) and dual (X, or $\ket{+}$/$\ket{-}$) bases.  Equivalently, the stabilizer generators for such codes can be split into a set of Z stabilizers (i.e., parity checks, tensor products of the identity I and Z), and a set of X stabilizers.  This facilitates implementation of encoded stabilizer operations, particularly the CNOT gate.  It also allows for error correction of X errors separately from Z error correction.  

A CSS code is based on two classical linear codes $C_1 \subset C_2 \subset \{0,1\}^n$ with, say, dimensions $k_1$ and $k_2$, respectively.  Let stabilizer code $C$ be stabilized by 
\begin{equation*}
\mathcal{S} = C_2^\perp(Z) \cup C_1(X) \enspace ,
\end{equation*}
Here, $C_1(X)$ represents the set of stabilizers $\{ \bigotimes_{i : x_i = 1} X_i : x \in C_1\}$; i.e., replace all 1s with Xs and all 0s with the identity.  Similarly for $C_2^\perp(Z)$, where $C_2^\perp = \{ y : \forall x \in C_2, x \cdot y \equiv 0 \mod 2 \}$.
Altogether, there are $(n-k_2) + k_1$ independent stabilizer elements, so $C$ encodes $k_2-k_1$ qubits.  

The logical X operations are $(C_2\setminus C_1)(X)$.  (That is, the logical Z basis states -- encoded 0/1 strings -- are superpositions over cosets of $C_1$ in $C_2$.)  Logical Zs are $(C_1^\perp\setminus C_2^\perp)(Z)$.  If the minimum weight of an element in $C_1^\perp$ (resp. $C_2$) is $d_1^\perp$ ($d_2$), then the distance of $C$ is $\min(d_1^\perp,d_2)$.  

A special case is if $C_1$ is self-orthogonal and $C_2 = C_1^\perp$.  Then, for example, application of transversal Hadamard gates, switching $X \leftrightarrow Z$ on each bit, is a transversal logical Hadamard gate -- switching $X_L$ and $Z_L$ within the code space.  (Indeed, then all single-qubit Cliffords can be applied transversally.)


The structure of CSS codes allows more efficient operations than for general stabilizer codes.  Importantly, transversal CNOT gates implement a transversal logical CNOT gate.  Indeed, logical Xs and logical Zs are correctly propagated, Xs copied forward and Zs back.  It remains to verify that each block remains in the code space -- is still stabilized by $\mathcal{S}$.  Initially the stabilizers are generated by 
$$
\begin{array}{r@{\otimes}l} 
P_Z&I^{\otimes n}\\
P_X&I^{\otimes n}\\
I^{\otimes n}&P_Z\\
I^{\otimes n}&P_X
\end{array}
$$
for all $P_Z \in C_2^\perp(Z)$ and $P_X \in C_1(X)$.  After applying transversal CNOTs from the first block into the second, the stabilizers become
$$
\begin{array}{r@{\otimes}l} 
P_Z&I^{\otimes n}\\
P_X&P_X\\
P_Z&P_Z\\
I^{\otimes n}&P_X
\end{array} \enspace .
$$
These are equivalent to the original set of stabilizers since, e.g., the fourth row can be used to cancel out the second $P_X$ in row two.

Fault-tolerant measurement is also easy for CSS codes; measuring each qubit in the Z (resp. X) eigenbases gives enough parity-check information to correct X (Z) errors and determine the values of $Z_L$ ($X_L$).  Indeed, if $P_Z$ is a Z stabilizer, $P_Z \ket{\psi} = \ket{\psi}$, and some set of X, Y or Z errors are applied to $\ket{\psi}$, then the noisy state will still be an eigenvector of $P_Z$.  The eigenvalue is called the \emph{syndrome}, and is $+1$ (resp. $-1$) if an even (odd) number of X or Y errors intersect the support of $P_Z$.  Projecting each qubit to be 0 or 1, $P_Z$'s syndrome is the parity of a subset of the qubits.

Nearly all fault-tolerance schemes use CSS codes, because of the ease of applying fault-tolerant CNOT gates and the convenience of measurements.  In particular, the presentations of the Steane- and Knill-type fault-tolerance schemes in Sec.~\ref{s:introfaulttoleranceschemes} make sense only for CSS codes.  On CSS codes, each of these schemes is based on the encoding of a circuit with trivial logical effect (Fig.~\ref{f:steaneknilltypeerrorcorrection}).\footnote{Knill's teleportation-based error correction in fact also works for more general stabilizer codes.}

The first fault-tolerant quantum error correction procedure, due to Shor, instead works by extracting syndromes of each stabilizer generator one at a time.  For example, Fig.~\ref{f:steaneft} extracts the syndrome of IIIZZZZ.  The ancilla cat state needs to be prepared fault tolerantly.  
The circuit
$$
\includegraphics{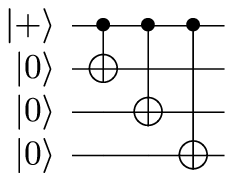}
$$
prepares a four-qubit cat state.  But here an XIIX error, giving $\tfrac{1}{\sqrt{2}}(\ket{0110}+\ket{1001})$, occurs with first-order probability (when the last CNOT fails), and kicks back two Z errors onto the data block.  This is not fault tolerant.  A verification procedure -- an easy special case of the techniques we will develop in Sec.~\ref{s:optimizedverificationprocedure} -- can be used to catch and remove correlated errors.  


Note that in Steane's X error correction procedure, based on transversal CNOTs into an encoded $\ket{+}$, X errors are copied forward to the ancilla, but Z errors are copied backward to the data!  If the ancilla is not fault tolerant against X errors, then repeating error syndrome extraction can be used to get more reliable syndrome information and make the error correction procedure fault tolerant anyway.  It is more serious if the ancilla is not fault tolerant against Z errors, although still sometimes tricks can make the error correction fault tolerant \cite{DiVincenzoAliferis06slow}.  


\comment{I like this paragraph, but it really belongs to the distance-three code threshold paper.  
What does quantum error correction consist of?  Roughly speaking, successful error correction should have two properties.  On a data block with no errors, or controlled errors (e.g., bounded in probability), there should be no logical effect and errors should remain controlled.  And an arbitrary state should be brought back into the codespace, except with controlled errors -- in this case, we can of course say nothing about a possible logical effect.}

\section{Quantum error-correcting codes} \label{s:quantumerrorcorrectingcodes}

An $[[n,k,d]]$ quantum error-correcting code encodes $k$ logical qubits into $n$ physical qubits with distance $d$ (meaning, $d$ is the minimum number of qubits touched by an nontrivial logical operation).  This section contains a brief review of some of the most important and popular quantum error-correcting codes used in fault-tolerance schemes.  They are all stabilizer codes, the codespace being the simultaneous $+1$ eigenspace of a set of $(n-k)$ independent, commuting Pauli operators.  

In this thesis, we will often also use for examples the classical repetition codes.  The $[n,1,n]$ repetition code maps one bit into $n$, via $0 \mapsto 0^n$, $1 \mapsto 1^n$, and has distance $n$ against bit flip (X) errors.  It has no protection -- distance one only -- against phase flip Z errors.  

\subsection{Four-qubit code} \label{s:fourqubitcode}

The $[[4,2,2]]$ quantum error-detecting code has distance only two, so can detect errors but not correct them.  It can be presented as\footnote{The basic construction generalizes to give $[[n,n-2,2]]$ codes for $n$ even \cite{Gottesman97}.} 
$$
\begin{array}{c c@{\,}c@{\,}c@{\,}c}
&X&X&X&X\\
&Z&Z&Z&Z\\
\cline{2-5}
X_{1,L} = &X&X&I&I\\
Z_{1,L} = &I&Z&I&Z\\
X_{2,L} = &X&I&X&I\\
Z_{2,L} = &I&I&Z&Z
\end{array}
$$
Because it can't correct errors, this code is typically used in schemes protecting erasure (detected) errors \cite{AlberBethCharnesDelgadoGrasslMussinger03erasure}.  It can also be used in fault-tolerance schemes based on postselection \cite{Knill03erasure,Knill05}, in which we condition the computation on not detecting any errors; see Sec.~\ref{s:unbiasednoise}.  Finally, it can allow for error correction if one relaxes the code concatenation structure slightly and keeps track of two different kinds of encoded errors \cite{Knill05} -- see Sec.~\ref{s:correctdisttwo}.

As an exercise in stabilizer algebra, one can check that the transversal logical Hadamard gate can be implemented with transversal physical Hadamard gates, followed by permuting bits two and three (by merely relabeling them, in a nonlocal gate model).  Indeed, $\text{XXXX} \leftrightarrow \text{ZZZZ}$.  Also, e.g., $X_{1,L} = \text{XXII} \mapsto \text{ZIZI} \approx \text{ZIZI} \cdot \text{ZZZZ} = \text{IZIZ} = Z_{1,L}$.

An encoding circuit is given in Fig.~\ref{f:fourfivesevenencoders}; fault-tolerant encoding of stabilizer states will be described in Sec.~\ref{s:encodingverification}.

\begin{figure}
\begin{center}
\includegraphics{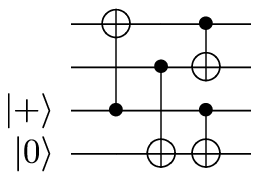}\qquad\quad
\includegraphics{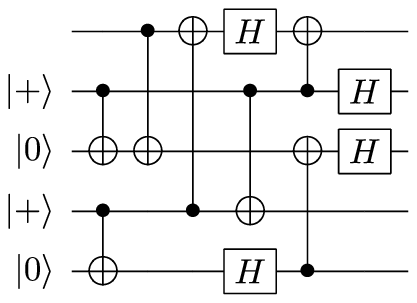}\qquad\quad
\includegraphics{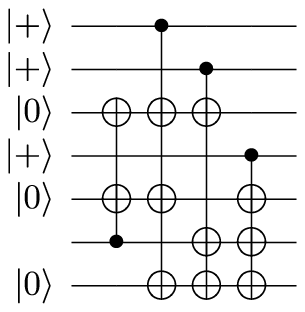}
\end{center}
\caption{Encoding circuits for the four-, five- and seven-qubit codes.  In the $[[7,1,3]]$ code encoder, a condensed notation is used for multiple NOTs with a single control.}\label{f:fourfivesevenencoders}
\end{figure}

This code has the interesting property that both ZZZZ and XXXX syndromes can be extracted fault-tolerantly, using just one ancilla qubit for each syndrome (Figure~\ref{f:fourqubitcodeft}).  
A single error in this circuit will either cause no error, one error, or an encoded/logical error on the \emph{second} of the two encoded qubits.  Never will a single error cause a logical error on the first encoded qubit.  Thus the circuit is fault tolerant, provided one is willing to sacrifice one of the two encoded qubits.  (Codes offering unequal protection to different encoded qubits are known as ``operator error-correcting" -- see Sec.~\ref{s:ninequbitbaconshorcode}.)

\subsection{Five-qubit code}

Five qubits are required for a distance-three quantum code.  The $[[5,1,3]]$ code \cite{LaflammeMiquelPazZurek96} can be presented to have stabilizer generators:
$$
XZZXI, IXZZX, XIXZZ, ZXIXZ \enspace .
$$
Logical X and logical Z can each be applied transversally.  The Clifford gate $T$, which conjugates X to Y to Z to X can also be applied transversally -- and this turns out to be useful in the ``magic states distillation" technique for achieving universality which we will discuss in Ch.~\ref{s:magicchapter}.)
However, the five-qubit code is inconvenient to work with in fault-tolerance applications because it is not CSS, and in particular the logical CNOT gate cannot be applied transversally \cite{Gottesman97}.


\subsection{Seven-qubit Steane code} \label{s:sevenqubitsteanecode}

The $[[7,1,3]]$ Steane code is based on the classical $[7,4,3]$ code, which has three parity checks:
$$
\begin{array}{c@{\,}c@{\,}c@{\,}c@{\,}c@{\,}c@{\,}c}
0&0&0&1&1&1&1\\
0&1&1&0&0&1&1\\
1&0&1&0&1&0&1
\end{array}
$$
The Steane code has the same parity checks (in stabilizer notation, the first row is, e.g., written IIIZZZZ).  It also has the dual stabilizers -- i.e., replace each Z with an X -- for altogether six independent stabilizers.  Logical Z and logical X are transversal.  Therefore, the Steane code has distance three, because an X error can be corrected in the computational basis, and a Z error can be corrected in the dual basis.

The $2^{4-1}$ even-weight classical codewords for the $[7,4,3]$ code are
$$
C_{\text{even}} = \left\{\begin{split} 
0000000, 1010101, 0110011, 1100110 \\
0001111, 1011010, 0111100, 1101001
\end{split}\right\} \enspace ,
$$
and the remaining codewords are the bitwise complements of these.  Encoded $\ket{0}$ can therefore be written out explicitly as $\ket{0}_L = \tfrac{1}{\sqrt{8}} \sum_{x \in C_{\text{even}}} \ket{x}$.  $\ket{1}_L = X^{\otimes 7} \ket{0_L}$.  This expression can also be derived directly from stabilizer algebra: $\ket{0}_L = \tfrac{1}{\sqrt{\abs{\mathcal{S}_X}}} \prod_{s \in \mathcal{S}_X} s \ket{0^7} = \prod_{s \in S_X} \left(\tfrac{1}{\sqrt{2}} (I + s)\right) \ket{0^7}$, where $\mathcal{S}_X$ is the set of X stabilizers, and $S_X$ is a generating set for $\mathcal{S}_X$.

The Steane code is CSS and self-dual, so in particular logical CNOT and also logical Hadamard can be applied transversally.  It is probably the most popular code used in fault-tolerance constructions, presumably because of its simplicity.

\subsection{Nine-qubit Bacon-Shor operator error-correcting code} \label{s:ninequbitbaconshorcode}

The $[[9,1,3]]$ Shor code is simply the classically three-bit repetition code ($\ket{0} \mapsto \ket{000}$, $\ket{1} \mapsto \ket{111}$), with its dual ($\ket{+} \mapsto \ket{\!+\!++}$, $\ket{-} \mapsto \ket{\!-\!--}$) concatenated on top, or vice versa.  The repetition code protects against bit flip errors to distance three, and its dual protects against phase flip errors to distance three.  The overall code therefore has distance three against arbitrary errors.  In stabilizer notation, the code's presentation is 
\begin{equation} \label{e:shorninequbitcode}
\begin{array}{c@{\,}c@{\,}c@{\,}c@{\,}c@{\,}c@{\,}c@{\,}c@{\,}c@{\,}c}
&Z & Z & \cdot & \cdot & \cdot & \cdot & \cdot & \cdot & \cdot \\
&\cdot & Z & Z & \cdot & \cdot & \cdot & \cdot & \cdot & \cdot \\
&\cdot & \cdot & \cdot & Z & Z & \cdot & \cdot & \cdot & \cdot \\
&\cdot & \cdot & \cdot & \cdot & Z & Z & \cdot & \cdot & \cdot \\
&\cdot & \cdot & \cdot & \cdot & \cdot & \cdot & Z & Z & \cdot \\
&\cdot & \cdot & \cdot & \cdot & \cdot & \cdot & \cdot & Z & Z \\
&X & X & X & X & X & X & \cdot & \cdot & \cdot \\
&\cdot & \cdot & \cdot & X & X & X & X & X & X \\
\cline{2-10}
X_L = &X & X & X & \cdot & \cdot & \cdot & \cdot & \cdot & \cdot \\
Z_L = &Z & \cdot & \cdot & Z & \cdot & \cdot & Z & \cdot & \cdot 
\end{array}
\end{equation}
(Here we have written $\cdot$ in place of I for clarity.)
This code has the property that adding the logical X operator as a stabilizer decouples the three groups of three qubits, i.e., $\ket{+}_L = (\ket{\!+\!++} + \ket{\!-\!--})^{\otimes 3}$.  This is handy in preparing $\ket{+}$ because it means that no error correlations need to be created between the different groups of three qubits.  In fact, there can be no error correlations at all; because on a group of three, any X error is either trivial or equivalent to another X error of weight one, and all nontrivial Z errors are equivalent to weight-one Z errors and to each other.  Encoding into a quantum error-correcting code typically creates correlated error events, and a verification procedure is typically required to remove these correlations -- as will be discussed in Sec.~\ref{s:encodingverification}.  However, encoding $\ket{+}$ for this code, no verification is required.  However, encoding $\ket{0}$ will create error correlations, requiring verification.  (Had the concatenation order been the reverse -- the repetition code placed on top of its dual -- then the situation for encoding $\ket{0}$ and $\ket{+}$ would have been reversed, as well.)

Note that the code as presented actually gives stronger protection against bit flip errors than against phase flip errors.  One bit flip error in each group of three can be corrected, whereas only one phase flip error can be corrected.\footnote{This kind of asymmetric protection property might be helpful in practice if the underlying noise model is known to be biased -- see Sec.~\ref{s:lowlevelft}.}  Intuitively, this means that there are more Z stabilizers than are really necessary.  Poulin \cite{Poulin05} considered combining some of the above Z stabilizers and removing others, while still preserving some bit-flip error protection.  A stabilizer code can be thought of as a decomposition of the Hilbert space into the codespace and its complement, here as $\mathcal{H}^9 = C \oplus B$ where one logical qubit is stored in $C$, the two-dimensional simultaneous $+1$ eigenspace of the eight stabilizer generators.  Removing a stabilizer generator, the stabilizer space doubles in dimension, so the decomposition becomes $\mathcal{H}^9 = (C \otimes A) \oplus B$, where $C$ and $A$ each contain a logical qubit.  But it is possible the logical register $C$ is protected against errors, while $A$ is unprotected.  This is not a particular problem -- just don't compute in $A$!  Such a code, in which encoded qubits have different amounts of protection, is known as an operator quantum error-correcting code \cite{KribsLaflammePoulin05operator}.  

Bacon gave an operator version of the Shor $[[9,1,3]]$ code which removed the asymmetry between X and Z stabilizers \cite{Bacon05operator}.  Simply multiply together all the odd Z stabilizers in Eq.~\eqref{e:shorninequbitcode} to give one Z stabilizer generator, multiply together all the even Z stabilizers to get another, and discard all the remaining Z stabilizers:
$$
\begin{array}{c@{\,}c@{\,}c@{\,}c@{\,}c@{\,}c@{\,}c@{\,}c@{\,}c@{\,}c}
&Z & Z & \cdot & Z & Z & \cdot & Z & Z & \cdot \\
&\cdot & Z & Z & \cdot & Z & Z & \cdot & Z & Z \\
&X & X & X & X & X & X & \cdot & \cdot & \cdot \\
&\cdot & \cdot & \cdot & X & X & X & X & X & X \\
\cline{2-10}
X_L = & X & X & X & \cdot & \cdot & \cdot & \cdot & \cdot & \cdot \\
Z_L = & Z & \cdot & \cdot & Z & \cdot & \cdot & Z & \cdot & \cdot 
\end{array}
$$
The code still protects the indicated logical qubit against X and Z errors.  (The other $9-4-1 = 4$ degrees of freedom, giving $A$ in the decomposition $\mathcal{H}^9 = (C \otimes A) \oplus B$, are not protected.)  It is also now symmetrical -- this is easiest to see by arranging the qubits in a $3 \times 3$ grid, so the X stabilizers are supported exactly on pairs of columns and the Z stabilizers on pairs of rows.  Logical CNOT can still be applied transversally, and now logical Hadamard can also be applied transversally, followed by relabeling (transposing the $3 \times 3$ grid of qubits).

In particular, Aliferis observed that $\ket{0}_L$ and $\ket{+}_L$ can each be split into tensor products of three qubits, $\ket{0}_L = (\ket{000} + \ket{111})^{\otimes 3}$ (after an appropriate qubit reordering), $\ket{+}_L = (\ket{\!+\!++} + \ket{\!-\!--})^{\otimes 3}$, when the extra degrees of freedom $A$ are set appropriately \cite{Aliferis05operator}.  This implies that encoding either of these states can be accomplished without requiring verification against correlated errors.

\subsection{15-qubit Reed-Muller code}

The $[[15,1,3]]$ Reed-Muller code has the interesting property that a non-Clifford gate, $\tfrac{X+Z}{\sqrt{2}}$, can be applied transversally \cite{KnillLaflammeZurek96,BravyiKitaev04}.  This is related to the magic states distillation technique for achieving universality which we will discuss in Ch.~\ref{s:magicchapter}, and has also been exploited in a scheme for fault-tolerant cluster state computing \cite{RaussendorfHarringtonGoyal05oneway}.

\subsection{$23$-qubit Golay code} \label{s:23qubitgolaycode}

The $[[23,1,7]]$ Golay code is a self-dual code based on the classical $[23,12,7]$ Golay code.  The parity checks of the classical code can be written:
$$
\begin{array}{c@{\,}c@{\,}c@{\,}c@{\,}c@{\,}c@{\,}c@{\,}c@{\,}c@{\,}c@{\,}c@{\,}c@{\,}c@{\,}c@{\,}c@{\,}c@{\,}c@{\,}c@{\,}c@{\,}c@{\,}c@{\,}c@{\,}c}
1&.&1&.&.&1&.&.&1&1&1&1&1&.&.&.&.&.&.&.&.&.&.\\
1&1&1&1&.&1&1&.&1&.&.&.&.&1&.&.&.&.&.&.&.&.&.\\
.&1&1&1&1&.&1&1&.&1&.&.&.&.&1&.&.&.&.&.&.&.&.\\
.&.&1&1&1&1&.&1&1&.&1&.&.&.&.&1&.&.&.&.&.&.&.\\
.&.&.&1&1&1&1&.&1&1&.&1&.&.&.&.&1&.&.&.&.&.&.\\
1&.&1&.&1&.&1&1&1&.&.&1&.&.&.&.&.&1&.&.&.&.&.\\
1&1&1&1&.&.&.&1&.&.&1&1&.&.&.&.&.&.&1&.&.&.&.\\
1&1&.&1&1&1&.&.&.&1&1&.&.&.&.&.&.&.&.&1&.&.&.\\
.&1&1&.&1&1&1&.&.&.&1&1&.&.&.&.&.&.&.&.&1&.&.\\
1&.&.&1&.&.&1&1&1&1&1&.&.&.&.&.&.&.&.&.&.&1&.\\
.&1&.&.&1&.&.&1&1&1&1&1&.&.&.&.&.&.&.&.&.&.&1
\end{array}
$$
and the quantum code has the same parity checks in both the Z and X eigenbases.\footnote{Note that each parity check has weight eight, and every pair of checks has inner product zero, mod two -- therefore the X and Z stabilizers commute, as required.  The checks are found by cyclically permuting the first parity check, then performing a sort of Gaussian elimination.}  The Golay code is more complicated than the Steane code -- for example, its higher distance requires some extra care in encoding and error correction (Sec.~\ref{s:encodingverification}).  However, Steane has compared a number of different quantum error correcting codes, using simulations of smaller codes to develop a heuristic analytical model to predict the performance of larger codes (including overhead considerations).  He found that the Golay code performed among the best of the codes considered, with respect both to overhead and the noise threshold \cite{Steane03}.  That a fault-tolerance scheme based on the Golay code can tolerate more noise than a similar scheme using the seven-bit code has also been confirmed rigorously, in a particular model \cite{Reichardt06Golay}.  However, neither analysis considered locality constraints, which may favor smaller codes, and the Golay code may end up being more useful for protecting the memory of a quantum computer than for protecting its computations (Sec.~\ref{s:architectures}).

\subsection{Larger codes, and asymptotically large codes} \label{s:largecodes}

The majority of these codes encode just a single qubit.  Fault-tolerance constructions are known for larger codes encoding multiple, even many, logical qubits \cite{Gottesman97}, and there is some evidence that these larger codes may give more efficient fault-tolerance constructions \cite{Steane03,Steane04computer}.  (However, our practical understanding of these larger codes is more limited, because of the combinatorial difficulty of simulating them.)  The problem with large codes seems to be that it is difficult to encode into them even to start the computation (Sec.~\ref{s:encodingverification}).  (For a general $n$-qubit stabilizer code, we only know how to encode using $O(n^2/\log n)$ basic stabilizer operations like the CNOT gate \cite{AaronsonGottesman04}.)  One strategy that might be very helpful in practice is to start with a fairly small code to reduce the effective error rate just enough to bootstrap into a larger, more efficient code \cite{Steane03,Knill03erasure}.  These layers of different codes can be maintained through the computation, or the smaller code can be removed (decoded directly or teleported out of the encoding, possibly with postselection -- see Sec.~\ref{s:bottomupdecoding}) once the larger code has been set up.  For example, to encode into the 23-qubit Golay code, one might start with 23 states prepared into the seven-qubit Steane code.  Then apply the Golay encoding circuit, error-correcting the Steane code after each level-one gate; and then decode the Steane code layer.  This has two potential advantages.  First of all, while encoding into an $n$-qubit code might take $\sim n^2$ gates to finish before the code can correct errors; once the encoding is finished, only $n$ gates (the number of gates in a logical operation) are usually required between error correction steps.  Secondly, encoding is perhaps the part of a fault-tolerance scheme most fraught with dangers -- for only during encoding do qubits within the same code block interact with each other, possibly causing correlated errors.  The independent errors introduced in decoding out the smaller code are less dangerous to fault tolerance (and also easier to analyze).  

Asymptotically large codes can have very good error-correction properties, expressed either in terms of distance (with the achievability of the Gilbert-Varshamov bound for nondegenerate codes \cite{EkertMacchiavello96gilbertvarshamovbound, CalderbankShor96}) or in terms of their correction of random errors \cite{Hamada04codes,GottesmanPreskill01codes}.  For quantum fault-tolerance, however, the difficulty of encoding limits the use of these codes.  Also, efficient classical algorithms are needed for decoding measurement results, or memory errors can accumulate in the quantum computer faster than they are corrected.


\section{Encoded ancilla preparation and verification} \label{s:encodingverification}

Both Steane- and Knill-type error correction rely crucially on the ability to prepare reliable ancillary encoded stabilizer states.  This is particularly the case for Knill's scheme, in which even computation takes place through careful ancilla preparation.  (Shor's scheme requires reliable stabilizer states which are not encoded.)  
High-fidelity ancillas do \emph{not} suffice for error correction; as described in Sec.~\ref{s:introductionancillapreparation}, error correlations must also be controlled.  

Let us remark immediately that for the two- and four-qubit codes used in the remainder of this thesis, fault-tolerant ancilla preparation is easy.  The reader unconcerned with fault tolerance using larger codes may certainly skip to the next chapter.  


Ancilla preparation circuits will typically lead to correlated errors (although not always; see Sec.~\ref{s:ninequbitbaconshorcode}).  Therefore, preparation typically must be followed by a verification procedure to reduce correlations.  While computation can often continue without waiting for error correction measurement results to come in, an ancilla cannot be used in error correction until verification has fully completed.  In terms of a fault-tolerance architecture, we can think of an ``ancilla factory" which continually prepares and verifies encoded ancillas, in large enough numbers that one will always be ready when needed for error correction.\footnote{With the restriction of local gates, ancillas must be prepared close enough to where they will be used, or a good transportation procedure must be devised.}  Efficient verification procedures are hence important to reduce delays and overhead.\footnote{The optimized verification procedures we present here can also be applied to the fault-tolerance scheme of \Ref\cite{DiVincenzoAliferis06slow}.}

In this section, we start by giving a general method for preparing a stabilizer state.  We then give a na{\"i}ve verification procedure.  Finally we optimize the verification procedure for specific codes, taking advantage of the fact that the preparation circuit does not create \emph{arbitrary} correlations.  

\subsection{Stabilizer state preparation} \label{s:stabilizerstatepreparation}

We describe Steane's time- and space-efficient procedure for preparing CSS-type encoded stabilizer states \cite{Steane02}.  (This procedure does not use the fewest number of CNOT gates or the minimum time [circuit depth] asymptotically -- see \Refs\cite{MooreNilsson98,AaronsonGottesman04} -- but performs well on small codes and does not require any extra qubits.)

Begin by using Gaussian elimination, and by rearranging qubits, to put the $n$-qubit target state's X (or Z) stabilizer generators in a standard form $\matrix{I&M}$, where $I$ is a $k$-dimensional identity matrix, and $M$ is a $k \times (n-k)$ binary matrix.  (Each bit $b$ corresponds to the Pauli $X^b$.)  For example, the X stabilizers for $\ket{0}$ encoded into the Steane code (Sec.~\ref{s:sevenqubitsteanecode}) are:
$$
\begin{array}{c@{\,}c@{\,}c@{\,}c@{\,}c@{\,}c@{\,}c}
I&I&I&X&X&X&X\\
I&X&X&I&I&X&X\\
X&I&X&I&X&I&X
\end{array} 
$$
Moving the fourth qubit into the third position gives
$$
\begin{array}{c@{\,}c@{\,}c}
X&\cdot&\cdot\\
\cdot&X&\cdot\\
\cdot&\cdot&X
\end{array}
\!\Bigg\vert\!
\underset{M}{\underbrace{
\begin{array}{c@{\,}c@{\,}c@{\,}c}
X&X&\cdot&X\\
X&\cdot&X&X\\
\cdot&X&X&X
\end{array}}} 
$$
Now, starting with $\ket{+^k0^{n-k}}$, use CNOT gates from the first $k$ qubits into the last $n-k$ qubits to generate each stabilizer.  For example, 
\begin{eqnarray*}
\text{initial X stabilizers:}
\overset{\substack{\text{control}\\ \text{qubits}}}{
\begin{array}{c@{\,}c@{\,}c}
X&\cdot&\cdot\\
\cdot&X&\cdot\\
\cdot&\cdot&X
\end{array}}
\!\Bigg\vert\!
\overset{\substack{\text{target}\\ \text{qubits}}}{
\begin{array}{c@{\;\,}c@{\;\,}c@{\;\,}c}
\cdot&\cdot&\cdot&\cdot\\
\cdot&\cdot&\cdot&\cdot\\
\cdot&\cdot&\cdot&\cdot
\end{array}}
&\overset{\Lambda_1(X_4 X_5 X_7)}{\longrightarrow}&
\overset{\substack{\text{control}\\ \text{qubits}}}{
\begin{array}{c@{\,}c@{\,}c}
X&\cdot&\cdot\\
\cdot&X&\cdot\\
\cdot&\cdot&X
\end{array}}
\!\Bigg\vert\!
\overset{\substack{\text{target}\\ \text{qubits}}}{
\begin{array}{c@{\,}c@{\;\,}c@{\;\,}c}
X&X&\cdot&X\\
\cdot&\cdot&\cdot&\cdot\\
\cdot&\cdot&\cdot&\cdot
\end{array}} \\
&\overset{\substack{\Lambda_2(X_4 X_6 X_7)\\ \Lambda_3(X_5 X_6 X_7)}}{\longrightarrow}&
\begin{array}{c@{\,}c@{\,}c}
X&\cdot&\cdot\\
\cdot&X&\cdot\\
\cdot&\cdot&X
\end{array}
\!\Bigg\vert\!
\begin{array}{c@{\,}c@{\,}c@{\,}c}
X&X&\cdot&X\\
X&\cdot&X&X\\
\cdot&X&X&X
\end{array} 
\end{eqnarray*}
The Z stabilizers are generated automatically.  Note that the number of CNOT gates required is exactly the number of nontrivial elements of $M$ -- different code presentations can affect this.

Since all the CNOT gates here commute, we can rearrange them to maximize parallelism.  A schedule corresponds to filling in nontrivial entries of $M$ with round numbers -- for example:
\begin{eqnarray*}
\begin{array}{c@{\,}c@{\,}c@{\,}c}
1&2&\cdot&3\\
3&\cdot&1&2\\
\cdot&3&2&1
\end{array}
&\longleftrightarrow&
\begin{split}
\text{round 1:}\;\, \Lambda_1(X_4), \Lambda_2(X_6), \Lambda_3(X_7) \\
\text{round 2:}\;\, \Lambda_1(X_5), \Lambda_2(X_7), \Lambda_3(X_6) \\
\text{round 3:}\;\, \Lambda_1(X_7), \Lambda_2(X_4), \Lambda_3(X_5) 
\end{split}
\end{eqnarray*}
In each time step, each control qubit can be used at most once, and each target qubit can be targeted at most once.  Therefore, no round number can appear twice in one row, nor twice in one column.  This implies that the number of rounds must be at least the maximum number of nontrivial entries in a row or column of $M$.  By Hall's marriage theorem (a bipartite graph of maximum degree $m$ can be covered with $m$ matchings), equality suffices.

Putting the qubits back in their original order, a three-round Steane-code encoding circuit for $\ket{0}_L$ is 
\begin{equation} \label{e:steaneencodingcircuit}
\raisebox{-4.5em}{\includegraphics[scale=.5]{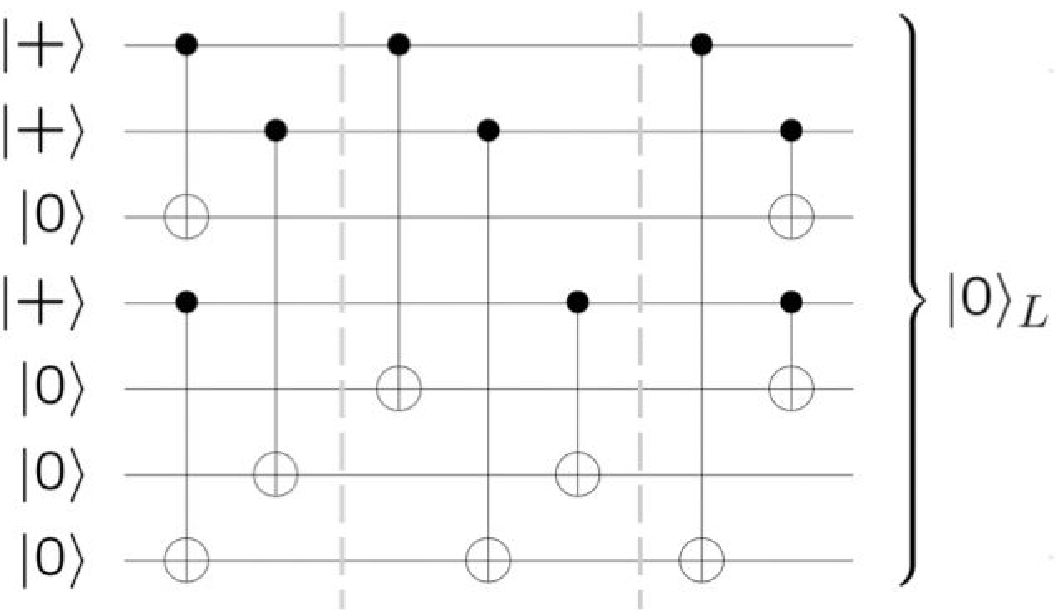}}
\end{equation}

For the Golay code, whose $M$ matrix was given in Sec.~\ref{s:23qubitgolaycode}, a seven-round schedule is given by:
\begin{equation} \label{e:golayencodingcircuitschedule}
\begin{array}{c@{\,}c@{\,}c@{\,}c@{\,}c@{\,}c@{\,}c@{\,}c@{\,}c@{\,}c@{\,}c@{\,}c}
1&.&2&.&.&4&.&.&3&5&6&7\\
2&4&3&5&.&6&7&.&1&.&.&.\\
.&2&4&3&5&.&6&7&.&1&.&.\\
.&.&5&6&7&1&.&2&4&.&3&.\\
.&.&.&7&1&3&4&.&5&6&.&2\\
4&.&7&.&2&.&1&5&6&.&.&3\\
3&5&6&2&.&.&.&1&.&.&7&4\\
5&1&.&4&6&7&.&.&.&3&2&.\\
.&7&1&.&3&5&2&.&.&.&4&6\\
6&.&.&1&.&.&3&4&7&2&5&.\\
.&6&.&.&4&.&.&3&2&7&1&5
\end{array}
\end{equation}

\subsection{Na{\"i}ve verification procedure} \label{s:naiveverificationprocedure}

The circuits from Eqs.~\eqref{e:steaneencodingcircuit} and~\eqref{e:golayencodingcircuitschedule} each give correlated errors -- they are not fault tolerant.  
In each case, a verification procedure is required, to catch correlated errors, before the states can be used.  

For example, in Eq.~\eqref{e:steaneencodingcircuit}, weight-two X errors occur with first-order probability.  Verification against X errors is therefore required for fault tolerance.  Z errors on the other hand are not correlated, so no Z verification is required.  Indeed, the maximum weight of any Z error on $\ket{0}_L$, up to stabilizer equivalences, is one; e.g., ZZIIIII is equivalent to IIZIIII.  

One standard method of ancilla verification is purification.  Prepare two ancillas, and check one against the other using transversal CNOTs followed by transversal measurement of the second ancilla.  Postselect on no detected errors in the second ancilla.  Verification of X errors can be shown schematically as 
$$
\includegraphics{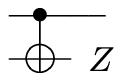}
$$
(Here, each wire denotes a code block, or logical qubit, the CNOT means transversal CNOT, and the Z at the right end of a wire means transversal measurement in the computational basis, postselecting on no detected X errors.)
The probability of any X error of weight $\geq 2$ on the output, conditioned on acceptance, is $O(\eta^2)$, where $\eta$ is the gate failure rate.  Indeed, if there is only a single failure, then it can occur either in preparing one of the ancillas, or in the transversal operations interacting the two ancillas.  But if only one ancilla has an error in it, then that error will be caught; and the transversal operations are fault tolerant.  

Verification of X and Z errors, when required, is schematically
$$
\includegraphics{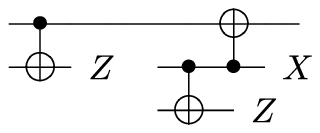}
$$
(Note that the ancilla used to check for Z errors must itself first be checked for X errors, or these errors will be copied upward.)

In his simulations, Steane finds that one round of purification -- outputting a state so that errors of weight $>1$ are second-order events -- works well enough.\footnote{Steane actually uses a slightly different procedure which uses fewer qubits at a time.}  However, although one round of purification suffices for the existence of a constant threshold, it is not strictly fault tolerant.  With a distance-$d$ code, we expect the effective error rate should drop like $\eta \rightarrow \eta^{(d+1)/2}$, but with only one round of purification we get just quadratic decay $\eta \rightarrow \eta^2$, giving up an advantage of using a large, high-distance code.  

For strict fault tolerance, we require that weight-$k$ errors are at most order-$k$ events, for $k \leq (d+1)/2$.  For codes of distance greater than three, one purification round does not generally suffice for strict fault tolerance.  Repeated purification is required.  For the Golay code, it is easy to see that three rounds of X error verification and two rounds of Z error verification are required to get a strictly fault-tolerant Golay code ancilla:
\begin{equation} \label{e:xzerrorverificationtables}
\textrm{
\begin{tabular}{r c c c c c c} 
& X error weight: & 0 & 1 & 2 & 3 & 4 \\
\cline{2-7}
X error order with $\quad$ $\raisebox{.25em}{\Qcircuit @C=.5em @R=.45em @!R {
&\qw&\qw&\qw&\qw
}}\smallskip \smallskip$& 0 verifications: & 0 & 1 & 1 & 1 & 1 \\
$\raisebox{1em}{\Qcircuit @C=.5em @R=.45em @!R {
&\ctrl{1}&\qw&\qw&\qw\\
&\targ   &\rstick{Z}\qw&&
}}\smallskip \smallskip$& 1 verification:  & 0 & 1 & 2 & 2 & 2 \\
$\raisebox{1em}{\Qcircuit @C=.5em @R=.45em @!R {
&\ctrl{1}&\qw          &\qw&\qw&\qw&\qw&\qw&\qw&\ctrl{1}&\qw          &\qw&\qw\\
&\targ   &\rstick{Z}\qw&   &   &   &   &   &\qw&\targ   &\rstick{Z}\qw&   &   \\
}}\smallskip \smallskip$& 2 verifications: & 0 & 1 & 2 & 3 & 3 \\
$\raisebox{1em}{\Qcircuit @C=.5em @R=.45em @!R {
&\ctrl{1}&\qw          &\qw&\qw&\qw&\qw&\qw&\qw&\ctrl{1}&\qw          &\qw&\qw&\qw&\qw&\qw&\qw&\ctrl{1}&\qw          &\qw&\qw\\
&\targ   &\rstick{Z}\qw&   &   &   &   &   &\qw&\targ   &\rstick{Z}\qw&   &   &   &   &   &\qw&\targ   &\rstick{Z}\qw&   &   \\
}}$& 3 verifications: & 0 & 1 & 2 & 3 & 4 \\
&&&&&&\\
&&&&&&\\
& Z error weight: & 0 & 1 & 2 & 3 & \\
\cline{2-6}
Z error order with $\quad$ $\raisebox{.25em}{\Qcircuit @C=.5em @R=.45em @!R {
&\qw&\qw&\qw&\qw
}}\smallskip \smallskip$& 0 verifications: & 0 & 1 & 1 & 1 &  \\
$\raisebox{1em}{\Qcircuit @C=.5em @R=.45em @!R {
&\targ    &\qw&\qw&\qw\\
&\ctrl{-1}&\rstick{X}\qw&&
}}\smallskip \smallskip$& 1 verification:  & 0 & 1 & 2 & 2 &  \\
$\raisebox{1em}{\Qcircuit @C=.5em @R=.45em @!R {
&\targ    &\qw          &\qw&\qw&\qw&\qw&\qw&\qw&\targ    &\qw          &\qw&\qw\\
&\ctrl{-1}&\rstick{X}\qw&   &   &   &   &   &\qw&\ctrl{-1}&\rstick{X}\qw&   &   \\
}}\smallskip \smallskip$& 2 verifications: & 0 & 1 & 2 & 3 &   
\end{tabular}}
\end{equation}
Here, each row of the table is derived from the previous rows.  For example, for a weight-four error to survive two X verifications, the worst case is that a weight-four error survived one verification (order-two event) and it exactly canceled out the same error in the third ancilla (first-order event) -- overall the probability is $O(\eta^3)$.

There are many different verification procedures outputting strictly fault-tolerant Golay code states $\ket{0}_L$.  For example, one can first check for Z errors, then X errors:
\begin{equation} \label{e:golaypurificationcircuit}
\includegraphics[scale=.7]{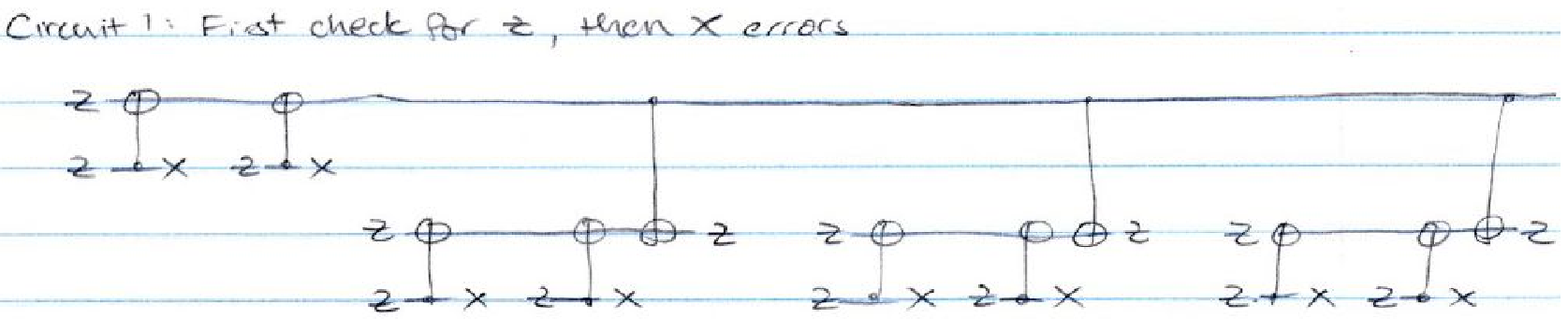}
\end{equation}
But one can also first check for X errors, then Z errors, or interlace the verifications in various ways:
\begin{equation*}
\includegraphics[scale=.7]{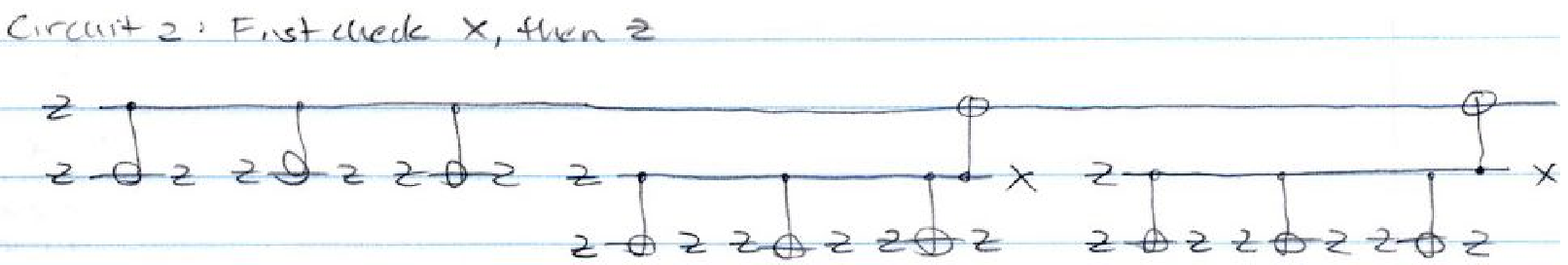}
\end{equation*}\begin{equation*}
\includegraphics[scale=.7]{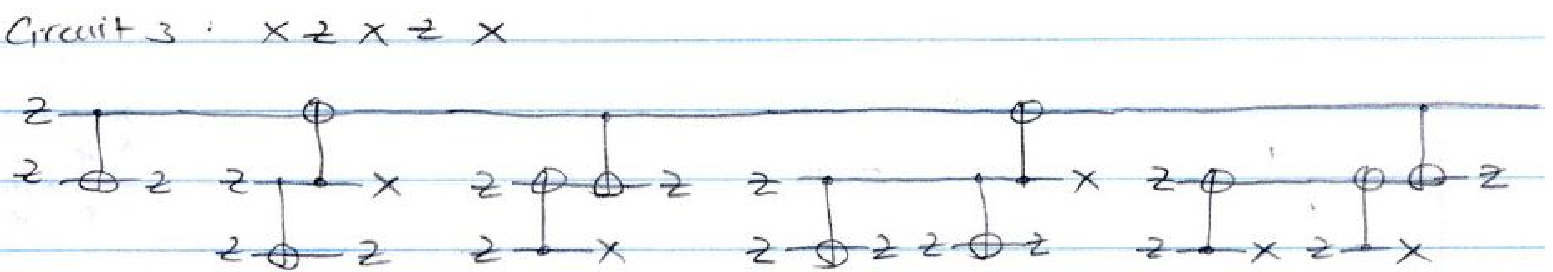}
\end{equation*}\begin{equation*}
\includegraphics[scale=.7]{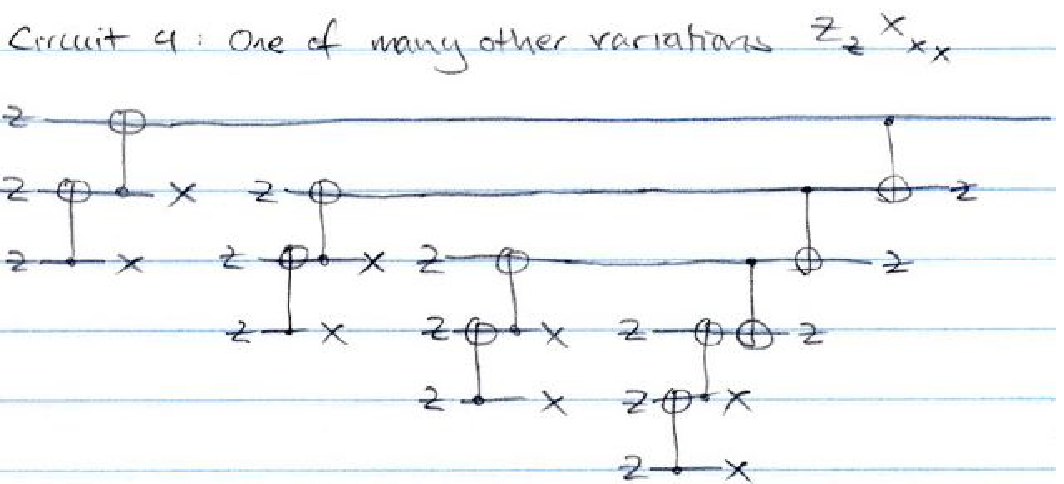}
\end{equation*}
Each of these procedures requires preparing twelve copies of an unverified $\ket{0}_L$.  

\subsection{Optimized verification procedure} \label{s:optimizedverificationprocedure}

The above verification procedures work, but are somewhat na{\"i}ve.  We can improve their efficiency if we notice, following Aliferis \cite{Aliferis05operator}, that while preparation creates correlated errors, it does not create \emph{arbitrary} error correlations.  

\subsubsection*{Steane code}
For example, consider the Steane code encoding circuit of Eq.~\eqref{e:steaneencodingcircuit}.  Assuming a single X error during preparation, (of course) arbitrary single-bit errors are possible on the output.  But the only possible two-bit errors are $X_1 X_7$, $X_2 X_3$ and $X_4 X_5$ -- these are the correlations created in the last round.  To see this, first notice that errors on the target qubits (qubits 3, 5, 6, 7) cannot spread.  And an X error, say on qubit one just before round two, will spread to $X_1 X_5 X_7 \sim X_3$ a one-bit error modulo the stabilizer element XIXIXIX.  

Therefore, if we prepare two states $\ket{0}_L$ with different preparation circuits -- say reversed -- then the correlated errors created will be different, as shown in Fig.~\ref{f:differentsteanepreparations}.
\begin{figure*}
\centering
\subfloat[][]{
\includegraphics[scale=.7]{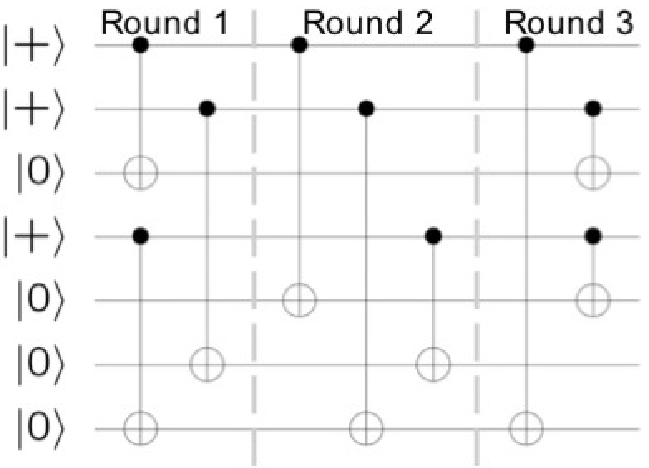}
}
\qquad
\subfloat[][]{
\includegraphics[scale=.7]{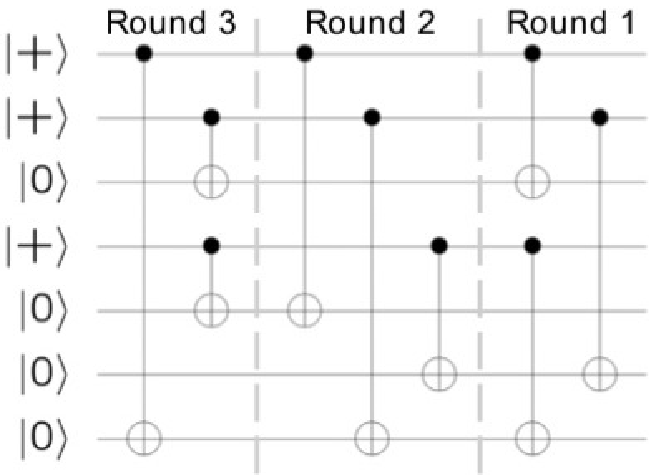}
}
\caption{Two different preparation circuits for the Steane $\ket{0}_L$ giving different correlated errors.  Assume at most one X error occurs during preparation.  The possible errors on the output in (a) are arbitrary single-qubit errors (of course), as well as the correlated errors $X_1 X_7$, $X_2 X_3$ and $X_4 X_5$ -- created by failures in the last round.  In (b), the round order has been reversed.  The only output weight-two X errors that can be caused by a single X error during preparation are $X_1 X_3$, $X_2 X_6$ and $X_4 X_7$.  
} \label{f:differentsteanepreparations}
\end{figure*}
Apply transversal CNOTs from one copy to the other, and measure the second.  On detecting a weight-one error, no action is required.  
On detecting a two-qubit error, correct it iff it could have been created by a single failure in the first state.
(Do not correct two-qubit errors which could have been created by a single failure in preparing the second state.)

Aliferis's optimized procedure still uses two preparations of $\ket{0}_L$, but it allows one to \emph{correct} detected errors, instead of postselecting on no detected errors.  This eliminates the overhead of an ``ancilla factory" readying multiple ancillas in parallel so that at least one will have survived verification -- particularly important in models with locality restrictions (like physical systems in low dimensions) in which moving ancillas into place is difficult.  The optimized procedure also allows for an ancilla to be used for error correction even before the verification measurements complete -- any necessary corrections can be propagated and applied later -- which is useful when physical measurements are slow.  

\subsubsection*{Golay code} \label{s:optimizedverificationproceduregolay}
Now consider the Golay code.  Encoding $\ket{0}_L$ for the Golay code, every X stabilizer we create has weight eight; and again, only certain correlated X errors can be created from a single X failure.  For example, possible output X errors from a single X failure on the control wire in the circuit 
\begin{equation} \label{e:eightqubitexample}
\begin{array}{c} 
\includegraphics{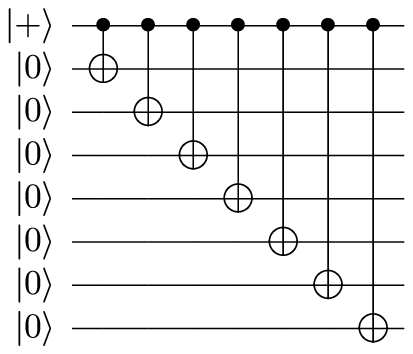}
\end{array}
\end{equation}
are only, writing $X_S \equiv \bigotimes_{i \in S} X_i$, $X_{\{1,\ldots,8\}} \sim I$, $X_{\{1,3,\ldots,8\}} \sim X_2$, $X_{\{1,4,\ldots,8\}} \sim X_{\{2,3\}}$, $X_{\{1,5,\ldots,8\}} \sim X_{\{2,3,4\}}$, $X_{\{1,6,7,8\}} \sim X_{\{2,\ldots,5\}}$, $X_{\{1,7,8\}} \sim X_{\{2,\ldots,6\}}$, $X_{\{1,8\}} \sim X_{\{2,\ldots,7\}}$.  Up to the stabilizer, these are each equivalent to a Xs on a consecutive sequence of qubits starting at 2.  Reversing the round order, as for the Steane code, only Xs on consecutive sequences of qubits ending at 8 can be created by a single X failure.  These two sets of errors are disjoint, as was the case for the Steane code.  

However, with two X failures, either the circuit of Eq.~\eqref{e:eightqubitexample} or the reversed-round circuit can output errors which are consecutive sequences $X_{\{i,i+1,\ldots,j\}}$.  Therefore, verifying one Golay ancilla against another prepared the same except with reversed rounds, weight-four X errors will survive with third-order probability.  
(Eq.~\eqref{e:eightqubitexample} is a subcircuit of the full Golay preparation circuit, showing only gates with some single control.)
Even postselecting on no detected errors, this is not strictly fault tolerant.

Still, some savings is possible.  
Define four round permutations: $A = (1,2,3,4,5,6,7)$, $A^r = (7,6,5,4,3,2,1)$, $B = (6,2,7,3,5,4,1)$, $B^r = (1,4,5,3,7,2,6)$.  Then the verification circuit
\begin{equation} \label{e:optimizedgolayverificationcircuit}
\begin{array}{c}
\includegraphics[scale=.5]{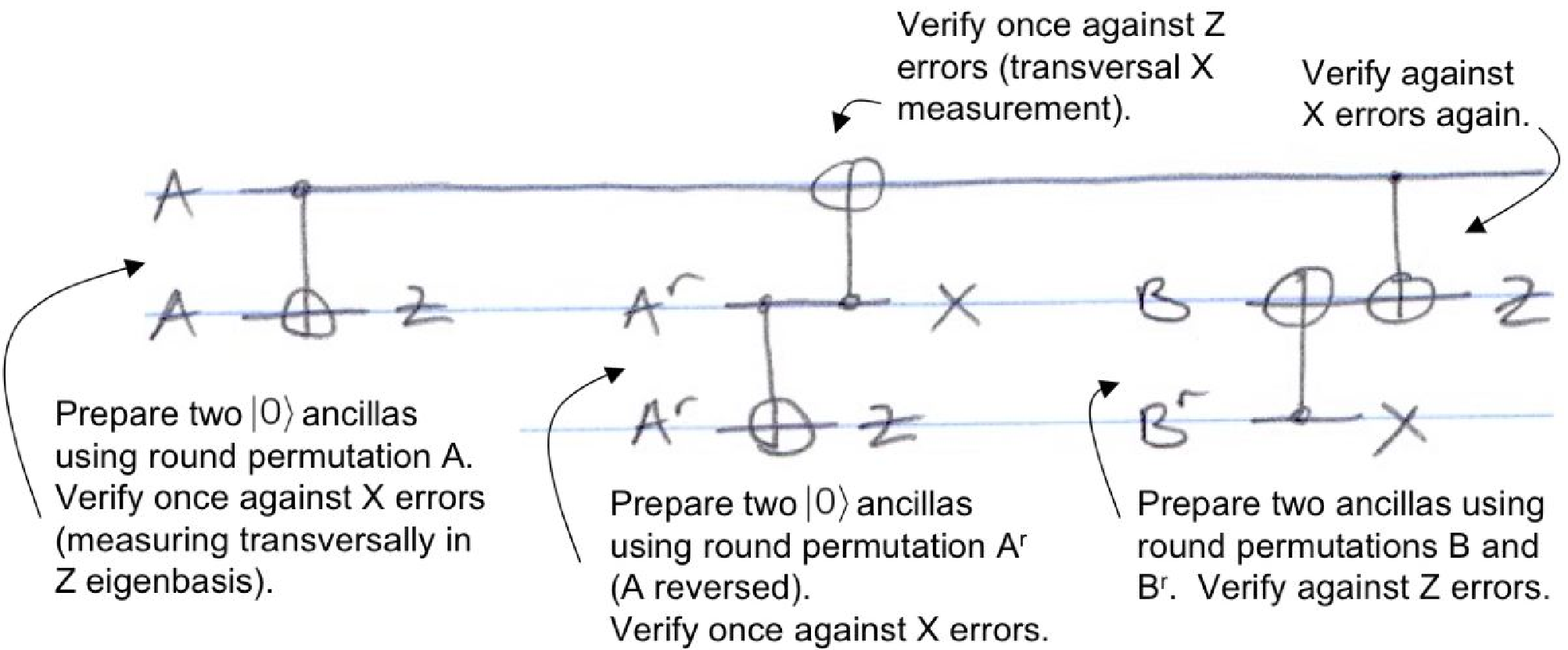}
\end{array}
\end{equation}
outputs a strictly fault-tolerant Golay code ancilla.  Here, each wire represents a $23$-qubit block, and CNOT gates represent transversal CNOTs.  At the left ends of the wires are preparations of $\ket{0}_L$ using the four different round permutations.  Z and X at the right end of a wire denote transversal measurement in the Z or X eigenbases, postselecting on no detected errors.

That the output is strictly fault tolerant does not follow immediately from an argument about Eq.~\eqref{e:eightqubitexample}, because different stabilizer elements can and do interact.  It requires a computer check of the different possibilities.  Checking X error fault tolerance reduces to checking X fault tolerance for the four circuits:
$$
\includegraphics{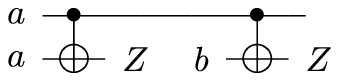}
$$
for $a \in \{A, A^r\}$ and $b \in \{B, B^r\}$.  This follows since we are concerned with preparations with up to three X failures.  (We don't have to consider the $A$, $B$ and $B^r$ permutations at once, for example, because that would use four failure locations.)  Checking Z error fault tolerance reduces to checking Z fault tolerance for the two circuits:
$$
\includegraphics{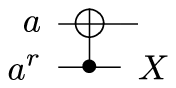}
$$
for $a \in \{A, B\}$, since we are concerned with preparations with up to two Z failures.  

Compared to the circuit of Eq.~\eqref{e:golaypurificationcircuit}, Eq.~\eqref{e:optimizedgolayverificationcircuit} requires half as many ancillas prepared.  
It is quite possible, though, that there are even better preparation procedures; one idea is to consider permutations of the CNOT gates which don't just reorder the rounds, but move gates between different rounds.  But note that although complexity and overhead is reduced, the noise threshold might also decrease, at least in a nonlocal gate model.  Intuitively, for the highest noise threshold, one would like the most reliable ancillary states possible, and so more verification rounds are better -- there is a direct tradeoff between overhead and reliability.  

\subsection{Definitions of fault-tolerant preparation} \label{s:differentftdefinitions}

Depending on how encoded ancilla states are to be used, and particularly with codes of distance $d >3$, different definitions of fault-tolerant preparation may be required.  

For example, a threshold proof along the lines of Aliferis-Gottesman-Preskill \cite{AliferisGottesmanPreskill05} or Reichardt \cite{Reichardt05distancethree} will work provided errors of weight $> 1$ are second-order events, as can be achieved with a single verification round.  However, the effective error rate will drop just quadratically $\eta \rightarrow \eta^2$ instead of $\eta \rightarrow \eta^{(d+1)/2}$.  Let us call a preparation procedure offering second-order protection \emph{weakly} fault tolerant.  

For \emph{strict} fault tolerance, on the other hand, an error of weight three should be a third-order event, weight-four errors should be fourth order, and so on. 

For an ancilla encoded into a distance-three code, to be used for error correction, the natural definitions of weak and strict fault tolerance coincide.  But, because X and Z failure events are correlated, there are still two variations in the definition.  A single Y failure, e.g., could cause a weight-one X error and weight-one Z error on different bits, for a weight-two error overall.  This still suffices for some threshold proof or estimation methods -- e.g., if one is separately estimating effective logical X and Z failure rates, then using the union bound to recombine them.  When one only requires bounds on the maximum of the weights of the X and Z error parts separately, call this \emph{CSS-type} fault tolerance.\footnote{This distinction between considering the total weight of an error and the maximum of the weights of the X and Z parts separately also arises in defining a perfect quantum code.  The $[[5,1,3]]$ code is perfect, because an error of total weight two is equivalent to a logical error plus a single bit error.  The $[[7,1,3]]$ code is perfect for X and Z errors separately, because it is based on the classical, perfect Hamming code.}
The two Golay-code verification procedures of Secs.~\ref{s:naiveverificationprocedure} and~\ref{s:optimizedverificationproceduregolay} above each give strict CSS-type fault tolerance.

Strict fault tolerance generally needs to be parameterized by an upper bound on the order of the events to be considered.  For weak fault tolerance, one is only concerned with showing that some events are second-order.  In Eq.~\eqref{e:xzerrorverificationtables}, we were only concerned with error orders up to four, allowing arbitrary errors to occur with fourth-order probability.  For an ancilla to be used for error correction, allowing arbitrary errors to occur as order $(d+1)/2$ events is often natural, since $(d+1)/2$ failures can cause an logical error (correction in the wrong direction).  However, if the ancilla is to be used for error \emph{detection}, then we'll want bounds on error weights caused by failures of orders going up at least to $d$.  

The required fault-tolerance definition also depends on the ancilla state being prepared: cat states for Shor-type error correction, $\ket{0}_L$ and $\ket{+}_L$ for Steane-type error correction, encoded Bell pairs $\tfrac{1}{\sqrt{2}}(\ket{00}_L+\ket{11}_L)$ (possibly with encoded unitaries applied) for Knill-type error correction/detection plus teleported computation.  In Chs.~\ref{s:postselectchapter} and~\ref{s:postselectdetailschapter}, we will define an encoding procedure for $\ket{\psi}_L$ as being suitably fault tolerant if it can be written
\def\incl #1#2{\makebox{\raisebox{#1}{\includegraphics[scale=.8]{images/#2}}}}
$$
\incl{0em}{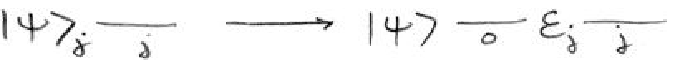}
$$
This notation will be explained later, but it roughly means that a noisy preparation of encoded $\ket{\psi}_L$ is the same as a perfect preparation of $\ket{\psi}$ followed by a particular kind of noisy encoder $\mathcal{E}$.  We will use several definitions for $\mathcal{E}$, including a version of CSS-type fault tolerance (in Sec.~\ref{s:unbiasednoise}).  With the strongest fault-tolerance definition, $\mathcal{E}(\ket{\psi})$ gives $\ket{\psi}_L$ with only bitwise-independent errors.  ($\mathcal{E}$ is not a physical operation, but a mathematical tool for the analysis.)

\chapter{Overview of postselection noise threshold proof} \label{s:overviewchapter}
\section{Fault tolerance}

Fault tolerance allows for computing reliably using noisy gates.  It is particularly important for quantum computers because of the fragility of entanglement.  Remarkably, though, classical fault-tolerance techniques carry over to the quantum case.  This chapter gives an overview of the postselection noise threshold proof, using primarily classical examples.  

How does fault tolerance work?  Consider the circuit we would like to run, perhaps to factor a large number.  It looks something like
\begin{equation} \label{e:overviewlinearcircuit}
\raisebox{3.5em}{
\Qcircuit @C=.5em @R=.5em @!R {
&\qw      &\qw     &\ctrl{1} &\qw     &\qw\\
&\targ    &\ctrl{2}&\targ    &\ctrl{1}&\qw\\
&\ctrl{-1}&\qw     &\targ    &\targ   &\qw\\
&\qw      &\targ   &\ctrl{-1}&\gate{X}&\qw
}}
\end{equation}
with time going from left to right.
The circuit has controlled-NOT gates and maybe some NOT gates (marked X).  The CNOT is defined by 
$$
\Qcircuit @C=.5em @R=.85em @!R {
\lstick{a}&\ctrl{1}&\rstick{a}\qw\\
\lstick{b}&\targ   &\rstick{a \oplus b}\qw
}
$$
It flips the target if and only if the control bit is set to one.

These gates are not yet enough; classically, we need something like an AND gate to compute nonlinear functions.  The reversible, quantum analog is called a Toffoli gate.  However, there is a general reduction to the case of gates which propagate Pauli operators linearly:
$$
\textrm{universal fault tolerance} \overset{\textrm{``magic"}}{\longrightarrow} \textrm{stabilizer operation fault tolerance}
$$
There is no sorcery here; the reduction technique is actually known as magic states distillation (Ch.~\ref{s:magicchapter}).  It allows us to consider initially just stabilizer operations, like the CNOT.  

What is the error model?  For now, assume perfect preparation and measurement of $\ket{0}$ and $\ket{1}$.  Model a noisy gate as a perfect gate, followed by \emph{independent}, probabilistic bit flip errors:
$$
\includegraphics{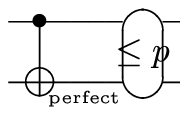}
$$
The circled noise location has three parameters, giving the probabilities of the three possible bit-flip errors on two bits:
$$
p_1 : \mathrm{XI} \qquad p_2 : \mathrm{IX} \qquad p_3 : \mathrm{XX}
$$
satisfying $p_1 + p_2 + p_3 \leq p$ for some small constant $p$.

To get around this noise, compute the whole circuit on top of an error-correcting code.  For example, we can use the three-bit repetition code, taking 
\begin{eqnarray*}
0 &\mapsto& 000 \\
1 &\mapsto& 111
\end{eqnarray*}
Then each ideal CNOT in Eq.~\eqref{e:overviewlinearcircuit} we'd like to apply is compiled into CNOTs across the code block:
\begin{equation} \label{e:cnotcompiledinto3bitrepetitioncode}
\raisebox{-3em}{\includegraphics{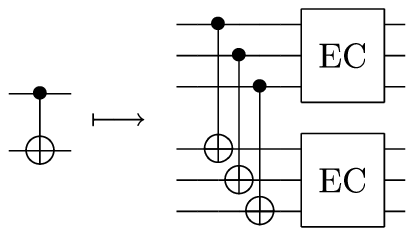}}
\end{equation}
The transversal gates are followed with a (noisy) error correction on each output block in order to prevent errors from accumulating.  (Error correction can be accomplished in the classical model with, e.g., majority gates, but different procedures are needed quantumly.)
By encoding into the repetition code, we improve reliability.  Previously, the CNOT failed with probability up to $p$.  Now, a single failure is correctable, and it takes two failures to circumvent the protection of a distance-three code.  

More formally, say we apply this circuit, and then perfectly decode each half, by taking the majority.  If there is at most one gate failure, then the same effect will be had by first decoding perfectly each block, and then applying a \emph{perfect} CNOT:
\begin{equation} \label{e:tutorialdecodecommute}
\raisebox{-5em}{\includegraphics{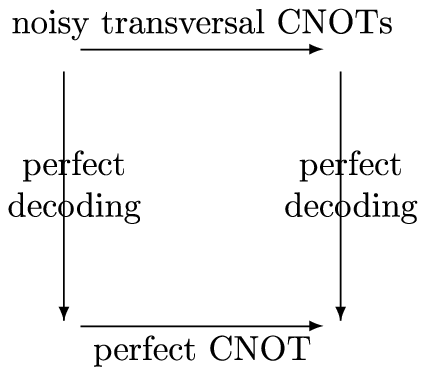}}
\end{equation}
If there are two or more failures, on the other hand, then the diagram may not commute.  
The probability of it not commuting 
is plotted below.  
Notice that beneath a constant error rate, there is improvement; the probability of a ``logical failure" is less than the initial failure rate.  
$$\includegraphics[scale=.8]{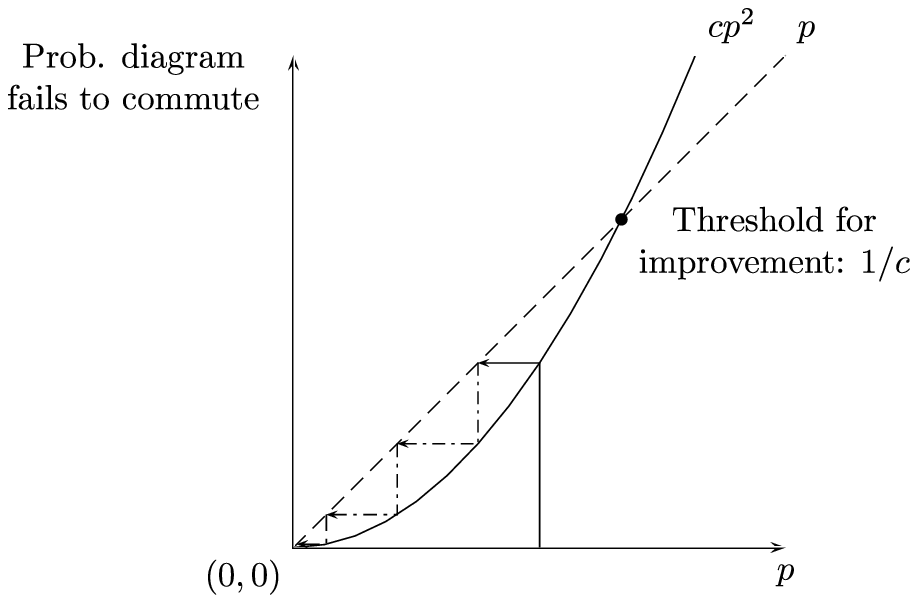}\qquad\qquad$$
Intuitively, below the threshold for improvement, effective error rates should drop rapidly, quadratically with each extra level of code concatenation (dot-dashed arrows above). 
It is difficult to make this intuition rigorous, though, because the error model is not preserved by gate compilation.  Still, Aharonov and Ben-Or, and independently Kitaev, in 1997 did manage to overcome the recursive analysis difficulty to prove the existence of a positive constant tolerable noise rate, or noise threshold~\cite{AharonovBenOr99, Kitaev96b}. 

Since 1997, researchers have worked to improve fault-tolerance schemes, and also the analysis of these schemes, in order to obtain higher proven or estimated tolerable noise rates.  Two results are particularly significant:
\begin{itemize}
\item Steane developed an optimized fault-tolerance scheme in 2002 to 2003~\cite{Steane02, Steane03}.  Using simulations, he \emph{estimated} that his scheme could tolerate noise rates as high as $3 \times 10^{-3}$ per gate.  
\item In 2005, Aliferis, Gottesman and Preskill, 
and this author independently, 
used more efficient proof techniques, with less worst-case slack, to prove the first rigorous noise lower bounds~\cite{AliferisGottesmanPreskill05, Reichardt05distancethree}.  The initial bounds were around $10^{-6}$ to $10^{-5}$ per gate, but unpublished calculations are now pushing a $10^{-4}$ lower bound.
\end{itemize}

\section{Postselection-based fault tolerance scheme}

Meanwhile, though, Knill developed a very different fault-tolerance scheme, which he estimated could tolerate noise rates as high as 3-6\% \cite{Knill05}, a breakthrough.  The scheme incorporates a number of new ideas, but probably the most important one is that it is based on using error \emph{detection} in Eq.~\eqref{e:cnotcompiledinto3bitrepetitioncode} instead of error correction.  When an error is detected, give up and start over -- or in other words condition, or postselect, on no detected errors.  

It is easy to see how this technique might improve the tolerable noise rate.  For example a distance-three code can only correct one error, and two errors can be corrected in the wrong direction.  But using error detection, two errors will be caught -- it takes three errors for a logical error to get past.  

On the other hand, one might think that an extensive reliance on error detection would cause an exponential overhead, because of all the times the computation is restarted.  This would be true classically.  But some purely quantum tricks, based on teleportation, allow the overhead to be efficient at least in theory (i.e., polynomial, or even polylogarithmic).  The overhead is still quite daunting, though, particularly at higher error rates.

Despite Knill's high threshold \emph{estimates}, however, there had been no proof that his scheme tolerated \emph{any} positive amount of noise at all.  Even the newer proof techniques fundamentally could not handle fault-tolerance schemes based on error detection.  This is because proofs roughly worked by trying to control most of the computer most of the time, but they allowed rare events to fall out of control.  Schematically, at a given time we can divide the probability mass of the quantum computer into a portion which is under control -- i.e., for which we have good bounds on the errors, or perhaps conditional probability bounds -- and the remaining probability mass over which we have no control, e.g.:
\begin{equation} \label{e:uncontrolledrenormalization1}
\raisebox{-5em}{\includegraphics[scale=.6]{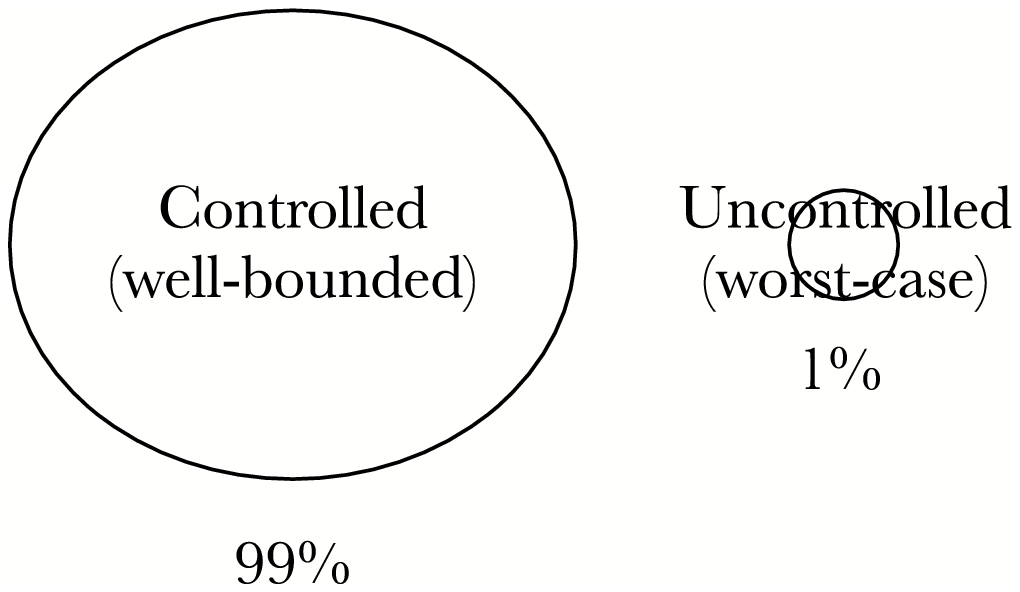}}
\end{equation}
The probability of surviving a round of error detection is rather small -- perhaps only a $1/99$ fraction of the controlled probability mass survives.  However, for the portion of probability mass which we lack any control, we must assume worst-case behavior.  Here, that means it could survive error detection with probability one:
$$
\qquad\quad\includegraphics[scale=.6]{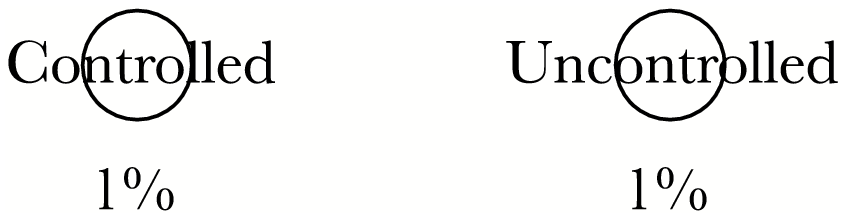}
$$
But then after renormalizing the distribution to condition on no detected errors, the fraction of probability mass which is out of control has exploded to half.  
$$
\quad\quad\includegraphics[scale=.6]{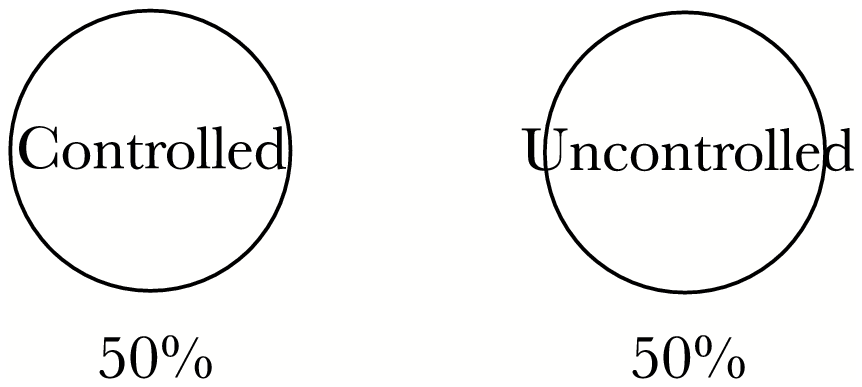}
$$
Intuitively, this maybe shouldn't happen.  The uncontrolled portion of the system ``should" be very bad, and should perhaps be even less likely to survive error detection than the controlled portion.  But in a proof, we have to assume worst-case behavior for uncontrolled events.
That these rare bad events can become exponentially more likely during renormalization frustrates previous proof techniques.  

I will sketch here a proof of a threshold for error-detection-based fault-tolerance schemes.  It gets around the difficulty described just above by maintaining control over \emph{all} the quantum computer, all the time.

\section{Sketch of postselection noise threshold analysis}

I will give just the main idea of the analysis.  Consider the CNOT gate, encoded using the two-bit repetition code, which can detect one error.  As in Eq.~\eqref{e:cnotcompiledinto3bitrepetitioncode}, it is compiled into transversal physical CNOT gates, followed by error detection on each output block.  
$$
\includegraphics{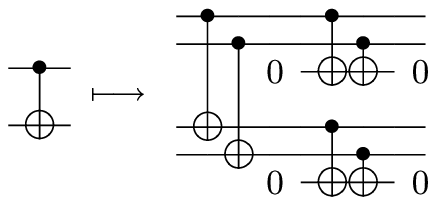}
$$
Error detection is implemented here by preparing a zero ancilla and applying CNOT gates into it.  If the control wires are the same -- both zero or both one -- then the ancilla will be unchanged.  But if the controls are not in the codespace, the ancilla will be flipped to one.  Therefore, condition on the ancilla ending up a zero.

To analyze this operation, we need some notation.  Define a noisy encoder as a perfect encoder followed by bitwise-independent noise.  (Such an encoder doesn't physically exist; it is a tool for the analysis.)
$$
\includegraphics[scale=.8]{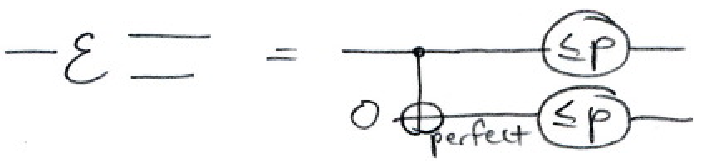}
$$
We analyze the logical CNOT gate on inputs with bitwise-independent errors.  
Ideally, we might like to say that if input errors are all independent, then so too will be the output errors:  
$$
\raisebox{-4em}{\includegraphics[scale=.8]{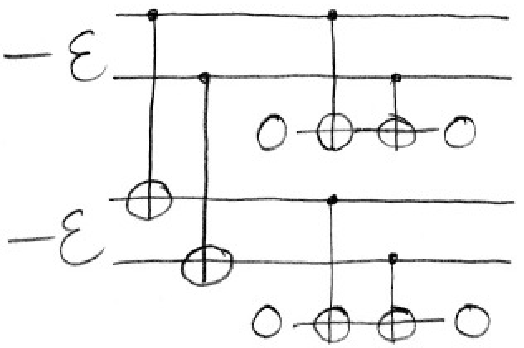}}
\quad\overset{?}{=}\quad
\raisebox{-1.8em}{\includegraphics[scale=.8]{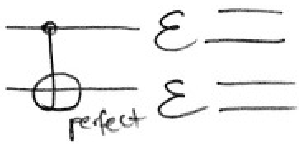}}
$$
This is not the case, however, because two errors in the first block will flip the whole block, therefore flipping the target block as well.  So the XXXX error occurs with probability $O(p^2)$, not the $O(p^4)$ probability that the right-hand side above asserts.  

Let us therefore revise our hypothesis to
\begin{equation} \label{e:independentbitandlogicalerrors}
\raisebox{-4em}{\includegraphics[scale=.8]{images/noisyencoderprecedingcircuit}}
\quad\overset{?}{=}\quad
\raisebox{-1.4em}{\includegraphics[scale=.8]{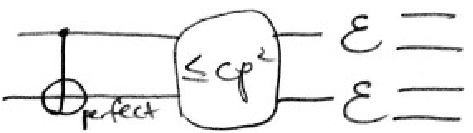}}
\end{equation}
That is, we allow for a correlated logical error on the two encoded output bits with $O(p^2)$ probability, independent of any further bit errors.  
This would still be very nice, as independent bit errors would make continuing the analysis to the next encoded CNOT gate easy; and independent logical errors would make extending the analysis to higher code concatenation levels easy.
But again, it is generically not the case.  Once the two blocks have interacted, they will be interdependent, and these dependencies cannot easily be removed.  (The right-hand side above has only seven free parameters -- four for the noisy encoders and three for the two-block logical error -- to describe a distribution on $2^4 = 16$ outcomes.)

It turns out, though, that the output distribution is very close to having this form:
$$
\includegraphics{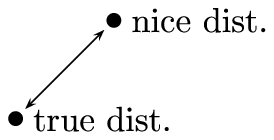}
$$
Eq.~\eqref{e:independentbitandlogicalerrors} holds approximately, and we may write
$$
\raisebox{-4em}{\includegraphics[scale=.8]{images/noisyencoderprecedingcircuit}}
\quad = \quad
\raisebox{-1.4em}{\includegraphics[scale=.8]{images/noisyencoderslogicalfollowingcircuit}}
\;+ \delta
$$
for some small error $\delta$.
But this sets up precisely the problematic situation we described in Eq.~\eqref{e:uncontrolledrenormalization1}.  The $\delta$ term is completely uncontrolled, so we have to assume worst-case behavior for it.  After the next postselection round, any upper bound on $\delta$ will be exponentially weaker.  

Fortunately, though, the output error distribution is quite close to \emph{many} of these nice error distributions, each satisfying the constraints on the right-hand side of Eq.~\eqref{e:independentbitandlogicalerrors}:
$$
\includegraphics{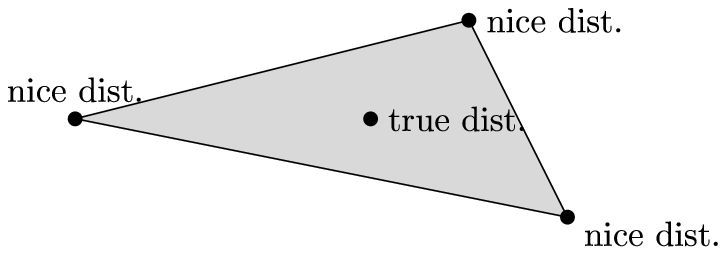}
$$
And it lies within their convex hull (in $2^4$ dimensions).  We write
\begin{equation} \label{e:overviewcnotmixingequality}
\raisebox{-4em}{\includegraphics[scale=.8]{images/noisyencoderprecedingcircuit}}
\quad = \quad
\left\{\raisebox{-1.4em}{\includegraphics[scale=.8]{images/noisyencoderslogicalfollowingcircuit}}\right\}
\end{equation}
meaning that the left-hand side can be rewritten as a mixture of distributions each having the form enclosed on the right-hand side.  

This condition is weaker than Eq.~\eqref{e:independentbitandlogicalerrors}, but sufficient.  The next encoded CNOT gate's inputs will not be bitwise independent -- but we can \emph{choose} one of the vertices of the mixture.  Then repeat the analysis, assuming bitwise-independent errors, to show that any possible output distribution can again be rewritten as a mixture of nice distributions.  
\begin{equation} \label{e:repeatedmixinganalysis}
\raisebox{-3.5em}{\includegraphics[scale=1]{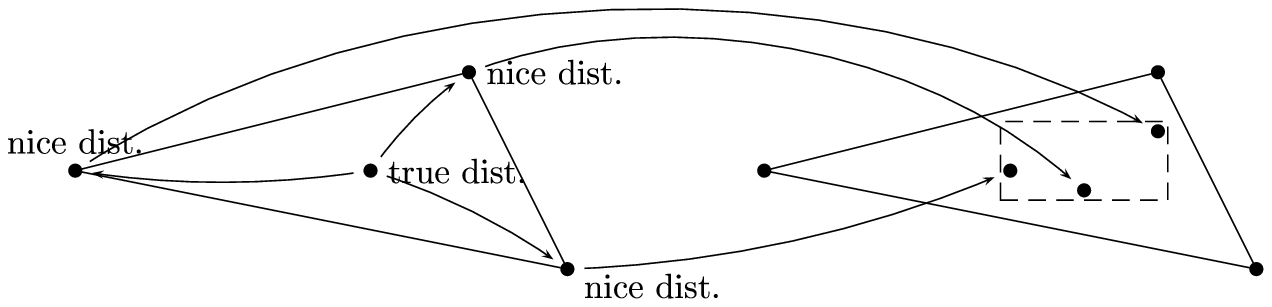}}
\end{equation}

How can we hope to prove this claim, though?  The output distribution is quite complicated, living in 16 dimensions.  It is easy to check that error orders are correct; see Table~\ref{f:overviewerrororders}.  
For example, the error IIIX (the last bit flipped) is a first-order event because there must be at least one failure for there to be an output error.  The circuit is also designed so that for two errors to survive on the output, there must have been two failures.  A single failure, for example in the first CNOT gates, can cause two errors, but such an event will be caught by error detection.

\def\IIII {\mathrm{IIII}}
\def\IIIX {\mathrm{IIIX}}
\def\IIXI {\mathrm{IIXI}}
\def\IIXX {\mathrm{IIXX}}
\def\IXII {\mathrm{IXII}}
\def\IXIX {\mathrm{IXIX}}
\def\IXXI {\mathrm{IXXI}}
\def\IXXX {\mathrm{IXXX}}
\def\XIII {\mathrm{XIII}}
\def\XIIX {\mathrm{XIIX}}
\def\XIXI {\mathrm{XIXI}}
\def\XIXX {\mathrm{XIXX}}
\def\XXII {\mathrm{XXII}}
\def\XXIX {\mathrm{XXIX}}
\def\XXXI {\mathrm{XXXI}}
\def\XXXX {\mathrm{XXXX}}
\begin{table}
\caption{} \label{f:overviewerrororders}
\begin{center}
\begin{tabular}{c|c}
\hline \hline
Error      & Probability        \\
\hline
IIII & $\Theta(1)$\\
IIIX, IIXI, IXII, XIII & $O(p)$\\
XXII, IIXX, XXXX, XIXI, XIIX, IXXI, IXIX & $O(p^2)$\\
IXXX, XIXX, XXIX, XXXI & $O(p^3)$ \\
\hline \hline
\end{tabular}
\end{center}
\end{table}

Aside from these constraints, however, the output error distribution is quite complicated.  
To work up to analyzing it 
let's
start by analyzing distributions on just two bits.  

\section{Two-bit mixing example} \label{s:tutorialmixingexample}

\def\II {\mathrm{II}}
\def\IX {\mathrm{IX}}
\def\XI {\mathrm{XI}}
\def\XX {\mathrm{XX}}
On two bits, there are three possible errors: IX, XI or XX.  Assume that $\pr[\IX] = \pr[\XI]$, perhaps by some symmetry.  Then, e.g., the distribution 
$$
\left(
\begin{array}{c}
\pr[\IX] = \pr[\XI] = 2p \\
\pr[\XX] = 5 p^2
\end{array}
\right)
$$
does not have bitwise independent errors -- there is a slight positive correlation.  But this distribution can be rewritten as 
$$
\left(
\begin{array}{c}
2p \\
5 p^2
\end{array}
\right)
= \frac{1}{2}
\left(
\begin{array}{c}
p \\
p^2
\end{array}
\right)
+ \frac{1}{2}
\left(
\begin{array}{c}
3p \\
9 p^2
\end{array}
\right) 
$$
a mixture of two distributions each with well-bounded, bitwise-independent errors.  (That the errors in each vertex of the mixture are appropriately bounded is important; \emph{any} distribution can be written as a mixture of the atoms $\pr[\II] = 1$, $\pr[\IX]=1$, $\pr[\XI]=1$ and $\pr[\XX] = 1$.)

Plotting the distributions, 
we see that $(2p,5p^2)$ could equally well have been rewritten as a mixture between $(0,0)$ and $(\tfrac{5}{2}p, (\tfrac{5}{2}p)^2)$:
$$
\includegraphics{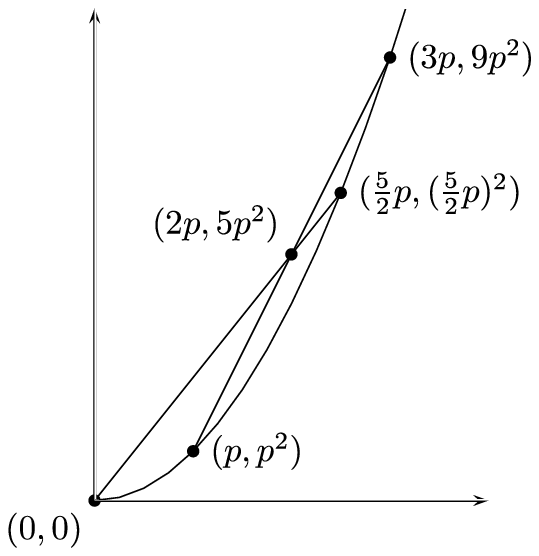}
$$
The latter mixture maintains tighter upper bounds on the first-order error probability, but worse lower bounds.  

Symmetrical bitwise-independent error distributions lie along the curve $y = x^2$ above.  Therefore, the convex hull of symmetrical bitwise-independent distributions in which $\pr[\IX] = \pr[\XI] \leq 3p$ is exactly the region between the curves $y = x^2$ and $y = 3 p x$:  
\begin{equation} \label{e:twobitexactmixing3}
\raisebox{-8em}{\includegraphics{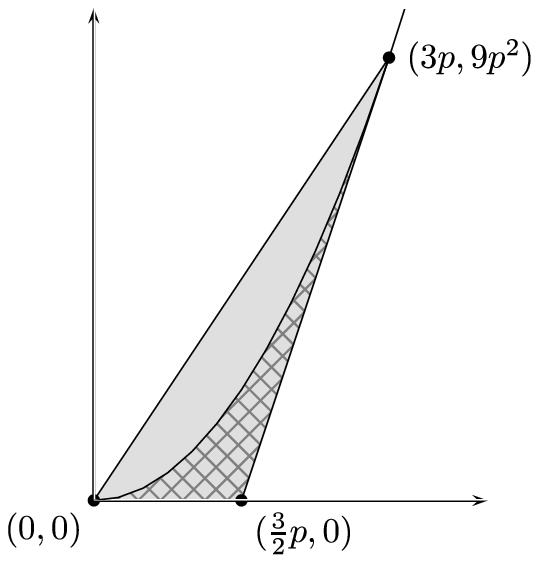}}
\end{equation}
Distributions \emph{beneath} the curve $y = x^2$ -- the crosshatched region above -- can be obtained by breaking the symmetry.  For example, mixing equally 
$$
\left(
\begin{array}{c}
\pr[\IX] = 3p, \pr[\XI] = 0 \\
\pr[\XX] = 0
\end{array}
\right)
\quad \text{and} \quad
\left(
\begin{array}{c}
\pr[\IX] = 0, \pr[\XI] = 3p \\
\pr[\XX] = 0
\end{array}
\right)
$$
gives the symmetrical distribution $(\pr[\IX] = \pr[\XI] = \tfrac{3}{2}p, \pr[\XX] = 0)$.

The region~\eqref{e:twobitexactmixing3} is a two-dimensional cross-section of the set of two-bit error distributions which can be rewritten as a mixture of bitwise-independent distributions with $\pr[\IX], \pr[\XI] \leq 3p$:
$$
\includegraphics{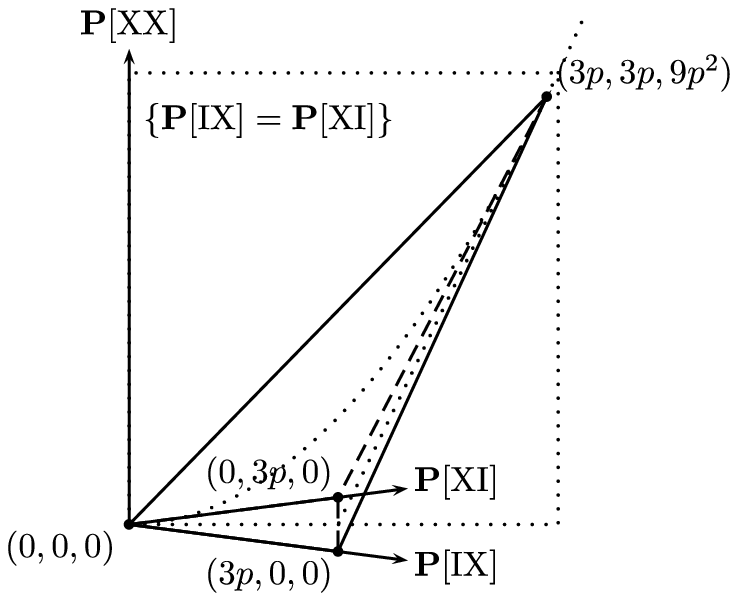}
$$

\section{Two approaches to mixing more complicated distributions}

Returning to our problem, proving Eq.~\eqref{e:overviewcnotmixingequality}, it is hard to imagine repeating analogous arguments.  In the two-bit example, we had to characterize the convex hull of just four distributions -- $(0,0,0)$, $(3p,0,0)$, $(0,3p,0)$, $(3p,3p,9p^2)$.  But now, we want to characterize the convex hull in $16$ dimensions of $2^4 \times 4$ extremal ``nice" distributions, satisfying the error bounds and independence constraints on the right-hand side of Eq.~\eqref{e:overviewcnotmixingequality}: For each of the four bits, we want to consider that bit error rate being either $0$ or $p$, and for the two-block logical failure location we want to consider the four possibilities $(\pr[\IX],\pr[\XI],\pr[\XX]) \in \{(0,0,0), (cp^2,0,0), (0,cp^2,0), (0,0,cp^2)\}$.  (It is sometimes easier to consider each of the eight possibilites, $\pr[\IX],\pr[\XI],\pr[\XX] \in \{0, \tfrac{c p^2}{3}\}$; the convex hull of the $2^4 \times 2^3$ distributions will then be smaller -- more conservative.)

Additionally, Table~\ref{f:overviewerrororders} puts only weak constraints on the distribution we are trying to rewrite as a mixture.  
 
Even worse, the quantum case is higher dimensional.  The smallest quantum error-detecting code uses four qubits, and even after various simplifications, the output error distribution will still be in 64 dimensions.  (Enforcing symmetries can reduce the dimensionality further, but only at the first level of encoding.)

There are two different approaches we follow to show that our distribution can be rewritten as a mixture of the extremal nice distributions:
\subsection{Low-dimensional numerical mixing}
Compute the output error distribution numerically -- or compute upper and lower bounds for each coordinate, as sketched in Eq.~\eqref{e:repeatedmixinganalysis}.  The output error distribution then lies somewhere inside an axis-aligned rectangular prism.  Verify that the entire prism lies within the convex hull of a set of nice error distributions.  
(Use linear-programming software to check that each vertex of the prism individually lies within the convex hull of extremal nice distributions 
-- e.g., $(0,0,0), (3p,0,0), (0,3p,0), (3p,3p,9p^2)$ in the two-bit mixing example above.)  
$$
\includegraphics[scale=1.5]{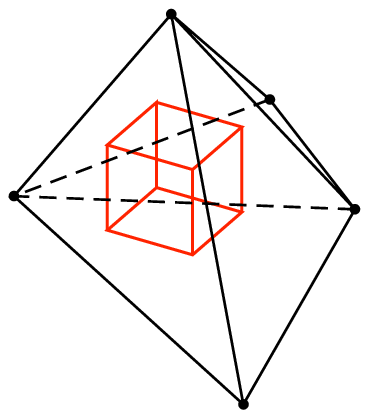}
$$
This is the approach we follow to analyze the first concatenation level in Ch.~\ref{s:numericalchapter}.  It is computationally infeasible in higher dimensions, and also is merely numerical.  
\subsection{Higher dimensions or existence proofs -- pull back to a simplex}
In order to prove that the analysis can be repeated indefinitely, to reduce effective error rates to \emph{arbitrarily} low levels, we need a more general analysis.  
In Sec.~\ref{s:tutorialmixingexample}, there were four different events -- II, IX, XI and XX -- and four extremal nice distributions.  Their convex hull was therefore just a simplex in three dimensions (probabilities in a distribution sum to one).  Similarly, characterizing the convex hull of, say, $2^7$ distributions over $\{0,1\}^7$ is straightforward; the Mixing Lemma we will present in Ch.~\ref{s:postselectchapter} gives simple, closed-form equations for the simplex's faces.  But the convex hull of many distributions over $\{\text{I}, \text{X}\}^4$ will have a more complicated structure, essentially because different errors can have the same effect.  (E.g., XXXX can occur with $O(p^2)$ probability as a logical error, but also with $O(p^4)$ probability as four separate bit errors, or as XIII + IXII + IIXX, etc.) 

To avoid this complication, we choose an injection into the probability space $\{\text{I}, \text{X}\}^4$ from the larger space $\{0,1\}^7$ in which errors caused by different sets of events are distinct and characterizing the convex hull is easy.  Extend linearly to all $\{0,1\}^7$ the map:
\begin{eqnarray*}
1000000 &\mapsto& \XIII \\
0100000 &\mapsto& \IXII \\
0010000 &\mapsto& \IIXI \\
0001000 &\mapsto& \IIIX \\
0000100 &\mapsto& \IIXX \\
0000010 &\mapsto& \XXII \\
0000001 &\mapsto& \XXXX 
\end{eqnarray*}
(The first four bits in $\{0,1\}^7$ make for bit errors in $\{\text{I},\text{X}\}^4$, and the last three bits give logical errors.)
Properly bounded, bitwise-independent distributions over $\{0,1\}^7$ pull forward to nice distributions over $\{\text{I}, \text{X}\}^4$.  
The map in linear, so convex combinations are preserved.  This lets us apply our understanding of mixtures of distributions over $\{0,1\}^7$ to characterize mixtures of nice distributions in $\{\text{I}, \text{X}\}^4$: A given error distribution is a mixture of the nice distributions if and only if it can be pulled back to $\{0,1\}^7$ and then rewritten as a mixture of bitwise-independent distributions.  These conditions are simple enough for easy noise threshold existence proofs, even in high dimensions, and can be checked with a linear program for numerical threshold lower bound calculations.

A more detailed description of this approach is given in Sec.~\ref{s:reliablebellstate}, and it is applied numerically in Ch.~\ref{s:numericalchapter}. 

\chapter{Postselection threshold for stabilizer operations against biased noise} \label{s:postselectchapter}
In this chapter, we prove the existence of a tolerable \emph{bit flip} noise threshold for the CNOT gate, using a scheme based on postselection.  Chapter~\ref{s:postselectdetailschapter} extends the proof to general Pauli noise, and Ch.~\ref{s:magicchapter} extends the gate set to be universal.

\section{Intuition} \label{s:intuition}

The intuitive difficulty in proving a threshold with postselection is possible negative correlations between logical errors (on the encoded state) and bit errors (away from the encoding).  (The state of the system, a distribution over pure states, can be specified by the ideal state plus a probability distribution of errors.)  For example, say in trying to prepare the $N$-bit encoded/logical state $\psi_L$, we get a logical error, $(E\psi)_L$, with some small probability.  Now postselect on no bit errors.  The good case $\psi_L$ survives with probability at least $(1-\eta)^N$ if the bit error rate is $\leq \eta$.  But if we lack any lower bounds on the bit error rate in $(E\psi)_L$, then it is possible that the logical error survives with probability one, becoming exponentially more likely after renormalizing the probability distribution.\footnote{If $\psi_L$ is encoded with $k$ levels of concatenation of an $n$-bit, $t$-error-correcting code, so $N = n^k$, then the probability of $(E \psi)_L$ should be $\sim\!(c \eta)^{(t+1)^k}$ for $c$ some constant determining the threshold for improvement.  But the renormalization penalty of $\sim\!(1-\eta)^{n^k}$ overwhelms this advantage.}

If we could prove that physical errors were completely uncorrelated from logical errors -- i.e., that errors within the codespace were independent of errors going outside the codespace, so $\psi_L$ and $(E \psi)_L$ had identical bit error rates -- then the above-described problem could never occur.  Postselecting on no bit errors would improve bit reliability without affecting the distribution of logical errors.  However, this is certainly not the case.  The true error distribution has all sorts of correlations, both between different code-concatenation levels (so postselecting on no bit errors can increase the probability of logical errors, as above), and between different code blocks (so postselection in one part of the computer can harm the state of the rest of the computer).

In fact, though, the true error distribution can be written as a \emph{mixture} of error distributions which have uncorrelated errors.  By itself, that is a trivial statement, as every probability distribution can be so written -- the set of distributions is a simplex whose (deterministic) vertices have strong independence properties.  However, the mixture can be written just over nice error distributions, in which errors are not only independent, but also bounded in probability.  As we carry out the analysis, then, at every step we simply condition on a certain nice error distribution from this mixture -- it doesn't matter which one!  After implementing, say, a logical CNOT gate, the error distribution loses its independence properties, but it can again be rewritten as a mixture of distributions with bounded-probability independent errors -- this strong inductive hypothesis is restored.  

Rewriting probability distributions with small correlations as mixtures of probability distributions with bounded-probability independent events is our main technical tool.  
The Mixing Lemma tells us exactly when a distribution $\pr[\cdot]$, with correlations between $n$ events, can be rewritten as a mixture of nice distributions in which those events are independent:

A point $(q_1, \ldots, q_n) \in [0,1]^n$ corresponds to a bitwise-independent distribution over $\{0,1\}^n$, in which the probability of $x$ is $\prod_{i=1}^n q_i^{x_i} (1-q_i)^{1-x_i}$.  Define the lattice ordering $y \preceq x$ for $x,y \in \{0,1\}^n$ if considered as indicators for subsets of $[n]$, $x \subseteq y$.  
\theoremstyle{theorem}
\newtheorem*{mixinglemma}{Mixing Lemma}
\begin{mixinglemma}
The convex hull, in the space of distributions over $n$-bit strings, of the $2^n$ bitwise-independent distributions $\{0,p_1\}\times\{0,p_2\}\times\cdots\times\{0,p_n\}$ is given exactly by those $\pr[\cdot]$ satisfying the inequalities, for each $x \in \{0,1\}^n$: 
\beq \label{e:mixinglemma}
\sum_{y \preceq x} (-1)^{\abs{x\oplus y}} \frac{\pr[\{z \preceq y\}]}{p(\{z \preceq y\})} \geq 0 \enspace ,
\eeq
where $p(\{z \preceq y\}) = \prod_{i=1}^n \delta_{y_i,1} p_i$, i.e., the probability of $\{z : z \preceq y\}$ in the distribution $(p_1,\ldots,p_n)$.
\end{mixinglemma}
\noindent  Note that this key lemma is completely classical, and so therefore is the essence of our argument.  The lemma's proof is deferred to Sec.~\ref{s:mixinglemmaproof}.

We illustrate the mixing technique in this chapter by applying it to a simple toy problem: fault-tolerance for CSS-type stabilizer operations against bit flip errors, using the concatenated two-bit repetition code with a postselection-based scheme.  
We prove that there exists a constant positive threshold for this postselection-based scheme.  The technique generalizes further, to full universality with arbitrary Pauli errors, but most of the key insights already appear from considering just this simple example.  (Section~\ref{s:extensions} briefly describes the tricks used to extend the technique; following chapters will explain them in detail.)

\section{Proof overview}

We introduce a simple independent bit flip noise model.  We then give several lemmas each roughly saying that an encoded circuit element has the correct logical effect -- except for rare logical errors -- and outputs blocks with only weakly correlated errors (ready for the next logical gate).  Applying these lemmas at a high enough level of concatenation, logical errors will be vanishingly rare, so the encoded circuit accurately simulates the initial ideal circuit.

The key lemma required is for the encoded CNOT gate, which implemented naively would create strong bit error correlations across different blocks.  Preventing such correlations reduces to preparing an encoded Bell pair with bit errors independent across its two halves.  It is probably impossible to prepare such a state, but we can prepare a encoded Bell pair such that the error distribution can be rewritten as a mixture of nearby distributions in each of which errors are independent across the two halves.

\section{Error model} \label{s:postselecterrormodel}

Assume perfect preparation and measurement of $\ket{0/1}, \ket{\pm} = \tfrac{1}{\sqrt{2}}(\ket{0}\pm\ket{1})$ qubits, but noisy physical controlled-NOT (CNOT) gates.  Each physical CNOT gate applies an ideal CNOT gate, then fails probabilistically and independently with an error rate $\leq \eta_0$, giving bit flip (X) errors on one or both of the affected qubits.\footnote{That is, one of II, IX, XI or XX is applied to the output each with some probability, with the total probability of the IX, XI and XX events bounded by $\eta_0$.  This is not an adversarial error model.}  (The ideal CNOT gate is defined by $\mathrm{CNOT}\ket{a,b} = \ket{a,a\oplus b}$, $a,b \in \{0,1\}$.)

In circuit diagram notation, we write
\begin{center}
\includegraphics[scale=.85]{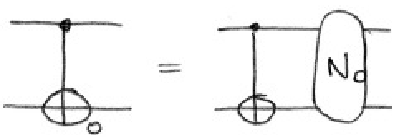}
\end{center}
Here, the left CNOT$_0$ is a physical, noisy CNOT gate, while the right CNOT is ideal.  The circled $N_0$ denotes introduction of IX, XI or XX errors with total probability at most $\eta_0$.

\section{Goal} \label{s:postselectgoal}

Fault tolerance is concerned with simulating ideal circuits using unreliable components.  
Say we have an inputless ideal circuit $\mathcal{C}$ which merely prepares $\ket{0/1}$ and $\ket{\pm}$ qubits and applies CNOT gates to output some quantum state $\ket{\psi}$.  We construct a fault-tolerant version of $\mathcal{C}$, $\mathrm{FT}\mathcal{C}$, by computing on top of the repeatedly concatenated two-bit repetition code.  The two-bit repetition code maps $0$ to $00$ and $1$ to $11$, and detects one bit flip error.  Concatenated on itself $k$ times, it becomes the $2^k$-bit repetition code, mapping $b$ to $b^{2^k}$ for $b \in \{0,1\}$.

By assumption, $\ket{0}_k = \ket{0^{2^k}}$ and $\ket{1}_k = \ket{1^{2^k}}$ can be prepared perfectly.  We need to show how to prepare reliably $\ket{+}_k = \tfrac{1}{\sqrt{2}}(\ket{0}_k + \ket{1}_k) = \tfrac{1}{\sqrt{2}}(\ket{0^{2^k}}+\ket{1^{2^k}})$ (a $2^k$-bit GHZ or cat state) and how reliably to apply encoded CNOT gates.  

What does it mean to do these operations ``reliably?"  Denote by \raisebox{-1.3ex}{\includegraphics[scale=.75]{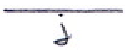}} a block of $2^j$ qubits.  Define a noisy encoding operator $\tilde{\mathcal{E}}_j$ recursively by 
\begin{gather} \label{e:noisye}
\includegraphics[scale=.85]{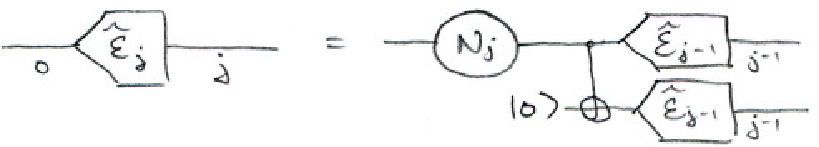} \\
\includegraphics[scale=.85]{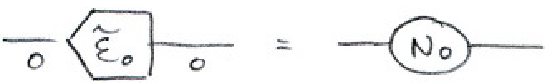} \nonumber 
\end{gather}
where the circled $N_j$ means 
independent introduction of a bit flip error with probability $\leq \eta_j = (c \eta_0)^{2^j}$ (some constant $c$).  $\tilde{\mathcal{E}}_j$ is not a physical operation, but is useful in our analysis.  

Reliable preparation of $\ket{+}_k$ means preparing $\tilde{\mathcal{E}}_j(\ket{+})$:
\begin{equation} \label{e:encplus}
\includegraphics[scale=.85]{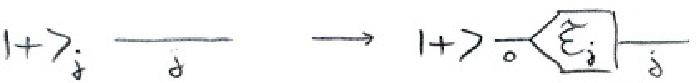}
\end{equation}
That is, noisy preparation of $\ket{+}_k$ should be the same as ideal preparation of $\ket{+}$, followed by a noisy encoding operator.  Here we write an arrow since error correlations mean we cannot enforce equality.  Our procedure for preparing $\ket{+}_k$ will produce a distribution over states which can be written as a mixture of noisy encodings of $\ket{+}$ with differing, but bounded, error parameters.  

Reliable application of a CNOT$_j$ gate means that we can commute noisy encoding operators past the encoded CNOT gate:
\begin{equation} \label{e:cnotcommute}
\includegraphics[scale=.85]{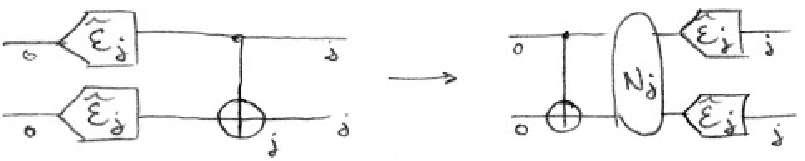}
\end{equation}
Once again, the left-hand side will be a mixture of diagrams of the type appearing on the right, all with bounded error parameters.  The right-hand CNOT gate is ideal, whereas the left-hand CNOT$_j$ indicates some implementation of an encoded gate.  The implementation will be specified below.

The simulating circuit, $\mathrm{FT}\mathcal{C}$, takes every preparation of $\ket{\phi}$ in $\mathcal{C}$ ($\phi \in \{0,1,+,-\}$) and replaces it with preparation of $\ket{\phi}_k$, and replaces every ideal CNOT in $\mathcal{C}$ with CNOT$_k$.  To analyze $\mathrm{FT}\mathcal{C}$, one repeatedly applies the above relationships to introduce noisy encoding operators $\tilde{\mathcal{E}}_k$ and then commute them past the CNOT$_k$s to the end of the circuit.  
One ends up with a mixture of diagrams, each looking like the ideal $\mathcal{C}$ with noise locations $N_k$ interspersed, and noisy encoding operators applied to the output qubits.\footnote{This formalism of commuting encoding operators through the circuit,
is similar to the commutative diagrams (with decoding operators) used in \Refs\cite{Reichardt05distancethree,AliferisGottesmanPreskill05} to define logical success or failure of an encoded gate.  Here, the noisy encoding operators do not commute past perfectly, for we have to take a probabilistic mixture of diagrams on the right-hand side.}  
This is our final goal; provided $k$ is large enough, so $\eta_k$ small enough (for $\eta_0 < 1/c$), it is unlikely that any of the errors $N_k$ actually occur, so we have a reliable simulation of $\mathcal{C}$.

More accurately, we want to guarantee that with high probability measurements at the end of $\mathrm{FT}\mathcal{C}$ give the same classical result as measurements at the end of $\mathcal{C}$.  It is straightforward to implement measurements; e.g., \raisebox{-1ex}{\includegraphics[scale=.75]{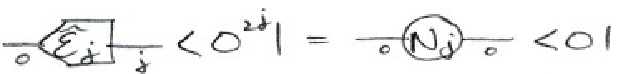}} (proof deferred to Sec.~\ref{s:postselectmeasurement}).
The more important extensions, beyond this toy error model and to full universality, are discussed in Sec.~\ref{s:extensions}.

\section{Proof overview}

\subsection{Reliable preparation of $\ket{+}_j$}

The proof of Eqs.~\eqref{e:encplus},\eqref{e:cnotcommute} is by induction.  The base cases, $j=0$, are immediate, by definition of the error model.  

We implement reliable preparation of $\ket{+}_j$ as a CNOT$_{j-1}$ from $\ket{+}_{j-1}$ into $\ket{0}_{j-1}$:
\def\incl #1#2{\makebox{\raisebox{#1}{\includegraphics[scale=.8]{images/#2}}}}
\begin{eqnarray*}
\incl{-1.5ex}{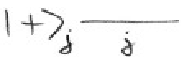} &=& \incl{-2.8ex}{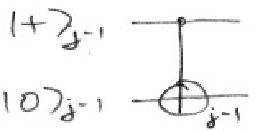} \\
&\rightarrow& \incl{-2.5ex}{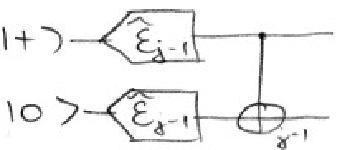} \\
&\rightarrow& \incl{-2ex}{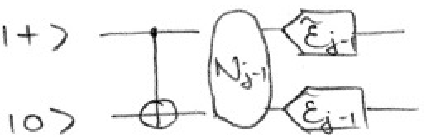} \\
&=& \makebox{\incl{-2ex}{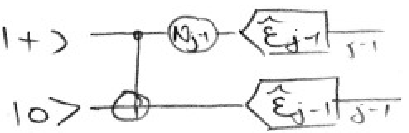}} \\
&=& \incl{-1.5ex}{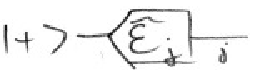}
\end{eqnarray*}
Here, the second and third lines follow from the level-$(j-1)$ versions of Eqs.~\eqref{e:encplus} and~\eqref{e:cnotcommute}, respectively.  
For the fourth line: Flipping both bits has no effect on $\tfrac{1}{\sqrt{2}}(\ket{00}+\ket{11})$, 
so XX is equivalent to II (no error) and IX is equivalent to XI.  
Thus set the probability of an error on bit two to zero, 
trivially independent of errors on bit one.  The last equality is by definition of a noisy encoder $\tilde{\mathcal{E}}_j$.  
(This requires adjusting the constant parameters of $\tilde{\mathcal{E}}_{j-1}$.  
A more careful analysis would track these parameters in order to determine the constant threshold, but to prove just the existence of a threshold, one merely has to check that the parameters stay under control.)

\subsection{CNOT gate implementation} \label{s:cnotteleport}

The fault-tolerant CNOT gate will be implemented by simultaneous teleportation and error-detection, similar to Knill's fault-tolerance scheme \cite{Knill03erasure, Knill04schemes,Knill05}.  
One can verify that \\
\beq\label{e:cnotreducestobellpairs}
\makebox{\raisebox{-10ex}{
\includegraphics[scale=.5]{images/cnotreducestobellpairs}
}}
\eeq
where $\ket{\psi} = \tfrac{1}{\sqrt{2}}(\ket{00}+\ket{11})$ a Bell pair, and $\bra{0}$, $\bra{+}$ denote postselected measurement of $0$ and $+$, respectively.\footnote{The success probability of this gadget is exactly $1/16$, although teleportation can be made deterministic.}  

In order to implement CNOT$_j$, then, it therefore suffices to create level-$j$ encoded Bell pairs $\ket{\psi}_j$ with independent errors across the two halves (using CNOT$_{j-1}$s, $\ket{+}_{j-1}$ and $\ket{0}_{j-1}$).  For then the two CNOT$_{j-1}$s used to implement the first logical CNOT in 
Eq.~\eqref{e:cnotreducestobellpairs}, 
between the two Bell pairs, will create correlations only in blocks about to be measured anyway, not in the output blocks.  The measurement $\bra{0}$ is implemented at level $j$ by transversal measurement $\bra{0^{2^j}}$ -- i.e., postselection on no detected X errors -- while measurement $\bra{+}$ can be implemented as $\bra{+^{2^j}}$.
((This argument can be made rigorous by pushing noisy encoders through, as we did to analyze reliable preparation of $\ket{+}_j$; see Sec.~\ref{s:cnotreductionanalysis}.)

That is, proving Eq.~\eqref{e:cnotcommute} reduces to giving a reliable preparation procedure for $\ket{\psi}_j$ satisfying:
\begin{equation} \label{e:ebellpair}
\includegraphics[scale=.95]{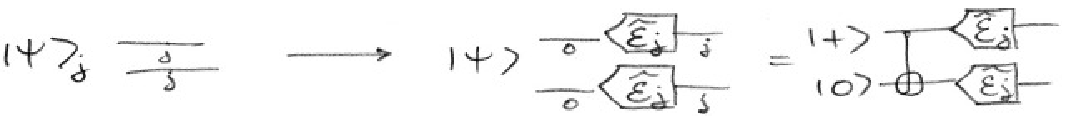}
\end{equation}

\subsection{Reliable preparation of $\ket{\psi}_j$} \label{s:reliablebellstate}

There are various ways of reliably preparing $\ket{\psi}_j$, and the choice of method has a large effect on the threshold for a particular scheme.  Here, we choose one of the simplest, shown on the left-hand side below (the boxed $\eta_{j-1}$s will be explained shortly):
\begin{eqnarray} \label{e:ebellprepproof}
\incl{-4em}{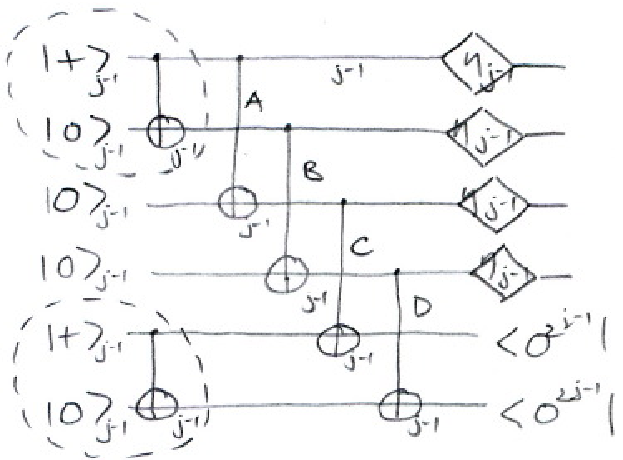} &\rightarrow& \incl{-3.7em}{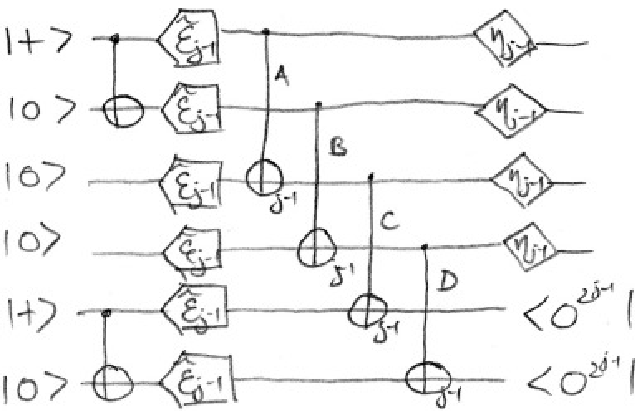} \\
&\rightarrow& \incl{-4em}{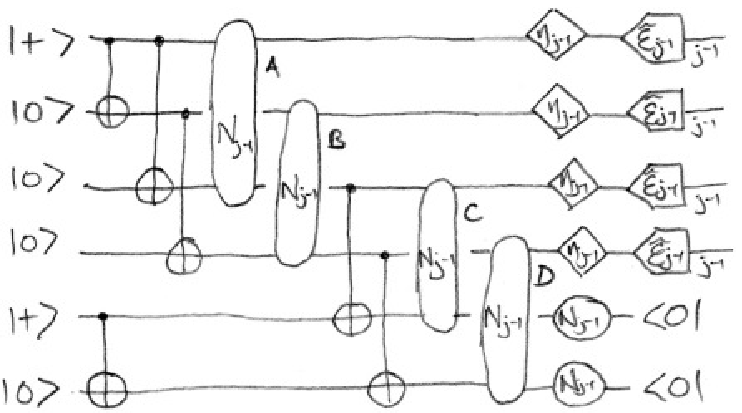} \nonumber \\ 
&\rightarrow& \incl{-2.5em}{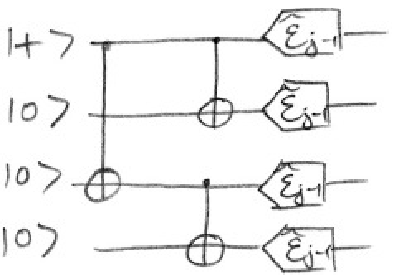} \nonumber \\
&=& \incl{-1.2em}{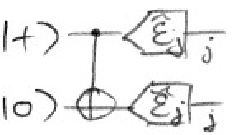} \nonumber 
\end{eqnarray}
The idea of this method is that CNOTs A and B prepare an encoded Bell pair with error correlations between its two halves.  CNOTs C and D are used to check for errors in the second half.  An error is caught if the measured block is out of the codespace, i.e., on outcomes $01$ or $10$.

In the first line above, we used Eq.~\eqref{e:encplus} twice, and in the second line used Eq.~\eqref{e:cnotcommute} at level $j-1$ as well as the measurement rule \includegraphics[scale=.75]{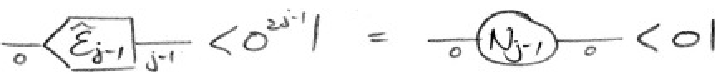}. 

\def\IIII {\mathrm{IIII}}
\def\XIII {\mathrm{XIII}}
\def\IXXX {\mathrm{IXXX}}
\def\IXII {\mathrm{IXII}}
\def\IIXI {\mathrm{IIXI}}
\def\IIIX {\mathrm{IIIX}}
\def\XXII {\mathrm{XXII}}
\def\XIXI {\mathrm{XIXI}}
\def\XIIX {\mathrm{XIIX}}
\def\IXXI {\mathrm{IXXI}}
\def\IXIX {\mathrm{IXIX}}
\def\IIXX {\mathrm{IIXX}}
\def\XXXX {\mathrm{XXXX}}

The third line, rewriting the distribution of level-$(j-1)$ errors after postselecting on acceptance as a mixture of independent error distributions, is the main step.  It can be checked directly by computing the convex hull of appropriately bounded bitwise-independent $X$ error distributions on the target encoded state: $16$ points $(0,\eta_{j-1}) \times \cdots \times (0,\eta_{j-1})$.  These points lie in only $8$ dimensions (\emph{not} $16$ dimensions labeled by $\{0,1\}^4$); since $\ket{0000}+\ket{1111}$ is unaffected by applying flipping all four bits, the different possible errors are $\IIII$ (no error), $\XIII$ (flip the first bit, equivalent to $\IXXX$), $\IXII$, $\IIXI$, $\IIIX$, $\XXII$, $\XIXI$ and $\XIIX$.\footnote{For explicit numerical calculations, linear programming software can be used to check that the convex hull of a given set of points contains a given distribution (or, all distributions satisfying certain coordinate-wise upper and lower bounds).}

The Mixing Lemma is therefore not required in this simple setting, if you are willing to get your hands dirty calculating the convex hull.  But more general error models require a larger error-detecting code and hence a larger ancilla state, and the symbolic calculation of the convex hull of a large number of points in high dimensions can be very difficult.  The Mixing Lemma gives a simple closed form for the convex hull of independent error distributions in $\{0,1\}^n$.  To illustrate its use in general, we apply it here (somewhat conservatively).

We would like to show that a distribution satisfying certain bounds lies in the convex hull of the distributions $(0,\eta_{j-1}) \times \cdots \times (0,\eta_{j-1})$, in the space $\{0,1\}^4 \mod \XXXX$.  But the Mixing Lemma only applies to points in $\{0,1\}^4$.  To apply it here, we need to \emph{linearly embed} $\{0,1\}^4/ \XXXX$ into $\{0,1\}^4$.  The simplest embedding is to evenly divide the probability mass of an error among those corresponding bit strings with minimum Hamming weight.  That is, map $\IIII$ to $0000$, $\XIII$ to $1000$, and divide the probability mass on the error $\XXII \sim \IIXX$ evenly between $1100$ and $0011$.

The Mixing Lemma gives $2^4$ inequalities to satisfy with $p_1 = p_2 = p_3 = p_4 = \eta_{j-1}$.  All except those for $x \in \{0,1\}^4$ with $\abs{x} \leq 1$ are automatic (since no probability mass has been put on strings of weight three or four).  These remaining inequalities are
\begin{gather*}
\begin{split}1 - \frac{1}{p_1}\pr[\{z \preceq 1000\}] - \cdots - \frac{1}{p_4}\pr[\{z \preceq 0001\}] \\ + \frac{1}{p_1 p_2}\pr[1100] + \cdots + \frac{1}{p_3 p_4} \pr[0011]  
\geq  0 \end{split}\\
\begin{split}\tfrac{1}{p_1}\pr[\{z \preceq 1000\}] - \tfrac{1}{p_1 p_2}\pr[1100] - \cdot\cdot - \tfrac{1}{p_1 p_4}\pr[1001]  \geq  0 \end{split}\\
\vdots \\
\begin{split}\tfrac{1}{p_4}\pr[\{z \preceq 0001\}] - \tfrac{1}{p_1 p_4}\pr[1001] - \cdot\cdot - \tfrac{1}{p_3 p_4}\pr[0011]  \geq  0 \end{split}
\end{gather*}
for $x = 0000, 1000, \ldots, 0001$, respectively.
It is sufficient to check instead the stronger inequalities
\begin{gather*}
\begin{split} \frac{1}{p_1}\pr[\{\XIII,\XXII,\XIXI,\XIIX\}] + \cdots \\+ \frac{1}{p_4}\pr[\{\IIIX,\XIIX,\IXIX,\IIXX\}] \leq 1 \end{split} \nonumber 
\end{gather*}\begin{gather*}
\begin{split}
\frac{1}{2 p_2}\pr[\XXII] + \frac{1}{2 p_3}\pr[\XIXI] + \frac{1}{2 p_4} \pr[\XIIX] & \leq \pr[\XIII] \\
\vdots \\
\frac{1}{2 p_1}\pr[\XIIX] + \frac{1}{2 p_2}\pr[\IXIX] + \frac{1}{2 p_3} \pr[\IIXX] & \leq \pr[\IIIX]
 \enspace .
\end{split}
\end{gather*}
The first inequality holds 
because any error at all occurring, and surviving error detection, is a first-order event (in $\eta_{j-1}$).  

The other inequalities are more interesting; they roughly require that 
conditional error events be first order (but note, e.g., that $\XXII$ is the same as $\IIXX$, so the first inequality can also be written as bounding $\pr[\IIXX]$ in terms of $\pr[\XIII]$).  This is where the boxed $\eta_{j-1}$s of Eq.~\eqref{e:ebellprepproof} come in: each represents the introduction of encoded bit flip errors (application of $\mathrm{X}^{2^{j-1}}$) \emph{by the experimentalist} with probability exactly $\eta_{j-1}$.  Probabilistically introducing errors to lower-bound the right-hand side enforces these inequalities.\footnote{In probabilistically adding logical errors, one has to maintain independence with bit errors.  
As discussed in Remark~\ref{t:randomness}, 
this is difficult -- unless physical NOT gates are perfect.  However, it can be done by only introducing the errors in your head, and tracking them with a classical computer.}

The Mixing Lemma now tells us that the embedded image of our error distribution can be rewritten as a mixture of bounded, bitwise-independent distributions, in $\{0,1\}^4$.  Therefore, the original error distribution over $\{0,1\}^4 / \XXXX$ is the same convex combination of the corresponding bitwise-independent error distributions over $\{0,1\}^4 / \XXXX$.

Indeed, generally, assume we are given distributions $\pi$ and $\pi_1, \ldots, \pi_k$ over state space $\Omega_1$.  (Here, $\pi$ is the actual error distribution over $\Omega_1 = \{0,1\}^4 / \XXXX$, and the $\pi_i$ are the bitwise-independent error distributions.)  Take $\Omega_2$ another state space, distributions $\rho_1, \ldots, \rho_{k'}$ over $\Omega_2$, and $f: \Omega_2 \rightarrow \Omega_1$.  (Here, $\Omega_2 = \{0,1\}^4$, the $\rho_i$ are bitwise-independent distributions, and $f$ is the map of an error string to its equivalence class.)  For a distribution $\sigma$ over $\Omega_2$, define $f(\sigma)$ a distribution over $\Omega_1$ by $f(\sigma)(\omega_1) = \sum_{\omega_2 : f(\omega_2) = \omega_1} \sigma(\omega_2)$.  Assume that $\{ f(\rho_i) : 1 \leq i \leq k' \} \subseteq \{ \pi_1, \ldots, \pi_k \}$.  If there is a distribution $\rho$ over $\Omega_2$ with $f(\rho) = \pi$, and lying in the convex hull of the $\rho_i$ -- say, $\rho = \sum_i p_i \rho_i$ pointwise -- then $\pi$ is in the convex hull of the $\pi_i$: 
\begin{eqnarray*}
\pi(\omega_1) 
&=& f(\rho)(\omega_1) \\
&=& \sum_{\omega_2 \in f^{-1}(\omega_1)} \rho(\omega_2) \\
&=& \sum_{\omega_2 \in f^{-1}(\omega_1)} \sum_i p_i \rho_i (\omega_2) \\
&=& \sum_i p_i \sum_{\omega_2 \in f^{-1}(\omega_1)} \rho_i (\omega_2) \\
&=& \sum_i p_i f(\rho_i)(\omega_1) \\
&=& \sum_j \left(\sum_{i : f(\rho_i) = \pi_j} p_i\right) \pi_j(\omega_1) \enspace .
\end{eqnarray*}
Note that there is significant freedom in choosing $\rho$ such that $f(\rho) = \pi$.  In the application above, recall we obtained $\rho$ from $\pi$ by evenly dividing an error's probability mass among minimum-weight bit strings in its preimage.  Different choices of $\rho$ can lead to better mixing; in this case, it should be advantageous to move a third-order probability mass away from weight-one bit strings and onto weight-three bit strings, and a fourth-order probability mass from $0^4$ to $1^4$.  While this optimization can affect numerical mixing calculations (see Ch.~\ref{s:numericalchapter}), it does not matter for proving the \emph{existence} of a threshold.  Different choices of $\Omega_2$, besides the obvious one, can sometimes make proving the existence of a threshold very easy, although probably at a cost in numerical efficiency; an example is given in Sec.~\ref{s:unbiasednoise}.

Deliberately introducing errors into the computation is counterintuitive, but is necessary for applying the Mixing Lemma.  For example, say that only failure locations A and C in Eq.~\eqref{e:ebellprepproof} are faulty, and the other locations are perfect; and moreover that A only fails as XX and C only fails as IX.  Then $\pr[\XIXI] > 0$ but $\pr[\XIII] = \cdots = \pr[\IIIX] = 0$ -- for acceptance, neither or both of A and C must fail.  This distribution cannot be written as a mixture of distributions with bounded, bitwise-independent X errors.  

\begin{remark} \label{t:thresholdpenalty}
Introducing errors might well hurt the threshold (presuming a postselection threshold even exists without introducing errors), but probably by no more than a small constant factor.  
If we are given lower bounds on CNOT gate failure rates, then it may not be necessary to deliberately introduce errors.  For example, if we are guaranteed that all physical CNOT$_0$ gates fail with the \emph{same} probability, $\leq \eta_0$, and on failure IX, XI and XX errors are equally likely, then it is not necessary to introduce any errors in constructing $\ket{\psi}_1$.\footnote{In Ch.~\ref{s:postselectdetailschapter}, we will also give a slightly more complicated ancilla preparation scheme in which the error rate is not deliberately increased.}
\end{remark}

\begin{remark}[Randomness] \label{t:randomness}
It is often useful to apply a gate (e.g., a swap gate or a Hadamard) with say probability $1/2$, in order to symmetrize a state.  This is more difficult in our encoded setting 
because noise will not be independent of whether or not the logical operation was applied.  For example, if one randomly swaps two blocks having independent errors, the output state will usually have correlations between the blocks.  
This is why we did not assume a symmetrical error model, $\pr[\XIII] = \cdots = \pr[\IIIX]$.  
\end{remark}

\begin{remark}
The proof's inductive structure is:
$$
\begin{diagram}[balance,width=3em,height=2em,tight]
\ket{+}_0 & \rTo & \ket{\psi}_0 & \rTo & \ket{\psi}_1 & \longrightarrow                    & \cdots & 
\longrightarrow                   & \ket{\psi}_{j-1} & \rTo  & \ket{\psi}_j & \longrightarrow & \cdots \\
          &      & \uTo         & \ruTo & \dTo & \rotatebox{30}{$\longrightarrow$} &        & 
\rotatebox{30}{$\longrightarrow$} & \dTo         & \ruTo      & \dTo   &   \rotatebox{30}{$\longrightarrow$} & \\
          &      & CNOT_0       & & CNOT_1 &                                   & \cdots & 
                                  & CNOT_{j-1}   &       & CNOT_j   & & \cdots 
\end{diagram}
$$
(Reliable preparation of $\ket{+}_j$ is equal to reliable preparation of $\ket{\psi}_{j-1}$ for this code, so was not actually needed.)
\end{remark}

\section{Extensions} \label{s:extensions} 

We briefly sketch some of the more important extensions to this basic proof.  These will be fleshed out further in Chs.~\ref{s:postselectdetailschapter} and~\ref{s:magicchapter}.

\subsection{Biased X noise model} 

This analysis can be extended to more interesting noise models, including both bit flip (X)  and phase flip (Z) errors.  The smallest quantum error-detecting code detecting both kinds of errors uses four qubits -- for example, concatenate the repetition code $b \mapsto bb$, $b \in \{0,1\}$ onto the dual repetition code $\ket{\pm} \mapsto \ket{\pm\pm}$.  The encoded Bell pair has eight qubits, and the Mixing Lemma can be applied with $n=24$ (three error possibilities, X, Z and Y$=i$XZ, for each bit) and nontrivial inequalities only for $x$ with $\abs{x} \leq 2$.  (The Mixing Lemma also generalizes to the natural lattice on $\{I, X, Y, Z\}^8$.)

By itself, the bit flip noise model presents interesting challenges.  For example, there are likely better ways of preparing large cat states -- e.g., adding a single qubit at a time instead of doubling -- but these can be difficult to analyze.  Actually, even reliably implementing all the stabilizer operations requires tricks in the biased noise case; because the Hadamard gate sends bit flip to phase flip errors, which the repetition code does not protect against.

\subsection{Universality} \label{s:universality}

We have not shown how to implement reliably a universal set of quantum gates; the CNOT gate, together with preparation and measurement of $\ket{0/1}$, $\ket{\pm}$, is a subset of the set of ``stabilizer operations" which are efficiently simulatable classically (Sec.~\ref{s:gottesmanknilltheorem}).  The extension to universality is via the technique of magic states distillation (although one needs to be careful about randomization -- see Remark~\ref{t:randomness}).  Magic states distillation lets us obtain universality at level $k$ using only level-$k$ stabilizer operations and certain unencoded noisy ancilla preparations.  

\subsection{Asymptotic efficiency} \label{s:postselectionasymptoticefficiency}

The error rate with this scheme drops doubly exponentially fast in $k$ the number of levels of code concatenation, meaning $k$ must be $\log \log N$ to reliably simulate an $N$-gate circuit $\mathcal{C}$.  The overhead is growing as $\exp(N \exp(k))$.  
Following Knill \cite{Knill03erasure}, the overhead can be made only polynomial or polylogarithmic by teleporting into the first of two levels of large random codes (Sec.~\ref{s:largecodes}).

\section{Numerics} \label{s:postselectchapternumerics}

We have proved the existence of a constant noise threshold, but no explicit threshold lower bound.  In his simulations, Knill found that the error distribution was quite close to having independent errors \cite{Knill04analysis}.  Therefore, writing the true error distribution as a mixture of nearby distributions with independent errors has the potential to give good threshold lower bounds.  (See Remark~\ref{t:thresholdpenalty}.)  
Some calculations, with more careful tracking of parameters, are given in Ch.~\ref{s:numericalchapter}.

There are many ways of optimizing the presented fault-tolerance scheme.  For example, it is probably better to verify against errors the full ancillary state used in implementing encoded CNOT gates (Sec.~\ref{s:cnotteleport}).  But rather than rediscover optimizations, it makes sense to analyze Knill's scheme, which has been optimized already using simulations.

In Ch.~\ref{s:correctionchapter}, we apply the mixing technique to a fault-tolerance scheme which does not use postselection.  We have not undertaken threshold calculations for this scheme, though.  Might the mixing technique be useful even for obtaining nonrigorous threshold estimates?

\section{Open problems} \label{s:postselectopenproblems}

Our threshold calculations can undoubtedly be improved, with optimized analysis.  Aside from this, we briefly sketch a concern with this model, and the main open problem.

\subsection{Local gates}

Physical constraints typically dictate that only neighboring or nearby qubits can interact with each other \cite{Gottesman00local,SvoreTerhalDiVincenzo04,MetodievCrossThakerBrownCopseyChongChuang04}.  We have however assumed that CNOT gates can be applied between arbitrary qubits.  Locality is not a particular problem for our proof technique, but may hamper postselection-based fault-tolerance schemes.

\subsection{Discrete noise}

We have assumed that the noise is discrete, with each gate failing with a Pauli error independently of the others.  Threshold results, for fault-tolerance schemes not based on postselection, exist for more general and more physically-realistic error models \cite{KnillLaflammeZurekScience98,AliferisGottesmanPreskill05,AharonovKitaevPreskill05}.  
Some noise correlations can be dealt with by applying the Mixing Lemma to the physical noise itself.  
However, this proof technique may be constrained to discrete Pauli noise models.

(In certain cases, more general noise can be converted to discrete Pauli noise via randomization techniques \cite{BennettDivincenzoSmolinWootters96, KernAlberShepelyansky04randomize, AschaurerDurBriegel05purify}.  For example, given the ability to apply perfect Pauli operations, an arbitrary single-qubit non-Markovian channel can be converted to a discrete Pauli error channel by conjugating the channel with a uniformly random Pauli.)

We have shown that the overall distribution of errors can be written as a mixture of distributions in each of which errors are independent and still of bounded strength.  In particular, this implies that in the full error distribution (before splitting it up), there can't be strong correlations between errors at very different levels, as described as problematic in Sec.~\ref{s:intuition}.  However, there can be some correlations in the full error distribution, especially between nearby levels.  For more general error models, this suggests that we look for a local condition which similarly implies fast decay of correlations.

Note that ``more general error models" can include effective error models arising from application of a different fault-tolerance scheme.  We ask, that is, when can a postselection-based fault-tolerance scheme be concatenated on top of another fault-tolerance scheme?  This question is well motivated practically, since switching schemes can give improved efficiency \cite{Steane03, Knill05}.
\chapter{Postselection threshold for stabilizer operations against general noise} \label{s:postselectdetailschapter}
\def\incl #1#2{\makebox{\raisebox{#1}{\includegraphics[scale=.8]{images/#2}}}}

In Ch.~\ref{s:postselectchapter}, we gave an overview of the proof of a positive threshold for stabilizer operations, for a scheme based on postselection on no detected errors, in the presence of biased X probabilistic noise.  The proof was based on maintaining an inductive invariant on the state of the system, that it can be written as the ideal state plus errors applied according to a mixture of distributions each with bounded failure rates and satisfying strong independence properties.  

In this chapter, we will fill in the remaining details required to prove the existence of a threshold for stabilizer operations.  (The extension to universality will be given in Ch.~\ref{s:magicchapter}.)  
We start by analyzing measurements, and prove the reduction from encoded CNOT gates to preparation of reliable Bell pairs that was sketched in Sec.~\ref{s:cnotteleport}.  

We extend the postselection threshold analysis to noise models with X (bit flip), Z (phase flip) and Y (both bit and phase flips) noise, and prove a threshold for the four-qubit code used as an \emph{operator} error-detecting code (introduced in Sec.~\ref{s:quantumerrorcorrectingcodes}).  (In Ch.~\ref{s:correctionchapter}, we will extend the analysis to schemes based on error correction, and to higher-distance quantum error-correcting codes.)

There are many ways of preparing reliable encoded Bell pairs.  The method presented in Sec.~\ref{s:reliablebellstate} required the experimentalist to deliberately introduce errors into the computation (since these errors are not actually applied to the data, but rather tracked in a classical computer, we call them ``virtual" errors).  The reason this was necessary was that, even though CNOT gates fail with bounded probability, the ratios of failure rates of different gates can be unbounded.  It is possible that some gates are even perfect, and this can give very strong correlations in the distribution of errors.  Introducing independent errors weakens the correlations sufficiently to regain the induction hypothesis, by applying the Mixing Lemma.  

At the very first level of encoding, it is not necessary for the experimentalist to introduce errors if he is guaranteed that all physical gates fail with exactly the same probability (or at least that there is a constant lower bound on their failure rates).  Such symmetry may be a reasonable assumption/definition for the physical noise model.  However, at higher levels of encoding, selecting an error distribution out of a mixture will break such a symmetry.\footnote{This is particularly the case since we are mixing between distributions in which an event occurs with probability either $p$ or zero.  But even if we looked for mixtures of distributions between say $p$ and $p(1-\epsilon)$ -- where $\epsilon$ is maybe $O(p^2)$ (the exponent should be the minimum weight of a stabilizer element) -- the difference still can increase doubly exponentially fast in the number of code concatenation levels.}  Therefore, even with a very careful analysis of the errors, it seems that at best a constant number of levels of concatenation could be analyzed without having to introduce errors to drown out correlations.

Having to introduce independent errors at rate $\eta_j$ is disappointingly conservative, because it could be that the true error rate is much lower than the upper bound of $\eta_j$.  In Sec.~\ref{s:alternativebellprep}, we give a different (but more complicated) ancilla preparation procedure which only requires introducing more errors after an error is already detected.  Therefore, the rate of introduced errors will be comparable to the true (unknown) error rate, possibly much less than $\eta_j$.  

Finally, we give the proof of the Mixing Lemma.  

\section{Measurement rule} \label{s:postselectmeasurement}

Here we explain more carefully the properties of reliable measurements.  We show that a level-$j$ measurement, conditioned on no detected errors, has the same behavior as an ideal measurement, except giving the wrong answer with probability $\leq \eta_j$.  

Define a computational-basis decoding operation by 
\begin{gather*}
\incl{0em}{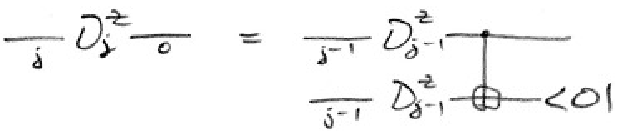} \\
\incl{0em}{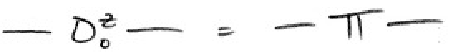}
\end{gather*}
That is, we project transversally into the computational basis (i.e., measure) -- represented by $\Pi$ -- then use parity checks to postselect on no detected errors.  The CNOTs used in the parity checks are perfect, because they are running in a classical computer.  

It is formally equivalent to delay all measurements to the end, commuting the ideal CNOTs to before the projections.  That is, 
\begin{gather*}
\incl{0em}{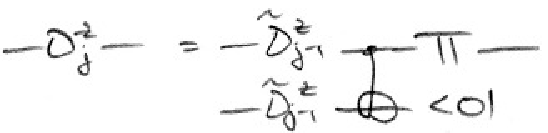} \\
\incl{0em}{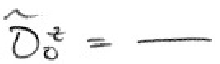}
\end{gather*}
Leading with a noisy encoder, we obtain
\begin{eqnarray*}
\incl{-.5em}{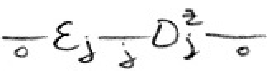} &=& \incl{-1em}{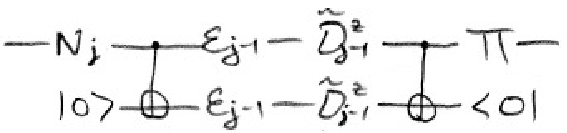} \\
&=& \incl{-1em}{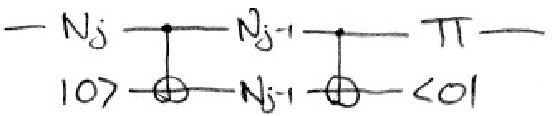} \\
&=& \incl{-.5em}{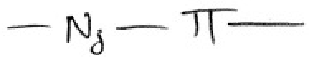}
\end{eqnarray*}
the desired behavior.  (Z errors, if present, will not be suppressed; however they have no effect on 0/1 basis states, so can be ignored.)  In particular, this implies that at the end of the fault-tolerant computation in $\text{FT}\mathcal{C}$, the probability of measuring zero (resp. one), conditioned on no detected errors, will be nearly the same as the probability of measuring zero (one) in the ideal circuit $\mathcal{C}$. 

\section{Reduction of CNOT$_j$ gate implementation to reliable preparation of Bell pair $\ket{\psi}_j$} \label{s:cnotreductionanalysis}

Here we explain in more detail how to implement reliable CNOTs using reliable Bell pairs (Sec.~\ref{s:cnotteleport}).  It is an exercise in pushing through noisy encoders.  

We start by preparing two level-$j$ Bell pairs, and applying transversal CNOT$_{j-1}$s between their first halves (which will be measured later on):
\begin{eqnarray*}
\incl{-2em}{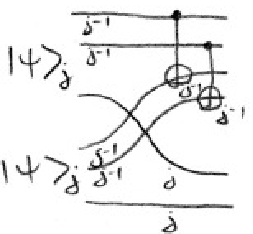} &\rightarrow& \incl{-2.5em}{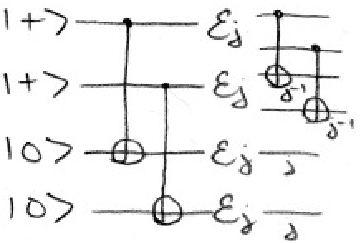} \\
&\rightarrow& \incl{-2.5em}{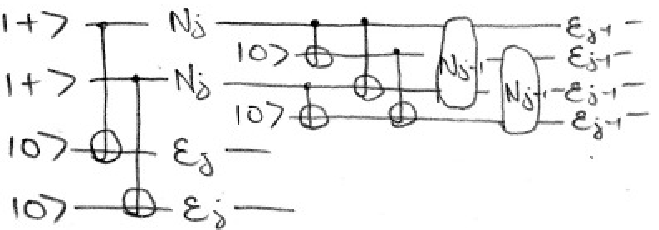} \nonumber \\
&=& \incl{-2.3em}{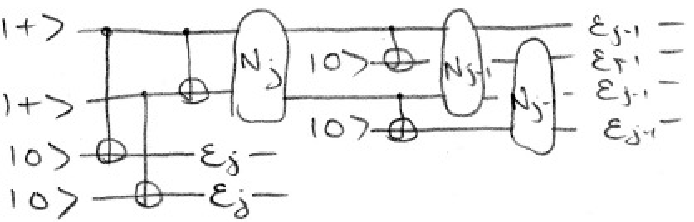} \nonumber
\end{eqnarray*}
Here we have used Eq.~\eqref{e:ebellpair} twice, pushed through level-$(j-1)$ encoders, and then pulled back an ideal CNOT gate.  Effectively, since 
$$
\incl{-2em}{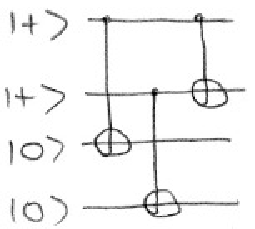} = \incl{-2em}{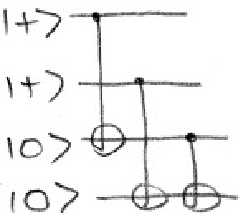}
$$
we have accomplished a CNOT between the second halves of the two Bell pairs, while correlating level-$(j-1)$ errors only on the first halves.  

The level-$j$ CNOT is then implemented by teleporting into this state, using CNOT$_{j-1}$s.  Substitute the whole above diagram as $\Gamma$ into
\begin{eqnarray*}
\incl{-4em}{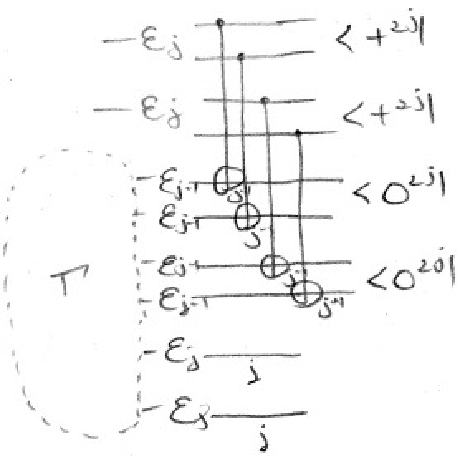} &\rightarrow& \incl{-2.5em}{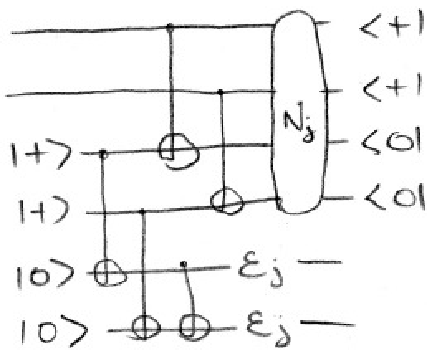} \\
&\rightarrow& \incl{-1em}{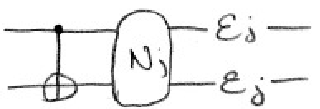} \nonumber
\end{eqnarray*}
Here we have again pushed through the noisy encoders $\mathcal{\tilde{E}}_{j-1}$, and used the measurement rule.  Because of postselection, two level-$(j-1)$ errors are required to cause a mistaken measurement.  The final step is a consequence of teleportation.  

Note that in fact it was not necessary to use CNOT$_{j-1}$s to accomplish the teleportation -- since these blocks are about to be measured anyway, it does not matter if correlations are created in errors below level $j-1$.  Instead, transversal level-0 CNOT gates can be used for an efficiency savings, because:
\begin{eqnarray*}
\incl{-1.5em}{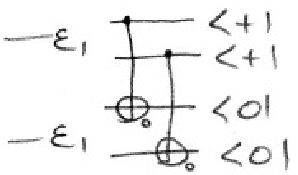} &\rightarrow& \incl{-1.8em}{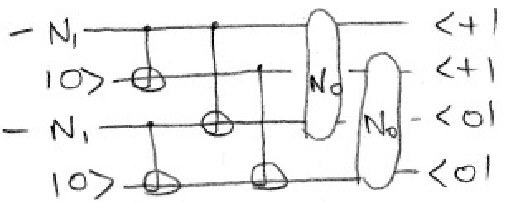} \\
&\rightarrow& \incl{-1em}{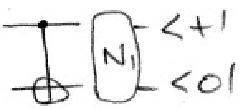} \nonumber
\end{eqnarray*}

\section{X and Z noise} \label{s:unbiasednoise} 

We have shown how to prepare reliable Bell pairs in the presence of biased X noise.  What about the more general case, where there is both X, Z and $\text{Y} = i\text{X}\text{Z}$ noise?  

The argument is a straightforward generalization of the biased X noise analysis.  
A fairly immediate extension can be derived by assuming X and Z errors occur independently, and then alternating levels of concatenation of the two-qubit repetition code and its dual.  However, obtaining a noise model with independent X and Z errors requires introducing errors at rate $\sqrt{\eta_0}$ and applying the Mixing Lemma.  This is extremely inefficient -- one expects it to reduce the threshold quadratically -- so instead we will give a direct argument.  

To make the proof slightly more interesting, let us use the four-qubit code in an ``operator" error-detecting fashion.  We will only be concerned with protecting one of the two encoded qubits against noise.

\subsection{Noise model}

For each physical CNOT gate, assume that a perfect CNOT is always applied, and then an independent biased die is tossed, causing one of the sixteen two-qubit Pauli products to be applied to the outputs.  
Each physical CNOT gate has sixteen parameters, $\eta_{0,\text{II}}, \eta_{0,\text{IX}}, \eta_{0,\text{IY}}, \eta_{0,\text{IZ}}, \ldots, \eta_{0,\text{ZZ}}$, specifying the probabilities of which two-qubit Pauli $\sigma$ is applied after the gate.  The total probability of the fifteen nontrivial two-qubit Paulis, $\sum_{\sigma \neq \text{II}} \eta_{0,\sigma}$, is upper-bounded by $\eta_0$.  Graphically, 
$$
\incl{0em}{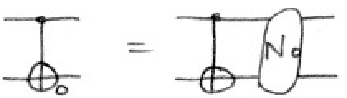}
$$

Other gates or operations can also be assumed to be faulty.  For example, assume that a physical preparation of $\ket{0}$ (resp. $\ket{+}$) fails probabilistically and independently, with probability $\leq \eta_0$, giving in fact $\ket{1}$ ($\ket{-}$).  Graphically, 
$$
\incl{0em}{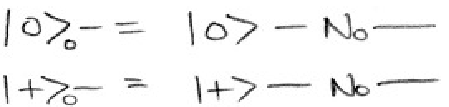}
$$
Physical measurements fail similarly, giving the wrong outcome.  Single-qubit Clifford gates are also faulty, and even including rest (memory) errors gives no more complications:
\begin{gather*}
\incl{0em}{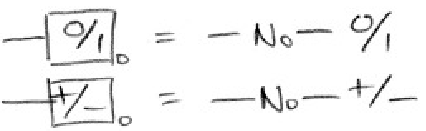}\\
\incl{0em}{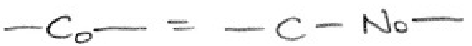}\\
\incl{0em}{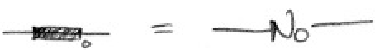}
\end{gather*}

\subsection{Four-qubit operator error-detecting code}

Recall the presentation of the $[[4,2,2]]$ error-detecting code from Sec.~\ref{s:fourqubitcode}:
$$
\begin{array}{c c@{\,}c@{\,}c@{\,}c}
&X&X&X&X\\
&Z&Z&Z&Z\\
\cline{2-5}
X_{L} = &X&X&I&I\\
Z_{L} = &I&Z&I&Z\\
X_{S} = &X&I&X&I\\
Z_{S} = &I&I&Z&Z
\end{array}
$$
The code's distance is two, meaning that it can detect any one-bit error.  
There are two encoded qubits.  Call the first encoded qubit, operated on by $X_L$ and $Z_L$, the \emph{logical} qubit.  Call the second encoded qubit the \emph{spectator} qubit.  Using this code as an operator error-detecting code (Sec.~\ref{s:ninequbitbaconshorcode}) means that we will only be concerned with protecting the logical qubit from errors, and will not worry about errors on the spectator.  

Concatenate this code using the logical qubits.  So the level-two-encoded logical and spectator qubits are encoded using sixteen physical qubits.  The four level-one spectator qubits are not used in the level-two code.  Therefore, there are now five total spectator qubits.  At $j$ levels of encoding, there will be one level-$j$-encoded spectator qubit, four level-$(j-1)$ spectators, sixteen level-$(j-2)$ spectators, and so on, for a total of 
$\tfrac{4^j-1}{4-1}$
spectator qubits -- and one logical qubit.

\begin{remark}[Different ways of using the four-qubit code] 
One different approach would be to fix the spectator qubit always to say $\ket{0}$, or in other words add $Z_S$ to the stabilizer to get a $[[4,1,2]]$ code.  Then we wouldn't have to worry about $Z_S$ errors, because these would preserve the codespace.  But an $X_S$ error would move out of the codespace, so would still be a concern.  By using the code in an operator fashion, we need concern ourselves about neither $Z_S$ or $X_S$ errors.  Also, as will become clear below, we can set the spectator's state to whatever is convenient, allowing for more efficient state preparation.  

Another different approach would be to use the second encoded qubit, too, in the concatenation procedure.  So for example, two level-two encoded qubits could be obtained using just eight physical qubits (two blocks of four, each block encoding two qubits).  This is more qubit-efficient, but it can be difficult to access separately the different encoded qubits within a block.  Knill has, however, analyzed a scheme in which the first level of encoding uses the $[[4,2,2]]$ code, and later levels use a particular $[[6,2,2]]$ code, using all the encoded qubits \cite{Knill05}.  (E.g., the first six-qubit code level uses twelve physical qubits, three blocks of four.)  This sort of scheme should be straightforward to analyze with mixing techniques.
\end{remark}

\subsection{Fault-tolerant Bell pair preparation}

In order to avoid having to track the state of the spectator qubits, which would complicate the notation, we will maintain these encoded qubits in a completely depolarized state, by introducing (virtual) errors.
Define the level-$j$ noisy encoder by
\begin{equation} \label{e:fourqubitnoisyencoder}
\incl{-2.5em}{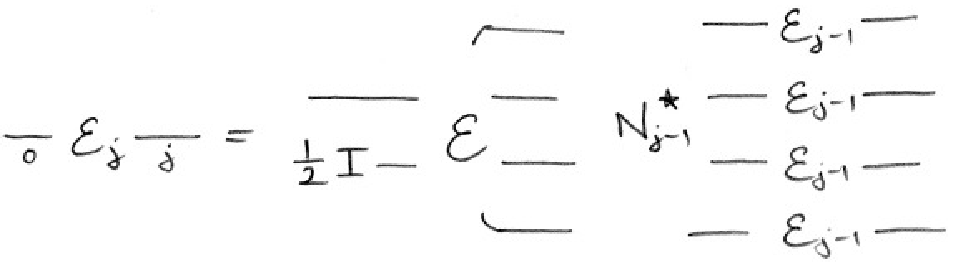}
\end{equation}
Here $\tfrac{1}{2}I$ means a depolarized input for the spectator, and $\mathcal{E}$ is the perfect encoder of a logical and spectator qubit:
$$
\incl{0em}{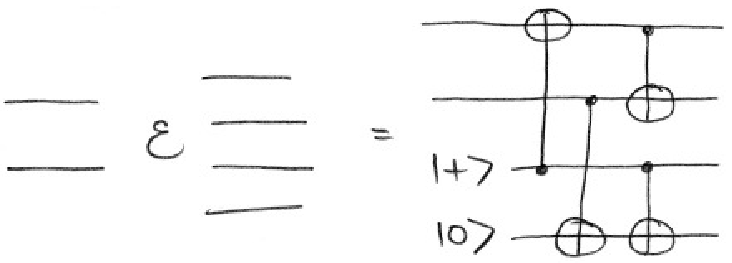}
$$
The $N_{j-1}^\star$ in fact represents twelve independent noise locations, each failing at rate at most $\eta_{j-1}$, and applying on failure one of \\
\begin{equation} \label{e:detectableerrors}
\text{\begin{tabular}{c c c c c c}
IIIZ & IIIX & IIIY & IIXI & IIXZ & IIXY \\ 
IZII & IZIX & IZIY & IZXI & IZYI & IYXI
\end{tabular}}
\end{equation}
Two errors are equivalent if their difference is a product of code stabilizers, encoded errors on the spectator qubit(s), and logical stabilizers (if any).  Hence there are $\tfrac{4^4}{2^4}$ inequivalent errors on a general four-qubit block (there are $4^4$ four-qubit Pauli operators, two independent code stabilizers and two spectator operators, but no logical stabilizers).  Here, we have allowed each detectable error to occur with first-order probability; removing the logical operators XXII, IYYI, ZIZI, and also IIII, there are $16-4=12$ inequivalent detectable errors.  Undetectable errors are second-order events.  

The base case $\mathcal{E}_0$ is trivial:
$$
\incl{0em}{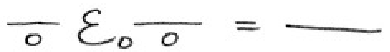}
$$

\begin{remark}
The most natural generalization of Eq.~\eqref{e:noisye} might be to define the noisy encoder by 
\begin{equation} \label{e:fourqubitsimultaneousnoisyencoder}
\incl{-2.5em}{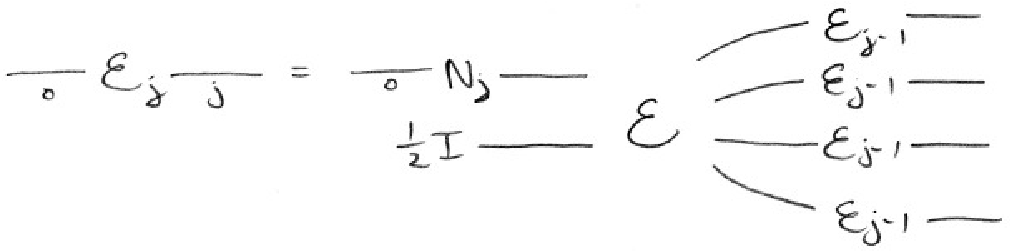}
\end{equation}
with a base case $\mathcal{E}_0$ as in Eq.~\eqref{e:noisye}.
But it turns out that the fault tolerance of Eq.~\eqref{e:fourqubitnoisyencoder} is sufficient, and easier to obtain than strict simultaneous X and Z error fault tolerance Eq.~\eqref{e:fourqubitsimultaneousnoisyencoder}.
(For simultaneous fault tolerance, in preparing an encoded Bell pair, \emph{three} verification rounds -- e.g., against X, Z, then X errors -- instead of just two, would be required on the second block in Fig.~\ref{f:eightqubitbell} below.)  Sec.~\ref{s:differentftdefinitions} discusses different definitions of fault tolerance.  By defining the noisy encoder according to Eq.~\eqref{e:fourqubitnoisyencoder}, we are using a fairly weak definition: second-order failures can be arbitrary and first-order failures must only be detectable.
\end{remark}

The fault-tolerance rules we aim to prove are nearly the same as before, 
except the noise locations now introduce Pauli errors and the noisy encoder is defined differently.  E.g., we'd like to show 
\begin{gather*}
\incl{0em}{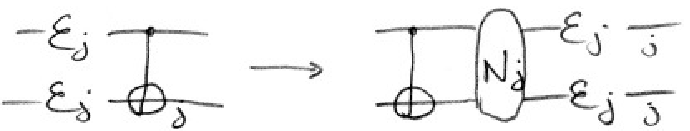} \\
\incl{0em}{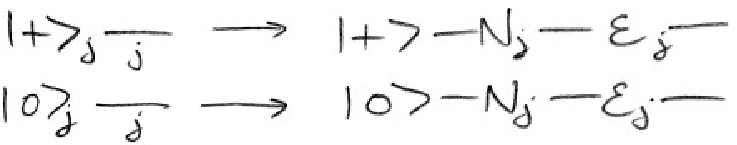}
\end{gather*}

Once again, the key step is implementing a reliable CNOT gate, which follows from reliable preparation of an encoded Bell pair $\ket{\psi} = \tfrac{1}{\sqrt{2}}(\ket{00} + \ket{11})$:
$$
\incl{0em}{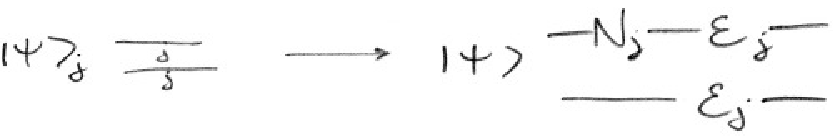}
$$
State verification is now slightly more complicated because there are both phase and bit flip errors to check for.  Several verification methods work, and a method which uses only one additional ancilla bit -- sacrificing some parallelism -- is given in Fig.~\ref{f:eightqubitbell}.  Not shown is the depolarization of the two spectator qubits at the end, and also the rate-$\eta_{j-1}$ independent introduction of each of the twelve detectable errors on each block, as in Eq.~\eqref{e:ebellprepproof}.  (Again, above the physical level, the qubits used are eight protected logical qubits, not any spectators.)  

\begin{figure}
\begin{gather*}
\incl{0em}{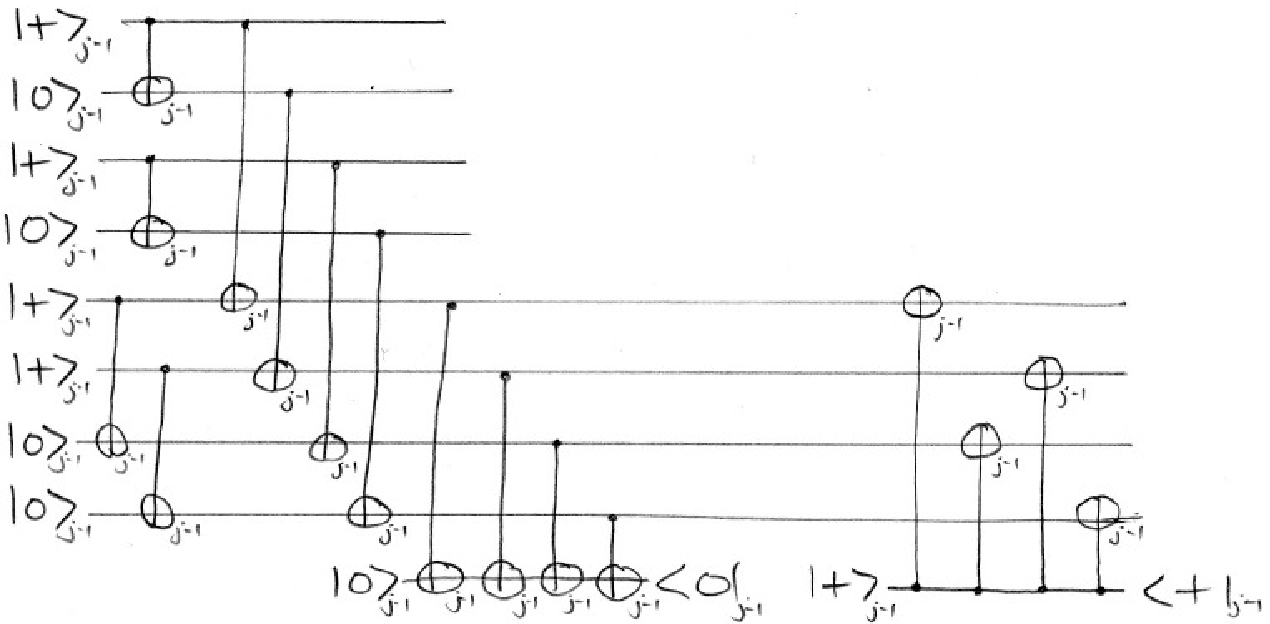} \\
\incl{0em}{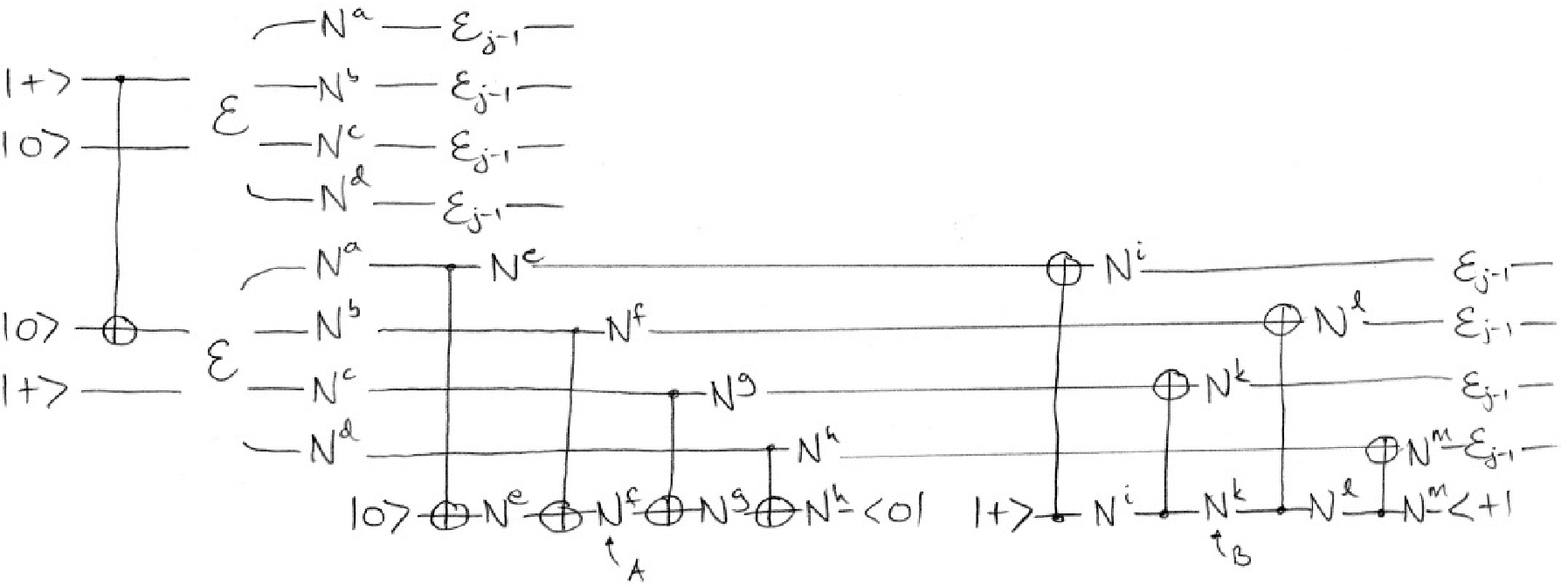}
\end{gather*}
\caption{Encoding and verification circuit for a level-$j$ Bell pair (top).  Not shown are depolarization of the spectator qubits and deliberate introduction of level-$(j-1)$ errors.  Bottom: The same circuit, with the noisy encoders pushed through, by induction, and with some noise locations combined.  All noise locations have failure rates upper-bounded by $\eta_{j-1}$.  Noise locations sharing the same superscript (e.g., the two $N^a$s) are a single location across multiple wires -- correlated failures.} \label{f:eightqubitbell}
\end{figure}

The first two CNOT gates on top encode $\ket{+}_L \ket{0}_S$ and the bottom two CNOT gates encode $\ket{0}_L \ket{+}_S$.  The spectator qubits' states are chosen so that in each case, only Bell pairs at the level $j-1$ are created -- so errors are independent leaving these first four gates.  

The order of the verification gates into or from the ancilla $\ket{0}$ or $\ket{+}$ is important, and chosen so that a Z error at $A$ kicks back into an error $Z_S$ on the spectator, and an X error at $B$ is copied into an $X_S$ error.  (Had the gates for Z error verification been ordered as those for X verification, an X error at $B$ would be copied into $X_L$, a first-order failure event causing a second-order logical error.)  

We will apply the Mixing Lemma with $n= 12 + 12 = 24$ -- for each block, the twelve failure locations of Eq.~\eqref{e:detectableerrors}.  Since we are deliberately introducing errors, applying mixing requires only upper bounding the probability of error events.  Make the two following simple observations:
\begin{itemize}
\item The maximum weight of an error supported entirely on a single code block is two.
\item The maximum weight of any error on the encoded Bell pair is three.  
\end{itemize}
(The weight of an error is the minimum Hamming weight of an equivalent error, modulo stabillizers and the spectator encoded operations.)  Moreover, for an error of weight three, choose a representative with minimum Hamming weight -- this equivalent error has two bit errors on one of the blocks (by pigeonhole).  These bit errors must together be detectable.  (Indeed, otherwise, they comprise a logical error, and since $X_L X_L$, $Z_L Z_L$ and $-Y_L Y_L$ are stabilizers, this logical error can be pulled onto the other block.  Therefore the error can be supported on a single block, so has weight at most two -- a contradiction.)  

Thus, there are no errors which should be third-order events -- we need only show that logical errors and errors supported across the two blocks are second-order events.  And indeed, a single failure event cannot cause an error supported across the two blocks -- the error-free verifications will catch any error in the second block coming from a failure in locations $a$ through $d$.  A single failure event also cannot cause a nontrivial undetectable error.

\subsection{Fault-tolerant Hadamard gate} \label{s:fthadamard}

A logical Hadamard can be applied by transversal Hadamards, followed by permuting bits two and three (by relabeling).  Therefore, the reliable encoded Hadamard can be implemented by applying these operations to the second, output half of a Bell pair, and then teleporting into the gate.  (To avoid increasing the bit error rate in the output block, apply the transversal Hadamards to the first, input half of the Bell pair.)

CNOT, Hadamard, Pauli gates, preparation of $\ket{0}$, and measurement in the $\ket{0}$, $\ket{1}$, and in the $\ket{+}$, $\ket{-}$ bases do not generate the full set of stabilizer operations, but will suffice to apply Theorem~\ref{t:practicalmagicdeterministic} (Sec.~\ref{s:introduction}) to obtain universality.

\section{Alternative Bell pair preparation procedure} \label{s:alternativebellprep}

The Bell pair preparation method of Sec.~\ref{s:reliablebellstate} was somewhat unsatisfactory, because it involved the experimentalist deliberately introducing errors into the computation.  We argued in Remark~\ref{t:thresholdpenalty} that the noise threshold should not be harmed too much by introducing these errors, and also that 
it isn't always necessary at the first code level.  Still, deliberately adding faults may seem like too pat a solution -- a way of avoiding difficulties without necessarily fully understanding them.  

In this section, we give an alternative method for preparing reliable Bell pairs.  We also give examples of methods which do \emph{not} work.  The goal is to illustrate the limitations of the mixing technique, and to give other ways these limitations can be designed around.  

To start, consider the following preparation procedure, based on purification (Sec.~\ref{s:stabilizerstatepreparation}):
\begin{equation} \label{e:altbellprep1}
\incl{-3.5em}{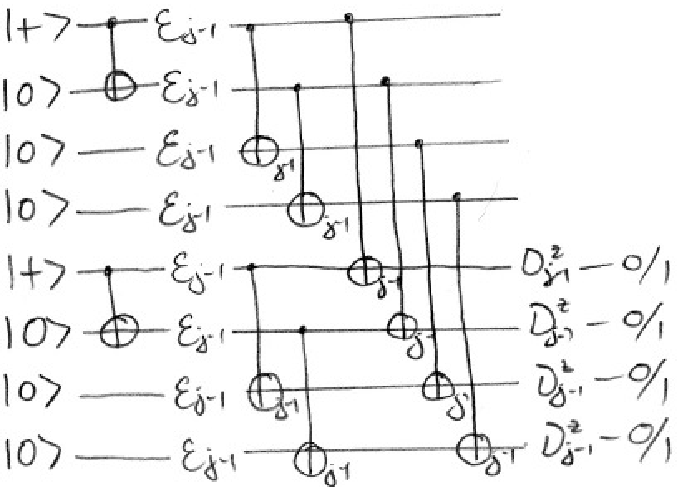} \rightarrow \incl{-3.5em}{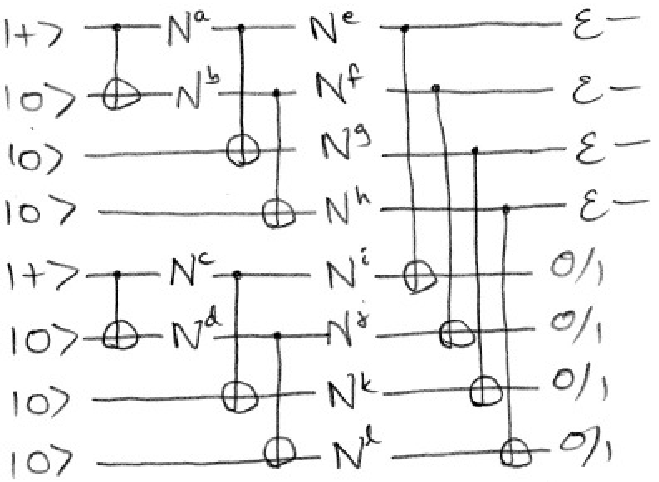}
\end{equation}
On the right, we have dropped some subscripts, and labeled noise locations for reference.  The markings 0/1 indicate measurement in the computational basis.  The preparation procedure requires too a verification rule describing how to use these measurement results; we'll give several below.  We have also used the rules 
\begin{gather*}
\incl{0em}{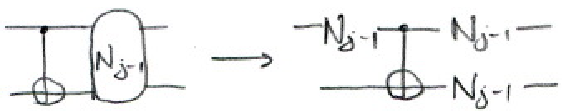} \\
\incl{0em}{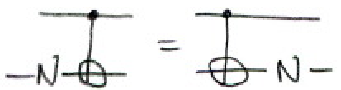}
\end{gather*}
in order to simplify.  (Note that these rules are only true for biased X noise.)  

To apply the Mixing Lemma (with $n=4$), we need that errors have the correct order and also that \emph{conditional} probabilities be well-bounded.  Here, every error has weight at most two, by symmetry (flipping all four bits with XXXX has no effect on the state).  Therefore, the two conditions are:
\begin{enumerate}
\item \label{e:condition1} Weight-two errors must have second-order probability.
\item \label{e:condition2} Conditioned on a bit being in error, the probability of another bit being in error must be first order.
\end{enumerate}

Generally, the first condition is easy to satisfy.  The second condition requires some care.  A sufficient combinatorially checkable criterion for it to hold is: For a subset $S$ of the error locations $\{a,\ldots, f\}$ which causes an output weight-two error mapped to $x \in \{0,1\}^4$ ($\abs{x} = 2$), to apply the Mixing Lemma, and for each $i \in x$, there exists a subset of $S$ causing the single error $i$ on the output.  

Indeed, for each $i,j \in [4] \equiv \{1,\ldots, 4\}$, we need upper bounds on the probabilities 
$$
\pr(i \vert j) = \frac{\pr(i,j)}{\pr(i)} \enspace .
$$
The numerator, $\pr(i,j)$, is a polynomial in the variables $p_a, \ldots, p_l$ the probabilities for failures at the respective locations.  Each term in this polynomial has degree at least two, by condition one.  The denominator, $\pr(i)$, is a first-order polynomial in the same variables.  We want to show that the ratio is $O(p_a + \ldots + p_l)$, i.e., that 
$$
\frac{\pr(i,j)}{\pr(i) \cdot (p_a + \ldots + p_l)} = O(1) \enspace .
$$
(The converse statement would be that there is some setting for the variables $p_a, \ldots, p_l$ making this ratio unbounded.)  
By continuity, it suffices to check that the denominator can't be made zero while the numerator is nonzero -- by setting zero the variables $p_\alpha \notin S$ and nonzero the variables in $S$.

Consider as a first attempt the very intuitive verification rule: Postselect on no detected errors.  This rule does \emph{not} work.  For, consider $S = \{a,c\}$, causing a weight-two error in the output.  No strict subset of $S$, $a$ or $c$, can cause any error at all in the output, for either error on its own will be caught.  Therefore, we can't guarantee that $\pr(1\vert 3)$ or $\pr(3 \vert 1)$ are first-order.  It is certainly possible that $p_\alpha = 0$ for $\alpha \neq a,c$, and $p_a$ and $p_c$ are nonzero; then a weight-two error sometimes occurs on the output, but never any weight-one errors.

Consider as a second attempt the verification rule: 
\begin{center}
\fbox{
\begin{minipage}[l]{4.25in}
\noindent { Verification rule:} (for Eq.~\eqref{e:altbellprep1}, and for the first round of Eq.~\eqref{e:altbellprepbig})
\begin{itemize}
\item On detecting one error, reject.
\item On detecting two errors, apply a uniformly random correction to the four wires ($2^4$ possibilities, or eight modulo the XXXX symmetry).
\end{itemize} 
\end{minipage} }
\end{center}
The intuition behind this rule is to enforce that one-bit errors can survive verification when two-bit errors can -- to avoid the problem with our first attempt.  Since we introduce corrections only after first detecting an error, this verification rule may be more satisfactory than that in Sec.~\ref{s:reliablebellstate}.

However, this verification rule does not work.  It does not even satisfy condition one.  A single failure, say at location $a$, can cause two errors to enter the verification, so two errors leave it with constant probability.  It is tempting, therefore, to change this rule, perhaps to: On detecting a two-bit error, apply a correction to a random \emph{one} of the four wires.  Then condition one would be satisfied.  But let's leave the rule as it is, and consider condition two.  

Condition two also fails to hold for this verification rule, not surprisingly.  If all failure locations are in fact perfect (have zero error rate) and only $p_a > 0$, then we can only say that $\pr(i \vert j) = O(1)$ for any $i,j \in [4]$, $i \neq j$.  Still, this verification rule is promising, because in fact for all $i,j,k \in [4]$, $i \neq j$, 
\begin{equation} \label{e:altbellpreproundone}
\frac{\pr(i,j)}{\pr(\text{exactly $k$})} = O(1) \enspace .
\end{equation}
(By ``exactly $k$", we mean the event that the output error has weight one (not zero or two) and is on wire $k$.)
That is, for any set $S$ of failure locations causing two errors $i$, $j$ in the output with constant probability, some (not necessarily strict) subset of $S$ causes error $k$ with constant probability.  
\\ Proof: 
\begin{itemize}
\item If no error was detected, then there must have been two errors in each encoded Bell pair entering verification.  The subset of $S$ causing say the error in the top pair, i.e., $S \cap \{a,b,e,f,g,h\}$, would on its own cause two errors to be detected, and a random correction to be applied.  The output error would be exactly $k$ with constant probability.  
\item If on the other hand two errors were detected, then $S$ itself causes a random correction to be applied, so again with constant probability the output error is exactly $k$. 
\end{itemize}

Let us now use the preparation procedure of, and verification rule for, Eq.~\eqref{e:altbellprep1}, as a subroutine for preparing encoded Bell pairs satisfying both conditions one and two.  Consider the following implementation:
\begin{equation} \label{e:altbellprepbig}
\incl{-11em}{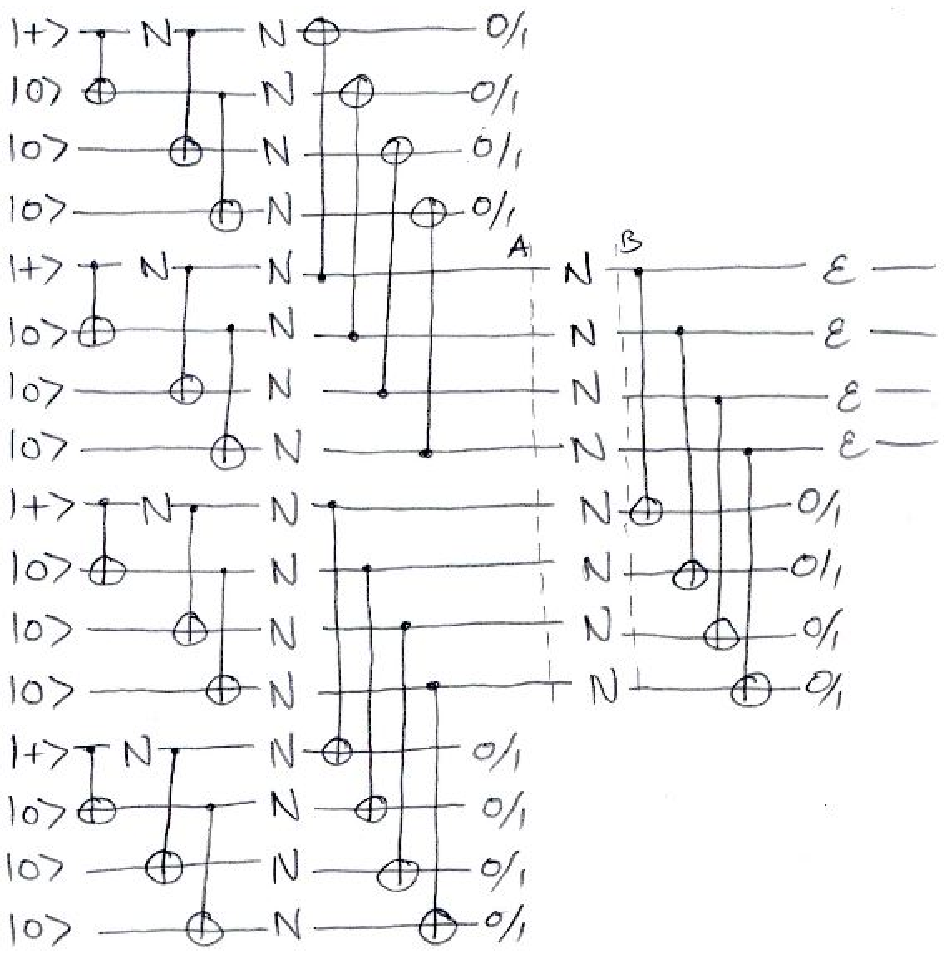}
\end{equation}
We start with four Bell pairs, and in the first round verify them pairwise according to the verification rule for Eq.~\eqref{e:altbellprep1}.  In the second round, we verify once more according to the rule:
\begin{center}
\fbox{
\begin{minipage}[l]{4.25in}
\noindent { Verification rule:} (for round two of Eq.~\eqref{e:altbellprepbig})
\begin{itemize}
\item On detecting one error, either correct it or not, according to a fair coin flip.
\item On detecting two errors, reject.  
\end{itemize} 
\end{minipage} }
\end{center}
(There are other verification rules which work; in formulating this rule, we have tried to use as much postselection as possible.)

This two-round preparation procedure indeed satisfies condition one.  For a weight-two error to survive on the output, there must be errors on both halves, so $\abs{S} \geq 2$.  

It also satisfies condition two.  Indeed, assume $S$ is a set of error locations causing a weight-two error on the output with constant positive probability.  
Note that at the end of round one, at the time marked A in Eq.~\eqref{e:altbellprepbig}, Eq.~\eqref{e:altbellpreproundone} holds for all $i,j,k \in [4]$, $i \neq j$.  At the time marked B, Eq.~\eqref{e:altbellpreproundone} holds for all $i,j \in [4]$, $i \neq j$, and either for $k \in \{i,j\}$ or $k \in [4] \setminus \{i,j\}$.  (The condition is slightly weakened because it is possible, e.g., for there to be two failures at $i$ and $j$ in the noise locations between A and B [i.e., immediately after round one]).
\\ Proof:
\begin{itemize}
\item{Case 1:} No errors detected in round two.  Then two errors must have entered the verification from either side (either canceling or complementing each other).  In particular, by Eq.~\eqref{e:altbellpreproundone}, a strict subset of $S$ just containing errors from one side will lead to weight-one errors on the output with constant probability.  
To apply the Mixing Lemma, there are two strings in $\{0,1\}^4$ this event can be mapped to (e.g., XXII is equivalent to IIXX, and can be mapped to either $1100$ or $0011$).  Map it to that string $x$ for which Eq.~\eqref{e:altbellpreproundone} holds for $k \in x$, on one of the sides.
\item{Case 2:} One-bit error detected in round two.  For the output error to have weight two when only one error was detected, errors from both sides must have entered verification.  And on exactly one side the errors must have had weight two.  (If the errors had weight one on both sides, or weight two on both sides, their combination would have even weight, not weight one.)  
The output error is symmetrical under XXXX to the weight-two error entering verification.  Therefore map this event to the $x \in \{0,1\}^4$ for which Eq.~\eqref{e:altbellpreproundone} holds for $k \in x$, on the side of the weight-two error.
\end{itemize}

We expect -- but have not checked -- that a similar procedure, except with four rounds of verification, will work for the more general case of X and Z noise, as considered in Sec.~\ref{s:unbiasednoise}.

\section{Proof of the Mixing Lemma} \label{s:mixinglemmaproof}

A point $(q_1, \ldots, q_n) \in [0,1]^n$ corresponds to a bitwise-independent distribution over $\{0,1\}^n$, in which the probability of $x$ is $\prod_{i=1}^n q_i^{x_i} (1-q_i)^{1-x_i}$.  Define the lattice ordering $y \preceq x$ for $x,y \in \{0,1\}^n$ if considered as indicators for subsets of $[n]$, $x \subseteq y$.  (E.g., $011 \preceq 010$; considered as subsets of $\{1,2,3\}$, $011$ is $\{2,3\}$ and $010$ is $\{2\}$.)  Equivalently, $y \preceq x$ iff $x \wedge y = x$, where $\wedge$ is the bitwise AND.

\begin{mixinglemma}
The convex hull, in the space of distributions over $n$-bit strings, of the $2^n$ bitwise-independent distributions $\{0,p_1\}\times\{0,p_2\}\times\cdots\times\{0,p_n\}$ is given exactly by those $\pr[\cdot]$ satisfying the inequalities, for each $x \in \{0,1\}^n$: 
\beq
\sum_{y \preceq x} (-1)^{\abs{x\oplus y}} \frac{\pr[\{z \preceq y\}]}{p(\{z \preceq y\})} \geq 0 \enspace ,
\eeq
where $p(\{z \preceq y\}) = \prod_{i=1}^n \delta_{y_i,1} p_i$, i.e., the probability of $\{z : z \preceq y\}$ in the distribution $(p_1,\ldots,p_n)$.
\end{mixinglemma}

\begin{proof}
For $w \in \{0,1\}^n$, let $w \cdot p$ denote the distribution $(w_1 p_1, \ldots, w_n p_n)$.  I.e., if $Z$ is drawn from $w \cdot p$, then the probability $Z = z$ is 
$
(w \cdot p)(z) 
= \prod_{i=1}^n (w_i p_i)^{z_i} (1- w_i p_i)^{1-z_i} 
$.  
In particular, 
$$
(w \cdot p)(\{z : z \preceq y\}) = \left\{\begin{array}{ll}\Pi_{i \in y} p_i = p(\{z \preceq y\}) & \textrm{if $w \preceq y$} \\  0 & \textrm{otherwise}\end{array}\right. \enspace .
$$

The convex hull of the distributions $\{w \cdot p : w \in \{0,1\}^n\}$ is contained in the set specified by the simultaneous inequalities~\eqref{e:mixinglemma}.  Indeed, $w \cdot p$ satisfies all the inequalities with equality, except that for $x = w$ for which it gives $1$:
\begin{equation*}
\sum_{y \preceq x} (-1)^{\abs{x \oplus y}} \frac{(w \cdot p)(\{z \preceq y\})}{p(\{z \preceq y\})} 
= \sum_{y : w \preceq y \preceq x} (-1)^{\abs{x \oplus y}} 
= \delta_{x,y} \enspace ,
\end{equation*}
since necessarily $x \succeq w$ for the sum over $y$ to be nonzero, and then $\sum_{k=0}^{\abs{w}-\abs{x}} \binomial{\abs{w}-\abs{x}}{k} (-1)^k = 0$ unless $\abs{x} = \abs{w}$.

Conversely, if a distribution $\pr[\cdot]$ satisfies inequalities~\eqref{e:mixinglemma}, then it lies in the convex hull of the distributions $\{w \cdot p : w \in \{0,1\}^n\}$.  Indeed, the $w \cdot p$ coordinate of $\pr[\cdot]$ is given by the value of the left-hand side of Eq.~\eqref{e:mixinglemma} for $x=w$.  These coordinates are nonnegative, and using these coordinates recovers the distribution $\pr[\cdot]$; for all $v \in \{0,1\}^n$,
\begin{eqnarray*}
\sum_{\substack{x,y\\ y \preceq x}} (-1)^{\abs{x\oplus y}} \frac{\pr[\{z \preceq y\}]}{p(\{z \preceq y\})} (x \cdot p)(\{z \preceq v\}) 
&=& \sum_{\substack{x,y\\ y \preceq x \preceq v}} \pr[\{z \preceq y\}] (-1)^{\abs{x\oplus y}} \frac{\prod_{i \in v} p_i}{\prod_{i \in y} p_i} 
\\ &=& \pr[\{z \preceq v\}] \enspace ,
\end{eqnarray*}
since again the sum over $x$ is zero unless $y=v$.
(The values $\pr[\{z \leq v\}]$ for different $v$ characterize $\pr[\cdot]$.)
\end{proof}

\chapter{Extension to universality via magic states distillation} \label{s:magicchapter}
\def\magic #1{\mathcal{U}(#1)}

\section{Introduction} \label{s:introduction}

Stabilizer operations -- consisting of Clifford group unitaries (generated by the Hadamard, phase and CNOT gates), preparation of $\ket{0}$ and measurement in the computational $\ket{0}, \ket{1}$ basis -- suffice for generating interesting, highly entangled quantum states.  
For example, cluster/graph states \cite{RaussendorfBriegel01} are stabilizer states -- i.e., can be prepared with stabilizer operations.
In terms of fault-tolerance, only stabilizer operations
are required for encoding/decoding quantum (stabilizer) codes and for applying error-detection or error-correction.\footnote{As always here, assuming adaptive classical control.}  Too, stabilizer operations are relatively easy to analyze rigorously, and to simulate -- because Clifford unitaries take Pauli errors just to different Pauli errors.\footnote{To classically simulate faults in encoded stabilizer operations on an arbitrary encoded state, with a probabilistic Pauli error model, one only needs to track the errors on each bit -- X, Y, Z or no error I -- not the encoded state itself.  Steane and Knill, and many others, have run extensive simulations of this type to determine threshold estimates for stabilizer operations \cite{Knill05,Steane03}.}
But of course the ease of simulating stabilizer operations (Gottesman-Knill theorem, Sec.~\ref{s:gottesmanknilltheorem}) is also their weakness; stabilizer operations certainly do not give a universal basis for quantum computation.  

What more do we need to achieve fault-tolerant encoded quantum universality?  For example, a fault-tolerant Toffoli gate would suffice.  Shor's original fault-tolerance construction gave a fault-tolerant Toffoli gate in terms of physical stabilizer operations and Toffoli gates, and adaptive classical control \cite{Shor96}.  That is, with code concatenation, the Toffoli gate can be implemented at $k$ levels of encoding in terms of the Toffoli gate and stabilizer operations at $k-1$ levels of encoding.  (Stabilizer operations at level $k$ are of course implemented in terms of stabilizer operations at level $k-1$.)

``Magic states distillation" \cite{BravyiKitaev04, KnillLaflammeZurekProcRSocLondA98, Knill04schemes} is a technique which instead allows for fault-tolerant universality at level $k$ from only level-$k$ stabilizer operations and some sort of universality operation (certain noisy ancilla preparations) at unencoded level zero \cite{Knill04analysis, Knill05}.  Magic states distillation lets us skip the fault-tolerance hierarchy for universal quantum computing operations, simplifying threshold proofs and estimates.  Roughly, this reduces proving a noise threshold for full universal quantum computation down to proving a threshold for stabilizer operations alone.  Moreover, this ``reduction" often works without affecting the maximum tolerable noise rate, or threshold, because the bottleneck is in achieving reliable stabilizer operations.

\subsection{Magic states distillation problem}

\theoremstyle{definition} 
\newtheorem*{magicstatesdistillation}{Magic states distillation problem}
\begin{magicstatesdistillation}
For which (single- or multi-) qubit (mixed) states $\rho$ does stabilizer operations plus repeated preparation of $\rho$ imply quantum universality?
\end{magicstatesdistillation}

If repeated preparation of $\rho$ and stabilizer operations gives universality, we say for short that $\rho$ ``gives universality," or $\magic{\rho}$.  
Single-qubit states $\rho$ can be parameterized by their Pauli coordinates, $\rho(x,y,z) = \tfrac{1}{2}{(I + x X + y Y + z Z)}$; see Fig.~\ref{f:blochsphere}.
Some of the main results on magic states distillation, from \Refs\cite{BravyiKitaev04, Reichardt04magic, Reichardt06magic}, are:

\begin{figure}
\begin{center}
\includegraphics[scale=.5]{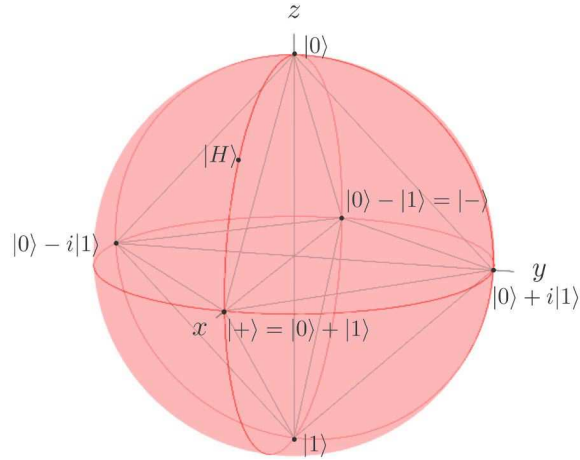}
\end{center}
\caption{{\bf Bloch sphere:} 
Single-qubit states are in one-to-one correspondence with points on or in the Bloch unit sphere in $\bf{R^3}$ (up to a global phase factor).  Coordinates $(x,y,z)$ correspond to the $2 \times 2$ density matrix $\rho(x,y,z) = \tfrac{1}{2}(I + x X + y Y + z Z)$, and these coordinates can be recovered from $\rho$ as $(\tr X \rho, \tr Y \rho, \tr Z \rho)$.   Pure states correspond to points on the surface of the sphere ($X$, $Y$ and $Z$ eigenstates along the $x$, $y$ and $z$ axes).  All points $\rho$ in the octahedron $\mathcal{O}$ the convex hull of the six single-qubit stabilizer states (Pauli eigenstates) give classically simulatable computations together with stabilizer operations.  To simulate preparation of $\rho$, randomly choose one of the vertices with the appropriate probability, and continue with the Gottesman-Knill simulation procedure. 
One-qubit unitaries correspond to rotations of the sphere \cite{NielsenChuang00,FujiwaraAlgoet99,RuskaiSzarekWerner01}, 
and one-qubit Clifford gates are exactly of the rotational symmetries of $\mathcal{O}$.
}
\label{f:blochsphere}
\end{figure}

\begin{theorem} \label{t:magicsummary}
$\magic{\rho(x,y,z)}$ if $$\max\{ \abs{x} + \sqrt{y^2+z^2}, \abs{y} + \sqrt{x^2+z^2}, \abs{z} + \sqrt{x^2+y^2}\} > 1 \enspace .$$  Also, $\magic{\rho(f x, f y, f z)}$ for $x = y = \tfrac{3 \sqrt{7} - 7}{7 (2 - \sqrt{2})}$, $z =  \tfrac{3 \sqrt{14} - 14}{7 (2 - \sqrt{2})}$ and $f = 0.9895$.  (Notice that $x+y+z = 3/\sqrt{7}$ and $\sqrt{x^2+y^2}+z = 1$.)
\end{theorem}

\noindent (See Figs.~\ref{f:oldnorotation_rotationsandt}, \ref{f:graph}.)  
Define $\ket{H}$ and $\ket{T}$ by $\ketbra{H}{H} \equiv \rho(\tfrac{1}{\sqrt{2}}, 0, \tfrac{1}{\sqrt{2}})$ and $\ketbra{T}{T} \equiv \rho(\tfrac{1}{\sqrt{3}}, \tfrac{1}{\sqrt{3}}, \tfrac{1}{\sqrt{2}})$, and let $\mathcal{E}_p$ be a depolarization channel with rate $p$: 
$$\mathcal{E}_p(\rho(x,y,z)) = (1-p) \rho(x,y,z) + p \rho(0,0,0) = \rho((1-p)x,(1-p)y,(1-p)z)\enspace .$$  
Then in particular $\magic{\mathcal{E}_p(\ketbra{H}{H})}$ holds if $p < 1-\tfrac{1}{\sqrt{2}} \approx 29.2\%$, and $\magic{\mathcal{E}_p(\ketbra{T}{T})}$ if $p < 1-\sqrt{\tfrac{3}{7}} \approx 34.5\%$.

\begin{figure}
\begin{center}
\includegraphics[scale=.5]{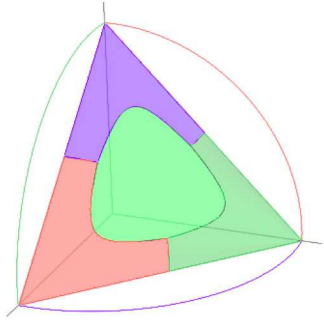}
\end{center}
\caption{One octant of the region of single-qubit states for which $\magic{\rho}$ is unknown (other octants are symmetrical).  The region is bounded by $1 < x+y+z \leq 3/\sqrt{7}$ and $\max\{ x + \sqrt{y^2+z^2}, y + \sqrt{x^2+z^2}, z + \sqrt{x^2+y^2}\} \leq 1$ (Theorem~\ref{t:magicsummary}).}
\label{f:oldnorotation_rotationsandt}
\end{figure}

\begin{figure}
\begin{center}
\includegraphics[scale=.5]{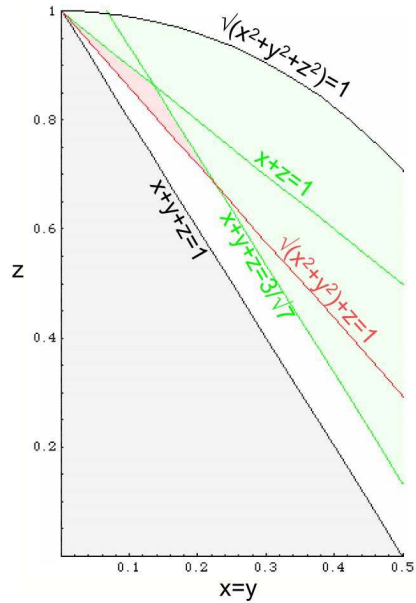}
\end{center}
\caption{A cross-section of Fig.~\ref{f:oldnorotation_rotationsandt} through the plane $x=y$.  Points within the octahedron $\mathcal{O}$ are shaded grey (lower-left), and points known to give universality are shaded green or red (upper right; not inclusive of the inner boundary).  (The point $\tfrac{0.9895}{7 (2 - \sqrt{2})}(3 \sqrt{7} - 7,3 \sqrt{7} - 7,3 \sqrt{14} - 14)$ is not shown; see \Ref\cite{Reichardt06magic}.)  There remains a gap between the planes $x+y+z=1$ and $x+y+z=3/\sqrt{7}$ where $\magic{\rho}$ is unknown.  (There is no gap in the $y=0$ cross-section of Fig.~\ref{f:oldnorotation_rotationsandt}.)}
\label{f:graph}
\end{figure}

\subsection{Applications to fault tolerance} \label{s:magicftapplications}

The main applications of magic states distillation are to fault-tolerant quantum computing.  In particular, it clearly addresses the problem of achieving universality using noisy ancilla preparation.  It does assume perfect stabilizer operations, though, which is certainly not realistic.  This assumption can be justified in two different contexts: 

1. Certain arguments for \emph{upper}-bounding the noise threshold assume that stabilizer operations are perfect and only the extra operation required for universality is noisy.  This is optimistic, but sufficient for an upper bound.  Theorem~\ref{t:magicsummary} is applied in Sec.~\ref{s:upperbounds} to prove that some of these recent upper bounds \cite{VirmaniHuelgaPlenio, BuhrmanCleveLaurentLindenSchrijverUnger06} are tight.

2. Assumptions, like perfect stabilizer operations, which are unrealistic at the physical level can sometimes be justified at higher levels of encoding in a fault-tolerant concatenated coding scheme.  In particular, it is possible that there are two different noise thresholds, one threshold for reliable stabilizer operations and a separate threshold for reliable universal quantum computation.  If the physical noise is below the threshold for reliable stabilizer operations, then we can assume that stabilizer operations are in fact perfect -- not at the physical level, but at some ``logical" level.  

Assuming perfect stabilizer operations, magic states distillation gives universality by using approximately prepared single-qubit ancillas.  However, these noisy ancillas need to be prepared not at the physical level, but at the same higher level of encoding at which the logical stabilizer operations are perfect.  

How can one reliably encode noisy ancillas?  Following Knill \cite{Knill04schemes,Knill05}, we use perfect encoded stabilizer operations to create an encoded Bell pair $\tfrac{1}{\sqrt{2}}(\ket{00}_L + \ket{11}_L)$ (the subscript $L$ denoting ``logical").  Then we decode one half from the bottom up, ideally obtaining $\tfrac{1}{\sqrt{2}}(\ket{0}\ket{0}_L + \ket{1}\ket{1}_L)$.  Finish by preparing a qubit in a ``magic" state like $\ket{H}$ or $\ket{T}$ and teleport it into the encoding, using a physical CNOT gate and two single-qubit measurements.  See Fig.~\ref{f:universalityreduction}.  If there is no noise, then the output state will be $\ket{H}_L$ or $\ket{T}_L$.  At that point, both stabilizer operations and ancilla preparation can be done at an encoded level, so we get encoded universality.  

\begin{figure}
\begin{center}
\includegraphics[scale=.5]{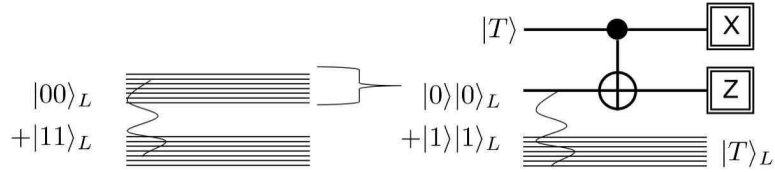}
\end{center}
\caption{Method for achieving universality by teleporting into an encoding.  If the X and Z measurements are not postselected, then a logical correction (not shown) might be required.}
\label{f:universalityreduction}
\end{figure}

In the presence of noise, the noise too will be teleported into the encoding, i.e., into logical noise.  As long as it is not too high, we can distill it out at the encoded level, using (perfect) encoded stabilizer operations.  Now the noise can come from three places: noise in the prepared single-qubit ancilla, noise in the physical teleportation circuit, and noise in the decoded half of the Bell pair.  As long as the total noise from these sources is not too large, magic states distillation will succeed.  Fortunately, magic states distillation can tolerate very high amounts of noise 
(Theorem~\ref{t:magicsummary}).
Therefore, it often turns out that the threshold for universal quantum computation is the same as that for just stabilizer operations.  The bottleneck is in achieving reliable stabilizer operations.  

This reduction, from universal fault tolerance to stabilizer operation fault tolerance, does not necessarily always work, because it requires that we be able to decode one half of an encoded Bell pair without introducing too much noise.  It is conceivable that we can prepare a perfect encoded Bell pair (because the noise is below the stabilizer operation threshold), but can't decode half of it without noise going out of control (or at least too high for magic states distillation).  

To date, though, the analysis of the decoding procedure, and therefore of encoded universality via magic states distillation, has not seemed to lead to any particular difficulties in fault-tolerance threshold analyses.  Here, the strong independence guaranteed by the Mixing Lemma will even let us postselect on no detected errors as we decode one half of the Bell state from the bottom up.  Ref.~\cite{Reichardt05distancethree} included a straightforward analysis of error correction during the decoding.  (In practice, perhaps hybrid decoding schemes might be useful; e.g., only postselect on no detected errors in decoding the last level.)  With an independent noise model, decoding cannot create correlated errors (decoding with postselection of course still can).  Therefore, decoding out of the code turns out to be a useful trick in various situations.  For example, to prepare a large stabilizer state with independent bit errors, one can first prepare the state on top of a fault-tolerance scheme, then decode out the bottom (see Sec.~\ref{s:postselectionasymptoticefficiency}).

Finally, note that while fault-tolerance schemes often use concatenated coding, and magic states distillation can also be phrased as projection into a certain code space (e.g., for the five-qubit code in the distillation scheme discussed in Sec.~\ref{s:fivequbit}), the two codes need bear no relationship to each other.

\subsection{Distillation stability conditions}

In the magic states distillation problem, it was assumed that the exact same state $\rho$ can be prepared repeatedly, and also that $\rho$ is known to the experimenter.  These are very strong assumptions, and not physically justified.  However, in considering noise threshold upper bounds (Sec.~\ref{s:upperbounds}), even delicate, artificial models can be of interest.  In applications to fault-tolerance lower bounds, though, additional stability conditions are required.  

Here, we require the ability to obtain universality from partially known states with only a known lower bound on their fidelity with $\ket{T}$.  We will prove:

\begin{theorem}[\cite{BravyiKitaev04,Reichardt05distancethree}] \label{t:practicalmagicrandom}
For any constant $\delta > 0$, perfect stabilizer operations with adaptive classical control, together with the ability to prepare states $\rho_i = \rho(x_i=y_i=z_i=f_i/\sqrt{3}) = \tfrac{1}{2}(I + \tfrac{f_i}{\sqrt{3}}(X+Y+Z)$ with each $f_i$ unknown but $\geq (1+\delta)/\sqrt{3}$, allows one to simulate universal quantum computation. 
\end{theorem}

\noindent The bound on the allowed error rate turns out to be the same for these nonidentical states as for the identical states assumed in Theorem~\ref{t:magicsummary}.  

The assumption of this theorem, that each prepared $\rho_i$ has equal $x$, $y$, $z$ coordinates, can be guaranteed by randomly applying $I$, $T \equiv \tfrac{1}{2}\left(\begin{smallmatrix}-1+i&1+i\\-1+i&-1-i\end{smallmatrix}\right)$, or $T^2$ each with probability $1/3$ independently to each prepared $\rho_i$.  ($T$ is a Clifford unitary which permutes the $x$, $y$, $z$ coordinates of the Bloch sphere: $T X T^\dagger = Y$, $T Y T^\dagger = Z$, $T Z T^\dagger = X$.  $\ket{T}$ is the $e^{2\pi i/3}$ eigenstate of $T$.)  
However, the ability to apply perfect Clifford unitaries may \emph{not} imply the ability to apply a random perfect Clifford unitary.  This symmetrization is not innocuous, particularly with a postselection-based fault-tolerance scheme like ours, or any scheme in which we are trying to maintain a strong error independence condition.  The problem is that the errors within the encoded $\rho$ will depend on whether logical $I$, $T$ or $T^2$ was applied.  

(Theorem~\ref{t:practicalmagicrandom} can suffice for error-correction-based fault-tolerance schemes using an analysis requiring only weaker control over errors \cite{Reichardt05distancethree}.  Also, recall from Sec.~\ref{s:postselectionasymptoticefficiency} that, for efficiency reasons, one expects to use postselection only for preparing codewords for an asymptotically large error-correcting code, and then achieve universality on top of this large code.  Theorem~\ref{t:practicalmagicrandom} might also suffice for this [speculative] application.)

Fortunately, the same distillation scheme is stable even up to off-axis state perturbations, so it is not absolutely necessary to symmetrize:

\begin{theorem} \label{t:practicalmagicdeterministic}
There exists a constant $\epsilon > 0$ such that perfect CNOT, Hadamard, preparation of $\ket{0}$ and measurement in the $\ket{0}$, $\ket{1}$ basis, with adaptive classical control, together with the ability to prepare (unknown) states $\rho_i$ each with fidelity $\geq 1-\epsilon$ with $\ket{T}$, allows one to simulate universal quantum computation. 

More explicitly, letting the Pauli coordinates of $\rho_i$ be $(x_i, y_i, z_i)$, universal quantum computation can be simulated provided $\max_i \max \{ \abs{\tfrac{1}{\sqrt{3}} - x_i}, \abs{\tfrac{1}{\sqrt{3}} - y_i}, \abs{\tfrac{1}{\sqrt{3}} - z_i} \} \leq 0.0527$.
\end{theorem}

In Theorem~\ref{t:practicalmagicdeterministic}, we have also relaxed the requirement for perfect stabilizer operations slightly, since CNOT and Hadamard do not generate the full Clifford group -- even with preparation of $\ket{0}$ and measurement in the $\ket{0}$, $\ket{1}$ basis, with adaptive control.  An additional operation such as $T$ or $\left(\begin{smallmatrix}1&0\\ 0&i\end{smallmatrix}\right)$ is required to get imaginary phases (e.g., the $Y$ $+1$ eigenstate is $\tfrac{1}{\sqrt{2}}(\ket{0} + i \ket{1})$).

\section{Encoding a noisy $\ket{T}$} \label{s:decodingandteleportedencoding}

\def\D {{\mathcal{D}}}
\def\E {{\mathcal{E}}}

\subsection{Decoding operation} \label{s:bottomupdecoding}

As sketched in Sec.~\ref{s:magicftapplications}, half an encoded Bell pair is decoded, from the bottom up, to the physical level, in order to allow teleporting into the encoding.  We start by defining and analyzing this decoding operation.  Define $\D_1$ as decoding one code level down to the physical level, postselected on no detected errors.  For the two-bit repetition code, 
$$
\includegraphics[scale=.8]{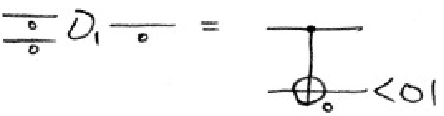}
$$
(The notation here is as in Ch.~\ref{s:postselectchapter}.)  One might instead define $\D_1$ to apply error-correction to the decoded qubit based on the extracted syndrome information, without affecting the analysis.  With the two-bit repetition code, though, it isn't possible to error-correct the output qubit (but see Ch.~\ref{s:correctionchapter}).  

The bottom-up decoding operation $\D_j$ is then defined recursively by
$$
\includegraphics[scale=.8]{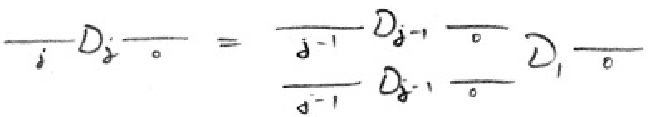}
$$
The behavior of $\D_j$ is characterized by 
\def\incl #1#2{\makebox{\raisebox{#1}{\includegraphics[scale=.8]{images/#2}}}}
\begin{eqnarray*}
\incl{-1.5ex}{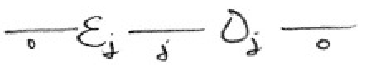} &=& \incl{-2.5ex}{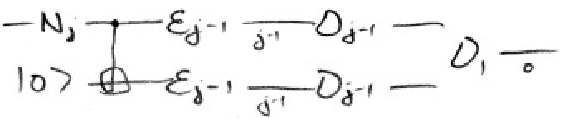} \\
&=& \incl{-2.2ex}{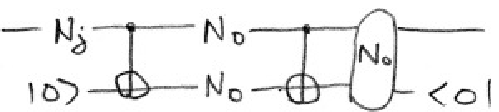} \\
&=& \incl{-.6ex}{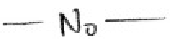}
\end{eqnarray*}
Here, the first line is by definition of the noisy encoder $\E_j$ and the second line by definition of the noisy CNOT$_0$ gate (Secs.~\ref{s:postselecterrormodel}-\ref{s:postselectgoal}), as well as by induction.  The content of the third line is that the noise rate remains bounded!  The reason the noise rate does not grow with $j$, say linearly, is that the first three error locations on line two above are quadratically suppressed by the error detection.  Below a constant threshold, the noise from the physical gates used in the final decoding dominates.  
Indeed, the recursion for the noise rate $p$ on the decoded qubit is dominated by 
$$
p \longrightarrow \eta_0 + c p^2
$$
for some constant $c$.
Initially, $p$ is bounded by a constant multiple of $\eta_0$, and it is easy to check that it stays so bounded.  
In \Ref\cite{Reichardt05distancethree}, this analysis of the decoding noise rates is carried out, and a numerical bound for the recursion's convergence is solved for.  There, it turns out that the threshold for decoding is above the threshold for reliable stabilizer operations, and so not a bottleneck -- and this is what we tend to expect.

\subsection{Teleportation encoding} \label{s:teleportedencoding}

To be able to apply Theorem~\ref{t:practicalmagicdeterministic}, we need first to prepare a reliable encoded states $\ket{\tilde{T}}_k$, i.e., reliable encodings of states each of which is a bounded distance from $\ket{T}$.  Start by extending our error model to allow preparation of (noisy) $\ket{T}$ at unencoded level zero:
$$
\includegraphics[scale=.8]{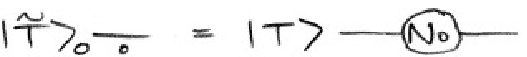}
$$
Assume we are below the threshold for stabilizer operations determined in Ch.~\ref{s:postselectchapter}, and also below the threshold for decoding determined above.  In particular, we can prepare reliable Bell states
$$
\includegraphics[scale=.8]{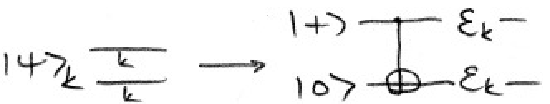}
$$
(see Sec.~\ref{s:reliablebellstate}).  Now let 
\def\incl #1#2{\makebox{\raisebox{#1}{\includegraphics[scale=.8]{images/#2}}}}
\begin{eqnarray} \label{e:teleportedencoding}
\incl{-1.5ex}{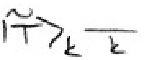} &\equiv& \incl{-4ex}{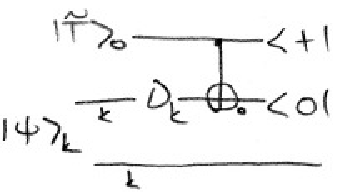} \\
&\rightarrow& \incl{-3ex}{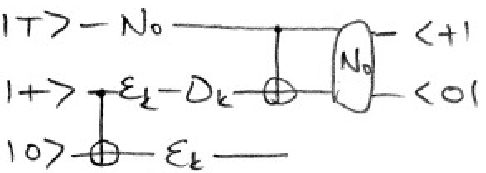} \nonumber \\
&=& \incl{-3ex}{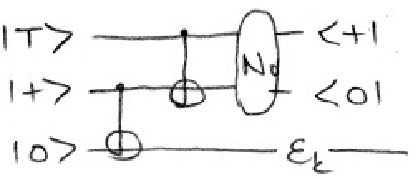} \nonumber \\
&=& \incl{-1ex}{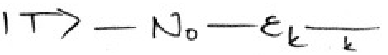} \nonumber
\end{eqnarray}
The last equality above is a consequence of (ideal) teleportation; the possible measurement errors lead to Pauli errors on the ideally teleported state.  

\section{Propagation of virtual errors} \label{s:virtualerrorpropagation}

Recall 
that the experimentalist probabilistically may himself have deliberately added errors into the system.  (In preparing a level-$j$-encoded Bell state, he probabilistically introduces errors at level $j-1$ in order to weaken error correlations within the system so the Mixing Lemma can be applied.)
As discussed in Sec.~\ref{s:reliablebellstate}, 
the method with which he introduces the errors, however, is important.  For example, to introduce a bit flip error at encoded level $j$, it is typically not okay simply to apply an X gate to each bit in the code block (i.e., transversally).  For, if X gates are not perfect, then doing so would create correlations between level-$0$ and level-$j$ errors.  (If X gates \emph{are} perfect, then transversal X application might be acceptable -- depending on other details of the error model, like the presence of memory errors.)

Instead, the experimentalist needs merely remember (in a classical computer) what errors should be added to each bit.  As Clifford gates are applied, this description can be updated using Clifford/Pauli commutation rules (known as tracking the ``Pauli frame").  Overall, the combined distribution of errors -- either ``virtual" errors or real errors -- can always be written as a mixture of error distributions with strong independence properties.  (There are possibly strong correlations within the virtual error distribution, and also within the physical error distribution; only for the combined distribution do we have independence properties.)

One might be concerned that the presence of universal gates will break this scheme.  After all, it is not possible to commute a Pauli error past a Toffoli gate, for example.\footnote{The Toffoli gate is not a Clifford operator (conjugating Paulis to Paulis), but is in so-called $C_3$, meaning it conjugates Pauli operators into the Clifford group \cite{GottesmanChuang99teleportation}.}  
However, in a scheme like ours, based on magic states distillation, any virtual error ends up in a Pauli measurement.  The error can be applied, perfectly, on the measurement outcome (in a classical computer).  Applying the virtual error to the measurement outcome is equivalent to applying a real error perfectly to the quantum state just before measurement.  This is the experimentalist's goal: not to track virtual errors through the circuit, but rather to introduce the errors -- make them real.  
It is never necessary to track a virtual error through a Toffoli gate.  Instead, virtual errors will enter our implementation of a Toffoli (using approximate ``magic" state preparation and application of Clifford operators classically adaptively based on measurement outcomes), and different virtual errors will leave -- and some virtual errors will become real during the implementation as well.

In particular, for our scheme, each qubit is touched by exactly one CNOT gate, and possibly a single-qubit Clifford unitary, before being measured and any virtual error made real.  The only exception is in the bottom-up decoding procedure of Sec.~\ref{s:bottomupdecoding}, in which $k$ CNOT$_0$s touch the final output qubit -- and then one more for teleportation.  For example, in Eq.~\eqref{e:teleportedencoding}, if we have tracked a virtual X error onto the decoded qubit, then we measure in the $\ket{0}$, $\ket{1}$ basis, and flip a 1 outcome to 0, and vice versa.

\section{Proofs of asymmetric distillation Theorems~\ref{t:practicalmagicrandom} and~\ref{t:practicalmagicdeterministic}}

This section briefly reviews the proofs of Theorem~\ref{t:practicalmagicrandom} \cite{Reichardt05distancethree, BravyiKitaev04} and Theorem~\ref{t:practicalmagicdeterministic}.  
The proofs are primarily just calculations.  
Throughout the section, we assume that stabilizer operations are \emph{perfect}.  

The idea is that many noisy copies of $\ket{T}$ are distilled with postselected stabilizer operations to one nearly perfect $\ket{T}$.  This can then be used to obtain universality.  Overall, despite using postselection, the procedure is efficient.  

\subsection{Efficient distillation of $\ket{T}$} \label{s:fivequbit}

\newcommand{\degrees}{\ensuremath{^\circ}}

Recall that $\ket{T}$ is defined, up to an unimportant phase, by $\ketbra{T}{T} \equiv \rho(\tfrac{1}{\sqrt{3}}, \tfrac{1}{\sqrt{3}}, \tfrac{1}{\sqrt{2}}) = \tfrac{1}{2}(I + \tfrac{1}{\sqrt{3}}(X + Y + Z)$.  
The polar coordinates of $\ket{T}$ are $(\theta,\phi) = (\cos^{-1} \tfrac{1}{\sqrt{3}}, \tfrac{\pi}{4})$, so, setting the phase, $\ket{T} = \cos \tfrac{\theta}{2} \ket{0} + e^{i \phi} \sin \tfrac{\theta}{2} \ket{1}$. 
$\ket{T}$ is the $e^{2\pi i/3}$ eigenstate of the $2\pi/3$ rotation $T = e^{i 2\pi/3} \rho(\tfrac{1}{\sqrt{3}}(1,1,1)) + e^{-i 2\pi/3} \rho(-\tfrac{1}{\sqrt{3}}(1,1,1)) =  \tfrac{1}{2}\left(\begin{smallmatrix}-1+i&1+i\\-1+i&-1-i\end{smallmatrix}\right)$, a $120\degrees$ rotation about $\ket{T}$ on the Bloch sphere ($T X T^\dagger = Y$, $T Y T^\dagger = Z$, $T Z T^\dagger = X$). 

Recall the five-qubit code, with stabilizer generators 
\begin{center}
\begin{tabular}{
c @{$\!\,$} c @{$\!\,$} c @{$\!\,$} c @{$\!\,$} c
@{$,\,$}
c @{$\!\,$} c @{$\!\,$} c @{$\!\,$} c @{$\!\,$} c
@{$,\,$}
c @{$\!\,$} c @{$\!\,$} c @{$\!\,$} c @{$\!\,$} c
@{$,\,$}
c @{$\!\,$} c @{$\!\,$} c @{$\!\,$} c @{$\!\,$} c
}
X&Z&Z&X&I&
I&X&Z&Z&X&
X&I&X&Z&Z&
Z&X&I&X&Z,
\end{tabular}
\end{center}
and logical X and logical Z being transversal X and Z, respectively.
Denote by $[f]^s$ the symmetric sum of $s$-tuples of the variables $f_1, \ldots, f_5$, i.e., 
$$
[f]^s \equiv \sum_{\substack{S \subseteq [5] \\ \abs{S} = s}} \prod_{i \in S} f_i \enspace .
$$

Take five prepared states, $\otimes_{i=1}^{5} {\rho_i}$, and use stabilizer operations to project into the codespace of the five-qubit code, then decode the logical qubit.  (This can be implementing by decoding the code, then measuring the extra bits and postselecting on no detected errors.)
On failure, try again.
A simple calculation gives that the probability of success is 
$$
p_\text{success} = \tfrac{1}{48}(3 + [f]^4) \enspace .
$$
The $x$, $y$, $z$ coordinates of the output state, conditioned on success, are equal, and equal to 
$$
\frac{1}{\sqrt{3}} \cdot \frac{1}{p_\text{success}} \frac{-1}{48}([f]^3 - 2 [f]^5) \enspace .
$$
These coordinates are negative, but one can rotate them back to be positive with a stabilizer operation.  Then the output state is $\rho(x=y=z=f_{out}/\sqrt{3})$, where 
$$
f_\text{out} = \frac{[f]^3 - 2[f]^5}{3+[f]^4} \enspace .
$$
In particular, considering the case when all $f_i$s are equal, one can easily check that $f_\text{out} > f$ when $f > \sqrt{3/7}$.  Recursively applying this procedure on the output states, the output fidelity converges to one.  A straightforward calculation shows that the convergence is fast, even accounting for the many discarded ancillas; the total overhead needed to reach fidelity $1-\theta$ scales as $(\log 1/\theta)^{\log_2 30}$ \cite{BravyiKitaev04}.  

What if the $f_i$ are not all equal?  The probability $p_\text{success}$ of decoding acceptance is clearly monotone in each $f_i$, so distillation remains efficient.  
Also, simple algebra gives that $\partial f_\text{out} / \partial f_i > 0$, so improving any of the input fidelities can only improve the output fidelity.  
Indeed, differentiate $f_ \text{out}$ with respect to $f_5$ -- other derivatives are related by symmetry.  Use the quotient rule $d (a/b) = \tfrac{1}{b^2}(b\, da - a\, db)$.  
The numerator, which does not involve $f_5$, is, after simplifications, 
$$
 f_1 f_2 \left( 3 - f_1 f_2 f_3 f_4 - f_3 f_4 - \tfrac{1}{3} f_1 f_2 f_3^2 f_4^2 - \tfrac{1}{3} f_1 f_2(f_3^2 + f_4^2) \right) + \textrm{symmetrical terms} . 
$$
Each term is nonnegative when the $f_i \in [0,1]$, implying that $f_\text{out}$ is monotone in each $f_i$ separately. 

To prove Theorem~\ref{t:practicalmagicdeterministic}, one can check that the same distillation procedure works, still with only polylogarithmic total overhead.  For the existence of an $\epsilon > 0$ such that states within $\epsilon$ of $\ket{T}$ are distilled to $\ket{T}$, it suffices to Taylor-expand the output state around the inputs being close to $\ket{T}$ (using three different small parameters for each of the five inputs) -- the difference of the output state from $\ket{T}$ is then second-order.  

Indeed, projecting 
\beq \label{e:practicalmagicstartstate}
\bigotimes_{i=1}^5 \tfrac{1}{2}\left(I + \tfrac{1}{\sqrt{3}}((1-\epsilon_{x,i}) X + (1-\epsilon_{y,i}) Y + (1-\epsilon_{z,i}) Z)\right)
\eeq
into the code space, and dropping quadratic or smaller terms in the $\epsilon_{x/y/z,i}$, gives
$$
\frac{1}{36}\left(6 - \sum_{i=1}^5 (\epsilon_{x,i} + \epsilon_{y,i} + \epsilon_{z,i})\right) \cdot \frac{1}{2} \left(I-\tfrac{1}{\sqrt{3}}(X+Y+Z)\right) \enspace ,
$$
or just $\frac{1}{2}(I-\tfrac{1}{\sqrt{3}}(X+Y+Z))$ after renormalization.

To determine an explicit numerical distillation sufficiency condition, project the state of Eq.~\eqref{e:practicalmagicstartstate} into the code space, but do not drop higher-order terms.  Renormalize, and compute the $\epsilon_{x/y/z}$ values of the resulting state.  Each is a ratio of two degree-five, multilinear polynomials.  Upper-bound the numerator by summing the absolute values of each monomial term.  (Note that the $\epsilon_{x/y/z,i}$ need not all be nonnegative.)  Lower-bound the denominator by subtracting from the constant term the sum of the absolute values of the nonconstant terms.  The resulting ratio is maximized by setting the $\epsilon_{x/y/z,i}$ as large as possible, say to $p$ -- giving 
$$
\frac{180 p^2 + 150 p^3 + 45 p^4 + 6 p^5}{24 - 60 p - 90 p^2 - 60 p^3 - 15 p^4} \enspace .
$$
Plotting this function, one observes that the cutoff for improvement over $p$ is above $0.0913$.  Divide by $\sqrt{3}$, giving $0.0527$, to match the notation in the statement of Theorem~\ref{t:practicalmagicdeterministic}.  This bound is probably very conservative, but it is still quite good.

An encoding circuit for the five-qubit code is given in Fig.~\ref{f:fivequbitencoder}.  The circuit for distillation is the same, except run in reverse with postselected measurements instead of preparations -- it only uses CNOTs, Hadamards, and measurement in the $\ket{0}$, $\ket{1}$ basis.  Therefore, this restricted set of operations suffices to distill $\ket{T}$.  From $\ket{T}$, the $Y$ $+1$ eigenstate can be prepared (Sec.~\ref{s:universalityfromt}), allowing by simple stabilizer algebra the completion of the Clifford group \cite{Gottesman97,KnillLaflammeZurek96}.

\begin{figure}
\begin{center}
\includegraphics[scale=1]{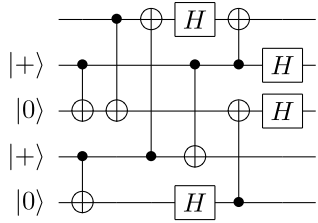}
\end{center}
\caption{Encoding circuit for the five-qubit code.}
\label{f:fivequbitencoder}
\end{figure}

\subsection{Reduction to distillation of $\ket{T}$: $\magic{\ket{T}}$} \label{s:universalityfromt}

Now let us prove $\magic{\ket{T}}$ \cite{BravyiKitaev04} -- i.e., that repeated preparation of $\ket{T}$ (perhaps via distillation), together with perfect stabilizer operations and adaptive classical control, gives universality.  

First, Shi proved the universality of the two gates CNOT and $R$, for $R$ any single-qubit real gate such that $R^2$ does not preserve the computational basis \cite{Shi02}.  In particular, this implies that stabilizer operations together with application of $\exp(i \tfrac{\phi}{2} X)$ gives universality for $\phi \notin \{i \tfrac{\pi}{2} : i \in \bf{Z}\}$.  (This fact can likely also be directly proved following \Ref\cite{BoykinMorPulverRoychowdhuryVatan00}.)  

Now in general, $\exp(i \tfrac{\phi}{2} Z)$ -- which the Hadamard conjugates to $\exp(i \tfrac{\phi}{2} X)$ -- can be implemented from prepared copies of $\exp(i \tfrac{\phi}{2} Z) \ket{+}$, together with stabilizer operations and adaptive control.  (Recall $\ket{+} = \tfrac{1}{\sqrt{2}}(\ket{0}+\ket{1})$ is the X $+1$ eigenstate.)  Indeed, 
\begin{eqnarray*}
(\alpha \ket{0} + \beta \ket{1})(\ket{0} + e^{i \phi} \ket{1}) &=& \begin{split}\alpha \ket{00} + \beta e^{i \theta} \ket{11} \\+ \alpha e^{i \phi} \ket{01} + \beta \ket{10}\end{split} ,
\end{eqnarray*}
by simply expanding out the tensor product.  
Then measure the parity, by applying a CNOT from the first qubit to the second.  
\begin{itemize}
\item{Even parity:} $\exp(i \tfrac{\phi}{2} Z)$ has been applied, as desired.
\item{Odd parity:} $\exp(-i \tfrac{\phi}{2} Z)$ has been applied, the opposite phase from what was desired.  In this case, repeat the procedure (using adaptive control), to carry out a random walk on phases which are integer multiples of $\phi$.  Terminate the walk when the phase is $+\phi$.  
\end{itemize}
For $\phi$ a rational multiple of $\pi$, the random walk is on a cyclic set, and the probability that more than $m$ steps are needed drops exponentially with $m$.  (For general $\phi$, this probability is $O(1/\sqrt{m})$.)

Finally, one can check that taking two copies of $\ket{T}$ and postselecting on even parity, the output is symmetric under Clifford gates to $\exp(i \tfrac{\phi}{2} Z) \ket{+}$, for $\phi = \tfrac{\pi}{6}$.

\begin{remark}
In Theorems~\ref{t:practicalmagicrandom} and~\ref{t:practicalmagicdeterministic}, we chose to distill $\ket{T}$ instead of $\ket{H}$ primarily for the simplicity of proving that distillation still works on asymmetrical inputs.  We have not checked the stability of any of the various distillation procedures for $\ket{H}$ \cite{BravyiKitaev04, Reichardt04magic, Reichardt06magic}.  However, the $\ket{H}$ distillation procedure, based on taking fifteen states close to $\ket{H}$ and projecting into the codespace of the fifteen-qubit Reed-Muller code \cite{KnillLaflammeZurek96,BravyiKitaev04,RaussendorfHarringtonGoyal05oneway}, is asymptotically more efficient than distilling towards $\ket{T}$.  Also, $\ket{H}$ is symmetrical to $\exp(i \tfrac{\pi}{8} Z) \ket{+}$, so a random walk to apply a phase of $\phi = \tfrac{\pi}{4}$ is not required -- on measuring odd parity, use a Clifford unitary to add a phase of $\tfrac{\pi}{2}$.  Finally, given a $\tfrac{\pi}{4}$ phase shift, a Toffoli gate can be implemented exactly \cite{KnillLaflammeZurekProcRSocLondA98}.  
\end{remark}

\section{Putting it together for encoded magic states distillation} \label{s:puttingmagictogether}

Finally, let us put it all together, as in Sec.~\ref{s:postselectgoal}.  Assume we are given an nonadaptive ideal quantum circuit $\mathcal{C}$, which takes as input only $\ket{0}$ states (we may assume any varying input is hardwired into the circuit), and whose output is measured in the 0,1 computational basis.  We wish to simulate the distribution of outputs of $\mathcal{C}$ using faulty quantum gates.  

Start by compiling $\mathcal{C}$ into the basis of Clifford group unitaries and application of $\exp(i \tfrac{\pi}{12} Z)$.  Assume there are $N$ gates total.  To simulate $\mathcal{C}$ fault-tolerantly, as $\mathrm{FT}\mathcal{C}$, set $k$ the number of code concatenation levels to be $\Omega(\log \log N)$.  Each ideal Clifford in $\mathcal{C}$ is implemented reliably as described in Ch.~\ref{s:postselectchapter}.  
The $\tfrac{\pi}{6}$ phase rotations are implemented as described in Sec.~\ref{s:universalityfromt} above, using reliable Clifford gates.  The needed encoded $\ket{T}$ states are prepared to an accuracy of $1/N$ using magic states distillation (Sec.~\ref{s:fivequbit}) with $\poly(\log N)$ preparations of $\ket{\tilde{T}}_k$ from teleportation encoding (Sec.~\ref{s:teleportedencoding}).

This completes the specification of $\mathrm{FT}\mathcal{C}$.  Overall, conditioned on passing any error-detection tests, 
we have some procedure taking as input $\ket{0}$s and noisy prepared $\ket{\tilde{T}}_0$s.  Using the rules of Sec.~\ref{s:decodingandteleportedencoding} and Ch.~\ref{s:postselectchapter}, noisy encoders $\tilde{\mathcal{E}}_k$ can be passed through the circuit, leaving an ideal circuit close to $\mathcal{C}$ interspersed with noise locations $N_k$ (independent failures at rate $\leq \eta_k = (c \eta_0)^{2^j}$).  None of these error events actually occur, with good probability, because $k$ has been chosen to be large enough.

\section{Fault-tolerance threshold upper bounds} \label{s:upperbounds}

Giving upper bounds for the fault-tolerance threshold (with a given set of operations and a given noise model) is difficult.  
There have been only a few approaches, and these tend to be tied delicately to a particular model.  
For example, Aharonov, Ben-Or, Impagliazzo and Nisan show that a useful noisy quantum circuit can only have logarithmic depth if fresh ancillas are not allowed to be introduced during the computation \cite{AharonovBenOr96,AharonovBenorImpagliazzoNisan96}.  But in practical quantum computing schemes, it is possible to initialize ancillas during the computation.  Razborov shows that the tolerable noise rate, of circuits with more than logarithmic depth, can be at most $1/2$ for a gate set with gates of fan-in two \cite{Razborov}.  But his approach does not allow for noiseless classical control based on measurement results.  In fact, too, interesting problems, including factoring, can be solved with log-depth quantum circuits, aided by classical computation \cite{CleveWatrous00}.  

Harrow and Nielsen \cite{HarrowNielsen} ask how much depolarizing noise can be tolerated by a two-qubit gate before it loses its power to generate entanglement; they find that the CNOT is the most resilient two-qubit gate, but does not tolerate independent depolarizing noise higher than 74\%. (Virmani, Huelga and Plenio improve this to $2/3$ with a more careful entanglement requirement \cite{VirmaniHuelgaPlenio}.)  Against simultaneous depolarizing noise, they find that the threshold is at most $8/9$, or $1/2$ for a somewhat-adversarial noise model (optimal noise process including correlated two-qubit noise).

Virmani et al. \cite{VirmaniHuelgaPlenio} assume that stabilizer operations, including the CNOT gate, are perfect, and ask how much noise can be tolerated in an additional gate used to achieve universality.  They show that the $\pi/8$ gate, $\exp(i \tfrac{\pi}{8}Z)$, with $(\sqrt{2}-1)/2\sqrt{2} \approx 14.6\%$ or more worst-case noise, or twice that amount of dephasing noise, becomes a convex combination of stabilizer operations and so this gate set can be simulated classically.  Among all the rotations $\exp(i \tfrac{\theta}{2}Z)$, the $\pi/8$ gate is the most resistant to dephasing noise according to their criterion.
The advantage of this approach, and also that of Harrow and Nielsen, is that it easily allows for the incorporation of noiseless classical control into the model.  

Buhrman, Cleve, Laurent, Linden, Schrijver and Unger extend these results to a depolarizing noise channel \cite{BuhrmanCleveLaurentLindenSchrijverUnger06}.  Again, assume that stabilizer operations are perfect, and assume that a noisy single-qubit gate is used to achieve universality.  They show that the $\pi/8$ gate with $(6-2\sqrt{2})/7 \approx 45.3\%$ or more depolarizing noise becomes a convex combination of stabilizer operations.  And again, the $\pi/8$ gate is the most noise-resistant single-qubit gate.  Therefore, $45.3\%$ is an upper bound on the noise threshold in this model.  

Magic states distillation shows the limit of the technique of Virmani et al., and of Buhrman et al.  
(This is perhaps not surprising, since according to the rough reduction of Sec.~\ref{s:introduction}, we expect the threshold bottleneck to be in achieving perfect stabilizer operations.)
Both their upper bounds are tight; with any less noise one gets universal quantum computation:

\begin{theorem} \label{t:pipereighttheorem}
Stabilizer operations, together with repeated application of a $\pi/8$ gate ($\exp(i \tfrac{\pi}{8}Z)$) subject to worst-case probabilistic noise at rate $p$ (or dephased at twice that rate), give universality if and only if $p < \tfrac{1}{\sqrt{2}}(1 - \tfrac{1}{\sqrt{2}})$.

Stabilizer operations, together with repeated application of a $\pi/8$ gate depolarized at rate $p$, give universality if and only if $p < (6-2\sqrt{2})/7$.
\end{theorem}

The $\pi/8$ gate with less than $(\sqrt{2}-1)/2\sqrt{2}$ worst-case probabilistic noise, or twice that amount of dephasing noise, takes $\ket{+}$ to a state $\rho(x,x,0)$ with $x > 1/2$, implying universality together with perfect stabilizer operations by Theorem~\ref{t:magicsummary}.  

The $\pi/8$ gate with $45\%$ depolarizing noise, however, takes $\ket{+}$ to a state well inside the octahedron $\mathcal{O}$, mixtures of one-qubit stabilizer states (Pauli eigenstates), of Fig.~\ref{f:blochsphere}.  Instead, inspired by the Jamiolkowski isomorphism, apply the noisy gate to the second half of a Bell pair (which can be prepared using stabilizer operations).  If the depolarizing noise rate is less than $(6-2\sqrt{2})/7$, then the output two-qubit state will lie outside the set of mixtures of two-qubit stabilizer states.  Moreover, there does exist a two-to-one-qubit stabilizer reduction giving a state outside $\mathcal{O}$; simply apply the parity-check procedure of Sec.~\ref{s:universalityfromt}.
Indeed, the renormalized output state at a depolarizing noise rate of $(6-2\sqrt{2})/7-\epsilon$ is computed to have $x,y,z$ coordinates of 
$$
\frac{1}{10+6\sqrt{2}+21\epsilon}\left((1+2\sqrt{2})(2+7\epsilon),-2\sqrt{2}(1+2\sqrt{2}+7\epsilon),0\right) ,
$$
for which $\abs{x}+\abs{y} > 0$ for $\epsilon > 0$.  By Theorem~\ref{t:magicsummary}, this state gives universality -- proving Theorem~\ref{t:pipereighttheorem}.  

One can show that for a one-qubit, Markovian (noisy) operation $\mathcal{E}$, stabilizer operations and $\mathcal{E}$ is universal (with adaptive classical control) iff $\magic{({\boldsymbol 1} \otimes \mathcal{E})(\ketbra{\Psi}{\Psi})}$ for $\ket{\Psi} = \tfrac{1}{\sqrt{2}}(\ket{00}+\ket{11})$ \cite{Reichardt06magic}.  
Ref.~\cite{Reichardt06magic} has more results on two-qubit-state magic states distillation procedures, and their application to fault-tolerance threshold upper bounds.   

\chapter{Thresholds for fault-tolerance schemes based on error correction} \label{s:correctionchapter}
In this chapter, we apply the mixing technique to prove the existence of fault-tolerance thresholds for schemes based on error correction, instead of postselection on no detected errors.  

In Sec.~\ref{s:correctdisttwo}, we sketch the proof of an error-correction noise threshold for stabilizer operations using a concatenated distance-two code.  (We give the essential details, but not the complete proof, to avoid duplicating the arguments of Chs.~\ref{s:postselectchapter}-\ref{s:magicchapter}.)  
Previous error correction noise threshold proofs applied only to codes of distance at least three~\cite{AliferisGottesmanPreskill05, Reichardt05distancethree}.  
Although the code here has distance only two, error correction at each code concatenation level is possible provided we use the information of whether or not an error was detected at the previous level.  For example, using the two-bit repetition code, the decoding of a $01$ block is ambiguous -- it can equally well decode to $0_L = 00$ or to $1_L = 11$.  What we do is correct the block arbitrarily, say to $0_L$, but also remember that there is a good chance the block is in error.  
Then, decoding of a level-two block, say 
$
0111,
$
gives $01$ at level one.  This is again ambiguous, but since the first bit is detected possibly to be in error, correct it to $1$.  A fault-tolerance scheme based on this type of decoding gives effective logical error rates with exponents increasing like the Fibonacci sequence.

A fault-tolerance scheme based entirely on postselection, as described in Chs.~\ref{s:postselectchapter}-\ref{s:magicchapter}, will give a threshold for reliable simulation of ideal quantum circuits, but with an inefficient, exponentially large overhead.  The scheme can be made efficient by switching ultimately to a different scheme, based on error correction.  Starting with a postselection-based scheme, instead of using error correction throughout, may allow for higher tolerable error rates (trading off higher overhead), and also allows the use of larger, more efficient codes.  The difficulty of encoding into large error correcting codes is what prompted the use of concatenated constant-sized codes in the first place, for ``bootstrapping" fault-tolerant encoding.  But bootstrapping with a postselection-based scheme, we can immediately start error correction with a large code and only need two concatenation levels to achieve a small enough effective error rate.\footnote{Note that we assume fast classical computation -- that decoding an $o(\log N)$-sized code takes only constant time -- so memory errors do not accumulate during decoding.  Without this assumption, a multiply concatenated structure of constant-sized codes, instead of just two levels of asymptotically large codes, may still be necessary to manage classical computation costs.}
In Sec.~\ref{s:postselectefficient}, we describe just how to achieve efficiently a sub-constant, $o(1)$, effective error rate for stabilizer operations, by encoding into a single large code.  The scheme can be extended to efficient fault-tolerant universality as in Ch.~\ref{s:magicchapter} (except the proof can be made even simpler now since, not using postselection, even \emph{random} stabilizer operations can be applied fault-tolerantly).

\section{Fibonacci-type threshold for a distance-two code with error correction} \label{s:correctdisttwo}

We start by proving a threshold for a Knill-type scheme based on correcting errors during the teleportation step, instead of merely postselecting on not detecting any.  Using a concatenated distance-two code, it isn't possible to decode a single level on its own.  Instead, we keep around hints from decoding the previous level, as to whether or not an error was \emph{detected} \cite{Knill05}.  
\begin{itemize}
\item A detected logical error can occur at level $j$ if there is an undetected level-$(j-1)$ error.  (E.g., using the two-qubit repetition code, if we measure a block as $01$, then we correct the block arbitrarily either to $0_L = 00$ or to $1_L = 11$, and mark the block detected.)
\item An undetected error can occur at level $j$ if there is a detected error and an undetected error at level $j-1$.  (E.g., say we measure $10$, where the second qubit has been detected to be possibly in error.  Of course we'll correct the second bit to get $1_L = 11$.  But if the first bit had an undetected error, we should have corrected to $0_L = 00$.)
\end{itemize}
Therefore the level-$j$ undetected error rate $\eta_j$ will be $O(\eta_{j-1} \eta_{j-2})$, instead of $O(\eta_{j-1}^2)$, so its exponent of $c \eta_0$ will be increasing like the Fibonacci sequence, instead of just as $2^j$.  That is, redefine
$$
\eta_j \equiv (c \eta_0)^{F(j+2)} \enspace ,
$$
where $F(n) = \tfrac{1}{\sqrt{5}}(\gamma^n + \gamma^{-n}) = F(n-1)+F(n-2)$ with $\gamma$ the golden ratio $(1+\sqrt{5})/2 \approx 1.62$.

For maximum simplicity, again assume the biased X noise error model of Sec.~\ref{s:postselecterrormodel}.

The main new ideas in proving a threshold for this scheme arise in considering the recursive decoding rule.  
\def\incl #1#2{\makebox{\raisebox{#1}{\includegraphics[scale=.8]{images/#2}}}}
Define a level-$0$ decoding operation in the computational basis ($Z$ eigenbasis) by
$$
\incl{-1em}{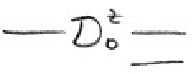} = \incl{-1em}{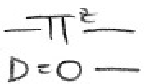} 
$$
It takes the input, projects it into the computational basis (with a measurement), and also appends an additional bit set to zero.  The second bit is a flag to indicate detected errors, and no errors can be detected at the physical level.

Level-$j$ decoding, of a code block of $2^j$ qubits, is defined as 
$$
\incl{-1em}{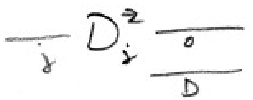} = \incl{-1.5em}{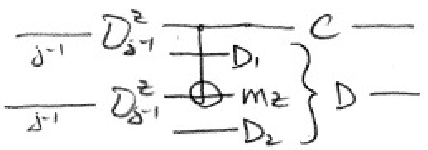} 
$$
It decodes each block, recursively, then decodes the two remaining bits, with just a single CNOT gate for the two-bit repetition code.  This CNOT gate is \emph{perfect}, since it is acting in the computational basis, in a classical computer.  A correction $C = X$ is applied if and only if an error is detected here and an error was detected in the first subblock at the previous level.  I.e., $C=X$ iff $D_1 \wedge (m_Z = 1)$; otherwise $C=I$.  The detected flag $D$ is set if an error is detected at this level, but not in either subblock, or if both subblocks have detected errors:
$$
D = ((m_Z = 1) \wedge \neg(D_1 \vee D_2)) \vee (D_1 \wedge D_2) \enspace .
$$
The point here is that when a block error is detected, and exactly one subblock has been marked detected, we can easily apply an appropriate correction.  If we detect an error and neither or both subblocks have been marked detected, then we don't know which subblock was actually bad.  We fix a correction rule anyway, but mark the output as detected to indicate that it could easily be in error; we could have corrected the wrong way. 

The rule characterizing level-$j$ decoding is
\begin{equation} \label{e:disttwocorrectdecoding}
\incl{-1em}{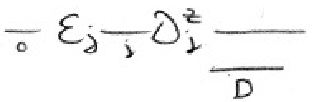} \rightarrow \incl{-1.3em}{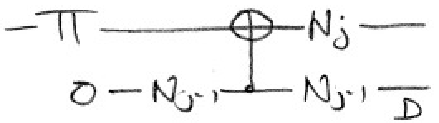}
\end{equation}
That is, an undetected error can occur at rate $\leq \eta_j$, while a detected error can occur at rate $\leq \eta_{j-1}$.  There is also the possibility that the detected-error flag is set, but there is in fact no error.

The proof is straightforward from the definitions.
$$
\incl{-1.5em}{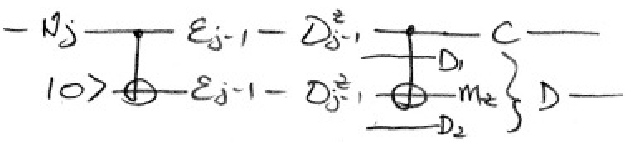} = \incl{-2.5em}{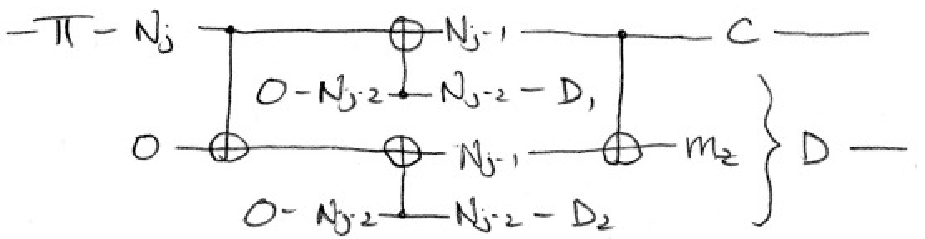}
$$
This then gives the right-hand side of Eq.~\eqref{e:disttwocorrectdecoding}.  
Indeed, let $E_1$ and $E_2$ mark the presence of errors on the two subblocks, so $m_Z = E_1 \oplus E_2$.  Let $E = E_1 \oplus C$ denote the presence of an error on the final output.  In Table~\ref{f:fibdecoderulederive}, all sixteen possible settings of the $E_i$ and $D_i$ are considered, together with the resulting $E$ and $D$ settings, and the order of the probability of each event.  Eq.~\eqref{e:disttwocorrectdecoding} follows, from the Mixing Lemma with $n=3$, since the error orders are correct.

\begin{table*} 
\caption{Exhaustive enumeration of the different decoding error possibilities.  $E_i$ indicates whether or not subblock $i$ is in error, and $D_i$ is the detected-error flag for that subblock, $i=1, 2$.  $m_Z = E_1 \oplus E_2$, $C = D_1 \wedge m_Z$.  $E = E_1 \oplus C$, $D = (m_Z \wedge \neg (D_1 \vee D_2)) \vee (D_1 \wedge D_2)$.  The error orders are set by induction on Eq.~\eqref{e:disttwocorrectdecoding}.}
\label{f:fibdecoderulederive}
\begin{center}
\begin{tabular}{c c c c | c c | c c | c}
\hline \hline
$E_1$ & $D_1$ & $E_2$ & $D_2$ & $m_Z$ & $C$ & E & D & order\\
\hline

0&0&0&0 & 0&0 & 0&0 & 1 \\
0&0&0&1 & 0&0 & 0&0 & $\eta_{j-2}$ \\
0&0&1&0 & 1&0 & 0&0 & $\eta_{j-1}$ \\
0&0&1&1 & 1&0 & 0&0 & $\eta_{j-2}$ \\
0&1&0&0 & 0&0 & 0&0 & $\eta_{j-2}$ \\
0&1&0&1 & 0&0 & 0&1 & $\eta_{j-2}^2 < \eta_{j-1}$ \\
0&1&1&0 & 1&1 & 1&0 & $\eta_{j-2}\eta_{j-1} = \eta_j$ \\
0&1&1&1 & 1&1 & 1&1 & $\eta_{j-2}^2 < \eta_{j-1}$ \\
1&0&0&0 & 1&0 & 1&1 & $\eta_{j-1}$ \\
1&0&0&1 & 1&0 & 1&0 & $\eta_{j-1}\eta_{j-2} = \eta_j$ \\
1&0&1&0 & 0&0 & 1&0 & $\eta_{j-1}^2 < \eta_j$ \\
1&0&1&1 & 0&0 & 1&0 & $\eta_{j-1}\eta_{j-2} < \eta_j$ \\
1&1&0&0 & 1&1 & 0&0 & $\eta_{j-2}$ \\
1&1&0&1 & 1&1 & 0&1 & $\eta_{j-2}^2 < \eta_{j-1}$ \\
1&1&1&0 & 0&0 & 1&0 & $\eta_{j-2}\eta_{j-1} = \eta_j$ \\
1&1&1&1 & 0&0 & 1&1 & $\eta_{j-2}^2 < \eta_{j-1}$ \\
\hline \hline
\end{tabular}
\end{center}
\end{table*}

The scheme is otherwise very similar to that described in Ch.~\ref{s:postselectchapter}.  

Reliable CNOT gates are implemented by teleporting into CNOT gates applied between Bell pairs.   The rule for commuting noisy encoders past level-$j$ CNOT gates needs to be changed, to allow for the possibility of a detected logical CNOT failure:
\begin{equation} \label{e:detectedcnot}
\incl{0em}{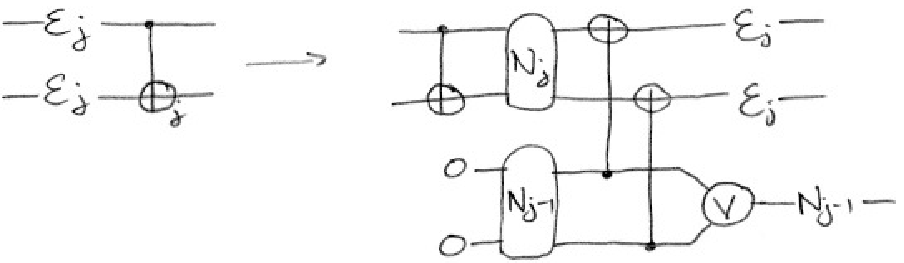}
\end{equation}
That is, there is probability $\leq \eta_j$ of an undetected correlated error on the output blocks.  A classical bit is also set to one with probability $\leq \eta_{j-1}$, to indicate the possible presence of a detected error on one or both of the output blocks.  
(The noisy encoder definition is unchanged from Eq.~\eqref{e:noisye} -- there will never be detected errors \emph{within} the output blocks.)

Reliable level-$j$ Bell pairs can be prepared using level-$(j-1)$ operations just as in Sec.~\ref{s:reliablebellstate}.  We postselect on not detecting any level-$(j-1)$ errors, as before, and also postselect on no detected failures of level-$(j-1)$ CNOT gates (Eq.~\eqref{e:detectedcnot}).  Since this postselection is only at the top level, it does not cause any efficiency concerns.  

The analysis of the teleportation to prove Eq.~\eqref{e:detectedcnot}, then, is nearly the same as in Sec.~\ref{s:cnotreductionanalysis}.  We only need some way of reliably measuring the logical state of a block based on error correction of lower levels, and this is provided by Eq.~\eqref{e:disttwocorrectdecoding}.  The difference is that, now, if any decoding of the measured blocks detects an error, the encoded CNOT is flagged as having a detected failure.

\section{Correction above postselection for asymptotic efficiency} \label{s:postselectefficient}

A fault-tolerance scheme based entirely on postselection cannot be used for efficient computation at a constant noise rate.  The error rate $\eta_k = (c \eta_0)^{2^k}$ with a postselection-based fault-tolerance scheme drops doubly exponentially fast in $k$ the number of levels of code concatenation.  To reliably simulate an $N$-gate circuit, $k$ can be set to $\log_2 \log_{1/c\eta_0} N$ so $\eta_k \leq 1/N$.  The overhead, though, grows as $\exp(N \exp(k))$, when we postselect on no detected errors in $N$ logical gates each using $\exp(k)$ physical qubits.

However, postselection can be used efficiently to bootstrap into a large quantum error-correcting code.  (Large quantum error-correcting codes cannot be used directly because of the difficulty of encoding -- see Sec.~\ref{s:classicalversusquantum}.)  
One would like to encode into an $n$-qubit code, with $n = \Theta(\log N)$, since against bitwise-independent errors at a rate beneath a constant threshold, large deviation bounds imply that the probability of incorrect decoding drops exponentially with $n$.  Encoding is accomplished on top of $k$ levels of concatenation of a distance-two code -- the repetition code for bit-flip error protection.  Then the $k$ concatenation levels are decoded out from the bottom up, as was done to create a Bell pair with one half encoded in Sec.~\ref{s:decodingandteleportedencoding} -- or alternatively teleported out.  This will introduce constant-rate errors onto each bit, but these errors occur \emph{independently}.  

For protection against bit flip errors, encoding into an $n$-qubit repetition code requires roughly $2n$ basic operations (preparation of $\ket{+0^{n-1}}$ followed by $n-1$ CNOT gates).  Therefore, if we do this encoding on top of $k$ levels of concatenation of the two-qubit repetition code, the overhead per encoding grows as $\exp(n \exp(k))$.  For this overhead to be efficient, $\poly(N)$, $k$ can only be a constant.  But then the probability of an uncontrolled failure will be $\sim n \cdot \eta_k = \omega(1)$, super-constant.  Therefore one level of a large quantum error correcting code does not suffice, even just for protection against bit flip errors.  For bit and phase flip error correction, the situation is even worse, since encoding into an $n$-qubit code typically requires $\sim n^2$ gates (Secs.~\ref{s:largecodes} and~\ref{s:stabilizerstatepreparation}, and Ref.~\cite{AaronsonGottesman04}), so the overhead per encoding grows as $\exp(n^2 \exp(k))$.

Even if we can't get a $1/N$ effective error rate efficiently, by encoding into an $n$-qubit code, we can achieve sub-constant effective error rates with efficient overhead.  We give the details below; the main novelty is the need to consider a sort of ``leakage" error.  Using this as a primitive, it is straightforward to prove an efficient fault-tolerance threshold.  The postselection fault-tolerance scheme can be used again on top of the $n$-qubit code (detecting logical errors, not bit errors within the code); or an error-correction-based scheme can be used, perhaps as in Sec.~\ref{s:correctdisttwo}.  Or, operations on top of the $n$-qubit code can be used to reliably encode into a $\Theta(\log N)$-qubit code, then the $n$-qubit code can be teleported out.  
We will not give the details of these various extensions.

Let $\mathcal{E}$ now denote an ideal (perfect) encoder into an $n$-qubit CSS-type quantum error-correcting code.  By computing over $k$ levels of the postselection-based scheme, then finally decoding out these levels, we can prepare ancilla states for teleporting into a Clifford unitary $U$:
\begin{equation} \label{e:largeencodingsclifford}
\incl{-2.5em}{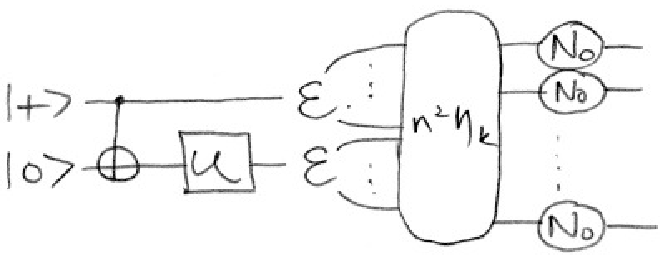}
\end{equation}
That is, the Bell pair with $U$ applied to one half is encoded perfectly, then an arbitrary correlated Pauli error is introduced with probability $O(n^2 \eta_k)$, then independent rate $O(\eta_0)$ Pauli errors are added to the outputs.  The correlated, rate $O(n^2 \eta_k)$ error is there because a level-$k$ logical error in the encoding circuit can spread out of control, and there are $O(n^2)$ encoding gates.  The $O(\eta_0)$ errors come from decoding.  

For teleporting into two-qubit Clifford gates like the CNOT gate, we use the same technique to prepare ancillas like
$$
\incl{0em}{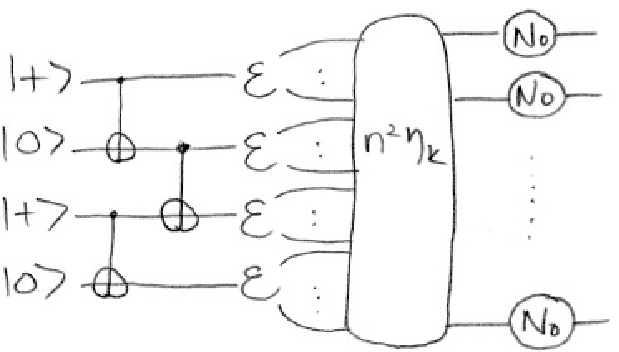}
$$

Let $\tilde{\mathcal{E}}$ denote $\mathcal{E}$ followed by independent introduction of Pauli errors with probability at most $\eta_0' = O(\eta_0)$ on each output bit.  Compiling an ideal circuit on top of the $n$-qubit code, we require two analysis rules for each Clifford unitary $U$:
\begin{equation} \label{e:largeurules}
\begin{split}
\incl{-.5em}{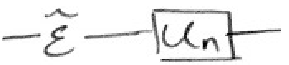} &\rightarrow \incl{-.5em}{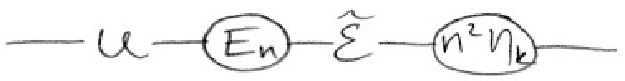} \\
\incl{-.5em}{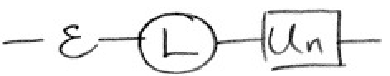} &\rightarrow \incl{-.5em}{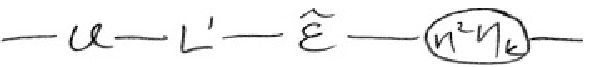}
\end{split}
\end{equation}
The first rule says that a noisy encoder followed by teleporting into a logical $U$ (using the ancilla from Eq.~\eqref{e:largeencodingsclifford}, as well as physical CNOT gates and measurements), is the same as applying $U$, followed by a logical Pauli error at rate $\exp(-n)$, then a noisy encoder -- and also arbitrary \emph{correlated} Pauli noise with probability $O(n^2 \eta_k)$.  This correlated noise is a sort of leakage error event, since its occurrence means we have lost control of the output's relation to the code space.  It arises from failures in the postselection-based scheme, which spread out of control at level $k$ during preparation of the ancilla of Eq.~\eqref{e:largeencodingsclifford}, before that scheme was decoded out.

The second rule describes the effect of the reliable $U$ implementation on a ``leaked" input.  $U$ is applied, followed by an arbitrary Pauli (marked $L'$) and reliable encoding.  That is, the leakage is removed, turned into a possible logical error, and the output state is restored to again lie ``close to" the codespace.  With probability $O(n^2 \eta_k)$, the output leaks.
Choosing $k$ and an $n$-qubit code appropriately, the rules of Eq.~\eqref{e:largeurules} give efficient reliable operations with sub-constant error rates.  

The rules for preparation and measurement are straightforward.  Universality can be obtained using magic states distillation, for which by previous arguments the key step is preparing an ancilla 
$$
\incl{0em}{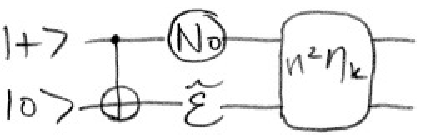}
$$
allowing for teleportation into the encoding.  This ancilla is a particular $(n+1)$-qubit stabilizer state, and its preparation uses the same technique as for Eq.~\eqref{e:largeencodingsclifford}.

\chapter{Numerical noise threshold calculations} \label{s:numericalchapter}
We here calculate rigorous numerical noise threshold lower bounds, up to some caveats listed below.  The purpose of this chapter is to show that the mixing technique to practical to apply, and efficient enough to give good threshold lower bounds.  We do not simply plug numbers into the analysis of Chs.~\ref{s:postselectchapter} and~\ref{s:postselectdetailschapter} used to prove the existence of some positive noise threshold.  A practical and efficient calculation requires a few new tricks.  Investigating them should give more insight into the behavior of fault-tolerance schemes, as well as into the mixing technique.  

Our calculations have not been optimized, and we fully expect that they can be significantly improved.

\section{Results}

We carry out two calculations, both using the concatenated four-qubit, operator error-detecting code (Sec.~\ref{s:fourqubitcode}).  

First, consider the case in which the experimenter exactly knows the noise model, and can tailor the fault-tolerance scheme to the model.  This is unrealistic, since at best noise model parameters can only be estimated.  However, this assumption makes it quite simple to prove a high tolerable noise rate, $> 7 \times 10^{-3}$ (in a model described below).  

Does the threshold remain so high when the noise model is not known exactly?  If the true noise model is slightly better than one thinks, it seems that should only help things.  But we do not know any rigorous argument behind this intuition, and a straightforward application of the mixing technique does not necessarily bear it out.  

Therefore, next consider the more reasonable situation, in which the experimenter is only given noise rate upper bounds.  Assume a symmetry in the noisy operations: every CNOT gate, for example, fails at exactly the same (unknown) error rate $p$, upper-bounded by $\eta_0$ (known).  We provide strong evidence that the threshold noise rate in this case is at least $10^{-3}$.  

A threshold exists even without the symmetry assumption -- i.e., if different CNOT gates fail at different rates between $0$ and $\eta_0$.  In fact, the mixing analysis does not preserve the symmetry assumption, so logical CNOTs at encoding levels one or higher \emph{will} effectively fail at different rates.  However, removing the symmetry assumption at the physical level would certainly reduce the threshold lower bound computed by a mixing argument, and would probably reduce the true threshold.  

We show that starting with a $10^{-3}$ noise rate (exactly), strict improvement in all noise parameters is achieved after a few levels of concatenation.  Even though the base noise model is exactly fixed, only the analyst knows this model and not the experimenter -- the analyzed fault-tolerance scheme does not depend on the noise model.   The analysis will not show that the entire continuous range $(0,10^{-3})$ of noise rates is tolerable.  Although we have checked a number of discrete points, $10^{-3}, 9 \times 10^{-4}, 8 \times 10^{-4}, \ldots, 10^{-4}$, the missing ingredient for a proof is a monotonicity result for mixing of level-one encoded Bell pairs.  (A threshold can be computed without such a monotonicity result, but then it is difficult to take advantage of the symmetry of CNOT failure rates.)

There is room for improvement in both calculations.  In particular, in the second case we use a suboptimal mixing criterion so that the calculations run quickly.  Also, fault-tolerance schemes which possibly are more efficient can be analyzed -- for example Knill's scheme using six-qubit codes concatenated on top of the four-qubit code \cite{Knill05}.  

All calculations have been run in Mathematica; code will be made available from the author.  

\section{Caveats}

\begin{enumerate}
\item We compute a threshold just for preparation and measurement in the X and Z eigenbases, and for the CNOT gate.  The Hadamard gate is not considered, but should not be a limiting factor (see Sec.~\ref{s:fthadamard}).  
\item Calculations are made numerically to 45-digit precision.  We do \emph{not} maintain rigorous upper and lower bounds on each number.  However, we don't feel this is a serious problem with the results, particularly since full precision can be regained at each new concatenation level.  The precision maintained is probably overkill.
\item We do not finish the calculations to show that effective error rates can be made arbitrarily small.  Instead we stop, and assume we are below the threshold, once it is shown that every error model parameter has strictly improved from the previous encoding level, via an analysis that can be applied repeatedly at different levels.  This is only a minor monotonicity assumption; but it can be removed by iterating the analysis here until error rates are very small, and then applying a separate, cruder analysis along the lines of Chs.~\ref{s:postselectchapter} and~\ref{s:postselectdetailschapter}.
\end{enumerate}

\section{Base (level-zero) noise model} \label{s:numericalbasenoisemodel}

Assume that each physical CNOT gate fails with the same probability $p$.  On failure, the two involved qubits are simultaneously depolarized.  Calculations can easily be run for different error models, but simultaneous depolarizing noise is fairly generic, and allows comparison with Knill's simulation- and model-based threshold estimates~\cite{Knill05}.  (Writing two-qubit depolarization as application of an independently and uniformly random Pauli II, IX, IY, IZ, $\ldots$, ZY, ZZ; with $1/16$ probability the identity II is applied.  The CNOT failure rate in this model is also therefore reasonably parameterized by $\tfrac{15}{16}p$, as in Knill's estimates.)  

Noise in preparation and measurement isn't as important as noise in the CNOT gate.  If for example the noise rate in preparing $\ket{0}$ is very high, then we can prepare two copies, and \emph{purify} one against the other:
$$
\includegraphics{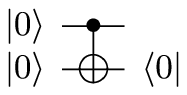}
$$
We apply a CNOT between the two and measure the second qubit.  If we measure 1, then we have detected an error, so start again -- i.e., postselect on measuring 0.  Assuming preparation and measurement failure rates are $O(p)$ -- where a failed preparation gives $\ket{1}$ instead of $\ket{0}$, and a failed measurement gives the wrong outcome.  Then the probability of $\ket{1}$ leaving this purification is $\tfrac{1}{4}p + O(p^2)$, since a failed CNOT will be detected unless either it doesn't flip the second bit, or there are at least two failures.  (This purification technique can be thought of as the unencoded-state analog of encoded-state purification described in Sec.~\ref{s:stabilizerstatepreparation}.)

Similarly, if the measurement error rate is very high, one can apply a CNOT into a prepared $\ket{0}$ and measure both qubits (in the 0/1 basis).  
$$
\includegraphics{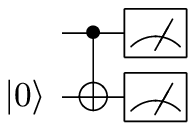}
$$
Output 0 on measuring 00 and 1 on measuring 11.  An error is detected on measuring 01 or 10, but detected errors are easier to manage than undetected errors (Sec.~\ref{s:detectederasureerrors}).

Therefore, take the preparation and measurement error rates each to be $p/4$.  (This is same assumption used by Knill~\cite{Knill05}.)  Equivalently, one can assume that following preparation or prior to measurement, the qubit is depolarized independently with probability $p/2$.  

For proving, or estimating, explicit threshold lower bounds, the different parameters need to be tracked carefully.  Often one effective error rate -- for e.g. an $\mathrm{XI}$ error following a CNOT gate -- might decrease with code concatenation, while another effective error rate -- e.g. for preparation of encoded $\ket{0}$ -- might increase.  Estimating a threshold for physical noise based only on improvement in the total effective CNOT error rate, and neglecting the other failure parameters, would give a too-optimistic ``pseudothreshold" \cite{SvoreCrossChuangAho}.

\section{Threshold calculation for a known noise model} \label{s:numericalexact}

We first assume that the noise rate $p$ is known exactly to the experimenter, and he can adjust the fault-tolerance scheme accordingly.  He will use this information in order to introduce precisely errors (or ``virtual" errors -- see Sec.~\ref{s:virtualerrorpropagation}) so that the error distribution has exactly the form we like (independent, bounded), instead of having to rewrite the error distribution as a mixture of such nice distributions.  A threshold proof in this setting has no need for the mixing technique.  We will show that $p=7 \times 10^{-3}$ can be tolerated.

\subsection{Level one}

\subsubsection{Level-one encoded Bell pair preparation and mixing}

We start by computing the error distribution on a level-one encoded Bell pair.  This state has six independent stabilizers, and two spectator qubits:
\beq \label{e:encodedbellpairstabilizers}
\begin{array}{c c@{\,}c@{\,}c@{\,}c@{\,}c@{\,}c@{\,}c@{\,}c}
&X&X&X&X&I&I&I&I\\
&Z&Z&Z&Z&I&I&I&I\\
&I&I&I&I&X&X&X&X\\
&I&I&I&I&Z&Z&Z&Z\\
&X&X&I&I&X&X&I&I\\
&Z&I&Z&I&Z&I&Z&I\\
\cline{2-9}
X_{1,S} = &X&I&X&I&I&I&I&I\\
Z_{1,S} = &Z&Z&I&I&I&I&I&I\\
X_{2,S} = &I&I&I&I&X&I&X&I\\
Z_{2,S} = &I&I&I&I&Z&Z&I&I
\end{array}
\eeq  The preparation circuit is shown in Fig.~\ref{f:numericalencodedbellpairprep}.  We compute the output error model using Knill's code for introducing and tracking errors on stabilizer states \cite{Knill04analysis}.  

\begin{figure*}
\begin{center}
\includegraphics[scale=1]{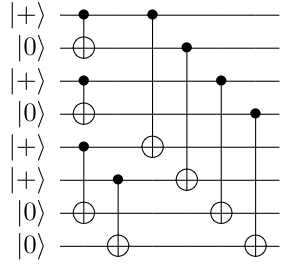}
\end{center}
\caption{
Preparation circuit for an unverified encoded Bell pair.
}
\label{f:numericalencodedbellpairprep}
\end{figure*}

After preparing the state, purify the state against X errors, then Z errors, then X errors, then Z errors.  To purify 
the state against X errors, prepare two identical copies and apply transversal CNOTs into the second copy.  Then measure the second copy in the Z eigenbasis, postselecting on no detected errors (i.e., if an error is detected, try again from the beginning).  To purify against Z errors, apply the CNOTs in the opposite direction and postselect on no detected Z errors after measuring in the X eigenbasis.  Running two rounds of purification against both kinds of errors works significantly better than just one round, although one round does suffice for fault tolerance (Sec.~\ref{s:naiveverificationprocedure}).  (The advantage of using three rounds of purification turns out to be small.)

We obtain a vector with $2^8 = 64$ dimensions, one for each error equivalence class.  (Error equivalence classes correspond to different syndromes $\pm 1$ for the eight stabilizers.)  To simplify the calculation, we now \emph{symmetrize} the state as much as possible.  This means:
\begin{enumerate}
\item Depolarize the two spectator qubits.  
\item With probability $1/2$ apply a perfect transversal Hadamard and swaps bits 2 and 3, and 6 and 7 (indexed top to bottom in Fig.~\ref{f:numericalencodedbellpairprep}).  This symmetrizes X and Z errors in the state.  Of course, we can't actually apply a perfect transversal Hadamard.  However, the exact same effect can be achieved, because of symmetries in the error models.  Putting a perfect Hadamard after a noisy $\ket{0}$ preparation is the same as a noisy $\ket{+}$ preparation, and similarly for measurement.  Also, surrounding a noisy CNOT with perfect Hadamards is the same as a noisy CNOT in the other direction.  Together, these symmetries imply that the effect of applying a perfect transversal Hadamard can be duplicated by going back and switching $\ket{0}$ and $\ket{+}$ preparations, X and Z eigenbasis measurements, and the direction of all CNOT gates.  (Note that the renormalization factor for postselection on no detected errors is the same is both cases.)
\item Uniformly at random apply one of the $32$ permutation symmetries of the encoded Bell pair in which spectators have been depolarized.  These symmetries are generated by: swapping the two halves, switching qubits 1 and 2 with 3 and 4, respectively, and switching qubits 1 and 3 with 2 and 4.
\end{enumerate}
Under the above symmetries, there remain only eleven inequivalent classes of errors.  Representatives for these equivalence classes, and all syndromes belonging to each class, are listed in Table~\ref{f:numericalexactlevelonesymmetrization}.  Total probabilities of each equivalence class, conditioned on acceptance, are given in Table~\ref{f:numericalexactlevelone}.
Note that Y errors are less likely than X or Z errors, since Y errors are caught by both X and Z error purification.  Also, probabilities scale roughly according to the minimum weight of an error in the equivalence class -- e.g., the single equivalence class with minimum-weight three is least likely, and looks like a third-order event.

\begin{table*} 
\caption{State symmetrization: For each of eleven equivalence classes, we list its member syndromes.  For $i =1, \ldots, 6$, the $i$th bit of the syndrome tells whether the error commutes (0) or anticommutes (1) with the $i$th stabilizer of Eq.~\eqref{e:encodedbellpairstabilizers}.  Error equivalence classes are labeled by a minimum-weight representative error.} \label{f:numericalexactlevelonesymmetrization}
\begin{center}
\begin{tabular}{c c} 
\hline \hline
Equiv. class& Member syndromes        \\
\hline
IIIIIIII & 000000\\
IIIIIIIZ& 000100, 000101, 001000, 001010, 010000, 010001, 100000, 100010\\
IIIIIIIY& 001100, 001101, 001110, 001111, 110000, 110001, 110010, 110011\\
IIIIIIXX& 000001, 000010\\
IIIIIIXY& 000110, 000111, 001001, 001011, 010010, 010011, 100001, 100011\\
IIIIIYYI& 000011\\
IIIZIIIZ& 010100, 010101, 101000, 101010\\
IIIZIIIX& 011000, 011001, 011010, 011011, 100100, 100101, 100110, 100111\\
IIIZIIIY& 011100, 011101, 011110, 011111, 101100, 101101, 101110, 101111,\\ &110100, 110101, 110110, 110111, 111000, 111001, 111010, 111011\\
IIIZIIXY& 010110, 010111, 101001, 101011\\
IIIYIIIY& 111100, 111101, 111110, 111111\\
\hline \hline
\end{tabular}
\end{center}
\end{table*}

\begin{table*} 
\caption{Probabilities of the eleven error equivalence classes for a twice-purified and symmetrized encoded Bell pair, with $p=\tfrac{7}{1000}$.} \label{f:numericalexactlevelone}
\begin{center}
\begin{tabular}{c|r@{.}l@{$\times$}l} 
\hline \hline
Equiv. class& \multicolumn{3}{c}{Probability}        \\
\hline
IIIIIIII & 9&64404&$10^{-1}$\\
IIIIIIIZ& 2&80860&$10^{-2}$\\
IIIIIIIY& 7&03928&$10^{-3}$\\
IIIIIIXX& 8&21862&$10^{-5}$\\
IIIIIIXY& 6&35893&$10^{-5}$\\
IIIIIYYI& 3&23595&$10^{-6}$\\
IIIZIIIZ& 1&27658&$10^{-4}$\\
IIIZIIIX& 7&76903&$10^{-5}$\\
IIIZIIIY& 1&02828&$10^{-4}$\\
IIIZIIXY& 3&31238&$10^{-7}$\\
IIIYIIIY& 1&28595&$10^{-5}$\\
\hline \hline
\end{tabular}
\end{center}
\end{table*}

This error distribution is probably not equivalent to one generated by bitwise-independent errors.  Even in the limit, taking the number of purification rounds to infinity, one does not expect the error model to converge to a bitwise-independent model.  Stabilizers couple the bits together.  (For example, consider purifying the cat state $\tfrac{1}{\sqrt{2}}(\ket{000}+\ket{111})$ against X errors.  Three X errors are acceptable, since XXX is a stabilizer, and this will create a higher-order correlation even in the limit.)  
			
However, by deliberately introducing errors at a total error rate less than $2.449\%$, we can make the overall distribution exactly equal to a bitwise-independent error distribution.  Indeed, Table~\ref{f:numericalexactconvolve11d} shows how two distributions over the eleven equivalence classes compose to give a third distribution.  Substitute for $p_i$ the actual error distribution (from Table~\ref{f:numericalexactlevelone}).  
(This table is derived by rewriting the two input distributions as distributions over syndromes [$2^6$ dimensions], composing in syndrome space, then moving back into the eleven symmetrized coordinates.)
We write down the target, a bitwise-independent error distribution in which each bit fails with the same rates: equal X and Z error rates (from the Hadamard symmetry) and smaller Y error rate.  Using Mathematica's NSolve function, we try to find a distribution $q_i$ which composes with $p_i$ to equal to the target.  We try to minimize the X, Y and Z error rates in the target distribution while maintaining that virtual errors are introduced with nonnegative probabilities.  The resulting virtual error distribution is given in Table~\ref{f:numericalexactdeliberateintroduce}.  The X and Z error rates on each qubit of the combined distribution are $2.8 \times 10^{-3}$, while the Y error rates are all $2.0 \times 10^{-3}$:
\beq \label{e:numericalexacterrorrates}
\text{X:}\; 2.8 \times 10^{-3} \quad \text{Y:}\; 2.0 \times 10^{-3} \quad \text{Z:}\; 2.8 \times 10^{-3} \enspace .
\eeq

\begin{table*} 
\caption{Composition of two distributions over the 11 equivalence classes.  
} \label{f:numericalexactconvolve11d}
\begin{center}
\begin{tabular}{c c c c} 
\hline \hline
Equiv. class& $p_i$ & $q_i$ & $(p \circ q)_i$        \\
\hline
IIIIIIII& $p_1$   &$q_1$   & $\left(\substack{p_{1} q_{1} + \tfrac{1}{8}p_{2}q_{2} + \tfrac{1}{8}p_{3}q_{3} + \tfrac{1}{2}p_{4}q_{4} +  \tfrac{1}{8}p_{5}q_{5} + p_{6} q_{6} \\+ \tfrac{1}{4}p_{7}q_{7} + \tfrac{1}{8}p_{8}q_{8} +  \tfrac{1}{16}p_{9}q_{9} + \tfrac{1}{4}p_{10}q_{10} + \tfrac{1}{4}p_{11}q_{11}}\right)$\\
IIIIIIIZ& $p_2$   &$q_2$   &$\left(\substack{p_{2} q_{1} +  p_{1} q_{2} + \tfrac{1}{4}p_{3}q_{2} + \tfrac{1}{2}p_{4}q_{2} + \tfrac{1}{2}p_{7}q_{2} + \tfrac{1}{4}p_{8}q_{2} + \tfrac{1}{4}p_{2}q_{3} + \tfrac{1}{4}p_{5}q_{3} \\+ \tfrac{1}{4}p_{9}q_{3} + \tfrac{1}{2}p_{2}q_{4} + \tfrac{1}{2}p_{5}q_{4} + \tfrac{1}{4}p_{3}q_{5} + \tfrac{1}{2}p_{4}q_{5} + p_{6} q_{5} + \tfrac{1}{4}p_{8}q_{5} + \tfrac{1}{2}p_{10}q_{5} \\+ p_{5} q_{6} + \tfrac{1}{2}p_{2}q_{7} +  \tfrac{1}{4}p_{9}q_{7} + \tfrac{1}{4}p_{2}q_{8} + \tfrac{1}{4}p_{5}q_{8} + \tfrac{1}{4}p_{9}q_{8}  + \tfrac{1}{4}p_{3}q_{9} + \tfrac{1}{4}p_{7}q_{9} \\+ \tfrac{1}{4}p_{8}q_{9} + \tfrac{1}{4}p_{10}q_{9} + \tfrac{1}{2}p_{11}q_{9} + \tfrac{1}{2}p_{5}q_{10} + \tfrac{1}{4}p_{9}q_{10} + \tfrac{1}{2}p_{9}q_{11}}\right)$\\
IIIIIIIY& $p_3$   &$q_3$   &$\left(\substack{p_{3}q_{1}+\tfrac{1}{4}p_{2}q_{2}+\tfrac{1}{4}p_{5}q_{2}+\tfrac{1}{4}p_{9}q_{2}+p_{1}q_{3}+p_{4}q_{3}+p_{6}q_{3}+p_{11}q_{3}\\+p_{3}q_{4}+\tfrac{1}{4}p_{2}q_{5}+\tfrac{1}{4}p_{5}q_{5}+\tfrac{1}{4}p_{9}q_{5}+p_{3}q_{6}+p_{8}q_{7}+p_{7}q_{8}+p_{10}q_{8}\\+\tfrac{1}{4}p_{2}q_{9}+\tfrac{1}{4}p_{5}q_{9}+\tfrac{1}{4}p_{9}q_{9}+p_{8}q_{10}+p_{3}q_{11}}\right)$\\
IIIIIIXX& $p_4$   &$q_4$   &$\left(\substack{p_{4}q_{1}+\tfrac{1}{8}p_{2}q_{2}+\tfrac{1}{8}p_{5}q_{2}+\tfrac{1}{4}p_{3}q_{3}+p_{1}q_{4}+p_{6}q_{4}+\tfrac{1}{8}p_{2}q_{5}+\tfrac{1}{8}p_{5}q_{5}\\+p_{4}q_{6}+\tfrac{1}{4}p_{7}q_{7}+\tfrac{1}{4}p_{10}q_{7}+\tfrac{1}{4}p_{8}q_{8}+\tfrac{1}{8}p_{9}q_{9}+\tfrac{1}{4}p_{7}q_{10}+\tfrac{1}{4}p_{10}q_{10}+\tfrac{1}{2}p_{11}q_{11}}\right)$\\
IIIIIIXY& $p_5$   &$q_5$   &$\left(\substack{p_{5}q_{1}+\tfrac{1}{4}p_{3}q_{2}+\tfrac{1}{2}p_{4}q_{2}+p_{6}q_{2}+\tfrac{1}{4}p_{8}q_{2}+\tfrac{1}{2}p_{10}q_{2}+\tfrac{1}{4}p_{2}q_{3}+\tfrac{1}{4}p_{5}q_{3}\\+\tfrac{1}{4}p_{9}q_{3}+\tfrac{1}{2}p_{2}q_{4}+\tfrac{1}{2}p_{5}q_{4}+p_{1}q_{5}+\tfrac{1}{4}p_{3}q_{5}+\tfrac{1}{2}p_{4}q_{5}+\tfrac{1}{2}p_{7}q_{5}+\tfrac{1}{4}p_{8}q_{5}\\+p_{2}q_{6}+\tfrac{1}{2}p_{5}q_{7}+\tfrac{1}{4}p_{9}q_{7}+\tfrac{1}{4}p_{2}q_{8}+\tfrac{1}{4}p_{5}q_{8}+\tfrac{1}{4}p_{9}q_{8}+\tfrac{1}{4}p_{3}q_{9}+\tfrac{1}{4}p_{7}q_{9}\\+\tfrac{1}{4}p_{8}q_{9}+\tfrac{1}{4}p_{10}q_{9}+\tfrac{1}{2}p_{11}q_{9}+\tfrac{1}{2}p_{2}q_{10}+\tfrac{1}{4}p_{9}q_{10}+\tfrac{1}{2}p_{9}q_{11}}\right)$\\
IIIIIYYI& $p_6$   &$q_6$   &$\left(\substack{p_{6}q_{1}+\tfrac{1}{8}p_{5}q_{2}+\tfrac{1}{8}p_{3}q_{3}+\tfrac{1}{2}p_{4}q_{4}+\tfrac{1}{8}p_{2}q_{5}+p_{1}q_{6}\\+\tfrac{1}{4}p_{10}q_{7}+\tfrac{1}{8}p_{8}q_{8}+\tfrac{1}{16}p_{9}q_{9}+\tfrac{1}{4}p_{7}q_{10}+\tfrac{1}{4}p_{11}q_{11}}\right)$\\
IIIZIIIZ& $p_7$   &$q_7$   &$\left(\substack{p_{7}q_{1}+\tfrac{1}{4}p_{2}q_{2}+\tfrac{1}{8}p_{9}q_{2}+\tfrac{1}{2}p_{8}q_{3}+\tfrac{1}{2}p_{7}q_{4}+\tfrac{1}{2}p_{10}q_{4}+\tfrac{1}{4}p_{5}q_{5}+\tfrac{1}{8}p_{9}q_{5}\\+p_{10}q_{6}+p_{1}q_{7}+\tfrac{1}{2}p_{4}q_{7}+\tfrac{1}{2}p_{11}q_{7}+\tfrac{1}{2}p_{3}q_{8}+\tfrac{1}{8}p_{2}q_{9}+\tfrac{1}{8}p_{5}q_{9}+\tfrac{1}{8}p_{9}q_{9}\\+\tfrac{1}{2}p_{4}q_{10}+p_{6}q_{10}+\tfrac{1}{2}p_{11}q_{10}+\tfrac{1}{2}p_{7}q_{11}+\tfrac{1}{2}p_{10}q_{11}}\right)$\\
IIIZIIIX& $p_8$   &$q_8$   &$\left(\substack{p_{8}q_{1}+\tfrac{1}{4}p_{2}q_{2}+\tfrac{1}{4}p_{5}q_{2}+\tfrac{1}{4}p_{9}q_{2}+p_{7}q_{3}+p_{10}q_{3}+p_{8}q_{4}+\tfrac{1}{4}p_{2}q_{5}\\+\tfrac{1}{4}p_{5}q_{5}+\tfrac{1}{4}p_{9}q_{5}+p_{8}q_{6}+p_{3}q_{7}+p_{1}q_{8}+p_{4}q_{8}+p_{6}q_{8}+p_{11}q_{8}\\+\tfrac{1}{4}p_{2}q_{9}+\tfrac{1}{4}p_{5}q_{9}+\tfrac{1}{4}p_{9}q_{9}+p_{3}q_{10}+p_{8}q_{11}
}\right)$\\
IIIZIIIY& $p_9$   &$q_9$   &$\left(\substack{p_{9}q_{1}+\tfrac{1}{2}p_{3}q_{2}+\tfrac{1}{2}p_{7}q_{2}+\tfrac{1}{2}p_{8}q_{2}+\tfrac{1}{2}p_{10}q_{2}+p_{11}q_{2}+\tfrac{1}{2}p_{2}q_{3}\\+\tfrac{1}{2}p_{5}q_{3}+\tfrac{1}{2}p_{9}q_{3}+p_{9}q_{4}+\tfrac{1}{2}p_{3}q_{5}+\tfrac{1}{2}p_{7}q_{5}+\tfrac{1}{2}p_{8}q_{5}+\tfrac{1}{2}p_{10}q_{5}\\+p_{11}q_{5}+p_{9}q_{6}+\tfrac{1}{2}p_{2}q_{7}+\tfrac{1}{2}p_{5}q_{7}+\tfrac{1}{2}p_{9}q_{7}+\tfrac{1}{2}p_{2}q_{8}+\tfrac{1}{2}p_{5}q_{8}\\+\tfrac{1}{2}p_{9}q_{8}+p_{1}q_{9}+\tfrac{1}{2}p_{3}q_{9}+p_{4}q_{9}+p_{6}q_{9}+\tfrac{1}{2}p_{7}q_{9}+\tfrac{1}{2}p_{8}q_{9}\\+\tfrac{1}{2}p_{10}q_{9}+\tfrac{1}{2}p_{2}q_{10}+\tfrac{1}{2}p_{5}q_{10}+\tfrac{1}{2}p_{9}q_{10}+p_{2}q_{11}+p_{5}q_{11}}\right)$\\
IIIZIIXY& $p_{10}$&$q_{10}$&$\left(\substack{p_{10}q_{1}+\tfrac{1}{4}p_{5}q_{2}+\tfrac{1}{8}p_{9}q_{2}+\tfrac{1}{2}p_{8}q_{3}+\tfrac{1}{2}p_{7}q_{4}+\tfrac{1}{2}p_{10}q_{4}+\tfrac{1}{4}p_{2}q_{5}+\tfrac{1}{8}p_{9}q_{5}\\+p_{7}q_{6}+\tfrac{1}{2}p_{4}q_{7}+p_{6}q_{7}+\tfrac{1}{2}p_{11}q_{7}+\tfrac{1}{2}p_{3}q_{8}+\tfrac{1}{8}p_{2}q_{9}+\tfrac{1}{8}p_{5}q_{9}+\tfrac{1}{8}p_{9}q_{9}\\+p_{1}q_{10}+\tfrac{1}{2}p_{4}q_{10}+\tfrac{1}{2}p_{11}q_{10}+\tfrac{1}{2}p_{7}q_{11}+\tfrac{1}{2}p_{10}q_{11}}\right)$\\
IIIYIIIY& $p_{11}$&$q_{11}$&$\left(\substack{p_{11}q_{1}+\tfrac{1}{4}p_{9}q_{2}+\tfrac{1}{2}p_{3}q_{3}+p_{11}q_{4}+\tfrac{1}{4}p_{9}q_{5}+p_{11}q_{6}\\+\tfrac{1}{2}p_{7}q_{7}+\tfrac{1}{2}p_{10}q_{7}+\tfrac{1}{2}p_{8}q_{8}+\tfrac{1}{4}p_{2}q_{9}+\tfrac{1}{4}p_{5}q_{9}\\+\tfrac{1}{2}p_{7}q_{10}+\tfrac{1}{2}p_{10}q_{10}+p_{1}q_{11}+p_{4}q_{11}+p_{6}q_{11}}\right)$\\
\hline \hline
\end{tabular}
\end{center}
\end{table*}

\begin{table*} 
\caption{Deliberately introducing errors with these probabilities, composed on the error distribution of Table~\ref{f:numericalexactlevelone}, according to the rules of Table~\ref{f:numericalexactconvolve11d}, gives a bitwise-independent error distribution in which each qubit has the same X, Y, Z failure rates $(2.8 \times 10^{-3}, 2.0 \times 10^{-3}, 2.8 \times 10^{-3})$.
} \label{f:numericalexactdeliberateintroduce}
\begin{center}
\begin{tabular}{c|r@{.}l@{$\times$}l} 
\hline \hline
Equiv. class& \multicolumn{3}{c}{Probability}        \\
\hline
IIIIIIII&9&75515&$10^{-1}$\\
IIIIIIIZ&1&56263&$10^{-2}$\\
IIIIIIIY&8&68035&$10^{-3}$\\
IIIIIIXX&1&33010&$10^{-7}$\\
IIIIIIXY&2&25246&$10^{-5}$\\
IIIIIYYI&4&82278&$10^{-6}$\\
IIIZIIIZ&5&59683&$10^{-6}$\\
IIIZIIIX&5&66308&$10^{-5}$\\
IIIZIIIY&6&95205&$10^{-5}$\\
IIIZIIXY&2&40975&$10^{-7}$\\
IIIYIIIY&1&92951&$10^{-5}$\\
\hline \hline
\end{tabular}
\end{center}
\end{table*}

No mixing is required!  Since by assumption the experimenter knows \emph{exactly} the base error model, he can compute precisely what virtual errors he needs to introduce, in order that the total distribution is exactly bitwise independent.  (While we have only carried out the calculations to a limited precision, we believe the results to be robust.)

\subsubsection{Level-one encoded CNOT, measurement and preparation} \label{s:numericalexactlevelonecnot}

Figure~\ref{f:cnotreducestobellpairssymmetry} recalls the implementation of a (logical) CNOT by teleportation.  Let us now compute the level-one encoded CNOT error model.  To do so, we need to understand the circuit of Fig.~\ref{f:encodedcnot}, which shows four encoded blocks from Fig.~\ref{f:cnotreducestobellpairssymmetry}.  The output halves of the two encoded Bell pairs are not shown, because their errors are independent of the input halves.  

\begin{figure*}
\begin{center}
\includegraphics[scale=.5]{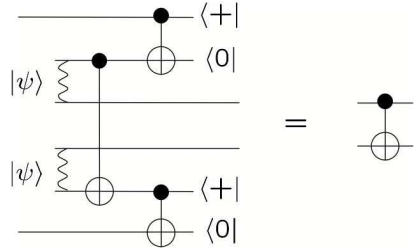}
\end{center}
\caption{Recall from Sec.~\ref{s:cnotteleport} that an encoded CNOT gate is implemented by teleporting into a CNOT between two encoded Bell pairs.  (Teleporting in the manner shown here, instead of using the circuit of Eq.~\eqref{e:cnotreducestobellpairs}, maintains the symmetry between X and Z errors under switching the direction of CNOT gates.)}
\label{f:cnotreducestobellpairssymmetry}
\end{figure*}

\begin{figure*}
\begin{center}
\includegraphics[scale=1]{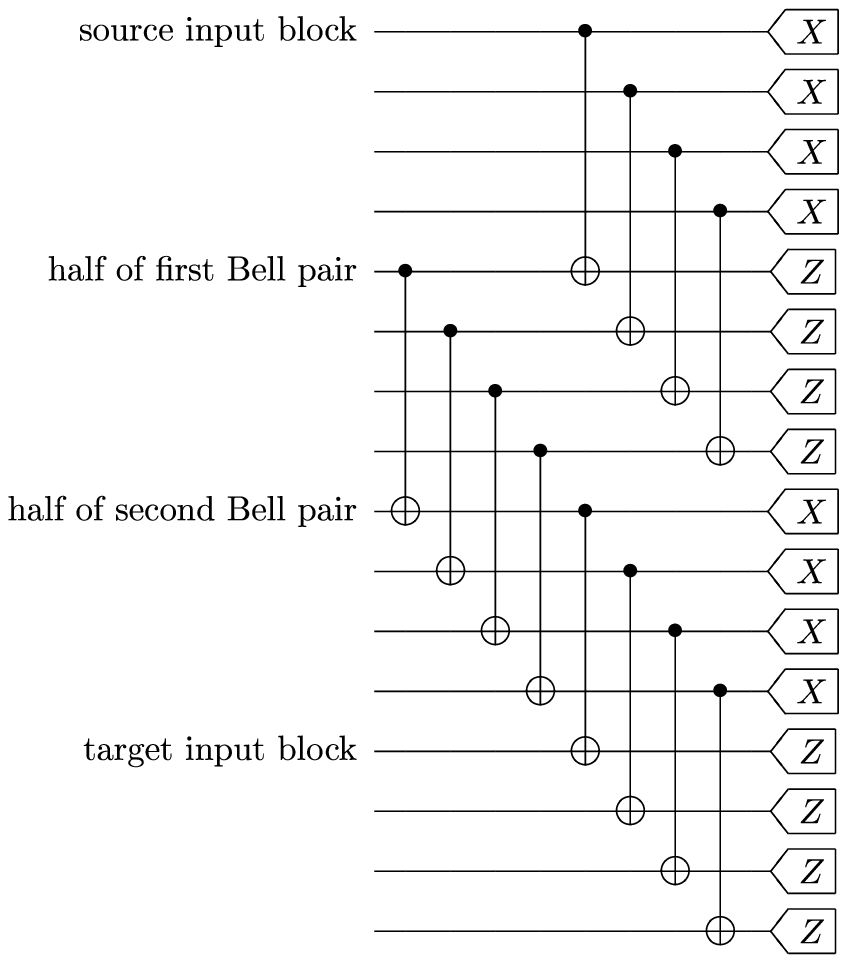}
\end{center}
\caption{
Having prepared the two encoded Bell pairs each with bitwise-independent errors, we are not concerned with the output halves (the middle two blocks above).  In order to compute the exact rate of introducing logical errors, conditioned on no detected errors, we need only analyze the circuit on the four input blocks, shown here.  There are forty-four total failure locations in this circuit (including the bitwise-independent errors on the input blocks), and we compute the exact logical CNOT error model they produce.
}
\label{f:encodedcnot}
\end{figure*}

There are forty-four failure locations in Fig.~\ref{f:encodedcnot}, comprising sixteen failures coming from the input wires, 12 CNOTs, and 16 measurements.  Errors on the input wires are at 
the
$(X, Y, Z)$ rates 
given exactly in Eq.~\eqref{e:numericalexacterrorrates}.
This is because the middle two input blocks are each one half of an encoded Bell pair, and the top and bottom input blocks are either coming from a CNOT gate output (in which case they are again one half of an encoded Bell pair), or coming from fresh preparation of encoded $\ket{0}$ or $\ket{+}$.  We will describe in a moment how to prepare encoded $\ket{0}$ and $\ket{+}$ with the exact same bitwise-independent error rates as on an encoded Bell pair -- not surprisingly, since they are simpler operations.

We compute the exact probability of all $2^{16}$ measurement outcomes.  First, assume there are no errors.  Then add in the effect of each error location, one at a time -- by summing scaled permutations of the vector of probabilities so far.  This calculation takes a few minutes.  When it is finished, we delete entries which have detected errors and renormalize -- postselect on acceptance.  The remaining entries are indexed by possible logical errors, which tell directly what the encoded CNOT error model is.  For example, a logical (X) error in the second block and no others implies that we have teleported into a CNOT followed by an XX error.  

Table~\ref{f:numericalexactencodedcnoterrormodel} gives the encoded CNOT logical error model computed.  The probability of acceptance is $80.581\%$.  Note that every nontrivial error is less likely than the original failure parameter $p/16 = \tfrac{1}{16}\cdot \tfrac{7}{1000}$.  Also, the error model remains symmetrical when the qubit blocks are swapped and conjugated by an ideal Hadamard; e.g., encoded XY errors have the same probability as YZ errors (but XZ, ZX and YY errors are only symmetrical to themselves).

\begin{table*} 
\caption{Level-one encoded CNOT error model.  The first column gives the measured (undetectable) logical error.  The second column gives the logical error following the encoded CNOT gate that the measured logical error implies, after teleportation (X errors are copied forward, and Z errors copied back).  The third column gives the probability of that event, conditioned on acceptance.} \label{f:numericalexactencodedcnoterrormodel}
\begin{center}
\begin{tabular}{c c|r@{.}l@{$\times$}l} 
\hline \hline
Measured & Encoded & \multicolumn{3}{c}{}\\
logical errors & CNOT error & \multicolumn{3}{c}{Probability}\\
\hline
\tt{IIII}&II&9&98449&$10^{-1}$\\
\tt{IIIX}&IX&4&26071&$10^{-4}$\\
\tt{IIZI}&ZZ&2&28605&$10^{-4}$\\
\tt{IIZX}&ZY&4&10378&$10^{-5}$\\
\tt{IXII}&XX&2&28605&$10^{-4}$\\
\tt{IXIX}&XI&6&14955&$10^{-5}$\\
\tt{IXZI}&YY&1&28423&$10^{-6}$\\
\tt{IXZX}&YZ&1&00200&$10^{-6}$\\
\tt{ZIII}&ZI&4&26071&$10^{-4}$\\
\tt{ZIIX}&ZX&3&34103&$10^{-6}$\\
\tt{ZIZI}&IZ&6&14955&$10^{-5}$\\
\tt{ZIZX}&IY&1&44506&$10^{-5}$\\
\tt{ZXII}&YX&4&10378&$10^{-5}$\\
\tt{ZXIX}&YI&1&44506&$10^{-5}$\\
\tt{ZXZI}&XY&1&00200&$10^{-6}$\\
\tt{ZXZX}&XZ&8&25795&$10^{-7}$\\
\hline \hline
\end{tabular}
\end{center}
\end{table*}

The preparation logical error rate should be zero, since preparing an encoded $\ket{0}_L$ or $\ket{+}_L$ is easier than preparing an encoded Bell pair $\tfrac{1}{\sqrt{2}}(\ket{00}_L+\ket{11}_L) = \tfrac{1}{\sqrt{2}}(\ket{++}_L+\ket{--}_L)$ (recall $\ket{\pm} = \tfrac{1}{\sqrt{2}}(\ket{0}\pm\ket{1})$).  It should therefore be possible to prepare $\ket{0}_L$ or $\ket{+}_L$ and then deliberately introduce errors so as the error distribution is bitwise-independent with 
the 
$(X,Y,Z)$ parameters 
of Eq.~\eqref{e:numericalexacterrorrates}.
We have not checked this, though.  Regardless, the preparation logical error rate is certainly less than the measurement logical error rate, since one way of preparing $\ket{0}_L$, e.g., is to prepare an encoded Bell pair then measure the first half, postselecting on measuring a logical $0$ with no detected errors.  

The measurement logical error rate can be conservatively upper-bounded by 
$$
\frac{\left(\begin{smallmatrix}8\\ 2\end{smallmatrix}\right) (4.8 \times 10^{-3})^2}{1 - 8(4.8 \times 10^{-3})} < 6.7 \times 10^{-4} < p/4 \enspace .
$$
This follows since the total probability of an X or Y error (or of a Z or Y error) is $4.8 \times 10^{-3}$.  Physical measurements fail at rate $p/4 < 4.8 \times 10^{-3}$.  A logical measurement has eight failure locations, four for the inputs and four for the physical measurements.  The numerator above upper-bounds the probability of two or more failures, while the denominator lower-bounds the acceptance probability.  

Therefore, we have shown strict improvement in every error parameter, which is strong evidence that we are below the noise threshold.  

\subsection{Levels two and above} \label{s:numericalexactlevelstwoandabove}

The experimenter can now concatenate the entire procedure on itself, and the analysis can be repeated.  (To run the analysis, it is not sufficient, though, to upper-bound the preparation and measurement logical error rates; they should be computed exactly.)  In the next section, it will not be possible to use the code-concatenation level-one analysis at levels two and above, because full symmetrization of twice-encoded Bell pairs will be impossible.  In particular, it is not legal to apply a random permutation to blocks of qubits: If one has a bitwise-independent error distribution on two blocks, and flips a coin whether to swap the blocks or not, the resulting distribution will typically have correlations.  In this case, however, it is possible to randomly permute blocks of qubits, because every block of qubits has an identical error distribution.  So symmetrization of higher-level-encoded Bell pairs proceeds exactly as symmetrization for level-one encoded Bell pairs.

To summarize, the experimenter knowing the base error model exactly has led to two useful technical simplifications:
\begin{enumerate}
\item Mixing was not required, because the experimenter could deliberately add errors so as to make the error distribution exactly bitwise independent.  
\item It is possible to apply random permutations on blocks of qubits, at level one or above -- every block has identically the same error distribution within it, so randomly permuting blocks does not create correlations.
\end{enumerate}
Without these simplifications, the analysis becomes less efficient, as we shall see in Sec.~\ref{s:numericalbound}.

\subsection{Extension to universality} \label{s:numericalexactuniversality}

Now we apply the magic states distillation technique of Ch.~\ref{s:magicchapter} to extend the above calculations from mere stabilizer operations to full quantum universality.  Theorem~\ref{t:practicalmagicdeterministic} gives numerical conditions sufficient for distilling $\ket{T}$ from states $\rho_i$, assuming perfect stabilizer operations -- i.e., at a high enough code concatenation level.  The missing ingredient is a numerical analysis of the teleportation encoding procedure of Sec.~\ref{s:decodingandteleportedencoding}, which starts with bottom-up decoding of one half of an encoded Bell pair.  

Two possible decoding circuits are shown in Fig.~\ref{f:decodingcircuits}; we will use the first one.  We compute the exact error distribution on the output qubit, conditioned on no detected errors, starting with independent $(X,Y,Z)$ errors at the rates of Eq.~\eqref{e:numericalexacterrorrates}.  The resulting error distribution should be composed with the level-one logical error rates and the analysis repeated.  We have not computed the level-one logical error rates, although it would be simple to do so.  Instead, we can conservatively substitute the rates of Eq.~\eqref{e:numericalexacterrorrates} again.\footnote{This is conservative assuming that the output error distribution is monotone in the starting error rates.  This assumption can presumably be proven easily, since the output error distribution is a simple function of the three input error rates (for a fixed error model for the CNOT gates and measurements) -- but we have not done so.}

\begin{figure*}
\begin{center}
\includegraphics[scale=1]{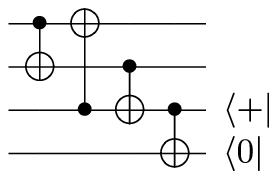} \qquad \quad
\includegraphics[scale=1]{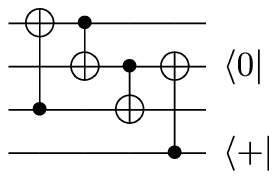}
\end{center}
\caption{Two possible decoding circuits for the four-qubit code.  The first output wire contains the result.  The second or third wire contains the decoded spectator qubit, which can be discarded.}
\label{f:decodingcircuits}
\end{figure*}

Rather than repeat the calculations over and over, though, it is easier to start by composing some error distribution with the rates of Eq.~\eqref{e:numericalexacterrorrates}, and show that the output error distribution is improved.  We find that starting with the composition of the distribution of Eq.~\eqref{e:numericalexacterrorrates} with bitwise-independent errors at $(X,Y,Z)$ rates $(7 \times 10^{-4}, 6 \times 10^{-4}, 1.5 \times 10^{-3})$, the output error rates are strictly better: $(6.223 \times 10^{-4}, 5.101 \times 10^{-4}, 1.477 \times 10^{-3})$.

The total error amount entering due to recursive bottom-up decoding is therefore at most $7 \times 10^{-4} + 6 \times 10^{-4} + 1.5 \times 10^{-3} = 2.8 \times 10^{-3}$ (the logical error rate at the top level of the encoded Bell pair is arbitrarily small, and in fact can be set to zero, so does not contribute).  The total error rate teleported into the encoding is this rate plus the total failure rates of the single physical CNOT gate and two measurements used for teleportation, plus the error rate on the initial noisy $\ket{T}$.  Theorem~\ref{t:practicalmagicdeterministic} allows for $\tfrac{1}{2} \cdot 0.0527$ worst-case noise, so the tolerable noise rate on the initial $\ket{T}$ is at least
$$
\tfrac{5.27 \times 10^{-2}}{2}-2.8 \times 10^{-3}-\tfrac{15}{16}\tfrac{7}{1000}-2\cdot \tfrac{1}{4}\tfrac{7}{1000} > 1.34 \times 10^{-2} \enspace .
$$
This number is probably very conservative, primarily because of the loose analysis of Theorem~\ref{t:practicalmagicdeterministic}.  We haven't fixed an noise model for preparing noisy $\ket{T}$ states -- however, assuming error rates no higher than those for measurement and preparation, we are well beneath the threshold for universality.  

\section{Threshold calculation with symmetrical, but only upper-bounded noise} \label{s:numericalbound}

\subsection{Noise model assumptions}

Let us now drop the assumption that the experimentalist knows the error model exactly; assume only that upper bounds on failure rates are known.  

The error model has various symmetries -- $\ket{0}$ and $\ket{+}$ preparations have the same failure rate, X and Z measurements have the same error rate, and conjugating a noisy CNOT by four Hadamard gates has an identical effect as a noisy CNOT in the opposite direction.  These assumptions are useful, but not essential, for an efficient analysis.  Without them, it turns out that there are seventeen error equivalence classes instead of eleven.  The analysis would run only slightly slower, and would probably give a similar threshold lower bound.  

An additional symmetry, which may be less apparent because it is so standard in the literature, is our assumption that every CNOT gate fails with exactly the same error parameters.  If CNOT gates instead fail at adversarially chosen rates (certain gates being nearly perfect while others failing at the upper bound rate), then any application of the mixing technique would likely give worse noise threshold bounds, because errors can have stronger correlations.  (A concrete example of how asymmetrically better gates can harm us was given in Sec.~\ref{s:reliablebellstate}.)  Below, we will use a very efficient analysis for encoding level one, and a less efficient analysis for levels two and above.  The more efficient analysis uses exact gate failure rates, while the less efficient analysis works with upper and lower bounds.  Without the assumption that CNOT gates are symmetrical to each other, the less efficient analysis would have to be used for level one, too.  

\subsection{Level one}

\subsubsection{Level-one encoded Bell pair preparation and mixing} \label{s:numericalboundedlevelonemixing}

Take the CNOT failure rate to be exactly $p=\tfrac{1}{1000}$.  Preparation and measurement error rates are each $p/4$, as described in Sec.~\ref{s:numericalbasenoisemodel}.  (We remark once again that the fault-tolerance scheme does not depend on $p$, unlike in the previous section.  Assuming that \emph{we} -- not the experimenter -- know $p$ exactly allows a more efficient analysis.)

Start by preparing a fully symmetrized encoded Bell pair, as in the previous section.  Table~\ref{f:numericalboundedlevelone} gives the resulting equivalence class probabilities; compare to Table~\ref{f:numericalexactlevelone}.  

\begin{table*} 
\caption{Probabilities of the eleven error equivalence classes for a twice-purified and symmetrized encoded Bell pair, with $p=\tfrac{1}{1000}$.} \label{f:numericalboundedlevelone}
\begin{center}
\begin{tabular}{c|r@{.}l@{$\times$}l} 
\hline \hline
Equiv. class& \multicolumn{3}{c}{Probability}        \\
\hline
IIIIIIII & 9&94988&$10^{-1}$\\
IIIIIIIZ& 4&00175&$10^{-3}$\\
IIIIIIIY& 1&00076&$10^{-3}$\\
IIIIIIXX& 1&63212&$10^{-6}$\\
IIIIIIXY& 1&25644&$10^{-6}$\\
IIIIIYYI& 6&29737&$10^{-8}$\\
IIIZIIIZ& 2&51403&$10^{-6}$\\
IIIZIIIX& 1&51138&$10^{-6}$\\
IIIZIIIY& 2&01314&$10^{-6}$\\
IIIZIIXY& 8&86856&$10^{-10}$\\
IIIYIIIY& 2&51645&$10^{-7}$\\
\hline \hline
\end{tabular}
\end{center}
\end{table*}

Next, rewrite this probability distribution as a mixture of bitwise-independent distributions.  To do so, we write down a list of $4^8$ distributions in which each of the eight bits is either:
\begin{itemize}
\item perfect, or 
\item has an X error with probability $q$ (but never Y or Z errors), or
\item has a Y error with probability $q$ (but never X or Z errors), or 
\item has a Z error with probability $q$ (but never X or Y errors).
\end{itemize}
Each of these distributions has eleven symmetrized coordinates, polynomials in $q$.\footnote{Only one type of error per bit, X, Y or Z, is allowed merely in order to simplify the routine which writes down the distribution's coordinates.  Better mixing (with a lower value of $q$) might be possible using a different set of bitwise-independent distributions -- for example, bitwise $(X,Y,Z)$ error rates of $(q/2, q/4, q/4)$, $(q/4, q/2, q/4)$, or $(q/4, q/4, q/2)$.  Too, mixing bitwise-independent distributions is convenient but not needed; mixing distributions which are independent only across the two halves of the Bell pair would suffice (see, e.g., Sec.~\ref{s:unbiasednoise}).}  
Removing duplicates, there are only $1507$ distinct points.  We wish to find the smallest value of $q$ such that the point given in Table~\ref{f:numericalboundedlevelone} lies in their convex hull.  (This implies that, as an error distribution, it lies in the convex hull of the $4^8$ bitwise-independent error distributions, because applying a fixed permutation and/or transversal ideal Hadamards takes one of the bitwise-independent distributions to another.)  For any fixed $q$, a linear programming routine with trivial objective function can check if one point lies inside the convex hull of some others; using this as a subroutine for a binary search lets us quickly minimize $q$.  We find that $q= 62798/10^8$ suffices.\footnote{This is very close to the first-order estimate of $\tfrac{5}{8}p = 0.625 p$.  Here, $p/8$ is the first-order Y error rate, and $p/8$ and $3p/8$ are the first-order rates for X and Z errors, depending on which error type was most recently tested for.}

\subsubsection{Level-one encoded CNOT, measurement and preparation}

The analysis of the level-one encoded CNOT gate is similar to that in Sec.~\ref{s:numericalexactlevelonecnot}.  However, we now do not know what the error distributions on the inputs are.  We only know that the distribution on each block of four qubits is one of $4^4$ bitwise-independent error distributions, since each block is half of an encoded Bell pair.  

In fact, in the above mixing, only $117$ of the $1504$ different vertices (including the perfect distribution, with no errors at all) have a positive coefficient in the mixture.  And there are only $34$ different distributions (closed under permutation and Hadamard symmetries) acting on half an encoded Bell pair.  These distributions are:
\begin{equation} \label{e:numericalboundedlevelonemixing}
\begin{array}{c@{,\;}c@{,\;}c@{,\;}c@{,\;}c@{,\;}c@{,\;}c@{,\;}c@{,\;}c@{,\;}c@{,\;}c}
\text{IIII}&\multicolumn{10}{c}{}\\
\text{XXXX}&\text{XYXY}&\text{XYXZ}&\text{XYYX}&\text{XYYZ}&\text{XYZY}&\text{XYZZ}&\text{XZXY}&\text{XZYY}&\text{XZZX}&\text{YXXY}\\
\text{YXYX}&\text{YXYZ}&\text{YXZX}&\text{YXZY}&\text{YXZZ}&\text{YYXZ}&\text{YYYY}&\text{YYZX}&\text{YYZZ}&\text{YZXY}&\text{YZYX}\\
\text{YZZY}&\text{ZXXZ}&\text{ZXYX}&\text{ZXYY}&\text{ZYXY}&\text{ZYYX}&\text{ZYYZ}&\text{ZZXY}&\text{ZZYX}&\text{ZZYY}&\text{ZZZZ}
\end{array}
\end{equation}
where an X on a qubit means the X error rate is $q$ but Y and Z error rates are zero -- with similar notation for Y and Z -- and an I on a qubit means that qubit is perfect.
The encoded CNOT error model of course depends on the bit error distributions of the inputs.  We therefore cannot compute the exact error model, but only upper bounds on its parameters, from the worst-case inputs.

To compute upper bounds on logical failure rates, we count all second- and third-order terms, as a polynomial in the variables $c$ and $m$, the CNOT and measurement failure parameters, respectively (here, $p/16$ and $p/4$), and also the input failure rates labeled by block ($1$ through $4$), bit ($1$ through $4$) and type (X, Y or Z).  The resulting polynomials are too long to display here, but Table~\ref{f:numericalboundedencodedcnoterrormodel} shows them with the substitutions $c=p/16$, $m=p/4$ and all input failure rates $q$.  

These polynomials can be evaluated with $p=1/1000$ and all input failure rates $q$.  But a savings of $> 17\%$ in the total failure rate upper bound can be had by instead maximizing each error parameter while varying each of the four input blocks over the $34$ different supported distributions ($\sim 34^4$ possibilities).\footnote{\emph{Lower} bounds on logical failure parameters can be obtained by minimizing over the different supported input distributions -- minimums in fact are achieved with perfect input blocks.  However these lower bounds are not very good, so we do not use them.}

\begin{table*} 
\caption{Level-one encoded CNOT error model second- and third-order terms, with substitutions.  Note, e.g., that  the YY error polynomial has no $q^2$ term, since no two failures in the inputs can by themselves lead to such a logical error.} \label{f:numericalboundedencodedcnoterrormodel}
\begin{center}
\begin{tabular}{c c|c} 
\hline \hline
Measured & Encoded &\\
logical errors & CNOT error & Probability\\
\hline
\tt{IIIX}&IX&$\tfrac{189}{128} p^2 + \tfrac{21}{2} p q + 20 q^2 + \tfrac{4183}{1024} p^3 + \tfrac{1381}{32} p^2 q + \tfrac{257}{2} p q^2 + 88 q^3$\\
\tt{IIZI}&ZZ&$\tfrac{189}{128} p^2 + \tfrac{23}{4} p q + 6 q^2 + \tfrac{4183}{1024} p^3 + \tfrac{2155}{64} p^2 q+ \tfrac{347}{4} p q^2 + 68 q^3$\\
\tt{IIZX}&ZY&$\tfrac{25}{128} p^2 + \tfrac{5}{4} p q + 2 q^2 + \tfrac{2019}{1024} p^3 + \tfrac{1177}{64} p^2 q + \tfrac{209}{4} p q^2 + 44 q^3$\\
\tt{IXII}&XX&same as ZZ\\
\tt{IXIX}&XI&$\tfrac{5}{128} p^2 + \tfrac{3}{4} p q + 6 q^2 + \tfrac{775}{1024} p^3 + \tfrac{895}{64} p^2 q + \tfrac{271}{4} p q^2 + 68 q^3$\\
\tt{IXZI}&YY&$\tfrac{1}{128} p^2+ \tfrac{603}{1024} p^3+ \tfrac{23}{8} p^2 q + 4 p q^2$\\
\tt{IXZX}&YZ&$\tfrac{1}{128} p^2+ \tfrac{235}{1024} p^3+ \tfrac{17}{8} p^2 q + 4 p q^2$\\
\tt{ZIII}&ZI&same as IX\\
\tt{ZIIX}&ZX&$\tfrac{1}{128} p^2+ \tfrac{603}{1024} p^3 + \tfrac{61}{8} p^2 q + 28 p q^2 + 32 q^3$\\
\tt{ZIZI}&IZ&same as XI\\
\tt{ZIZX}&IY&$\tfrac{1}{128} p^2+ \tfrac{1}{4} p q + 2 q^2 + \tfrac{235}{1024} p^3 + \tfrac{309}{64} p^2 q+ \tfrac{117}{4} p q^2 + 44 q^3$\\
\tt{ZXII}&YX&same as ZY\\
\tt{ZXIX}&YI&same as IY\\
\tt{ZXZI}&XY&same as YZ\\
\tt{ZXZX}&XZ&$\tfrac{1}{128}p^2 + \tfrac{123}{1024} p^3 + \tfrac{11}{8} p^2 q + 4 p q^2$\\
\hline \hline
\end{tabular}
\end{center}
\end{table*}

Add in a conservative remainder term, 
\begin{multline*}
\sum_{i=0}^{4} \sum_{j=0}^{4} \binomial{12}{i} \binomial{16}{j} \binomial{16}{4-i-j} \left(\tfrac{15}{16}p\right)^i q^j \left(\tfrac{1}{4}p\right)^{4-i-j} \\= \tfrac{108008495}{2^{16}} p^4 + \tfrac{518585}{2^6} p^3 q + \tfrac{212175}{2^4} p^2 q^2 + 8540 p q^3 + 1820 q^4 \enspace ,
\end{multline*}
to account for fourth- and higher-order terms.  Then renormalize -- divide by a conservative lower bound on the acceptance probability, $1-12 (\tfrac{15}{16}p) - 16q - 16(\tfrac{1}{4}p)$ -- to finish our calculation of upper bounds on each failure parameter of the level-one encoded CNOT gate.  The resulting failure probability upper bounds are given in Table~\ref{f:numericalboundedencodedcnoterrormodellevelone}, with symmetrical entries combined.\footnote{Note that these probability upper bounds need to be converted to bounds on likelihoods to apply Knill's stabilizer manipulation code.}  
The error model is now quite nonuniform, with the most likely errors (IX or ZI) being more than 500 times as the least likely error, XZ.  

\begin{table*} 
\caption{Level-one encoded operation error model upper bounds.} \label{f:numericalboundedencodedcnoterrormodellevelone}
\begin{center}
\begin{tabular}{c c|r@{.}l@{$\times$}l} 
\hline \hline
\multicolumn{2}{c}{Encoded} & \multicolumn{3}{c}{Probability}\\
\multicolumn{2}{c}{error parameter} & \multicolumn{3}{c}{upper bound}\\
\hline
&IX, ZI&1&39080&$10^{-5}$\\
&ZZ, XX&6&12551&$10^{-6}$\\
&ZY, YX&1&87585&$10^{-6}$\\
&XI, IZ&2&04980&$10^{-6}$\\
CNOT&YY&2&68270&$10^{-8}$\\
&YZ, XY&2&59751&$10^{-8}$\\
&ZX&4&77280&$10^{-8}$\\
&IY, YI&1&01937&$10^{-6}$\\
&XZ&2&53796&$10^{-8}$\\
\hline
\multicolumn{2}{r|}{Preparation/Measurement}&3&09425&$10^{-6}$\\
\hline \hline
\end{tabular}
\end{center}
\end{table*}

The calculation for encoded measurements is very similar, but with simpler counting.  Again, maximize the second- and third-order terms over the supported input distributions, add a remainder term, and renormalize.  For logical X measurement, the second-order count is 
\begin{multline*}
4 m^2 + 2 m y_1 + 2 m y_2 + 2 m y_3 + y_1 y_3 + y_2 y_3 + 2 m y_4 + y_1 y_4 + y_2 y_4 + 2 m z_1 + y_3 z_1 + y_4 z_1 + 2 m z_2 \\+ y_3 z_2 + y_4 z_2 + 2 m z_3 + y_1 z_3 + y_2 z_3 + z_1 z_3 + z_2 z_3 + 2 m z_4 + y_1 z_4 + y_2 z_4 + z_1 z_4 + z_2 z_4 \enspace ,
\end{multline*}
where again $m$ is the bit measurement error rate and, e.g., $z_3$ is the probability of a Z error on the third bit of the input block.  The result is an upper bound on the level-one logical X measurement error rate, of $3.09425 \times 10^{-6}$.  Logical Z measurement is symmetrical, so of course has the same upper bound on its error rate. 

\begin{figure*}
\begin{center}
\includegraphics[scale=1]{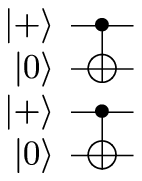} $\qquad\quad$
\includegraphics[scale=1]{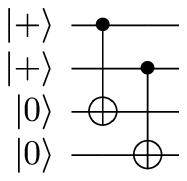}
\end{center}
\caption{
Preparation circuit for an encoded $\ket{+}$ (left), and encoded $\ket{0}$.  No verification is required.  After preparing the state, the spectator qubit is depolarized.  
In the former case, in which $\ket{+}_L\ket{0}_S$ is prepared, this means applying the virtual error XIXI with probability $1/2$.  In the latter case, in which $\ket{0}_L\ket{+}_S$ is prepared, apply IIZZ with probability $1/2$.
}
\label{f:numericalencodedprep}
\end{figure*}

We have checked that for encoded preparations as in Fig.~\ref{f:numericalencodedprep}, the logical preparation error rate can be set to zero.  The output error distribution lies in the convex hull of the same $34$ bitwise-independent error distributions of Eq.~\eqref{e:numericalboundedlevelonemixing}, with the same value of $q$ (actually, even with a slightly lower value).  (If one does not depolarize the spectator qubit, though, then the same value of $q$ suffices for a mixture, but \emph{not} with only the distributions of Eq.~\eqref{e:numericalboundedlevelonemixing}.)  Some numerical evidence suggests that these distributions being the only ones required is not a coincidence, but remains true even for lower values of $p$.  

However, restricting the set of bitwise-independent distributions allowed in a mixture makes it more -- if still not very -- plausible that $p=1/1000$ might be a tolerable noise rate, but some smaller noise values might not be tolerable (even with an equally symmetric error model).  Therefore, we will conservatively set the logical preparation error rate to equal the logical measurement error rate.  As discussed in Sec.~\ref{s:numericalexactlevelonecnot}, this error rate can be achieved by measuring one half of an encoded Bell pair.  (Preparing an encoded state by measuring one half of an encoded Bell pair also ensures that the the output's bit error distribution will be one of the distributions of Eq.~\eqref{e:numericalboundedlevelonemixing}, so the optimization over the CNOT inputs is valid.)

\subsection{Levels two and above}

Every error parameter, for level-one CNOTs, measurements and preparations, has strictly improved from the base error model.  However, this is \emph{not} yet strong evidence for being beneath the threshold, because two essential symmetries have been lost. 
\begin{enumerate}
\item Despite our symmetry assumption for physical CNOT gates, we cannot assume that different level-one CNOT gates fail with identical error parameters.  Symmetries are broken when one picks out a bitwise-independent distribution from a mixture.  
\item We similarly cannot assume that every level-one encoded block has the same distribution of bit errors within it. 
\end{enumerate}
The first asymmetry roughly means that we need to worry about adversarially placed \emph{perfect} logical CNOT gates, which can strengthen error correlations.  The second asymmetry means that we cannot randomly permute level-one encoded blocks of qubits (see Sec.~\ref{s:numericalexactlevelstwoandabove}).  Therefore, in analyzing the next level, we will not be able to work in an eleven-dimensional symmetrized space, but will have to keep track of sixty-four coordinates.  The difficulty of checking upper and lower bounds in each of sixty-four coordinates ($2^{64}$ points to check) will require us to use a less efficient (but computationally easy) analysis based on a slightly generalized Mixing Lemma.

\subsubsection{Generalized Mixing Lemma}

Let $\{p_i\}_{i=1}^m$ denote the distribution on $\{0,1,\ldots, m\}$: $\pr[i]=p_i$ for $i>0$ and $\pr[0] = 1- \sum_{i=1}^m p_i$.  Fix values $q_i^j$ for $i \in \{1, \ldots, m\}$ and $j \in \{1, \ldots, n\}$.  To each $w \in \{0,1,\ldots,m\}^n$ associate the product distribution in which the $j$th coordinate is distributed according to 
$$
\left\{q_{w_j}^j \delta_{w_j,i}\right\}_{i=1}^m \enspace ;
$$
that is, if $w_j=0$ then the $j$th coordinate is perfect, and otherwise the probability is divided between $0$ and $w_j$ with respective probabilities $1-q_{w_j}^j$ and $q_{w_j}^j$.  Define a lattice ordering $y \preceq x$ for strings $x,y \in \{0,1,\ldots, m\}^n$ if $x_i \in \{0,y_i\}$ for $i=1,\ldots,n$.

\begin{lemma} \label{t:generalizedmixinglemma}
The convex hull of the $(m+1)^n$ distributions associated to the strings $\{0,1,\ldots, m\}^n$ is given exactly by those $\pr[\cdot]$ satisfying, for each $x \in \{0,1, \ldots, m\}^n$:
$$
\sum_{y \preceq x} (-1)^{\abs{y}-\abs{x}} \frac{\pr[\{z \preceq y\}]}{y(\{z \preceq y\})} \geq 0 \enspace .
$$
\end{lemma}

\noindent We omit the proof, which follows the same steps as in Sec.~\ref{s:mixinglemmaproof}.

\subsubsection{Level-two encoded Bell pair preparation and mixing} \label{s:numericalboundedleveltwomixing}

We use the error rate upper bounds determined above in order to upper bound the error rates in a level-two encoded Bell pair.  Working with only upper bounds instead of an exact error model requires a little more care; for example the sum of probability upper bounds is $>1$.  

More seriously, we cannot symmetrize the level-two encoded Bell pair.  Because level-one blocks have different bit error distributions (depending on which distribution from Eq.~\eqref{e:numericalboundedlevelonemixing} was chosen), applying a random permutation to the level-one blocks would create correlations in the bit error distributions within those blocks.  Therefore, we need to consider all $2^6 = 64$ coordinates (a different coordinate for each syndrome of the six stabilizers of Eq.~\eqref{e:encodedbellpairstabilizers}; spectators are depolarized).  We have upper bounds for each coordinate, but the lower bounds are trivial (0 for nontrivial syndromes).  To obtain lower bounds for ensuring correlations aren't too strong, the experimenter now has to deliberately introduce errors into the state -- which was not necessary for level-one encoding.

But after introducing errors, then what?  The actual distribution is bounded above and below on each coordinate, so lies somewhere in an axis-aligned rectangular prism.  We'd like to check that every point in this prism lies in the convex hull of the $4^8$ bitwise-independent error distributions described in Sec.~\ref{s:numericalboundedlevelonemixing}.  For this, it suffices to check every vertex of the prism.  But there are $2^{64}$ vertices, far too many to check.  
Evidence from checking random vertices of the rectangular prism suggests that this mixing might allow for very good bounds (low $q$).  But even one bad vertex -- not lying in the convex hull of the bitwise-independent error distributions -- would invalidate the argument.  

We therefore instead turn to the Mixing Lemma of Ch.~\ref{s:postselectchapter}, generalized now to work on the natural lattice on $\{I, X, Y, Z\}^8$ (set $m=3$ and $n=8$ in the above formulation, I corresponds to $0$).  The Mixing Lemma sufficiency conditions are very easy to check; if a term is positive ($\abs{y}-\abs{x}$ even), we use our lower bound for the numerator, and we use the upper bound otherwise.  

Embed the $64$ syndromes into the lattice $\{I, X, Y, Z\}^8$ by evenly dividing the probability mass on a given syndrome among those errors of minimum weight generating that syndrome.  (This is suboptimal; better mixing can be obtained by moving probability mass, from both lower and upper bounds, onto higher-weight lattice elements.)  No error equivalence class has a minimum-weight error of weight greater than three, so the vast majority of the lattice is unused.  In fact, there are only 153 lattice elements with positive probability mass on a child -- i.e., only 153 inequalities to check to apply Lemma~\ref{t:generalizedmixinglemma}.  

We first try deliberately introducing errors according to a bitwise-independent error distribution which closely approximated the error upper bounds on nontrivial syndromes.  In fact, the distribution in which the $(X,Z,Y)$ error rates are (roughly)
$$
(2.41884 \times 10^{-5}, 2.07782 \times 10^{-6}, 1.04583 \times 10^{-6})
$$
on each of the first four qubits, and 
$$
(2.41907 \times 10^{-5}, 2.07653 \times 10^{-6}, 1.04560 \times 10^{-6})
$$
on each of the last four qubits, approximated the error upper bounds to better than $1.4 \times 10^{-10}$ in total variation distance.  However, using this distribution, the smallest value of $q$ for which all 153 mixing inequalities are satisfied is roughly $2.2 \times 10^{-3}$, rather disappointing.  

At $q=2.1 \times 10^{-3}$, only the inequalities corresponding to $x=\text{YIIIIIII}$ and its permutations fail.  Therefore try to adjust the introduced error distribution in order to satisfy these inequalities, without upsetting the others.  By an iterative optimization procedure, we settle on the following two adjustments:
\begin{enumerate}
\item Multiply the above X and Z error rates by $7/8$, and multiply the Y error rate by $13/2$.  
\item Then remove any probability mass on weight three errors.  Multiply the probability of each weight-two error by $4/5$, and the probability of any weight-one error by $7/5$.  Renormalize to obtain again a distribution.  
\end{enumerate}
With these adjustments, all inequalities are satisfied at $q=5.4 \times 10^{-4}$, a significant improvement.  

\subsubsection{Level-two encoded CNOT, measurement and preparation}

Obtaining upper bounds for the level-two encoded CNOT error model is very similar to the level-one CNOT case.  We again count second- and third-order terms, and add in a conservative fourth-order term.  The same polynomial as before can be reused, but for efficiency it is useful to redo the counts.  This time, keep different variables for each kind of CNOT failure, instead of just a single variable $c$, because the level-one CNOT error model is nonuniform, whereas all failure parameters of the physical CNOT were equal.  Also, we will not optimize over just those vertices supported in the mixture, so we can use a single variable $q$ for all input error rates, instead of $48$ different variables before (indexed by block, bit and X, Y or Z).  

The resulting upper bounds on level-two CNOT error rates are given in Table~\ref{f:numericalboundedencodedcnoterrormodelleveltwo}.  There is improvement in most, but not all, parameters.

\begin{table*} 
\caption{Level-two and level-three encoded operation error model upper bounds.} \label{f:numericalboundedencodedcnoterrormodelleveltwo} \label{f:numericalboundedencodedcnoterrormodellevelthree}
\begin{center}
\begin{tabular}{c c|r@{.}l@{$\times$}l|r@{.}l@{$\times$}l} 
\hline \hline
\multicolumn{2}{c}{Encoded} & \multicolumn{6}{c}{Probability upper bound}\\
\multicolumn{2}{c}{error parameter} & \multicolumn{3}{c}{Level two} & \multicolumn{3}{c}{Level three}\\
\hline
&IX, ZI&6&25510&$10^{-6}$&1&78923&$10^{-6}$\\
&ZZ, XX&1&90588&$10^{-6}$&5&47678&$10^{-7}$\\
&ZY, YX&6&07117&$10^{-7}$&1&72365&$10^{-7}$\\
&XI, IZ&1&80892&$10^{-6}$&5&14121&$10^{-7}$\\
CNOT&YY&2&21913&$10^{-10}$&1&82493&$10^{-11}$\\
&YZ, XY&2&21914&$10^{-10}$&1&82492&$10^{-11}$\\
&ZX&5&55412&$10^{-9}$&8&36132&$10^{-10}$\\
&IY, YI&6&00379&$10^{-7}$&1&70904&$10^{-7}$\\
&XZ&2&21917&$10^{-10}$&1&82492&$10^{-11}$\\
\hline
\multicolumn{2}{r|}{Preparation/Measurement}&4&71147&$10^{-6}$&1&36833&$10^{-6}$\\
\hline \hline
\end{tabular}
\end{center}
\end{table*}

The level-two measurement, and preparation, error models can be computed with the same polynomials as before.  We get an upper bound on the error rate of $4.71147 \times 10^{-6}$.

\subsubsection{Level-three encoded CNOT, measurement and preparation}

All CNOT error parameters have improved in going from one level of encoding to two levels.  However, the preparation and measurement error models have actually worsened slightly.  Repeating the exact same analysis (with the deliberately introduced errors determined with the same procedure) gives strict improvement in all error model parameters at level three.  The upper bound on the level-three preparation and measurement error rates is $1.36833 \times 10^{-6}$.  The upper bounds on the level-three CNOT error model are given in Table~\ref{f:numericalboundedencodedcnoterrormodellevelthree}.  This is strong evidence that $p=1/1000$ was below the noise threshold.  Repeating the analysis further should give rapid improvement in error rates.

(The calculations of Sec.~\ref{s:numericalexactuniversality}, carried out at error rates which were all higher, imply that once again the noise threshold for full universality is the same as that for stabilizer operations.)

\subsection{Areas for improved analysis}

This analysis has not been optimized, and can probably be significantly improved.  The greatest loss of efficiency seems to be in applying Lemma~\ref{t:generalizedmixinglemma} on level-two encoded Bell pairs, instead of mixing directly in the $64$-dimensional syndrome space.  A computationally efficient algorithm for mixing in the syndrome space would be useful.  Lacking that, significant improvements can probably be obtained by:
\begin{enumerate}
\item Choosing a better error distribution to introduce for lower bounds.  As we saw in Sec.~\ref{s:numericalboundedleveltwomixing}, optimizing this choice can give large improvements.
\item Better application of Lemma~\ref{t:generalizedmixinglemma} by varying the embedding into the lattice $\{I,X,Y,Z\}^8$.  We have also run these threshold calculations for the biased X noise error model used for convenience in Ch.~\ref{s:postselectchapter}, and in that case adjusting the embedding into the binary lattice yielded a $20\%$ improvement in the mixing parameter $q$.  The technique may be less promising with the more complicated lattice.  So far, a few experiments with different embeddings (but without varying the lower-bounding distribution) have only allowed 
improving $q$ from $5.4 \times 10^{-4}$ for each of $(X,Y,Z)$ errors, to $(5.3 \times 10^{-4}, 4.9 \times 10^{-4}, 5.3 \times 10^{-4})$.
\end{enumerate}
These improvements can be automated with a larger linear program.

Recall from Sec.~\ref{s:alternativebellprep} that there exist encoded Bell pair preparation procedures which do not require deliberate introduction of errors in order to prove the existence of a positive threshold.  We do not know how to obtain a good numerical analysis of such a scheme, however.  (The existence of a threshold followed from proving that the maximum of a ratio of two quadratic polynomials, under certain bounds, was finite.  It is more difficult to evaluate this maximum explicitly.)  In any case, we would expect any savings to be small since currently errors are only introduced at moderate rates, and only at levels two and above.

Modifying the fault-tolerance scheme itself should also give improved lower bounds.  For example, it is certainly beneficial to purify the four-encoded-qubit ancilla state of Fig.~\ref{f:cnotreducestobellpairssymmetry} (as mentioned in Sec.~\ref{s:postselectchapternumerics}).  Analyzing Knill's optimized scheme, which uses a six-qubit code at the second code concatenation level, should be worthwhile.  However, it is computationally difficult to compute exact error distributions in larger systems, and mixing is also more expensive.  

\clearpage

\bibliographystyle{bibtex/halpha}
\addcontentsline{toc}{chapter}{Bibliography}
\ssp	
\bibliography{bibliography}

\end{dissertationText}
\end{document}